\documentclass[10pt]{article}

\pdfoutput=1 

\usepackage[T1]{fontenc}
\usepackage[utf8]{inputenc}
\usepackage[english]{babel}

\usepackage{amsmath,amsfonts,amssymb,amscd,mathrsfs,amsthm}
\usepackage{bbold}

\usepackage[]{graphicx}
\usepackage{xcolor,dsfont}
\definecolor{darkblue}{rgb}{0.1,0.1,.7}
\usepackage[colorlinks, linkcolor=darkblue, citecolor=darkblue, urlcolor=darkblue, linktocpage]{hyperref} 
\usepackage[square, comma, compress,numbers]{natbib}
\usepackage{geometry}
\usepackage[percent]{overpic}
\usepackage[margin=10pt,font=small,labelfont=bf]{caption}
\usepackage{subcaption}
\usepackage{booktabs}
\numberwithin{equation}{section}

\usepackage{tikz}
\usetikzlibrary{decorations.pathmorphing}
\usetikzlibrary{decorations.markings}
\usepgflibrary{shapes.geometric}

\def\bop{\mathcal{O}}

\def\qaq{\quad \text{and} \quad}
\def\dag{\dagger}
\def\phd{{\phantom{\dagger}}}

\def\lra{\leftrightarrow}
\def\La{\Lambda}

\def\zb{\bar{z}}

\newcommand{\NO}[1]{{:\!#1\!:}}
\newcommand{\ud}[2]{^{#1}_{\phantom{#1}#2}}

\def\Red{\color [rgb]{0.9,0.1,0.1}}

\def\Blue{\color [rgb]{0.0,0.2,0.7}}

\newcommand{\floor}[1]{\left \lfloor #1 \right \rfloor }
\newcommand{\braket}[3]{\langle #1|#2|#3 \rangle}

\newcommand{\brakket}[2]{\langle #1|#2\rangle}

\newcommand{\ket}[1]{|#1\rangle}
\newcommand{\dket}[1]{|#1\rangle\!\rangle}
\newcommand{\bra}[1]{\langle #1|}
\newcommand{\expec}[1]{\langle #1 \rangle}
\def\normord#1{\mathop{:}\nolimits\!#1\!\mathop{:}\nolimits}

\def\dps{\displaystyle}
\def\ldef{\mathrel{\mathop:}=}
\def\rdef{=\mathrel{\mathop:}}
\newcommand{\limu}[1]{\mathrel{\mathop{\sim}\limits_{\scriptstyle{#1}}}}
\def\fns{\footnotesize}
\newcommand{\reef}[1]{(\ref{#1})}
\def\beq{\begin{equation}} 
\def\eeq{\end{equation}}
\def\nn{\nonumber} 
\def\bsub{\begin{subequations}}
\def\esub{\end{subequations}}

\def\mbb{\mathbb}
\def\mbf{\mathbf}
\def\mca{\mathcal}
\def\mfr{\mathfrak}
\def\mrm{\mathrm}
\def\msc{\mathscr}
\def\mtt{\mathtt}
\def\msf{\mathsf}
\def\ads2{AdS${}_2$}

\def\th{\tfrac{1}{2}}
\def\half{\frac{1}{2}}
\def\pd{\partial}
\def\a{\alpha}
\def\b{\beta}

\def\dd{\delta}

\def\la{\lambda}
\def\ga{\gamma}

\def\DD{\Delta}
\def\Oo{\mathcal{O}}
\def\sO{\mathrm{O}}
\def\l{\ell} 
\def\eps{\epsilon}
\def\vareps{\varepsilon}

\renewcommand{\Im}{\operatorname{Im}}

\newcommand{\bDelta}{{\bf \Delta}}
\newcommand{\ii}{\mathrm{i}}
\newcommand{\specialcell}[2][c]{%
  \begin{tabular}[#1]{@{}c@{}}#2\end{tabular}}

\begin{document}

\vspace*{-.6in} \thispagestyle{empty}
\begin{flushright}
CERN-TH-2021-065
\end{flushright}
\vspace{1cm} {\Large
\begin{center}
  {\bf  Hamiltonian truncation in Anti-de Sitter spacetime}
\end{center}}
\vspace{1cm}
\begin{center}
{\bf Matthijs Hogervorst${}^{a}$, Marco Meineri${}^{a,b}$, Jo\~{a}o Penedones${}^a$ and Kamran Salehi Vaziri${}^a$ }\\[2cm] 
{
  ${}^a$  Fields  and  Strings  Laboratory,  Institute  of  Physics,\\
  \'{E}cole  Polytechnique  F\'{e}d\'{e}rale  de  Lausanne, Switzerland\\
  ${}^b$ CERN, Department of Theoretical Physics, \\
CH-1211 Geneva 23, Switzerland
}
\\
\end{center}
\vspace{14mm}

\begin{abstract}
 Quantum Field Theories (QFTs) in Anti-de Sitter (AdS) spacetime are often strongly coupled when the radius of AdS is large, and few methods are available to study them. In this work, we develop a Hamiltonian truncation method to compute the energy  spectrum of QFTs in two-dimensional AdS. 
 The infinite volume of constant timeslices of AdS leads to divergences in the energy levels.
 We propose a simple prescription  to obtain finite physical energies and test it with numerical diagonalization in several models: the free massive scalar field, $\phi^4$ theory, Lee-Yang and Ising field theory. 
  Along the way, we discuss spontaneous symmetry breaking in AdS and derive a compact formula for perturbation theory in quantum mechanics at arbitrary order. Our results suggest that all conformal boundary conditions for a given theory are connected via bulk renormalization group flows in AdS.
\end{abstract}
\vspace{12mm}

\newpage

{
\setlength{\parskip}{0.05in}
\tableofcontents
}

\newpage

\section{Introduction}
\label{sec:intro}
%!TEX root = ../HTinAdS.tex
%%%%%%%%%%%%%%%%%%%%%%%%%%%%%%%%%%%%%%%%%%%%%

Strongly coupled Quantum Field Theories (QFT) are challenging.
In this work, we are interested in UV complete QFTs defined as relevant deformations of free or solvable Conformal Field Theories (CFT).\footnote{In practice, we can include Yang-Mills theory in this class although, strictly speaking, it cannot be formulated as a gauge-invariant relevant deformation of a free theory.}
For example, if the QFT has a mass gap, we would like to determine the masses of the stable particles and their scattering amplitudes.
Generically, this question cannot be addressed with perturbative methods because there is no natural parameter to expand in.
Decades of previous work have devised ingenious approaches like expanding  in the number of spacetime dimensions \cite{Wilson:1971dc} or in inverse powers of the rank of the gauge group \cite{tHooft:1973alw}. In many cases, one can also implement a lattice discretization and use (costly) numerical methods (like Monte Carlo, tensor networks, etc.) to recover the continuum limit. 
In this work, we explore an alternative approach by placing the QFT in a curved spacetime. For example, a QFT on a compact spatial manifold like a sphere of radius $R$ offers a new handle into the non-perturbative regime. In this case, one can study the energy spectrum as a function of $R$ to interpolate from the perturbative small-$R$ regime to the strongly coupled large-$R$ limit.\footnote{See \cite{Tilloy:2021yre, Tilloy:2021hhb} for recent progress setting up tensor networks directly in the continuum. }
This is the basis of the Hamiltonian truncation approach to QFT$_{d+1}$ placed on $\mathbb{R} \times S^{d}$, initiated in~\cite{Yurov:1989yu} for the case of 1+1 dimensions.\footnote{We refer to Ref.~\cite{James:2017cpc} for a recent review of the Hamiltonian truncation literature.}

Here, we consider QFT on hyperbolic or Anti-de Sitter (AdS) space of radius $R$.
More precisely, we shall study deformations of exactly solvable theories. The action is given by
\beq
\label{eq:Sin}
S = S_0 + \la \int_{\mrm{AdS}_{d+1}}\!dx \sqrt{g} \,\mca{V}(x)
\eeq
where $g$ is the metric of AdS${}_{d+1}$ and $\mca{V}$ is a bulk local operator  with mass dimension $\DD_{\mca{V}} < d+1$.  Let us emphasize that the metric is non-dynamical: this is a study of QFT in a fixed curved background.
The unperturbed theory described by $S_0$ can either be a free theory or a solvable Boundary CFT (BCFT).\footnote{Notice that AdS${}_{d+1}$ is conformally equivalent to $\mathbb{R}^d \times \mathbb{R}^+$.
This is obvious in Poincaré coordinates $$ds^2_{\rm AdS} = \frac{R^2}{z^2}\left( dz^2 + dx^idx_i \right)\,,\qquad \qquad x^i \in \mathbb{R}^d\,,\qquad z>0\,.$$
}
In this setup, physical observables depend on the dimensionless parameter $\bar{\lambda} \equiv \la
R^{d+1-\DD_\mca{V}}$.
This parameter allows us to continuously connect the perturbative regime of small $\bar{\la}$ with the strongly coupled regime of large $\bar{\la}$.
This is similar to the case mentioned above of QFT on $\mathbb{R} \times S^{d}$. 
The choice of the AdS${}_{d+1}$ background has two main advantages.
The first is that it preserves more symmetry. Indeed, AdS is a maximally symmetric space. The isometry group of (Lorentzian) AdS${}_{d+1}$ is the conformal group $SO(2,d)$. This means that the Hilbert space of the theory organizes in representations of the conformal algebra.
The second advantage is that AdS has a conformal boundary where we can place boundary operators and their correlation functions  solve conformal bootstrap equations. This means that the well established conformal bootstrap methods can be used to study non-conformal QFTs~\cite{Paulos:2016fap}.
Notice that the extrapolation to $\bar{\la} \to \infty$ corresponds to the flat space limit where the mass spectrum and scattering amplitudes of the QFT in flat space can be recovered.

In the present paper we choose to study the above setup in the Hamiltonian framework. To be precise, we study the theory \eqref{eq:Sin} in global coordinates, which endow AdS with the topology of a solid cylinder. The Hilbert space is that of the undeformed theory corresponding to $S_0$. The time evolution  is governed by a Hamiltonian of the form\footnote{The Hamiltonian is dimensionless because we measure energies in units of $1/R$, with $R$ the AdS radius.}
\beq
\label{eq:Hin}
H = H_0  + \bar{\la} \,V
\eeq
where $H_0$ is the Hamiltonian of the $\la = 0$ theory, and the interaction term $V$ corresponds to $\mca{V}$ integrated over a timeslice (this is explained in detail in section \ref{sec:Vmat}).

The key problem we address in our work is the diagonalization of $H$ for finite $\la$, which describes how the  states and their energies change after turning on the bulk interaction $\mca{V}$. Since the strength of the dimensionless coupling $\bar\la$ grows with $R$, the small-radius limit can be attacked in Rayleigh-Schr\"{o}dinger perturbation theory (see e.g.~\cite{Fitzpatrick:2010zm}), but when $R$ is sufficiently large the theory is in a strong-coupling regime, and perturbation theory breaks down.\footnote{See the recent work~\cite{Brower:2019kyh} for a lattice Monte Carlo approach to address the finite-coupling regime in AdS${}_2$.}

Nevertheless, we can attempt to study this diagonalization problem using the Hamiltonian truncation toolbox. This amounts to introducing a UV cutoff $\La$ and  diagonalizing $H$ in the finite-dimensional subspace of states with unperturbed energy less than $\La$. The continuum limit is then recovered by taking $\La \to \infty$. In this work, we study several two-dimensional QFTs on AdS$_2$ with this strategy. 

The UV cutoff, at least in the implementation explained above, breaks the conformal symmetry, hence Hamiltonian truncation does not take advantage of the full isometry group of AdS. On the other hand, the method is extremely versatile and straightforward to implement. It is worth stressing that the application of equal-time Hamiltonian truncation to infinite volume physics is new --- although lightcone conformal truncation~\cite{Katz:2016hxp,Anand:2020gnn} can be used to treat flat-space physics using Hamiltonian truncation technology.\footnote{In lightcone conformal truncation, the need to work with a finite basis of states leads to an effective IR cutoff, see for instance the discussion in appendix F of~\cite{Anand:2020qnp}.}
It presents conceptual challenges, which we will tackle in the first part of the paper. In this respect, AdS is an ideal playground: due to the UV/IR connection familiar from AdS/CFT, the conformal theory on the boundary provides a handle into the infrared behavior of the observables. We hope that the lessons drawn from this setup will be useful in other contexts.

This paper is organized as follows. In section~\ref{sec:prel} we review some basic features of the physics and symmetries of \ads2. In sectionx~\ref{sec:cutoff} we explain in detail the existence of UV divergences that arise in the  Hamiltonian truncation scheme, and we explain how to obtain the correct continuum limit in a large class of theories. This section contains one of the main results of the paper, \emph{i.e.} eq. \eqref{prescription}, which allows to obtain the correct energy gaps. Finally, in sections \ref{sec:massive} and \ref{sec:minimal} we present results for several theories. 
The massive free scalar deformed by $\phi^2$ and $\phi^4$ operators is discussed in detail in section \ref{sec:massive}. Next, in section \ref{sec:minimal}, we study relevant deformations away from the Lee-Yang and Ising minimal models. Future directions are discussed in section.~\ref{sec:disc}. The exactly solvable models -- \emph{e.g.} the massive free theories -- allow to test the formula \eqref{prescription}. On the contrary, figures \ref{fig:spectrumphi4}, \ref{fig:LYExtp}, \ref{fig:LYSpectrum}, \ref{fig:MIsingSpectrum} and \ref{fig:MNegIsingSpectrum} contain genuinely nonperturbative results on the spectrum of strongly coupled theories in AdS.

We also include several complementary appendices. We are especially proud of appendix \ref{app:QMPT}, where we derive a new simple formula for the order-$\la^n$ correction to the eigenvalues of the Hamiltonian in perturbation theory.

\section{Quantum Field Theory in AdS}
\label{sec:prel}
%!TEX root = ../HTinAdS.tex
%%%%%%%%%%%%%%%%%%%%%%%%%%%%%%%%%%%%%%%%%%%%%

As promised, we will study QFTs defined in the bulk of $(d+1)$-dimensional AdS spacetime, specializing to $d=1$. The generalization to higher $d$ is relatively straightforward, but will not be necessary in the rest of this work. In this section, we review some basic facts about the correspondence between bulk (non-gravitational) AdS physics and the conformal  theory that lives on the boundary of AdS. For a more extensive pedagogical treatment, we refer for instance to~\cite{KaplanBottom} and~\cite{Penedones:2016voo}.

\subsection{Geometry}

 Euclidean \ads2 with radius $R$  can be defined as a hyperboloid
\beq
\label{eq:ads2}
-(X^0)^2 + (X^1)^2 + (X^2)^2 = -R^2~, \qquad X^0>0
\eeq
inside $\mbb{R}^{1,2}$, which we will endow with global coordinates $\tau \in \mbb{R}$ and $r \in [-\th \pi,\th \pi]$ via
\beq
X^\mu(\tau,r) = \frac{R}{\cos r} \, (\cosh \tau,\sinh \tau,\sin r) \,.
\eeq
The coordinate $\tau$ (resp.\@ $r$) plays the role of a Euclidean time (resp.\@ a space) coordinate. In the above coordinates, the AdS metric is given by
\beq
ds^2 = \left(\frac{R}{\cos r}\right)^2 \left[d\tau^2 + dr^2\right].
\label{AdSmetric}
\eeq
The conformal boundary of AdS${}_2$ is the union of the two lines $r = \pm \th \pi$.

It follows from the definition~\reef{eq:ads2} that \ads2 is manifestly invariant under a group $SO(1,2) \sim SL(2,\mbb{R})$ of continuous isometries. The corresponding Lie algebra is described by three generators $\{H,P,K\}$ with commutators
\beq
\label{eq:comm}
[H,P] = P,
\quad
[H,K] = -K,
\quad
[P,K] = -2H.
\eeq
The generators act on local scalar bulk operators as
\beq
[G,\Oo(\tau,r)] = \mca{D}_G \cdot \Oo(\tau,r)
\eeq
where 
\bsub
\label{eq:genAct}
\begin{align}
  \label{eq:Dact} \mca{D}_H &= \pd_\tau, \\
  \label{eq:Pact} \mca{D}_P &= e^{-\tau}(\sin r \, \pd_\tau + \cos r\,\pd_r), \\
  \label{eq:Kact} \mca{D}_K &= e^{\tau}(\sin r \, \pd_\tau - \cos r\, \pd_r).
\end{align}
\esub
In addition, there is a parity symmetry $\mtt{P}$ (with $\mtt{P}^2 = 1$) that maps $r \mapsto -r$. Under parity, the generators transform as
\beq
\label{eq:parity}
\mtt{P} \begin{pmatrix} H \\ P \\ K \end{pmatrix}  \mtt{P}^{-1} = \begin{pmatrix} +H \\ -P \\ -K \end{pmatrix}.
\eeq
The symmetries of AdS put strong constraints on correlation functions of bulk operators. For instance, any one-point correlation function  $\expec{\Oo_i(\tau,r)} =  \braket{\Omega}{\Oo_i(\tau,r)}{\Omega}$ in the vacuum state $\ket{\Omega}$ must be constant.\footnote{By assumption, the vacuum state $\ket{\Omega}$ is annihilated by all $SL(2,\mbb{R})$ generators.} Vacuum two point correlation functions of scalar bulk operators --- at say $(\tau,r)$ and $(\tau',r')$ --- only depend on a single $SO(1,2)$ invariant
\beq
\label{eq:crossratio}
\xi = \frac{(X^\mu(\tau,r) -X^\mu(\tau',r'))^2}{R^2}=
\frac{\cosh(\tau - \tau') - \cos(r - r')}{2\cos r \cos r'} \geq 0.
\eeq
Vacuum higher-point correlators or correlators inside excited states depend on a larger number of $SL(2,\mbb{R})$ invariants. 

\subsection{The Hilbert space and the bulk-to-boundary OPE}\label{sec:BCFT}

The Hilbert space of a QFT in AdS$_2$ decomposes into multiplets of the isometry group $SL(2,\mbb{R})$. 
Such a multiplet is defined by a primary state $\ket{\psi_i}$, satisfying $H \ket{\psi_i} = \DD_i \ket{\psi_i}$ and $K \ket{\psi_i} = 0$. In a unitary QFT, all primary states must have $\DD_i \geq 0$. Associated to any primary state with $\DD_i > 0$ are infinitely many descendant states of the form
\beq
\label{eq:descDef}
\ket{\psi_i,n} \ldef \frac{1}{\sqrt{n!(2\DD_i)_n}} \, P^n \ket{\psi_i},
\quad
n=0,1,2,\ldots
\eeq
identifying $\ket{\psi_i,0} = \ket{\psi_i}$, and writing $(x)_n = \Gamma(x+n)/\Gamma(x)$ for the Pochhammer symbol. 
 The normalization in~\reef{eq:descDef} is chosen such that the descendant states are unit-normalized, as long as the primary state $\ket{\psi_i}$ is unit-normalized as well. The state $\ket{\psi_i,n}$ has energy $\DD_i + n$, and the generators $K$ and $P$ act as
 \beq
\label{eq:KPonstates}
K \ket{\psi_i,n} = \ga(\DD_i,n) \ket{\psi_i,n-1}
\qaq
P \ket{\psi_i,n} = \ga(\DD_i,n+1) \ket{\psi_i,n+1}
\eeq
using the shorthand notation $\ga(\DD,n) \ldef \sqrt{n(2\DD+n-1)}$. In addition, there is a unique vacuum state $\ket{\Omega}$ without descendants that is annihilated by all $SL(2,\mbb{R})$ generators, so in particular it has $\DD_\Omega = 0$. Moreover, if the theory is invariant under parity, all primary states have a definite parity, meaning that $\mtt{P} \ket{\psi_i} = \pi_i \ket{\psi_i}$ with $\pi_i = \pm 1$. Then from~\reef{eq:parity} it follows that descendant states transform as
\beq
\mtt{P} \ket{\psi_i,n} = (-1)^n \pi_i \ket{\psi_i,n}.
\eeq

In a QFT in AdS, one can define boundary operators as (rescaled) limits of local bulk operators pushed to the conformal boundary of AdS.
According to the bulk state -- boundary operator map \cite{Paulos:2016fap}, any primary state $\ket{\psi_i}$ corresponds to a boundary primary operator $ \bop_i$ with dimension $\DD_i$, according to the rule
\beq
\ket{\psi_i} = \lim_{\tau \to -\infty} e^{-\DD_i \tau} \Oo_i(\tau) \ket{\Omega}.
\eeq
Often, we will therefore use the notation $\ket{\Oo_i}$ for boundary states instead of $\ket{\psi_i}$. Likewise, descendant states $\ket{\psi_i,n}$ can be obtained from boundary descendants $[P,\dotsm[P,\Oo_i(\tau)]\dotsm]$. Sometimes the operator picture will be useful, but we will not always explicitly identify the local operators that generate specific states. For instance, a simple-looking state in the Fock space of a massive boson could originate from a complicated-looking boundary operator, and vice versa. 

In turn, the state-operator map implies that any bulk operator can be expanded in a basis of boundary scaling operators, placed at an arbitrary point in AdS. For instance, one can replace a bulk operator $\mca{V}(\tau,r)$ with its bulk-to-boundary operator product expansion (OPE) at the $r=\pi/2$ boundary:
\beq
\mca{V}(\tau,r) = \sum_{i} b_i \left(\frac{\pi}{2}-r\right)^{\Delta_i} \left[\mathcal{O}_i(\tau) + \text{descendants}\right]
\label{bulkToBound}
\eeq
where the sum runs over all primary operators $\Oo_i$ and the $b_i$ are so-called bulk-to-boundary OPE coefficients. 
As we will see shortly, the low-lying operators in the bulk-to-boundary OPE of the deforming operator in eq.~\eqref{eq:Sin} are important in establishing the UV completeness of the resulting interacting QFT. 

In particular, all the examples treated in sections \ref{sec:massive} and \ref{sec:minimal} belong to a special class of QFTs in AdS: the Hilbert space of the undeformed theory has a gap larger than one above the vacuum. In equations:
\beq
\Delta_i>1~, \quad \forall\ \ket{\psi_i}\neq \ket{\Omega}~.
\label{UVfiniteCond}
\eeq
The absence of relevant deformations on the boundary removes the necessity of fine tuning. Together with the absence of bulk UV divergences, which are the same as in flat space, this condition ensures that the action \eqref{eq:Sin} is well-defined without the addition of counterterms, in a neighborhood of $\lambda=0$.\footnote{Non-perturbatively, the scaling dimension may be pushed down to $\Delta=1$ at some value $\bar{\lambda}^*$. In this case the boundary condition becomes unstable and generically the theory cannot be defined beyond this value. In sections \ref{sec:massive} and \ref{sec:minimal} we shall see that this phenomenon is generic. Further discussion can be found in subsection \ref{sec:sbb} and section \ref{sec:disc}.} Concretely, we shall point out at various points in the next subsection and in section \ref{sec:cutoff} where this condition is important.

\subsection{The interacting Hamiltonian}
\label{sec:Vmat}

Suppose that we're given an exactly solvable QFT in AdS${}_2$; denote its Euclidean action by $S_0$. This can for instance be a two-dimensional rational BCFT, or a theory of free massive bosons and/or fermions.\footnote{See for example appendix A of~\cite{Aharony:2010ay} for a discussion of general free fields in AdS${}_{d+1}$.} We can then turn on a relevant interaction, by shifting
\beq
S_0 \mapsto S= S_0 + \lambda \int_{\text{AdS}_2}\!\sqrt{g}d^2x\, \mca{V}(x)
\label{Slambda}
\eeq
where $\mca{V}(x)$ is a bulk operator to which we can assign a scaling dimension $\DD_\mca{V} < 2$.\footnote{In the 2$d$ Landau-Ginzburg theory of a scalar $\phi$, all polynomial operators $\mca{V} = \phi^p$ have $\DD_\mca{V} = 0$.} The generalization to multiple relevant operators $\mca{V}_i$ is straightforward, but won't be discussed explicitly here. In the Hamiltonian picture, this translates to 
\beq
\label{eq:Ddef}
H_0 \mapsto H = H_0+ \bar{\la} V,
\qquad \qquad
V \ldef R^{\DD_\mca{V}} \int_{-\pi/2}^{\pi/2} \frac{dr}{(\cos r)^2}\, \mca{V}(\tau=0,r)
\eeq
writing $\bar{\la} \ldef \la R^{2-\DD_\mca{V}}$. The factor of $R^{\DD_\mca{V}}$ is absorbed into $V$ in order to make the matrix elements of $V$ dimensionless.

We would like to study the spectrum and eigenstates of $H$  by diagonalizing it inside the Hilbert space from subsection~\ref{sec:BCFT}. Before we do so, let us first remark that the Hamiltonian $H$ from eq.~\reef{eq:Ddef} is not necessarily well-defined,  since $V$ may not exist.
A sufficient condition for $V$ to be well-defined is that $\mca{V}(\tau,r)$ vanishes sufficiently fast near the boundary. In particular, we require that
\beq
\label{eq:as}
\lim_{r \to \pm \pi/2} \, \frac{1}{\cos r} \, \mca{V}(\tau = 0,r) = 0
\eeq
inside any normalizable state. This means that the bulk-to-boundary OPE \eqref{bulkToBound} of $\mca{V}$ should not contain operators with $\Delta_i \leq 1$. If this condition is \emph{not} satisfied, an IR regulator needs to be introduced and boundary counterterms corresponding to the offending operators must be added to $H$. In particular, we require that the bulk one-point function $\expec{\mca{V}}$, which corresponds to the identity on the r.h.s.\@ of eq.~\eqref{bulkToBound}, vanishes.\footnote{In fact, removing the identity operator does not require the introduction of an explicit IR cutoff. Since $\expec{\mca{V}(0,r)}$ does not depend on $r$, one can simply replace $\mca{V}\to \mca{V}-\expec{\mca{V}}$, which is equivalent to adding a cosmological constant counterterm to $S$.} From the point of view of the boundary, the condition \eqref{UVfiniteCond} corresponds to the fine tuning of relevant operators at leading order in $\bar{\lambda}$. Therefore, it is not surprising that it may not be sufficient for the theory to be well defined at finite $\bar{\la}$. Nevertheless, \eqref{eq:as} is special from the point of view of Hamiltonian truncation, since it ensures that the matrix elements of $V$ are finite. Notice also that it is special to AdS: for instance, it does not appear when one attempts to study renormalization group (RG) flows on the strip $\mbb{R} \times [0,\pi]$ using the Truncated Conformal Space Approach (TCSA)~\cite{Dorey:1997yg}. Although we will focus on theories which do not require an IR cutoff owing to the condition~\eqref{UVfiniteCond}, we will nevertheless introduce an IR cutoff $\eps$ in certain arguments in section~\ref{sec:cutoff}.

Even after ensuring that all the individual matrix elements of $V$ are well-defined, the operator $H$ may not be diagonalizable due to the presence of high-energy states in the UV theory. This is a familiar problem in Hamiltonian truncation: for instance, it is well-known in TCSA on $\mbb{R} \times S^d$ that any perturbation with $\DD_{\mca{V}} \geq \th(d+1)$ leads to a divergent cosmological constant~\cite{Hogervorst:2014rta}. As we mentioned, this is not surprising, since UV-completeness in perturbation theory may require fine tuning at all orders in the coupling. However, it turns out that \emph{any} AdS${}_2$ theory suffers from this problem, regardless of the unperturbed theory $S_0$ or the pertubation $\mca{V}$. We shall address this issue in detail in section~\ref{sec:cutoff}.

In actual Hamiltonian truncation computations, one diagonalizes the Hamiltonian~\reef{eq:Ddef} inside a subspace of the full Hilbert space. Typically, one fixes a UV cutoff $\Lambda$ and keeps all states with energy smaller than $\Lambda$ in the $\la = 0$ theory. Let us make this procedure explicit by introducing some notation. In the Hilbert space, the action of the operator $V$ can be represented by a matrix $\mbf{V}$, namely:
\beq
\label{eq:Vmatel}
V \ket{\psi_i,n} = \sum_{j} \sum_{m=0}^\infty \mbf{V}\ud{j;m}{i;n} \ket{\psi_j,m}.
\eeq
In turn, the matrix elements of $\mbf{V}$ can be computed by integrating wavefunctions over timeslices: 
\beq
\label{eq:matelt}
{\mbf V}\ud{j;m}{i;n} = R^{\DD_\mca{V}} \int_{-\pi/2}^{\pi/2} \frac{dr}{(\cos r)^2} \, \braket{\psi_j,m}{\mca{V}(0,r)}{\psi_i,n}.
\eeq
Such integrals always exist, owing to the condition~\reef{eq:as}.
Consequently, the operator $H$ acts as
\beq
\label{eq:Hfull}
H \ket{\psi_i,n} = (\DD_i + n) \ket{\psi_i,n} + \bar{\la} \sum_{j} \sum_{m=0}^\infty \mbf{V}\ud{j;m}{i;n} \ket{\psi_j,m}.
\eeq
In Hamiltonian truncation, one defines a truncated Hamiltonian $H(\Lambda)$ simply starting from~\reef{eq:Hfull} by discarding all states with energy strictly larger than $\La$.

Due to the $SL(2,\mbb{R})$ symmetry of the theory, matrix elements with different $m,n$ are linearly related --- see e.g.\@ Ref.~\cite{Fitzpatrick:2010zm} for a discussion of this phenomenon in AdS${}_3$. The resulting constraints on the matrix elements are discussed in appendix~\ref{app:constraints}.

Finally, let us comment on the choice of basis of the Hilbert space. In a unitary QFT, it is always possible to organize the states such that the primaries are orthonormal, that is to say $\brakket{\psi_i}{\psi_j} = \dd_{ij}$ for all $i,j$. This convention is assumed above in \reef{eq:Vmatel} and \reef{eq:matelt}. In practice, it is not necessary to organize the Hilbert space in terms of primaries and descendants, let alone to impose an orthonormality condition between different primaries of the same dimension $\DD$. Working in a completely general basis of states $\ket{\varphi_i}$, the eigenvalue equation $H\ket{\Psi} = \mca{E} \ket{\Psi}$ can be recast as
\beq
\label{eq:genev}
H_{ij}\,\beta^j = \mca{E} \, G_{ij}\,\beta^j,
\quad
\ket{\Psi} = \beta^i \ket{\varphi_i}
\eeq
where $H_{ij} = \braket{\varphi_i}{H}{\varphi_j}$ and $G_{ij} = \brakket{\varphi_i}{\varphi_j}$ are Hermitian; the matrix $G$ is known as the Gram matrix corresponding to the basis $\ket{\varphi_i}$. The energies $\mca{E}$ and eigenstates $\ket{\Psi}$ of $H$ are physical (up to a choice of normalization of $\ket{\Psi}$), although the coefficients $\beta^i$ are basis-dependent. This is particularly important when the unperturbed theory is non-unitary, in which case the Gram matrix is not positive definite, and there exists no orthonormal basis of states. We will encounter a non-unitary QFT in the form of the Lee-Yang model, discussed in subsection~\ref{sec:leeyang}.

\subsection{The energy spectrum}
\label{sec:pheno}

Let $\DD_i(\bar{\lambda})$ be the eigenvalues of the AdS Hamiltonian. The dimensionless coupling appearing in the Hamiltonian is $\bar{\lambda} = \lambda R^{2-\DD_\mca{V}}$ so in the limit $R \to 0$ the interaction vanishes and $\DD_i(0)$ is the dimension of the $i$-th state in the non-interacting theory.
At small radius we can write
\beq
\label{eq:expansion}
\DD_i(\bar{\lambda}) = \DD_i(0) + \omega_{i,1}  \bar{\lambda} + \omega_{i,2}  \bar{\lambda}^2 + \ldots
\eeq
for some coefficients $\omega_{i,n}$ that can be determined through Rayleigh-Schr\"{o}dinger perturbation theory (or alternatively, using Witten diagrams).   In the presence of multiple relevant interactions, Eq.~\reef{eq:expansion} is modified in a straightforward way. Depending on the UV theory and the perturbation in question, the perturbative expansion~\reef{eq:expansion} can be asymptotic, or it can have a finite radius of convergence. Crucially, every level $\DD_i(\bar{\lambda})$ corresponding to an $SL(2,\mbb{R})$ primary is accompanied by an infinite tower of descendants with energies $\DD_i(\bar{\lambda}) + \mbb{N}$. 

When AdS has a large radius, it is expected to approximate flat-space physics. In the current paper we are interested in \emph{massive} flows, where at low energies the Hamiltonian $H$ describes a theory of particles with masses $m_i > 0$. We then have
\beq
\label{eq:midef}
\DD_i(\bar\la)  \limu{\bar\la \to \infty}  m_i R = \kappa_i \bar{R}~, 
\eeq
for the lightest primary states, corresponding to single-particle states in the flat space limit. In eq. \eqref{eq:midef}, we defined the radius of AdS in units of the coupling:\footnote{The notation $\bar{R}$ becomes ambiguous when there is more than one coupling, but this will not play a role in the present work. }
\beq
\bar{R}  \ldef |\la|^{1/(d+1-\DD_\mca{V})} R = |\bar{\la}|^{1/(d+1-\DD_{\mca{V}})}.
\label{RbarDef}
\eeq
 A simple way to derive the linear growth in eq. \eqref{eq:midef} is by noticing that a rescaling $(\tau,r) \to (\tau,r)/R$ is needed to obtain the flat space metric from eq.~\eqref{AdSmetric} in the $R \to \infty$ limit.
The dimensionless coefficients $\kappa_i > 0$  can in general only be determined non-perturbatively.
More generally, the eigenvalues $\DD_i$ will grow like $E_i R$ when $ R \to \infty$, with $E_i$ the center of mass energy of the corresponding state in the flat space limit. 

\begin{figure}[!htb]
\begin{center}
\includegraphics[scale=0.43]{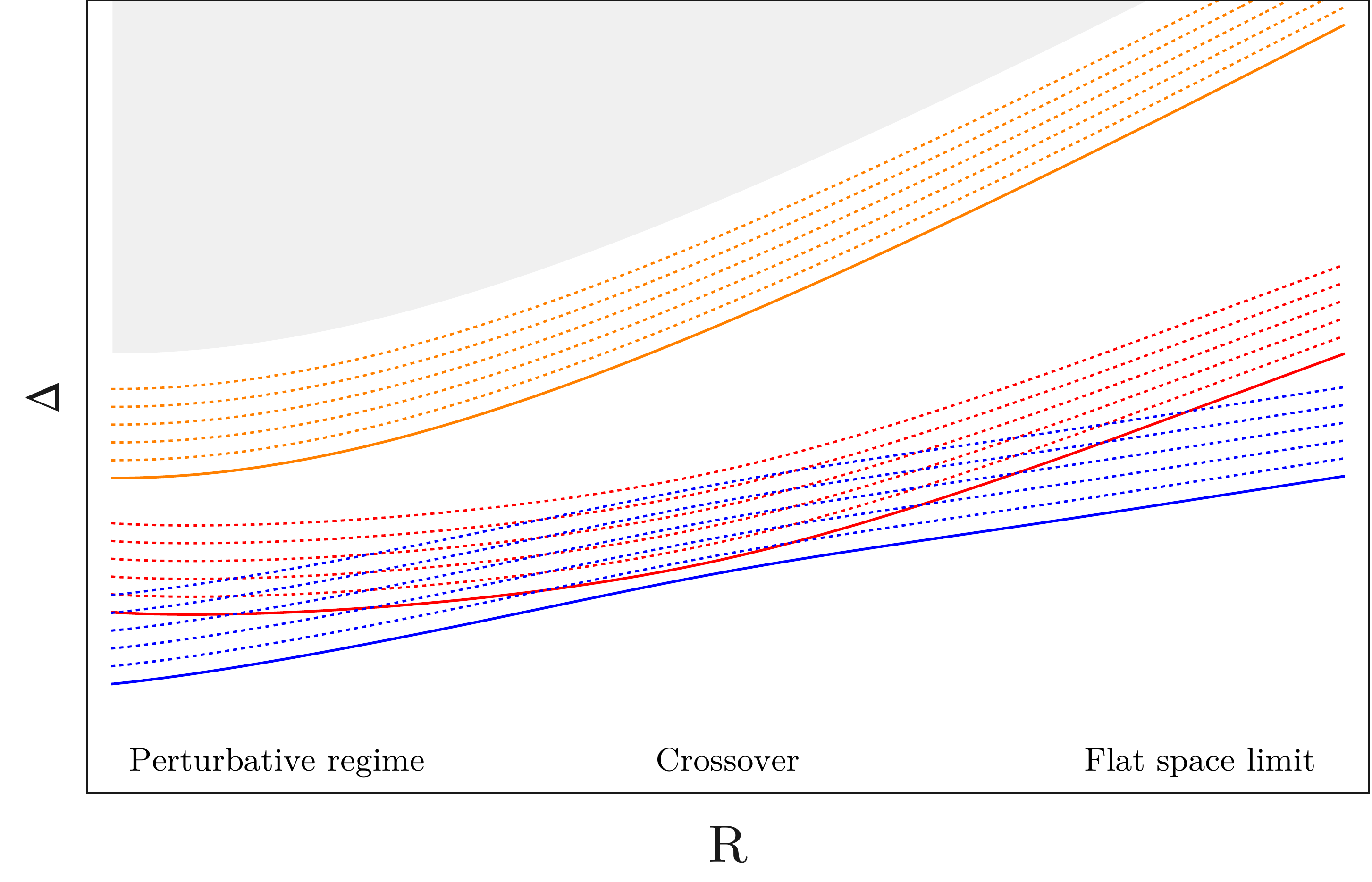}
\end{center}
\vspace{-4mm}
\caption{{ Sketch of the spectrum of the AdS Hamiltonian as a function of the AdS radius $R$.  The solid lines describe different primary states, and for every primary we have drawn its five lowest descendants (dotted). Additional states at higher energies are not shown (shaded region). The slope of the lines on the right side of the plot measures the masses of different particles in flat space. }}
\label{fig:phenoPlot}
\end{figure}

In figure~\ref{fig:phenoPlot} we have drawn a schematic plot of the first few energies of an AdS Hamiltonian as a function of $R$.
Notice that the structure of $SL(2,\mathbb{R})$ multiplets implies the existence of many level crossings. 
These must occur whenever two primary states asymptote to states of different (center of mass) energy in the flat space limit $R\to \infty$.
This is surprising because the interaction $V$ has non-zero matrix elements between different (undeformed) $SL(2,\mathbb{R})$ multiplets and generic Hamiltonians show level repulsion.
In fact, this is what we see at finite truncation $\Lambda$, since the cutoff breaks the $SL(2,\mathbb{R})$ symmetry.   As we shall see, exact level crossing only happens in the limit $\Lambda \to \infty$.  

Alternatively, it is possible to flow to a massless theory at large radius. In that case, the levels $\DD_i$ will have a finite limit as $R \to \infty$, coinciding with the spectrum of a specific boundary CFT. Such RG flows will not be encountered in the present work, hence we will not discuss this case in more detail.

\subsection{Spontaneous symmetry breaking}\label{sec:sbb}

In flat space and infinite volume, QFTs may spontaneously break symmetries. 
It is natural to ask if the same can happen in AdS. The answer is not obvious because AdS spatial slices of constant $\tau$ have infinite volume but behave like a box of finite volume, in the sense that they give rise to a discrete energy spectrum.
In addition, one may entertain the possibility that a symmetry may be spontaneously broken inside a finite region of AdS and restored near the boundary due to the effect of symmetry-preserving boundary conditions. 

In \cite{Carmi:2018qzm}, the phases of the $O(N)$ model in AdS where studied at large $N$. The effective potential was found to allow for symmetry preserving and symmetry breaking vacua, both stable under small fluctuations of the fields. Expanding the fields around the two vacua, the properties of the two phases where studied, but the existence of a phase transition and the possibility of phase coexistence were not clarified.

In order to address these questions, in appendix \ref{app:SSB}, we consider the following simple model  of a scalar field in AdS with Euclidean action 
\beq
S= \int d^{d+1}x \sqrt{g} \left[ \frac{1}{2} (\partial \phi)^2 + V(\phi) \right]\,,
\eeq
where $V(0)=0$ and $V(-\phi)=V(\phi)$ is $\mathbb{Z}_2$ symmetric.
Furthermore, we impose symmetry-preserving  boundary conditions $\phi \to 0$ at the AdS boundary.
 The global minimum of $V(\phi)$ is attained at $\phi=\phi_t \neq 0$ in order to favour spontaneous symmetry breaking.

We claim that there are only two possibilities:
\begin{enumerate}
\item[{\bf 1.}] The global minimum of the action is zero and it is attained by the constant solution $\phi=0$.
\item[{\bf 2.}] The action is not bounded from below and its value can always be decreased further by setting $\phi=\phi_t$ in a bigger region of AdS.
\end{enumerate}

The main consequence of this claim is that there is no choice of potential $V(\phi)$ for which the Euclidean path integral is dominated by a finite size bubble of the true vacuum surrounded by a region of false vacuum close to the  boundary of AdS.  
Notice that this scenario would conflict with the homogeneous property of AdS. Indeed, if we could decrease the action with a finite size bubble then we would decrease the action further with several of them placed far apart. But merging bubbles of true vacuum decreases the action further and we fall into case {\bf 2.}. 

In flat Euclidean space only {\bf 2.} is possible, except if $\phi=0$ is the global minimum of $V(\phi)$.
This is obvious because a large bubble of true vacuum gives a negative contribution  to the action (from the potential term) proportional to the volume of the bubble and a positive contribution (from the kinetic term) proportional to the surface area of the bubble. However, in AdS, area and volume both grow exponentially with the geodesic radius and it is not obvious which term dominates.
In appendix \ref{thinwall} we will use the thin wall approximation to get some intuition for the claim above. We  present a more general argument in appendix \ref{generalbubble}.
A quantitative conclusion of the argument presented there is that, at least at the classical level, the symmetry preserving vacuum is unstable if $V''(0)<-d^2/4$. We recognize this as the Breitenlohner-Freedman bound \cite{Breitenlohner:1982jf}. 

The analysis of this simple model suggest that spontaneous symmetry breaking can happen for QFT in AdS. Moreover, when it happens, it is very similar to infinite volume flat space time. The main novelty is that symmetry preserving AdS boundary conditions can make some stationary points of the potential fully stable even if they are not its global minimum. However, if these vacua become unstable, then the true vacua correspond to field configurations that break the symmetry all the way to the AdS boundary. Moreover, as in flat space, they give rise to superselection sectors. 

In this paper, we shall study the theory of a scalar field with a quartic potential in subsection \ref{sec:phi4}. The considerations presented above suggest that the theory in AdS has two phases, where the $\mathbb{Z}_2$ symmetry is respectively preserved or broken. While in the flat space limit the phase transition is continuous, we do not expect this to be the case at finite $R$, since, as discussed, the $\mathbb{Z}_2$ preserving boundary conditions may stabilize the theory for a negative value of the mass square mass. Although studying the symmetry breaking line will prove difficult in Hamiltonian truncation, we shall offer some more comments in subsection \ref{sec:phi4}.

\section{Hamiltonian truncation in AdS, or how to tame divergences}
\label{sec:cutoff}
%!TEX root = ../HTinAdS.tex
%%%%%%%%%%%%%%%%%%%%%%%%%%%%%%%%%%%%%%%%%%%%%

The computation of the spectrum of the interacting theory via Hamiltonian truncation is straightforward in principle. The Hamiltonian of the exactly solvable theory induces a grading of the Hilbert space, which is labeled by the discrete set of energies $e_i$. We can truncate the Hilbert space to the finite-dimensional space of states with energies $e_i \leq \La$, and diagonalize the interacting Hamiltonian \eqref{eq:Hin} in this subspace.
As the cutoff is raised, one expects to approximate the exact spectrum of the interacting energies with increasing accuracy.

However, a naive implementation of this procedure for any theory in AdS${}_2$ runs into trouble: the vacuum energy $\mca{E}_\text{vac}(\La),$ as well as the rest of the spectrum $\mca{E}_i(\La)$, diverge in the $\La \to \infty$ limit. Let us remark that these divergences do not arise from short-distance singularities in the bulk. Rather, they are related to the non-compact nature of space. Due to the UV-IR connection familiar from the AdS/CFT correspondence,  in Hamiltonian truncation the same divergences originate from the high-energy tail of the unperturbed spectrum.

As a matter of principle, it's possible to subtract $\mca{E}_\text{vac}(\Lambda)$ from the Hamiltonian (adding a cutoff-dependent counterterm proportional to the identity operator), which amounts to measuring energy gaps $\mca{E}_i(\La)-\mca{E}_\text{vac}(\La)$. Surprisingly, these gaps display a systematic shift away from their physical values: they oscillate and fail to converge to a definite value in the $\Lambda\to\infty$ limit --- see figure~\ref{fig:rowPlot}.  
Instead, we claim that the exact physical energy spectrum is recovered by means of the following prescription:
\beq
\boxed{\mca{E}_i-\mca{E}_\text{vac} = \lim_{\La\to\infty} \mca{E}_i(\La+e_i)-\mca{E}_\text{vac}(\La)~.}
\label{prescription}
\eeq
The aim of this section is to illustrate the origin of the divergences, describe their nature in Rayleigh-Schr\"{o}dinger perturbation theory, and give an argument for the validity of the simple prescription \eqref{prescription} to all orders. The examples in sections \ref{sec:massive} and \ref{sec:minimal} provide further evidence to our claim.

\subsection{An example}\label{sec:exNew}

To get some intuition for the prescription \eqref{prescription}, let us start with an example where we know the exact spectrum of the Hamiltonian.
Consider the case where $H_0$ describes a free scalar $\phi$ of mass $m$ in AdS$_2$.
The field $\phi$ admits the following mode expansion
\beq
\phi(\tau = 0,r) = \sum_{n=0}^\infty f_n(r) \left(a_n^\dag + a_n^\phd \right)~.
\eeq
The  functions $f_n(r)$ are given explicitly in eq. \eqref{eq:fnDef}, but their form will not be important here.
The creation and annihilation operators satisfy the usual commutation relations $\left[ a_n, a_m^\dag \right] = \delta_{nm}$. The Hilbert space of the theory is the Fock space generated by acting with creation operators on the vacuum $| \Omega \rangle$. In this basis, the non-interacting normal-ordered Hamiltonian reads
\beq
\label{eq:H0massive}
H_0 = \sum_{n=0}^\infty (\DD+n) a_n^\dag a_n^\phd\,,
\eeq
where $\DD$ is related to the mass $m$ of the scalar field via $\DD (\DD-1) = m^2 R^2$. 

We will study the following Hamiltonian:
\beq
\label{eq:Dmass}
H = H_0 + \bar{\la}  \,  V 
\eeq
where $\bar{\la}=\la R^2$ and
\beq
V  = \int_{-\pi/2 }^{\pi/2 }\!\frac{dr}{(\cos r)^2}\; \NO{\phi^2(\tau=0,r)}~.
\eeq
The new Hamiltonian $H$  describes a free massive scalar with mass squared $m^2+2\lambda$.
Therefore, its spectrum is the same as $H_0$ if we replace $\Delta$ by 
\beq
\label{eq:DcorrMain}
\DD( \bar{\la}) = \half + \sqrt{\left(\DD-\half\right)^2 + 2 \bar{\la}} = \DD + \frac{ \bar{\la}}{\DD - \th} - \frac{ \bar{\la}^2}{2(\DD-\th)^3} + O( \bar{\la}^3)~.
\eeq
Let us now see how this result emerges in the Hamiltonian truncation approach.
Firstly, we write $V$ in terms of the ladder operators
\beq
\label{Vaadagger}
V  = \sum_{m,n = 0}^\infty A_{mn}(\DD) (a_m^\dag a_n^\dag + 2a_m^\dag a_n^\phd + a_m^\phd a_n^\phd)~,
\eeq
where the coefficients $A_{mn}(\DD)$ are given explicitly in eq.~\eqref{eq:AmatP}, but their form will not be important here.
Secondly, we truncate the Hilbert space to the states with unperturbed energy less than $\Lambda$.
We will finally diagonalize $H$ inside this truncated Hilbert space.

\begin{figure}[htb]
\centering
   \includegraphics[scale=1]{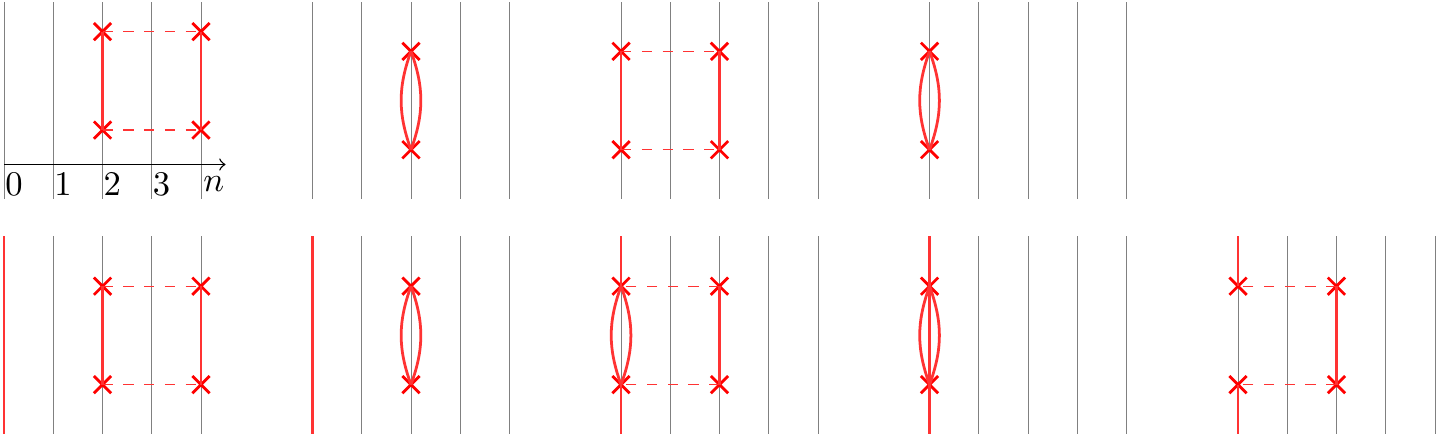}
  \caption{Diagrams contributing at second order in perturbation theory to the vacuum energy (top row) and to the energy of the first excited state $\ket{\chi} = a_0^\dagger |\Omega \rangle$ (bottom row).
 In the top row, we have intermediate two-particle  states, while in the bottom row, we have intermediate one and three particle  states.}
\label{fig:allfeynmanMain}
\end{figure}

It is instructive to study this problem perturbatively in $\lambda$, using Rayleigh–Schrödinger perturbation theory. 
We focus on  the vacuum $\ket{\Omega}$ and on the first excited state $\ket{\chi} \ldef \ket{0} = a_0^\dag \ket{\Omega}$.  Explicitly, we obtain 
\bsub
\label{RSpt}
\begin{align}
\mca{E}_\text{vac}(\Lambda) &= - \bar{\la}^2 \sum_{m \ge n \ge 0}^{m+n+2\DD < \Lambda} \frac{|\braket{\Omega}{V }{m,n}|^2}{2\DD+m+n} + O(\bar{\la}^3)
\label{RSpta}
 \\
\mca{E}_{\chi}(\Lambda) &= \DD + \bar{\la} \langle \chi | V |\chi \rangle   - \bar{\la}^2 \sum_{m \ge 1}^{m+\DD < \Lambda} 
\frac{|\braket{\chi}{V  }{m}|^2}{m}
 - \bar{\la}^2 \sum_{m \ge n \ge 0}^{m+n+3\DD < \Lambda} 
 \frac{|\braket{\chi}{V }{m,n,0}|^2}{2\DD+m+n} 
+O(\bar{\la}^3)~,
\end{align}
\esub
with the sums running over normalized single-particle states $|m\rangle = a_m^\dag |\Omega\rangle$, two particle states $|m,n\rangle \propto a_m^\dag a_n^\dag |\Omega\rangle$ and three-particle states $|m,n,0\rangle \propto a_m^\dag a_n^\dag a_0^\dag |\Omega\rangle$ with unperturbed energy less than $\Lambda$. 
It is possible to use a diagrammatic language to depict the terms in these expressions, as shown in figure  \ref{fig:allfeynmanMain}.
In these diagrams, time runs upwards and the horizontal axis corresponds to different mode numbers, \emph{e.g.} the state $\ket{1,2,2,4} \propto a_1^\dagger (a_2^\dagger)^2 a_4^\dagger \ket{\Omega}$ would be represented by single lines at $n=1$ and $n=4$ and a double line at $n=2$. The vertex $V$ is represented by crosses on the same dashed horizontal line and it can either lower or raise occupation numbers according to \eqref{Vaadagger}. 
The original state appears both at the bottom and at the top of the diagram.
At order $\bar{\la}^2$, the operator $V$ is applied twice and, in principle, all possible intermediate states need to be taken into account. However, working at finite cutoff $\La$ means that only a finite number of intermediate states are allowed, namely those with unperturbed energy smaller than $\Lambda$. 
This diagramatics is explained in more detail in appendix \ref{sec:feynman}. 
 
\begin{figure}[!htb]
\begin{center}
\includegraphics[scale=0.22]{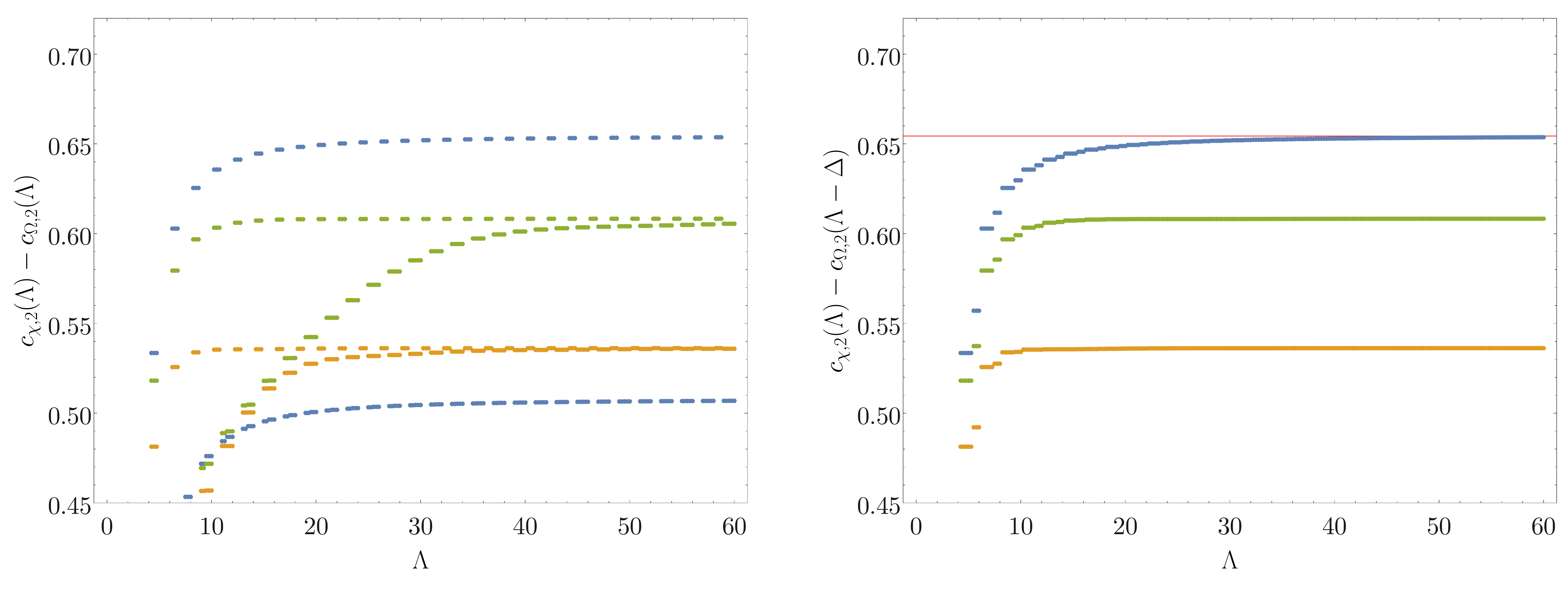}
\end{center}
\caption{Order $\lambda^2$ contribution to the energy gap of the massive boson with $\DD = \sqrt{2}$, comparing the naive prescription \eqref{naiveMain} (left) and the prescription \eqref{prescription} (right). The unregulated theory is shown in blue; the same theory regulated with $\epsilon=1/10$ (resp.\@ $\epsilon=1/5$) are shown in green (orange). For $\epsilon>0$ the two prescriptions agree as $\Lambda \to \infty$.
However, there are significant deviations (oscillations)  for $\Lambda \lesssim (\text{few})/\epsilon$.
In particular, for $\epsilon=0$, only the prescription  \eqref{prescription} converges to the exact result shown as a solid red line; there are $\sO(1)$ oscillations that persist up to arbitrarily large cutoff if the naive prescription is used.}
\label{fig:rowPlot}
\end{figure}

As shown in appendix \ref{sec:ExplicitSecondOrder},  both energies (at order $\bar{\la}^2$) diverge linearly with $\Lambda$,
\beq
\label{lindiv}
\mca{E}_\text{vac}(\Lambda) \approx \mca{E}_{\chi}(\Lambda) \approx 
- \sigma_\infty(\DD) \bar{\la}^2 \La~.
\eeq
This divergence emerges from the double sums in \eqref{RSpt}, or equivalently from the diagrams on the left most column of 
figure \ref{fig:allfeynmanMain}.
Naively, one could think that the   energy gap could be obtained from the difference
\beq\label{naiveMain}
\lim_{\Lambda \to \infty} \mca{E}_{\chi}(\Lambda)-\mca{E}_\text{vac}(\Lambda)~.
\eeq
However this does not reproduce the exact result \eqref{eq:DcorrMain} at $\mathcal{O}(\bar{\lambda}^2)$.
To find the solution of this puzzle, let us look more carefully at the structure of the perturbative formulas \eqref{RSpt}.
We can write 
\begin{align}
\label{Edif}
\mca{E}_{\chi}(\Lambda_1)-\mca{E}_\text{vac}(\Lambda_2) =&
 \DD + \bar{\la} \langle \chi | V |\chi \rangle   - \bar{\la}^2 \sum_{m\ge 1}^{m+\DD<\Lambda_1} 
\frac{|\braket{\chi}{V  }{m}|^2}{m} 
- \bar{\la}^2 \sum_{m\ge 0}^{m+3\DD<\Lambda_1} 
 \frac{|\braket{\chi}{V }{m,0,0}|^2}{2\DD+m}
 \\
  &
   +\bar{\la}^2 \sum_{m\ge 0}^{m+2\DD<\Lambda_2} 
 \frac{|\braket{\Omega}{V }{m,0}|^2}{2\DD+m}
 - \bar{\la}^2 \sum_{m\ge n\ge 1}^{m+n+3\DD<\Lambda_1 \atop m+n+2\DD>\Lambda_2} 
 \frac{|\braket{\Omega}{V }{m,n}|^2}{2\DD+m+n}
+\mathcal{O}(\bar{\la}^3)~, \nonumber
\end{align}
where we used that $\braket{\chi}{V }{m,n,0}=\braket{\Omega}{V }{m,n}$ for both $m$ and $n$ different from $0$. 
Notice that the last sum is suspicious because it only involves two-particle states with energy  between $\Lambda_2$ and $\Lambda_1-\Delta$.\footnote{
For simplicity, we assumed $ \Lambda_2< \Lambda_1-\Delta$.} 
Indeed, 
the exact result is obtained in the limit $\Lambda_1 , \Lambda_2\to \infty$ simply by dropping the contribution of the last term. This is what happens if we apply the prescription \eqref{prescription}.
Diagrammatically, this corresponds to the cancellation of 
the diagrams on the first two columns in figure \ref{fig:allfeynmanMain}. Although the diagrams in the first column give rise to a linear divergence, the prescription \eqref{prescription} ensures that they cancel exactly.
Notice that, if we keep $\Lambda_1 - \Lambda_2 $ fixed in the limit $\Lambda_1 , \Lambda_2\to \infty$, then the last sum in \eqref{Edif} gives a finite contribution, due to the growth of the overlaps $|\braket{\Omega}{V }{m,n}|$. This is the same growth responsible for the linear divergence \eqref{lindiv}. Furthermore, the discreteness of the unperturbed spectrum, labeled by integers $m,\,n,$ generates the order one oscillation visible in the left panel of figure \ref{fig:rowPlot}. We shall comment further on these oscillations in subsection \ref{subsec:second}.

\begin{figure}[!htb]
\begin{center}
\includegraphics[scale=0.25]{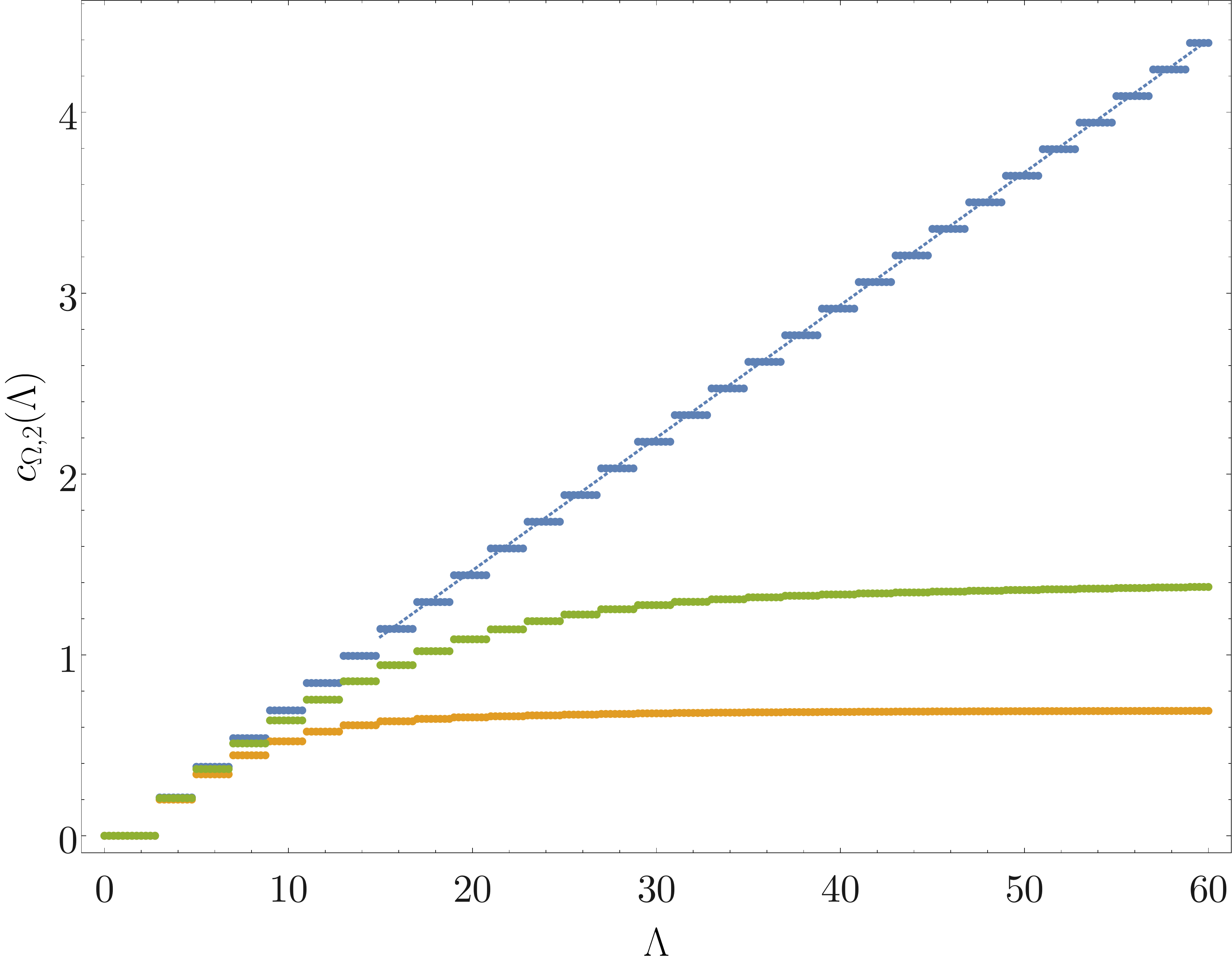}
\end{center}
\caption{Order ${\la}^2$ contribution to the Casimir energy of the $\phi^2$ flow, where the UV theory is a massive free boson with $\DD = \sqrt{2}$. The notation $c_{i,n}$ is defined in~\reef{Epertc}. Blue dots: the theory without a spatial cutoff as given by \eqref{RSpta}; green resp.\@ orange dots: the same theory cut off at $\eps = 1/10$ resp.\@ $\eps = 1/5$. Notice that the Casimir energy of the full theory diverges linearly with $\Lambda$, whereas the same interaction  with $\eps > 0$ asymptotes to a finite value. The asymptotic value appears to be reached when $\La \sim  1/\eps$. The dotted blue line shows the large-$\La$ behavior predicted using~\reef{lindiv}.}
\label{fig:vacPlot}
\end{figure}

The divergence of the ground state energy as the cutoff is removed is a consequence of the infinite volume of space. In fact, it is instructive to regulate the interaction term in the free boson example so that it affects only a finite volume:
\beq
V_\eps = \int_{-\pi/2+\eps}^{\pi/2-\eps}\!\frac{dr}{(\cos r)^2}\; \NO{\phi^2(\tau=0,r)}~.
\eeq
For $\eps > 0$ the eigenvalues of the Hamiltonian are finite as $\Lambda \to \infty$. In particular, in figure \ref{fig:vacPlot} we compare the order $\bar{\lambda}^2$ contribution to the vacuum energy, as given by \eqref{RSpt},  for the theory with $\eps = 0$ (in blue) with the same quantity for $\eps = 1/10$ (in green) and $\eps=1/5$ (in orange).
The Casimir energy of the $\eps = 0$ theory grows linearly with $\La$, whereas the regulated theories have a finite limit as $\La \to \infty$. 

We can also study the energy gap between the state $\ket{\chi} = a_0^\dag |\Omega \rangle$  and the vacuum.  
In the theory regulated with $\epsilon>0$, we can compute the energy gap as in \eqref{naiveMain} because the energy levels remain finite as $\Lambda \to \infty$. 
However, convergence only emerges for $\Lambda \gtrsim 1/\epsilon$. This explains why we cannot use the naive prescription \eqref{naiveMain} in the unregulated theory.
These issues are depicted in figure \ref{fig:rowPlot}. 

Let us summarize the general lesson. If we fix the vacuum energy to vanish for the unperturbed Hamiltonian, then the energy of each state receives an infinite contribution in the perturbed theory. The divergence of the vacuum energy is a general feature of any QFT in infinite volume, and is not an issue in the continuum limit, since in a fixed background the vacuum energy is not observable. On the other hand, this means that the spectrum of the Hamiltonian does not have a limit as the truncation level is increased, which is the source of the ambiguity in the determination of the energy gaps. We saw in a specific example that the divergence of the vacuum energy is linear with the truncation cut-off $\Lambda$. In subsection \ref{subsec:VacSecond}, we show that this is a general feature of second order perturbation theory. More generally, we expect this to be true at all orders, as we will explain in subsection \ref{sec:allorders}. In the same example, we also saw that the prescription \eqref{prescription} solves the ambiguity and leads to the correct energy gaps. 

It is worth noticing that, due to the mentioned UV/IR connection, the manifestation of the problem in perturbation theory, \emph{i.e.} the state-dependence of the contributions of intermediate states close to the cutoff, is analogous to the situation in flat space when authentic UV divergences are present \cite{EliasMiro:2020uvk,Anand:2020qnp}. Contrary to the latter, though, the infinite volume divergences we are concerned with all come from disconnected contributions. This makes it plausible that the simple prescription \eqref{prescription} has a chance of working at all orders. 
Indeed, let us now justify eq. \eqref{prescription} for a generic QFT in AdS$_2$.

\subsection{The strategy}
\label{subsec:pert_proof}

As remarked in subsection \ref{sec:Vmat}, the theory associated to the action \eqref{eq:Sin} may require a regulator to cutoff divergences close to the boundary of AdS. If the QFT is UV complete, as we shall assume in this paper, a set of counterterms exist such that all correlation functions are finite and invariant under the AdS isometries when the cutoff is removed \cite{deHaro:2000vlm}. This ensures that the energy gaps above the vacuum are well defined in the continuum limit, as it can be seen for instance via the state operator map. In the following, we assume the spectral condition \eqref{UVfiniteCond}, together with $\expec{\mca{V}}=0,$ so that such a procedure is not necessary. Furthermore, we also assume that local UV divergences in the bulk are absent. 

Nevertheless, our strategy will precisely consist in introducing a local cutoff at the AdS boundary, such that, in the regulated theory, the individual energy levels are finite as well. We shall find the regulated spectrum via Hamiltonian truncation, and analyze the convergence properties as the truncation cutoff $\Lambda$ is removed.  
Concretely, let us regulate the theory by cutting off all integrals over AdS at a coordinate distance $\epsilon$ from the boundary, \emph{i.e.} $r \in [-\pi/2+\epsilon,\pi/2-\epsilon]$. This means modifying the operator $V$ from~\eqref{eq:Ddef} as follows:
\beq
\label{eq:regV}
V_\eps \ldef R^{\DD_{\mca{V}}} \int_{-\pi/2+\eps}^{\pi/2-\eps}\!\frac{dr}{(\cos r)^2}\; \mca{V}(\tau = 0,r).
\eeq
We shall keep $\epsilon$ finite for the moment, and take the  limit  $\epsilon \to 0$ only when computing physical observables. Quantities computed in the regulated theory will contain $\epsilon$ as a subscript or a superscript.

Our argument in favor of the validity of eq.~\eqref{prescription} at all orders requires recasting the usual  Rayleigh-Schr\"{o}dinger perturbation theory in a convenient if unusual form. A reminder of perturbation theory in Quantum Mechanics can be found in appendix \ref{app:RSPT}, together with the proof of the formulas which we use below. The interacting energies admit an expansion of the form\footnote{We assume that the $i$th state is non degenerate. We will comment on the degenerate case in subsection \ref{sec:degen}.}
\beq
\mca{E}^\epsilon_i = e_i + \sum_{n=1}^\infty (-1)^{n+1} c^\epsilon_{i,n}  \, \bar{\la}^n~,
\label{EepPert}
\eeq
including the vacuum to which we associate the index $i=\Omega$. Notice that the bare energies $e_i$ do not need to be regulated. The coefficients $c^\epsilon_{i,n}$ admit a compact expression in terms of the spectral densities defined as follows:
\beq
\label{densityDef}
\braket{i}{V_\epsilon(\tau_1 + \ldots \tau_{n-1}) \dotsm V_\epsilon(\tau_1 + \tau_2)V_\epsilon(\tau_1)
V_\epsilon(0)}{i}_\text{conn} = \prod_{\l=1}^{n-1} \int_{0}^\infty \!d\a_\l\, e^{-(\a_\l - e_i)\tau_\l} \, \rho^\epsilon_{i,n}(\a_1,\ldots,\a_{n-1})~.
\eeq
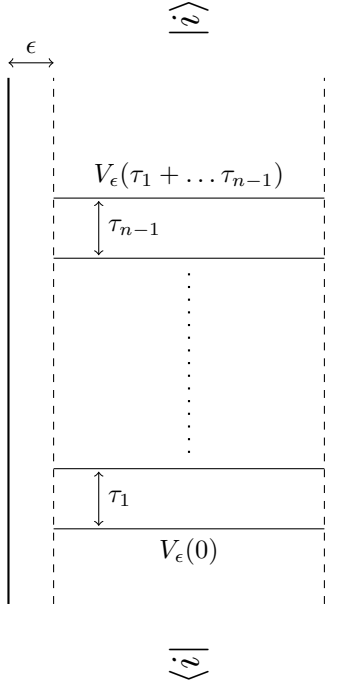
\begin{figure}[t]
\centering
\begin{tikzpicture}[scale=2]
\draw [thick]  (-1.2,-1.5) -- (-1.2,2);
\draw [thick]  (1.2,-1.5) -- (1.2,2);
\draw [dashed]  (-0.9,-1.5) -- (-0.9,2);
\draw [dashed]  (0.9,-1.5) --  (0.9,2);
\draw   (-0.9,1.2) -- node [above] {$V_\eps(\tau_1 +\dots \tau_{n-1})$} (0.9,1.2) ;
\draw [<->] (-0.6,0.82) -- node [right] {$\tau_{n-1}$} (-0.6,1.18);
\draw   (-0.9,0.8) -- (0.9,0.8);
\draw   (-0.9,-0.6) --  (0.9,-0.6);
\draw   (-0.9,-1) -- node [below] {$V_\eps(0)$} (0.9,-1);
\draw [<->] (-0.6,-0.98) -- node [right] {$\tau_1$} (-0.6,-0.62);
\draw [loosely dotted, thick] (0,0.7) -- (0,-0.5);
\draw [<->] (-1.2,2.1) -- node [above] {$\epsilon$} (-0.9,2.1);
\node at (0,2.4) [rotate = 90, scale =1.5] {$\ket{i}$};
\node at (0,-1.9) [rotate = 90, scale =1.5] {$\bra{i}$};
\end{tikzpicture}
\caption{A depiction of the correlation function on the l.h.s. of eq. \eqref{densityDef}. The vertical thick lines denote the boundary of AdS$_2$. The horizontal lines denote insertions of the regulated potential $V_\eps$, as defined in eq. \eqref{eq:regV}. The subtractions of disconnected pieces, which define the connected correlator in eq. \eqref{densityDef}, are not shown.}
\label{fig:Vs}
\end{figure}
The correlator on the left hand side is depicted in figure \ref{fig:Vs}. Crucially, it is connected with respect to the state $\ket{i}$, meaning that all the subtractions are evaluated in the same state. Explicit examples for the first few values of $n$ are given in appendix \ref{app:RSPT}, along with the expression of the spectral densities in terms of the matrix elements of $V$. Now, the general term in the perturbative expansion \eqref{EepPert} reads
\beq
\label{cinDensity}
  c^\epsilon_{i,n} =  \int_0^\infty \prod_{\l=1}^{n-1} \!\frac{d\a_\l}{\a_\l - e_i} \, \rho^\epsilon_{i,n}(\a_1,\ldots,\a_{n-1})~.
\eeq
This remarkably compact expression, compared to the increasingly convoluted textbook formulas, does not seem to be well-known in the QFT literature. The fact that the Casimir energy can be expressed in terms of connected Feynman diagrams appears (without proof) already in refs.~\cite{Bender:1968sa,Bender:1969si} by Bender and Wu. A proof of the formula for the Casimir energy is given in ref.~\cite{Klassen:1990dx} by Klassen and Melzer, specializing to the case of a perturbed $2d$ QFT. Finally, in~\cite[eq.~(2.25)]{Klassen:1991ze} by the same authors a formula is given (without proof) for the Rayleigh-Schr\"{o}dinger coefficients of the $i$-th state in the case of a perturbed 2$d$ CFT on the cylinder. The formula appearing in~\cite{Klassen:1991ze} is generically divergent and needs to be regulated --- in the time domain, its integrands generically blow up as $\tau_i \to \infty$. We prove \eqref{cinDensity} in appendix \ref{app:RSPT}. Notice that, as long as $\epsilon$ is finite, the integrals in eq.~\eqref{cinDensity} converge.

The coefficients of the perturbative expansion of the physical energy gaps can be written as follows: 
\beq
c_{i,n} -c_{0,n}  = \lim_{\epsilon\to 0}  \int_0^\infty\!\frac{d\vec{\a}}{\vec{\a} - e_i} \left[ \rho^\epsilon_{i,n}(\vec{\a}) - \rho^\epsilon_{0,n}(\vec{\a}-e_i)\right]~,
\label{gapsPertTheory}
\eeq
where the vector notation abbreviates the product in eq.~\eqref{cinDensity}. This formula is crucial and deserves an important comment: the $\epsilon\to 0$ limit is not guaranteed to commute with the integral. The fact that it does is equivalent to the validity of eq.~\eqref{prescription}. Indeed, \emph{if} the limit can be taken inside the integration, we obtain the following simple chain of equalities:
\begin{align}
c_{i,n} -c_{0,n} &= \lim_{\epsilon\to 0} \lim_{\Lambda\to \infty}  \int_0^\La\!\frac{d\vec{\a}}{\vec{\a} - e_i} \left[ \rho^\epsilon_{i,n}(\vec{\a}) - \rho^\epsilon_{0,n}(\vec{\a}-e_i)\right] \nonumber\\
& = \lim_{\Lambda\to \infty}  \int_0^\La\!\frac{d\vec{\a}}{\vec{\a} - e_i} \left[ \rho_{i,n}(\vec{\a}) - \rho_{0,n}(\vec{\a}-e_i)\right] \label{ProofChain}\\
&= \lim_{\Lambda\to \infty} \left(c_{i,n}(\La) -c_{0,n}(\La-e_i)\right)~, \nonumber
\end{align}
where we defined the expansion of the truncated energies:
\beq
\mca{E}_i(\La) = e_i+\sum_{n=1}^\infty (-1)^{n+1} c_{i,n}(\La)  \, \bar{\la}^n~.
\label{Epertc}
\eeq
Hence, we are left with the task of proving that the $\epsilon\to 0$ and the $\La\to\infty$ limits do commute. Of course, for this to be true it is not sufficient that the integral in the second line of eq. \eqref{ProofChain} converges. Indeed, intuitively we should rule out the existence of bumps which give a finite contribution to the integral, and run away to infinity as $\eps \to 0$. Formally, the dominated convergence theorem states that the limits commute if, for all $\eps<\eps_0$, the integrand is bounded by a \emph{fixed} function whose integral from 0 to $\infty$ converges. 

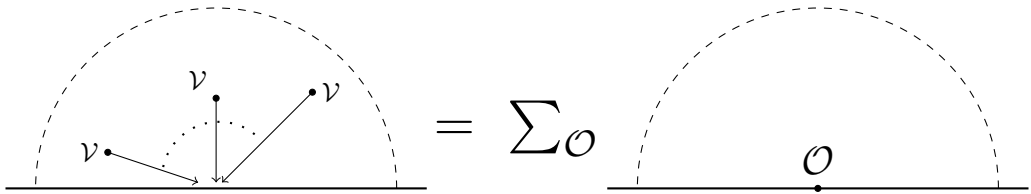
\begin{figure}[t]
\centering
\begin{tikzpicture}[scale=0.8]
\draw [thick]  (-3.5,0) -- (3.5,0);
\draw [dashed] (3,0) arc [start angle=0, end angle=180, radius=3];
\draw [<-] (0,0.1) -- (0,1.5); 
\filldraw (0,1.5) circle [radius=0.05] node [above left] {$\mca{V}$};
%\node at (-0.2,1.7) {$\mca{V}$};
\draw [<-] (0.1,0.1) -- (1.6,1.6); 
\filldraw (1.6,1.6) circle [radius=0.05] node [right] {$\mca{V}$};
\draw [<-] (-0.3,0.1) -- (-1.8,0.6); 
\filldraw (-1.8,0.6) circle [radius=0.05] node [left] {$\mca{V}$};
%\node at (1.4,1.7) {$\mca{V}$};
\draw [loosely dotted, thick] (-0.880,0.355) arc [start angle = 168, end angle=45, radius=0.95];
\node at (5,1) [scale=2] {$= \ \sum_\mathcal{O}$};
\draw [thick]  (6.5,0) -- (13.5,0);
\draw [dashed] (13,0) arc [start angle=0, end angle=180, radius=3];
\filldraw (10,0) circle [radius=0.05] node [above,scale=1.5] {$\mathcal{O}$};
\end{tikzpicture}
\caption{Sketch of the OPE channel relevant to establishing the singular behavior of the correlator in eq. \eqref{densityDef} as some of the $\tau_m \to 0$. The OPE is obtained, as usual, by projecting the insertions of the $\mca{V}$ operators on the l.h.s. onto a complete set of states on the dashed semicircle. The coefficients of the OPE are higher-point functions involving the insertions of $\mca{V}$ and the boundary operator $\mathcal{O}$. An explicit example is considered in appendix \ref{sec:twopgen}, see eq. \eqref{VVbOPE} and figure \ref{fig:VVboundary}.}
\label{fig:multipleOPEsketch}
\end{figure}

Before engaging in the details of the argument, let us describe the main ingredients, which are few and simple. Clearly, establishing the validity of the manipulations in eq. \eqref{ProofChain} requires to bound the large $\vec{\alpha}$ limit of the spectral densities $\rho_{i,n}^\eps$. It is intuitive, albeit hard to prove, that this limit is controlled by the small $\vec{\tau}$ limit of the correlator on the l.h.s. of eq. \eqref{densityDef}. As it can be seen in figure \ref{fig:Vs}, when one, or more, of the $\tau_m$ go to zero, the insertions of the potential collide. Non-analyticities in these limits may arise because a subset of the perturbing operators $\mca{V}$ collide in the bulk, or, if $\eps = 0$, on the boundary. Both contributions are important in establishing the rate of convergence of the energy gaps as we lift the cutoff $\Lambda$, which we will analyze in subsection \ref{sec:rate}. On the contrary, by assumption, the individual energy levels themselves are finite as long as $\eps>0$. Hence, in order to establish the prescription \eqref{prescription} we are only interested in the fusion of the insertions $\mca{V}$ with the boundary of AdS. As explained above, our assumptions also imply that the OPE \eqref{bulkToBound} is sufficiently soft to make the matrix elements of $V$ finite. We conclude that the relevant small $\vec{\tau}$ singularities are due to the simultaneous collision of \emph{multiple} local operators $\mca{V}$ on a \emph{unique} point on the AdS boundary. This prompts us to study the OPE depicted in figure \ref{fig:multipleOPEsketch}: the lightest operator appearing in the OPE, whose expectation value in the external state $\ket{i}$ does not vanish, is responsible for the small $\vec{\tau}$ singularity, and eventually for the growth of the spectral density at infinity.

In the following, we shall detail the above steps in the first non-trivial order in perturbation theory. Then, we will discuss the generalization of the procedure to all orders in perturbation theory. Finally, we will discuss the rate of convergence of Hamiltonian truncation in AdS.

\subsection{Divergence of the vacuum energy at second order}
\label{subsec:VacSecond}

Let us begin by showing that the linear divergence of the vacuum energy with the truncation level, which we observed in the example \ref{sec:exNew}, is a general feature at second order in perturbation theory. Since $\braket{\Omega}{\mca{V}}{\Omega}=0$, eq. \eqref{densityDef} reduces in this case to
\beq
\braket{\Omega}{V(\tau) V(0)}{\Omega}
 = \int_{0}^\infty \!d\a \, e^{-\a \tau} \, \rho_{0,2}(\a)~.
\label{rhoVacSecond}
\eeq
Since we are interested in characterizing the behavior of the first energy level as a function of the cutoff $\Lambda$, we set $\epsilon=0$ in this subsection. More general results can be found in appendix \ref{sec:twoptvac}. As explained in subsection \ref{subsec:pert_proof}, we need to extract the large $\alpha$ behavior of $\rho_{0,2}(\a)$. Reflection positivity ensures that $\rho_{0,2}(\a)>0$. This allows us to invoke a Tauberian theorem \cite{Pappadopulo:2012jk} to rigorously tie the large $\alpha$ limit of the spectral density to the small $\tau$ limit of the correlator on the l.h.s. of eq. \eqref{rhoVacSecond}.

The latter limit is computed in appendix \ref{sec:twoptvac}, and reads as follows:\footnote{As explained in appendix \ref{sec:twoptvac}, this result is valid if $\Delta_\mca{V}<3/2$, otherwise it must be replaced by eq. \eqref{VacSmallTauBulk}. In particular, if $\Delta_\mca{V}>3/2$, the leading contribution to the divergence of the vacuum energy comes from a bulk UV singularity, which we assumed to avoid in this work.}
\beq
\braket{\Omega}{V(\tau) V(0)}{\Omega} \overset{\tau \to 0}{\sim} \frac{c}{\tau^2}~.
\label{VVvacSmallTau}
\eeq   
The coefficient of the singularity is theory dependent, and, as explained in the same appendix, it can be computed as follows:
\beq
R^{2\Delta_\mca{V}}\braket{\Omega}{\mca{V}(\tau, r)\mca{V}(0, r')}{\Omega} \equiv f_\Omega(\xi)~,
\eeq
\beq
c = 8\int_0^\infty\! d\xi\, \textup{arcsinh} \left(2 \sqrt{\xi(\xi+1)}\right) f_\Omega(\xi) ~, 
\label{CsmallTauVac}
\eeq
where we recall that $\xi$ is defined in eq. \eqref{eq:crossratio}. In the language of the OPE described in figure \ref{fig:multipleOPEsketch}, the small $\tau$ limit corresponds to the contribution of the identity operator, the only one surviving when the r.h.s. of the OPE is evaluated in the vacuum.

As advertised, the Tauberian theorem reviewed in \cite{Pappadopulo:2012jk} implies the following asymptotics for the integrated spectral density:
\beq
\int_0^\Lambda \! d\a \,\rho_{0,2}(\a) \overset{\Lambda\to \infty}{\sim}
\frac{c}{2}\, \Lambda^2~.
\eeq
The contribution to the vacuum energy at second order can then be extracted from eq. \eqref{EepPert}, by truncating the integral in eq. \eqref{cinDensity} to the region $[0,\Lambda]$ and integrating by parts:
\beq
\mathcal{E}_\Omega(\Lambda) \overset{\Lambda\to \infty}{\sim} 
- c\, \Lambda\, \bar{\lambda}^2 + O(\bar{\lambda}^3)~.
\eeq
We will have opportunities to check this formula in the examples of sections \ref{sec:massive} and \ref{sec:minimal}.

\subsection{The argument at second order}
\label{subsec:second}

Let us now turn to our goal of establishing the prescription \eqref{prescription}.
At second order in perturbation theory, the object of interest is the two-point function of the potential:
\beq
 \braket{i}{V_\eps(\tau)V_\eps(0)}{i}_\text{conn} = \braket{i}{V_\eps(\tau)V_\eps(0)}{i} - \braket{i}{V_\eps}{i}^2~.
\eeq
Let us first point out the following simple but crucial fact. The difference of the spectral densities appearing in eq. \eqref{ProofChain}, \emph{with} the shift in their arguments, is the spectral density of the difference of the correlators in state $\ket{i}$ and in the vacuum: 
\beq
g_\eps(\tau)=\braket{i}{V_\eps(\tau)V_\eps(0)}{i}_\text{conn} - \braket{\Omega}{V_\eps(\tau)V_\eps(0)}{\Omega}_\text{conn}
= \int_{-e_i}^\infty \!d\a\, e^{-\a \tau} \left(\rho^\eps_{i,2}(\a+e_i)-\rho^\eps_{0,2}(\a)\right)~.
\label{gfromrho}
\eeq
We shall denote the new spectral density as
\beq
\bDelta \rho^\eps_{i,2}(\alpha)=\rho^\eps_{i,2}(\a+e_i)-\rho^\eps_{0,2}(\a)~.
\label{Dspec}
\eeq
We are interested in the large $\alpha$ behavior of the spectral density, and we would like to leverage our knowledge of the correlation functions in position space. While it is clear that singularities at small $\tau$ in $g_\eps(\tau)$ are related to the large $\alpha$ behavior of the spectral density, it is non-trivial to obtain a precise correspondence. Indeed, $\bDelta \rho^\eps_{i,2}(\alpha)$ is a non-positive distribution, and the Tauberian theorems which usually provide those estimates cannot be applied.\footnote{See however \cite{Pal:2019yhz}, where a lower bound on the average of a non-positive spectral density was obtained, subject to a condition on the amount of negativity allowed.} Instead, we shall tackle the question directly, and point out the additional assumptions necessary to obtain the needed asymptotics. Let us first recall the naive answer. \emph{If} the correlator has a singular power-like small $\tau$ limit, \emph{i.e.}
\beq
if\qquad g_\eps(\tau) \overset{\tau\to 0}{\sim}  \tau^{-\beta}~, \quad \beta>0~,
\eeq
and \emph{if} the spectral density had a power-like behavior at large $\alpha$ as well, then we would conclude that
\beq
\bDelta \rho^\eps_{i,2}(\alpha) \overset{\alpha\to \infty}{\sim} 
\alpha^{\beta-1}~.
\eeq
Since the spectral densities are in general infinite sums of delta functions, we 
have to interpret the last equation in an averaged sense.
Let us describe how this works in more detail. 

We focus our attention on the following averaged spectral density:
\beq
R_\eps(\Lambda)=\int_{-\infty}^\Lambda\! d\a\,\bDelta \rho^\eps_{i,2}(\alpha)
= \int_{\gamma-\ii \infty }^{\gamma+\ii \infty } \frac{d\tau}{2\pi\ii \tau} 
e^{\Lambda \tau} g_\eps(\tau)~, \qquad \gamma>0~.
\label{Raver}
\eeq
$R_\eps(\Lambda)$ is a finite sum which can be easily inferred from eq. \eqref{rho2states}. The right hand side is obtained by commuting the $\alpha$-integral with the inverse Laplace transform of $g_\epsilon(\tau)$, and one can explicitly check that eq. \eqref{Raver} gives the correct result.\footnote{This is how it goes: $g_\epsilon(\tau)=\sum_{n=0}^\infty a_n e^{-\a_n \tau}$. This sum converges absolutely for $\Re\tau>0$. Hence, we can commute the sum and the integral to get $R_\eps(\Lambda)=\sum_{n=0}^\infty a_n \int_{\gamma-\ii \infty }^{\gamma+\ii \infty } \frac{d\tau}{2\pi\ii \tau} e^{(\Lambda-\a_n) \tau}$. The last integral converges and the result is the expected  $R_\eps(\Lambda)=\sum_{\alpha_n<\Lambda} a_n $.} While the correlation functions in eq. \eqref{gfromrho} are analytic when $\Re \tau>0$, singularities may arise in the left half plane. In particular, as we show in appendix \ref{sec:twopgen}, $g_\eps(\tau)$ generically has a branch point at $\tau=0$,
\beq
g_\eps(\tau) \overset{\tau \to 0}{\sim} \tau^{\Delta_*-2}~,
\label{gepscaling}
\eeq 
where $\Delta_*$ is the dimension of the leading boundary operator above the identity in the OPE illustrated in figure \ref{fig:multipleOPEsketch} - see also figure \ref{fig:VVboundary} and the detailed explanation in appendix \ref{sec:twopgen}. 
If the exactly solvable Hamiltonian corresponds to a CFT, then $\Delta_*\leq 2$, due to the universal presence of the displacement operator, see e.g. \cite{Billo:2016cpy}. If $\Delta_* =2$, the singularity is logarithmic. 
More precisely, in appendix \ref{sec:twopgen},  we show that, if $\Delta_*\leq 2$, $g_\eps(\tau)$  is bounded as follows:\footnote{More precisely, this result assumes that the bulk OPE is non-singular. In appendix \ref{sec:twopgen} we also discuss what happens if this is not the case, in a restricted scenario -- see the paragraph around eq. \eqref{GepsBulk}. See also the discussion in subsection \ref{sec:rate}. In general, as long as the small $\tau$ limit is less singular than $\tau^{-1}$ and it is uniform in $\epsilon$, the rest of the argument goes through. The first condition is equivalent to $\Delta_{\tilde{\mca{V}}}>2\Delta_\mca{V}-2$, $\Delta_{\tilde{\mca{V}}}$ being the first operator above the identity in the $\mca{V}\times \mca{V}$ OPE. This is the same as requiring that bulk UV divergences are absent, which we assume in this work.}
\beq
|g_\epsilon(\tau)|<C(\tau_0) 
\left\lbrace
\begin{array}{ll}
|\tau|^{\Delta_*-2}~, & \Delta_* <2 \\
\log |\tau| ~,  & \Delta_*=2~,
\end{array}\right.
\quad |\tau|<\tau_0~.
\label{gepbound}
\eeq
Here, $\tau_0>0$ is an arbitrary fixed number. The crucial step in the derivation of this result is that, when replacing the OPE of figure \ref{fig:multipleOPEsketch} in the left hand side of eq. \eqref{gfromrho}, the identity operator cancels out, and with it the strongest singularity \eqref{VVvacSmallTau} at small $\tau$. As a consequence, the spectral density \eqref{Dspec} has a softer large $\a$ behavior than what would happen without the shift. It is important that the bound \eqref{gepbound} is uniform in $\epsilon$.

In section \ref{sec:massive}, we will consider deformations of free massive QFTs, for which $\Delta_*>2$ is possible. Although we have not analyzed this case in detail, we are confident that the bound \eqref{gepbound} still holds, up to analytic terms at $\tau=0$. In other words, a strict bound can be obtained when $\Delta_*>2$ by taking enough derivatives of $g_\eps(\tau)$. We offer some comments in this direction in appendix \ref{sec:twopgen}. This turns into a bound for a higher moment of $\bDelta \rho$, which still allows to proceed with the argument. The examples in section \ref{sec:massive} lend support to this conclusion.

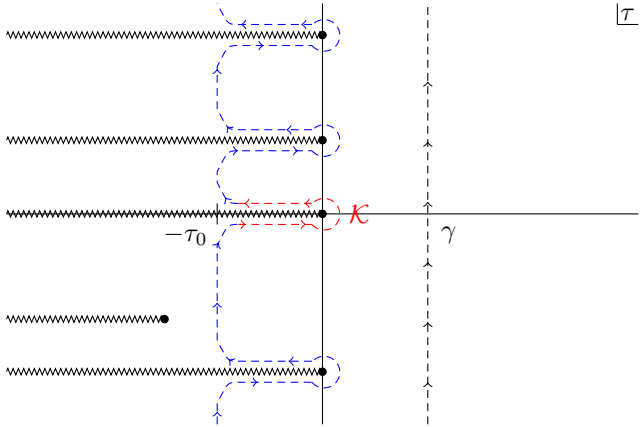
\begin{figure}[h]
\centering
\begin{tikzpicture}[scale=1.4]
\draw  (0,-2) -- (0,2);
\draw  (-3,0) -- (3,0);
\draw (2.8,2) -- (2.8,1.8);
\draw (2.8,1.8) -- (3,1.8);
\node at (2.9,1.9) {$\tau$};
\draw [decorate,decoration={zigzag,amplitude=1.2pt,segment length =2 pt}] (-3,0) -- (0,0);
\filldraw (0,0) circle [radius=1pt]; 
\draw [decorate,decoration={zigzag,amplitude=1.2pt,segment length =2 pt}] (-3,0.7) -- (0,0.7);
\filldraw (0,0.7) circle [radius=1pt]; 
\draw [decorate,decoration={zigzag,amplitude=1.2pt,segment length =2 pt}] (-3,1.7) -- (0,1.7);
\filldraw (0,1.7) circle [radius=1pt]; 
\draw [decorate,decoration={zigzag,amplitude=1.2pt,segment length =2 pt}] (-3,-1.5) -- (0,-1.5);
\filldraw (0,-1.5) circle [radius=1pt]; 
\draw [decorate,decoration={zigzag,amplitude=1.2pt,segment length =2 pt}] (-3,-1) -- (-1.5,-1);
\filldraw (-1.5,-1) circle [radius=1pt]; 
\node at (-1.3,-0.2) {$-\tau_0$};
\draw (-1,0.1) -- (-1,-0.1);
\node at (1.2,-0.2) {$\gamma$};
\draw [densely dashed,postaction = {decoration={markings, mark=between positions 0.1 and 0.95 step 0.8cm with {\arrow{>};}},decorate}] (1,-2) -- (1,2); % Upwards gamma contour
\draw [blue,densely dashed,postaction = {decoration={markings, mark=between positions 0.1 and 0.95 step 0.8cm with {\arrow{>};}},decorate}] (-1,-2) -- (-1,-1.8) .. controls (-0.9,-1.6) .. (-0.8,-1.6) -- (-0.1,-1.6); % First blue part
\draw [blue,densely dashed] (-0.1,-1.6) arc [start angle=-140,end angle=140,radius=0.15]; % First arc (blue)
\draw [blue,densely dashed,postaction = {decoration={markings, mark=between positions 0.1 and 0.95 step 0.8cm with {\arrow{>};}},decorate}] (-0.1,-1.4) -- (-0.8,-1.4) .. controls (-0.9,-1.4) .. (-1,-1.2) -- (-1,-0.3) .. controls (-0.9,-0.1) .. (-0.8,-0.1); % Second blue part
\draw [red,densely dashed,postaction = {decoration={markings, mark=between positions 0.1 and 0.95 step 0.8cm with {\arrow{>};}},decorate}] (-0.8,-0.1) -- (-0.1,-0.1); % Red horizontal part
\draw [red,densely dashed] (-0.1,-0.1) arc [start angle=-140,end angle=140,radius=0.15] node [midway,right] {$\mathcal{K}$}; % Second arc (red)
\draw [red,densely dashed,postaction = {decoration={markings, mark=between positions 0.1 and 0.95 step 0.8cm with {\arrow{>};}},decorate}] (-0.1,0.1) -- (-0.8,0.1); % Red horizontal part
\draw [blue,densely dashed,postaction = {decoration={markings, mark=between positions 0.1 and 0.95 step 0.8cm with {\arrow{>};}},decorate}] (-0.8,0.1) .. controls (-0.9,0.1) .. (-1,0.3) -- (-1,0.4) .. controls (-0.9,0.6) .. (-0.8,0.6)  -- (-0.1,0.6);; % Third blue part
\draw [blue,densely dashed] (-0.1,0.6) arc [start angle=-140,end angle=140,radius=0.15]; % Third arc (blue)
\draw [blue,densely dashed,postaction = {decoration={markings, mark=between positions 0.1 and 0.95 step 0.8cm with {\arrow{>};}},decorate}] (-0.1,0.8) -- (-0.8,0.8) .. controls (-0.9,0.8) .. (-1,1) -- (-1,1.4) .. controls (-0.9,1.6) .. (-0.8,1.6) -- (-0.1,1.6); % Fourth blue part
\draw [blue,densely dashed] (-0.1,1.6) arc [start angle=-140,end angle=140,radius=0.15]; % Fourth arc (blue)
\draw [blue,densely dashed,postaction = {decoration={markings, mark=between positions 0.1 and 0.95 step 0.8cm with {\arrow{>};}},decorate}] (-0.1,1.8) -- (-0.8,1.8) .. controls (-0.9,1.8) .. (-1,2);  % Fifth blue part
\end{tikzpicture}
\caption{A sketch of the generic singularity structure of $g_\eps(\tau)$ The singularities are generically branch points, and the cuts are oriented away from the region $\Re \tau>0$, where the correlator is analytic. The black dashed contour is the original one in eq. \eqref{Raver}, while the blue dashed one is shifted to the left in order to leverage the exponential suppression at large $\Lambda$. Finally, the red contour is $\mathcal{K}$ in eq. \eqref{Rfromtau0}. }
\label{fig:invLapCont}
\end{figure}

On the other hand, further singularities may appear on the imaginary $\tau$ axis. A generic example of the analytic structure in the $\tau$ plane is depicted in figure \ref{fig:invLapCont}.
Let us now shift the integration contour in \eqref{Raver} to the left, as shown in figure \ref{fig:invLapCont}. It is clear from eq. \eqref{Raver} that the part of the contour which runs in the left half plane is exponentially suppressed at large $\Lambda$. 
As for the $\Re \tau=0$ axis, if $\bDelta \rho^\eps_{i,2}(\alpha)$ was positive, the stronger singularity would be at $\tau=0$, as it is easy to deduce from eq. \eqref{gfromrho}. The large $\Lambda$ limit of $R_\eps(\Lambda)$ would then be controlled by the strength of this singularity, a result familiar from the Hardy-Littlewood Tauberian theorem. 
Since the spectral density is not necessarily positive, we will \emph{assume}
that the large $\Lambda$ behaviour of $R_\eps(\Lambda)$ is still dominated by the singularity at $\tau=0$. This assumption is not harmless, and we shall offer some comments about it at the end of this subsection.
 Keeping in mind figure \eqref{fig:invLapCont}, we then obtain
\beq
\label{Rfromtau0}
|R_\eps(\Lambda)| \sim 
 \left|\int_\mathcal{K} \frac{d\tau}{2\pi\ii} \frac{e^{\Lambda \tau}-1}{\tau} g_\eps(\tau) +\int_\mathcal{K} \frac{d\tau}{2\pi\ii \tau}  g_\eps(\tau) \right|~.
\eeq 
Here, $\mathcal{K}$ is a keyhole contour around the origin which extends up to $\Re\tau=-\tau_0$. With the benefit of hindsight, we added and subtracted 1. Indeed, we can now shrink to the real axis the first $\mathcal{K}$ contour, and we can use the inequality (\ref{gepbound}) to write
\beq
|R_\eps(\Lambda)|\leq 
2C(\tau_0)\int_{-\tau_0}^0\frac{d\tau}{2\pi}|e^{\Lambda \tau}-1|
|\tau^{\Delta_*-3}| 
+\left|\int_\mathcal{K} \frac{d\tau}{2\pi\tau}  g_\eps(\tau)\right|~.
\label{ineq}
\eeq
Notice that the first integral converges if $\Delta_*>1$, which is guaranteed by the condition \eqref{UVfiniteCond}. The second integral is finite and $\Lambda$ independent. 
Therefore, under the assumptions we have made, we conclude that\footnote{It should be noted that, when $\Delta_*=2$, one obtains $F(\Lambda) \sim \log^2\Lambda$ in eq. \eqref{boundAsymp}. This is sufficient to prove the prescription \eqref{prescription}, but is a weak bound. Indeed, the integrated spectral density $R(\Lambda)$ is expected to only grow like $\log \Lambda$. The weak point is the inequality \eqref{ineq}: when we shrink the contour $\mathcal{K}$ onto the real axis, the integral of the left hand side reduces to the discontinuity, which replaces $\log\tau$ with a constant and does lead to the expected asymptotics.}
\beq
|R_\eps(\Lambda)|\leq F(\Lambda) \overset{\Lambda\to\infty}{\sim} \Lambda^{2-\Delta_*}~,
\label{boundAsymp}
\eeq
with $F(\Lambda)$  independent of $\epsilon$.
We shall now assume \eqref{boundAsymp} to prove  \eqref{ProofChain}.  It is sufficient to integrate by parts once, to express the energy shift \eqref{gapsPertTheory} in terms of $R_\eps(\Lambda)$:
\beq
c_{i,2}-c_{0,2}=\lim_{\eps\to 0}\lim_{\Lambda\to\infty} \left(\frac{R_\eps(\Lambda-e_i)}{\Lambda-e_i}+
\frac{R_\eps(-e_i)}{e_i}+\int_0^\Lambda \frac{d\alpha}{(\alpha-e_i)^2}R_\eps(\alpha-e_i)\right)~.
\eeq
Thanks to eq. \eqref{boundAsymp}, we clearly can swap the small $\eps$ and the large $\Lambda$ limits in the first two addends. As for the last integral, we can fix a constant $\Lambda_*$ and write the following inequality:
\begin{align}
&\int_0^{\Lambda_*} \frac{d\alpha}{(\alpha-e_i)^2}R_0(\alpha-e_i)-\int_{\Lambda_*}^\infty \frac{d\alpha}{(\alpha-e_i)^2}F(\alpha-e_i) \notag \\
\leq&\lim_{\eps\to 0}\int_0^\infty \frac{d\alpha}{(\alpha-e_i)^2}R_\eps(\alpha-e_i)\\
\leq &\int_0^{\Lambda_*} \frac{d\alpha}{(\alpha-e_i)^2}R_0(\alpha-e_i)+\int_{\Lambda_*}^\infty \frac{d\alpha}{(\alpha-e_i)^2}F(\alpha-e_i)~. \notag
\end{align}
Here, we took the $\eps\to 0$ limit inside the integral over the finite domain $[0,\Lambda_*]$, and used the inequality \eqref{boundAsymp} to express the remaining integrand in terms of the positive and $\eps$ independent function $F(\Lambda)$.
 Taking $\Lambda_*\to \infty$, we finally obtain the sought result \eqref{ProofChain}. The prescription \eqref{prescription} at order $\lambda^2$ in perturbation theory immediately follows. 

Let us now come back to the additional singularities on the $\Im \tau$ axis in figure \ref{fig:invLapCont}. Contrary to the singularity at $\tau=0$, the contribution of each of them is not enhanced by the $1/\tau$ pole in \eqref{Raver}. Hence, if they are not enhanced with respect to eq. \eqref{gepbound}, they are subleading at large $\Lambda$. This assumption hides a remarkable series of cancellations, analogous to the cancellation of the identity block at $\tau=0$, which, as discussed, leads to the bound \eqref{gepbound}. Indeed, each of the correlators in eq. \eqref{gfromrho} is expected to have an infinite series of double poles along the imaginary axis. The reason for this is that such double poles lead to order one oscillations in the $c_{i,2}(\Lambda)$, the second order energy shifts in eq. \eqref{Epertc}. The oscillations, visible for instance in figure \ref{fig:vacPlot}, are a generic consequence of the discreteness of the spectrum. Without an exact cancellation, the oscillations affect the energy gaps as well, as in figure \ref{fig:rowPlot}. The absence of order one oscillations in the examples treated in sections \ref{sec:massive} and \ref{sec:minimal}, then, gives evidence for the correctness of the prescription \eqref{prescription} beyond the arguments presented so far. Analyzing these cancellations in detail requires studying the correlators of the deformation $V_\eps$ in Lorentzian kinematics: an interesting task for the future. For the moment, let us make a couple of simple remarks. If the unperturbed spectrum has integer scaling dimensions, then all correlators are periodic for $\tau \to \tau +2\pi \ii$. This is the case for the minimal models with the boundary condition we choose in section \ref{sec:minimal}. Then, barring additional singularities within the first period, the cancellation of all the double poles simply follows from the result we proved for $\tau=0$. On the other hand, the periodicity is broken in the general case, including the perturbation of massive free theories which we will treat in section \ref{sec:massive}. An example of this non periodic behavior can be seen in appendix \ref{app:periodicity}. Nevertheless, we can still envision a mechanism for the cancellation of the double poles. As we discussed in subsection \ref{subsec:VacSecond} and in appendix \ref{app:smalltau}, the double pole at $\tau=0$ in each correlator in eq.  \eqref{gfromrho} is due to a disconnected contribution. If the unperturbed theory is free, such contribution corresponds to a disconnected Witten diagram, where the boundary operators which create and annihilate the states $\ket{i}$ are spectators. Such diagrams exchange a fixed number of particles, and so they are periodic in $\Im(\tau)$ with period $2\pi$ up to a phase, which is crucially independent of the boundary state $\ket{i}$. Therefore, again, each correlator in eq. \eqref{gfromrho} has an infinite series of double poles, which all cancel in $g_\eps(\tau)$. The absence of the double poles can be exhibited in the example of a massive free scalar, and we do so in appendix \ref{app:periodicity}. More specifically, there we also check that the remaining non-analyticities on the imaginary $\tau$ axis are not enhanced with respect to the $\tau=0$ singularity.

A last comment is in order. Even if the leftover singularities in figure \ref{fig:invLapCont} are softer than the one at $\tau=0$, one could still worry that there is an infinite sum over all of them at $\tau = i t_n$ for real $t_n$. However, this sum is suppressed by the presence of the rapidly oscillating phases $e^{i\Lambda t_n}$. The convergence of the sum then depends on the density of the singularities along the imaginary axis, as $\Im \tau \to \infty$. One might expect this density to be approximately constant. If the boundary spectrum of the unperturbed theory is integer, this is obvious. In the general case, the exact periodicity may be replaced by some recurrence time, fixed by the scaling dimensions $\Delta$, which nevertheless should not change this picture. Notice, however, that it is conceivable that non-analyticities appear with a periodicity of some multiple of $\pi$ even in the general case, in keeping with the periodic structure of AdS in real time: this happens in the scalar example analyzed in appendix \ref{app:periodicity}.  All in all, if the density is at most constant, the sum converges at least conditionally.

\subsection{Generalization at all orders}
\label{sec:allorders}

The aim of this section is to sketch the generalization of the previous argument to all orders in perturbation theory. We shall not attempt to be detailed in this case, rather we will stress the ingredients that should make the argument follow through.

At $n$-th order in perturbation theory, we need to bound the (average of the) spectral density 
\beq
\bDelta \rho_{i,n}^\eps(\vec{\a})=\rho_{i,n}^\eps(\vec{\a}+e_i)-\rho_{0,n}^\eps(\vec{\a})~,
\eeq
as the vector $\vec{\a}$ becomes large in a generic direction of its $(n-1)$ dimensional space. The bound should be uniform in $\eps$ and allow the integration of 
\beq
\int_0^\La\!\frac{d\vec{\a}}{\vec{\a} - e_i}  \bDelta \rho_{i,n}^\eps(\vec{\a}-e_i)
\label{nthOrderInt}
\eeq
over a hypercube of size $\Lambda$, as $\Lambda \to \infty$: see eq. \eqref{ProofChain}. 
It is easy to see that, also in the general case, $\bDelta \rho_{i,n}^\eps(\vec{\a})$ is the Laplace transform of the following difference of connected correlators:
\begin{multline}
g^n_\eps(\vec{\tau})=\braket{i}{V_\epsilon(\tau_1 + \ldots \tau_{n-1}) \dotsm V_\epsilon(\tau_1 + \tau_2)V_\epsilon(\tau_1)
V_\epsilon(0)}{i}_\text{conn} \\
- \braket{\Omega}{V_\epsilon(\tau_1 + \ldots \tau_{n-1}) \dotsm V_\epsilon(\tau_1 + \tau_2)V_\epsilon(\tau_1)
V_\epsilon(0)}{\Omega}_\text{conn}~.
\label{gepsn}
\end{multline}
Recall that all the subtractions of disconnected pieces are computed in the state $\ket{i}$ for the first correlator and $\ket{\Omega}$ for the second. We maintain the assumption that the large $\vec{\alpha}$ limit of the spectral density is controlled by the limit of $g^n_\eps(\vec{\tau})$ when a subset of components of the vector $\vec{\tau}$ vanish. In turn, this limit is controlled by the OPE channel of figure \ref{fig:multipleOPEsketch}, at the level of the correlation function of the local operators $\mca{V}$.

Then, let us first consider the case where a proper subset of components of $\vec{\tau}$ vanish together. In this case, the contribution from the identity in the OPE in figure \ref{fig:multipleOPEsketch} vanishes separately in each connected correlator on the right hand side of eq. \eqref{gepsn}. This is a general property of cumulants, valid in any state: they vanish whenever a subset of random variables is statistically independent from the others.\footnote{In practice, it can be easily verified by noticing that the OPE can be applied at the level of the generating function. Schematically, for the case where one subset $\Pi$ of the insertions coalesce:
\begin{equation}
\mathcal{Z}_{\ket{i}} = \braket{i}{e^{\sum_m J(x_m)\mca{V}(x_m)}}{i} \to \braket{\Omega}{e^{\sum_{p\in \Pi} J(x_p)\mca{V}(x_p)}}{\Omega} \braket{i}{e^{\sum_{m\notin \Pi} J(x_m)\mca{V}(x_m)}}{i} \equiv \mathcal{Z}_{\Pi} \mathcal{Z}_{\bar{\Pi}}~.
\end{equation}
Since the generating function factorizes, its log splits into a sum, and the connected correlator immediately vanishes.}

On the other hand, if all the components of $\vec{\tau}$ vanish together, then the contribution of the identity does not cancel out at the level of the single connected correlator. Rather, it cancels out in the difference \eqref{gepsn}, as it can be checked again by inserting a complete set of states in the generating function.

The cancellation of the identity should be sufficient for the integral in eq. \eqref{nthOrderInt} to converge uniformly in $\eps$ and allow the chain of equalities \eqref{ProofChain} to hold. Let us sketch the argument at the level of power counting, and let us focus on the radial behavior, where all components of $\vec{\alpha}$ diverge with fixed ratios. In this case, the asymptotics of $\bDelta \rho_{i,n}^\eps$ is controlled by the small $|\vec{\tau}|$ limit, for which we expect
\beq
|g^n_\eps(\vec{\tau})| < C(\tau_0) |\vec{\tau}|^{\Delta_*-n}~, \qquad 
|\vec{\tau}|<\tau_0~.
\eeq
Here, $\Delta_*$ is the lowest state above the identity which couples to the $n$ insertions of $\mca{V}$, and in the absence of selection rule it is the same as in eq. \eqref{gepscaling}. The spectral density would then be bounded by a function with the following asymptotics:
\beq
|\bDelta \rho_{i,n}^\eps(\vec{\a})| \lesssim |\vec{\a}|^{1-\Delta_*}~.
\eeq
Of course, this equation is understood in an averaged sense, and a detailed computation would make use of the integrated density analogous to eq. \eqref{Raver}. Convergence of eq. \eqref{nthOrderInt} is then guaranteed in the limit under scrutiny as long as $\Delta_*>1$, which is enforced by the spectral condition \eqref{UVfiniteCond}.

Let us finally point out that, running the previous argument for each connected correlator in eq. \eqref{gepsn}, the presence of the identity in the OPE implies that each individual energy level diverges linearly with $\Lambda$ at all orders in perturbation theory.

\subsection{Degenerate spectra}
\label{sec:degen}

The main formula presented in~\reef{prescription} applies to \emph{any} AdS Hamiltonian regulated by a hard cutoff. However, the argument presented in section~\ref{subsec:second} and~\ref{sec:allorders} is based on the fact that in perturbation theory, energies $\mca{E}_i$ are closely related to integrated correlation functions. To be precise, the $n$-th order contribution of the energy $\mca{E}_i$ is related to the $n$-point connected correlator of $V(\tau)$ in the state $\ket{i}$, as is shown in appendix~\ref{app:RSPT}. This relation does not hold in {degenerate} perturbation theory, as we will briefly discuss here. Physically, theories with degenerate spectra are particularly important for Hamiltonian truncation in AdS${}_2$, hence they deserve special attention. In 2d BCFTs degeneracies arise because of the very nature of Virasoro modules, and in theories of free bosons or fermions, degeneracies naturally arise from multi-particle states in Fock space.

For concreteness, we have in mind a unitary quantum theory where $g \geq 2$ states have the same unperturbed energy $e_{*}$. There are three different scenarios for the spectrum at non-zero coupling $\la$:
\begin{itemize}
\item[(i)] the $g$-fold degeneracy is completely lifted at first order in perturbation theory;
\item[(ii)] the degeneracy is only fully lifted at some order $n \geq 2$ in perturbation theory;
\item[(iii)] the degeneracy is exact and present for any value of $\la$.
\end{itemize}
Let us focus on the first scenario, since it applies generically.\footnote{Case (ii) can happen when $V$ explicitly breaks a symmetry of the unperturbed Hamiltonian, and (iii) is the hallmark of an unbroken symmetry/integrability, e.g.\@ a boson or fermion perturbed by a mass term $\phi^2$ or $\bar{\psi}\psi$.} It is well-known that the appropriate eigenstates are eigenvectors $\ket{\a}$ of $V$ restricted to the degenerate subspace $\msc{D}$, that is to say states satisfying
\beq
\label{eq:keydegenerate}
H_0 \ket{\a} = e_{*} \ket{\a}
\qaq
\braket{\a}{V}{\b} = v_\a \dd_{\a\b}
\eeq
and the requirement that the degeneracy is lifted at first order means that all $v_\a \in \mbb{R}$ must be different.  The energies $\mca{E}_\a$ are then determined unambiguously, taking the form
\beq
\mca{E}_\a = e_* + \bar\la v_\a - {\bar\la}^2 e^{(2)}_\a + {\bar\la}^3 e^{(3)}_\a + \ldots
\eeq
for some coefficients $e^{(n)}_\a$ that can be determined using Rayleigh-Schr\"{o}dinger perturbation theory. Using the shorthand notation $V_{\phi \psi} = \braket{\phi}{V}{\psi}$, the first two coefficients can be expressed as
\bsub
\beq
\label{eq:deg2}
e_{\a}^{(2)} =  V_{\a \psi}\, \frac{1}{e_\psi - e_{*}} \, V_{\psi \a}
\eeq
and
\begin{multline}
\label{eq:deg3}
e_{\a}^{(3)} =  V_{\a\psi} \, \frac{1}{e_\psi - e_{*}}\, V_{\psi \phi}\, \frac{1}{e_\phi - e_{*}} \, V_{\phi \a} - v_\a \cdot V_{\a \psi}\, \frac{1}{(e_\psi - e_{*})^2} \, V_{\psi \a}  \\
- {\Red V_{\a \psi} \, \frac{1}{e_\psi - e_{*}} \, V_{\psi \b} \, \frac{1}{v_{\b} - v_\a} \, V_{\b \phi} \, \frac{1}{e_\phi - e_{*}} \, V_{\phi \a}}\,.
\end{multline}
\esub
In both~\reef{eq:deg2} and~\reef{eq:deg3} sums over intermediate states are implicit: sums over $\psi,\phi$ run over all states outside of $\msc{D}$, whereas the sum over $\beta$ runs over all states in $\msc{D}$ except the state $\ket{\a}$ itself. As a matter of fact, the second-order energy $e_\a^{(2)}$ can be extracted from the connected correlation function
\beq
g_{\a,2}(\tau) = \braket{\a}{V(\tau)V(0)}{\a} - \braket{\a}{V}{\a}^2
\eeq
exactly like in the case of non-degenerate quantum mechanics. However, the term in {\Red red} appearing in the second line of the third-order energy~\reef{eq:deg3} is  qualitatively different. For one, it features four insertions of the matrix $V$, hence $e_\a^{(3)}$ is at best related to a four-point function of $V(\tau)$ inside the state $\ket{\a}$. More importantly, one of the factors appearing on the second line of~\reef{eq:deg3} has $1/(v_\b - v_\a)$ as a denominator. Such a denominator cannot arise from a correlation function of $V(\tau)$ in the interaction picture, obeying $\pd_\tau V(\tau) = [H_0,V(\tau)]$. We conclude that in general, the $e_\a^{(n)}$ cannot be expressed in terms of connected correlators of $V(\tau)$.

Nevertheless, it appears that the prescription~\reef{prescription} correctly predicts the spectra of AdS theories with degenerate unperturbed spectra. It is not difficult to check the prescription up to low orders in perturbation theory (for example in the case of a massive scalar with integer $\DD$).\footnote{For generic $\DD$, the degeneracies present in the spectrum spectrum of the free massive scalar theory persist when a $\phi^2$ interaction is turned on. When $\DD$ is an integer there are additional degeneracies that are lifted by a $\phi^2$ deformation.} In section~\ref{subsec:IsingT}, the $\eps$ deformation of the $2d$ Ising model is studied; it is again degenerate at zero coupling, but its degeneracies are partially resolved at finite coupling. This RG flow is studied using Hamiltonian truncation, and the results agree well with analytic predictions (up to truncation errors), see figure~\ref{fig:TIsingSpectrum}. The other Virasoro theories featured in this work also have denegeracies, but since they aren't exactly solvable, one cannot explicitly check the prescription~\reef{prescription} \emph{a posteriori}.

In summary, there is no reason to doubt that the prescription~\reef{prescription} predicts correct spectra for the case of degenerate AdS spectra, but we have not found a simple modification of the proof from the non-degenerate case. We leave this problem open for future work.\footnote{In the presence of degeneracies, it is tantalizing to split the Hamiltonian $H = H_0 + \lambda V$ into two pieces as $H = \tilde{H}_0 + \lambda \tilde{V}$, where $\tilde{H}_0 = H_0 + \lambda \mrm{P} V \mrm{P}$ and $\tilde{V} = V - \mrm{P} V \mrm{P}$, where $\mrm{P}$ projects onto a degenerate subspace. The new unperturbed Hamiltonian $\tilde{H}_0$ is now non-degenerate when $\la \neq 0$. One can try to rederive the prescription by treating $\tilde{V}$ as a perturbation instead of $V$.}

\subsection{Truncation errors}
\label{sec:rate}

In the previous sections, we have discussed in detail a procedure that is needed to study AdS Hamiltonians using a hard energy cutoff. In particular, we showed that the standard TCSA prescription --- fixing a cutoff $\Lambda$ and subtracting the Casimir energy --- predicts wrong energy spectra.
Given the correct prescription~\reef{prescription}, Hamiltonian truncation in AdS is still an approximation, since one necessarily works at some finite cutoff $\Lambda$. In this section, we estimate the error on the energy levels due to this truncation, comparing the situation in AdS to the better-understood problem of Hamiltonian truncation on finite volume spaces. 

The basic result we will work towards is closely related to the discussion from subsection~\ref{subsec:second} and its generalization to higher orders in perturbation theory from subsection~\ref{sec:allorders}. Concretely, let's consider the second-order contribution of some interaction $\bar{\lambda} V$ to the $i$-th energy level, working at cutoff $\Lambda$:
\bsub
\begin{align}
\label{eq:Eian}
\bDelta \mca{E}_i(\Lambda)& \ldef \mca{E}_i(\Lambda) - \mca{E}_\text{vac}(\Lambda - e_i) \nn \\
&= e_i + \bar\lambda \braket{i}{V}{i} + \sum_{n=2}^\infty (-1)^{n+1}  \bar\lambda^n \bDelta c_{i,n}(\Lambda)~,
\end{align}
where the first cutoff-dependent term is given by
\beq
\bDelta c_{i,2}(\Lambda) = \int^{\Lambda}\!\frac{d\a}{\a-e_i}\, \bDelta \rho_{i,2}(\a-e_i)~.
\eeq
\esub
Here, we use the definition~\reef{Dspec} for the shifted spectral density $\bDelta \rho_{i,2}(\a)$.  Likewise, the higher-order coefficients $\bDelta c_{i,n}(\La)$ can be expressed in terms of spectral densities $\bDelta \rho_{i,n}(\vec{\a})$. In contrast to the previous sections, here we no longer consider a spatial cutoff $\eps$, working in full AdS${}_2$ with $\eps = 0$ throughout. In an otherwise UV-finite theory, the energy gap $\bDelta \mca{E}_i(\Lambda)$ has a finite limit as $\Lambda \to \infty$, so the cutoff error is given by tails of spectral integrals:
\beq
\label{eq:cutofferr}
\text{cutoff error} = \bDelta \mca{E}_i(\infty) - \bDelta \mca{E}_i(\Lambda) = - \bar\lambda^2  \int^{\infty}_\Lambda\!\frac{d\a}{\a-e_i}\, \bDelta \rho_{i,2}(\a-e_i) + \sO(\bar\lambda^3).
\eeq
If we can bound integrals like~\reef{eq:cutofferr}, we immediately obtain the desired error estimates for energy gaps in AdS${}_2$. But this is precisely what we did in subsection \ref{subsec:second}, using the crucial eq. \eqref{boundAsymp}. Recall that this equation follows from considering the leading non-trivial contribution to the OPE of figure \ref{fig:multipleOPEsketch}, coming from a boundary operator $\Psi(\tau)$ of dimension $\DD_{*}$. We should expect the bound to be saturated, generically, and therefore we obtain an error of the order
\beq
\label{eq:bdyerrors}
\bDelta \mca{E}_i(\infty) - \bDelta \mca{E}_i(\Lambda) \; \limu{\Lambda \to \infty} \; \frac{c_i}{\Lambda^{\DD_{*}-1}}~.
\eeq
In light of the discussion in subsection \ref{sec:allorders}, we expect eq. \eqref{eq:bdyerrors} to be true at all orders in $\bar{\lambda}$. 
The state-dependent coefficient $c_i(\bar{\lambda})$ encodes the large-$\a$ asymptotics of the densities $\bDelta \rho_{i,n}(\a)$. For instance, eq. \eqref{eq:chiError} gives $c_1$, the coefficient of the first excited state for a free massive boson, at second order in perturbation theory.  More generally, $c_i$ can be determined in perturbation theory following the computations of appendix \ref{sec:twopgen} -- see in particular formula \reef{VVbOPE}. For later reference, we mention that $c_i$ is proportional to the boundary OPE coefficient $C_{\Oo_i \Oo_i \Psi}$ (up to a number that depends on $\mca{V}$ and $\Psi$ but not on the state $\ket{i}$), where $\Oo_i(\tau)$ is the boundary operator corresponding to the state $\ket{i}$.

Now, by assumption there are no boundary states of dimension $\leq 1$ --- otherwise, the Hamiltonian is IR-divergent to begin with --- hence error terms of the form~\reef{eq:bdyerrors} always vanish as the cutoff is removed, but when the exponent $\DD_{*} - 1$ is small, they can be important. In particular, the identity module in Virasoro minimal models in AdS${}_2$ contains a state with $\DD_{*} = 2$ that describes the displacement operator, leading to a $1/\Lambda$ error. For theories of massive particles in the bulk of AdS, the dimension $\DD_{*}$ of the leading boundary state depends on the particle's mass (in units of the AdS radius): the heavier the bulk particle is taken to be, the smaller the error~\reef{eq:bdyerrors} will be.

We stress that eq.~\reef{eq:bdyerrors} is true asymptotically, that is to say for $\Lambda$ bigger than some cutoff scale $\Lambda_0$ which is \emph{a priori} unknown. In particular, if Hamiltonian truncation computations only probe cutoffs below $\Lambda_{0}$, the formula in question does not necessarily predict the order of magnitude of actual cutoff errors. Nonetheless, the formula~\reef{eq:bdyerrors} is consistent with all of the computations performed in the present work.

It is natural to ask whether the error~\reef{eq:bdyerrors} can be removed by adding so-called \emph{improvement terms} to the Hamiltonian. These would be proportional to either boundary or (integrated) bulk operators, with cutoff-dependent coefficients that vanish as $\La \to \infty$, and their role would be precisely to cancel large truncation errors in observables. Indeed, since $c_i$ only depends on the state $\ket{i}$ through the OPE coefficient $C_{\Oo_i \Oo_i \Psi}$, it should be possible to remove the error~\reef{eq:bdyerrors} by adding a term to the Hamiltonian of the form
\beq
\label{eq:bdyimprovement}
H \mapsto H + \frac{\xi_\Psi \bar{\lambda}^2}{\Lambda^{\DD_{*} - 1}} \Psi(\tau = 0) + \ldots
\eeq
for some computable coefficient $\xi_\Psi$. As follows from eq.~\reef{VVbOPE}, the coefficient $\xi_\Psi$ is theory-dependent: it's related to the integrated bulk-bulk-boundary correlator $\expec{\mca{V} \mca{V} \Psi}$, which is not fixed by conformal symmetry alone. We have not undertaken this exercise in the present work, but it looks like an important step to take with the goal of performing high-precision Hamiltonian truncation computations in the future, especially in the case of Virasoro minimal models.

As we emphasized, eq. \eqref{eq:bdyerrors} captures the contribution of the bundary of AdS to the truncation error. However, the large $\alpha$ behavior of the spectral density is also influenced by bulk contributions. Equivalently, the bulk OPE contributes non-analytic terms in $\tau$ to the correlation functions \eqref{densityDef}. Let us focus again on the second order in perturbation theory. The singularities in question are in one-to-one correspondence with operators appearing in the $\mca{V} \times \mca{V}$ OPE (taking the UV theory to be conformal for concreteness). Indeed, suppose that $\mca{V} \times \mca{V}$ contains an operator $\tilde{\mca{V}}$ of dimension $\DD_{\tilde{\mca{V}}}$. Then, as discussed in appendix \ref{sec:twopgen} -- see in particular eq. \eqref{GepsBulk} -- the subtracted two-point function of the potential is expected to receive at small $\tau$ a contribution of the following kind:
\beq
\braket{i}{V(\tau)V(0)}{i}_\text{conn} - \braket{\Omega}{V(\tau)V(0)}{\Omega}_\text{conn}
\limu{\tau \to 0}  \tau^{1-2\DD_\mca{V} + \DD_{\tilde{\mca{V}}}}~.
\label{VVrateBulk}
\eeq
Such terms have nothing to do with the geometry of AdS: they are also present if one  quantizes the same theory on the cylinder $\mbb{R} \times S^{1}$, see for example~\cite{Hogervorst:2014rta}. When eq. \eqref{VVrateBulk} gives the leading small $\tau$ behavior, the energy gaps have errors that go as
\beq
\label{eq:bulkerrors}
\bDelta \mca{E}_i(\infty) - \bDelta \mca{E}_i(\Lambda) \; \limu{\Lambda \to \infty} \; \frac{d_i}{\Lambda^h},
\quad
h = \DD_{\tilde{\mca{V}}} - 2\DD_\mca{V} + 2
\eeq
for some coefficient $d_i$ which is proportional to the matrix element $\braket{i}{\tilde{\mca{V}}(\tau = 0,r)}{i}$, integrated over a timeslice. Here the exponent $h$ only depends on the UV properties of the theory: it does not depend on the choice of boundary conditions, for instance. By definition, $h > 0$ in a UV-finite theory, as is the case for all theories discussed in the present work, so errors of the above type vanish as the cutoff is removed. 
Nevertheless, a word of caution is in order about eq. \eqref{VVrateBulk}: boundary effects might modify it when the integral of $\braket{i}{\tilde{\mca{V}}}{i}$ over a time slice diverges. In appendix \ref{sec:twopgen}, the precise conditions are discussed. 
In the rest of the work, the bulk OPE will cause the leading truncation error only for the $\phi^4$ deformation of a free massive scalar, see subsection \ref{subsec:phi4conv}. In that case, this boundary enhancement does not take place.\footnote{However, when $h$ is integer, $h=2$ in this case, the scaling~\reef{eq:bulkerrors} is typically enhanced by logarithms, and in fact we will find the error to go as $\ln(\Lambda)/\Lambda^2$.}

Truncation errors coming from bulk singularities can also be removed by adding local counterterms to the Hamiltonian, i.e.\@ by modifying $H$ as follows:
\beq
\label{eq:improvement}
H \mapsto H +  \frac{c_{\tilde{\mca{V}}} \bar\lambda^2}{\Lambda^h} \int_{-\pi/2}^{\pi/2}\frac{dr}{(\cos r)^2}\, \tilde{\mca{V}}(\tau=0,r) + \ldots
\eeq
for some computable coefficient $c_{\tilde{\mca{V}}}$. For CFTs quantized on $\mbb{R} \times S^{d-1}$, these counterterms were discussed in detail in~\cite{Hogervorst:2014rta} (and in the $d=2$ case before in~\cite{Giokas:2011ix} and~\cite{Watts:2011cr}). A detailed discussion for the massive scalar on $\mbb{R} \times S^1$ is presented in~\cite{Rychkov:2014eea} and related works~\cite{Rychkov:2015vap,Elias-Miro:2015bqk,Elias-Miro:2017tup}. In this paper, we will not actually add improvement terms of the form~\reef{eq:improvement}, so we will not discuss Hamiltonians of the above form in detail. The computation of the coefficient $c_{\tilde{\mca{V}}}$ would go along the same lines as computations from the literature, apart from an integration over an (infinite-volume) timeslice of AdS${}_2$, contrasting with the computation on $S^1$ from~\cite{Rychkov:2015vap}.\footnote{The computation on $S^1$ is simpler for an additional reason, namely the $U(1)$ global symmetry that acts by translations on the spatial circle, which is absent in AdS.}

\section{Deformations of a free massive scalar}
\label{sec:massive}
%!TEX root = ../HTinAdS.tex
%%%%%%%%%%%%%%%%%%%%%%%%%%%%%%%%%%%%%%%%%%%%%

After the general discussion of Hamiltonian truncation in the previous sections, we will now turn to a concrete quantum field theory: a massive scalar $\phi$, described by the action:
\beq
\label{eq:ScalarAction}
S=\int d^2x \sqrt{g} \left(\half(\pd_{\mu}\phi)^2 + \half m^2 \phi^2 +\frac{\kappa}{2} \mbf{R} \phi^2  + \text{local interactions}\right)~,
\eeq
where $\mbf{R} = -2/R^2$ is the scalar curvature of \ads2. The curvature coupling $\kappa$ is arbitrary, but since the curvature is constant, any shift in $\kappa$ can be reabsorbed into the definition of $m^2$, and henceforth we will set $\kappa = 0$. The free theory was briefly treated in section~\ref{sec:exNew}; here we will discuss the quantization in slightly more detail. We recall that $\phi(\tau,r)$ has a mode decomposition 
\beq
\phi(\tau = 0,r) = \sum_{n=0}^\infty f_n(r) (a_n^\phd  + a_n^\dag)
\eeq
with creation and annihilation operators that obey canonical commutation relations
\beq
[a_m^\phd,a_n^\dag] = \dd_{mn}~.
\eeq
The mode functions $f_n(r)$ are given by
\beq
\label{eq:fnDef}
f_n(r) =  \sqrt{\frac{4^\DD}{4\pi}\frac{\Gamma^2(\DD)}{\Gamma(2\DD)}}\,  \sqrt{\frac{n!}{(2\DD)_n}}\,  (\cos r)^{\DD} \, C^{{\DD}}_n(\sin r)~.
\eeq
Here $\DD$ is a positive root of $\DD(\DD-1) = m^2 R^2$ and $\dps{C_n^{\nu}(\cdot)}$ denotes a Gegenbauer polynomial. 
The free Hamiltonian $H_0$ is given in~\reef{eq:H0massive}, and from it we can deduce that the $n$-th mode $a_n^\dag$ has energy $\DD+n$.

As a consistency check of the above, we find that the Green's function of the field $\phi$ is given by
\beq
\label{eq:scalar2pt}
\braket{\Omega}{\mrm{T}\phi(\tau,r)\phi(\tau',r')}{\Omega} =  \frac{1}{4\pi} \frac{\Gamma^2(\DD)}{\Gamma(2\DD)}\, \xi^{-\DD}\, {}_2F_1\!\left[{{\DD,\DD}~\atop~{2\DD}} \Big| - \frac{1}{\xi}\right] \rdef G(\xi)
\eeq
where $\xi$ is the cross ratio from Eq.~\reef{eq:crossratio}. This result could also have been derived directly, by looking for ${SL}(2,\mbb{R})$-invariant solutions of the equation of motion with the correct boundary conditions.

In what follows, we will analyze the scalar theory in AdS${}_2$ with the following interaction terms turned on:
\beq
\label{eq:Smassdef}
S = S_0 + \int_{\mrm{AdS}_2}\!( \la_2  \phi^2 + \la_4  \phi^4 + \ldots)
\eeq
where the $\ldots$ can denote any local bulk interaction (not necessarily $\mbb{Z}_2$-even, and possibly containing derivatives). In the Hamiltonian language, these interactions read
\beq
\label{eq:LGham}
H = H_0 + \bar{\la}_2 V_2 + \bar{\la}_4 V_4 + \ldots
\eeq
where\footnote{Here normal ordering simply means moving annihilation operators to the right, e.g.\@ $\NO{a_1^\dag a_1 a_2^\dag} = a_2^\dag a_1^\dag a_1$.}
\beq
\bar{\la}_n \ldef \la_n R^2
\qaq
V_{n} \ldef \int\!\frac{dr}{(\cos r)^2} \, \NO{\phi^n(\tau=0,r)}.
\eeq
The operators $H$, $V_n$ etc.\@ act on the Hilbert space of the free theory, which is the Fock space generated by the modes $\{a^\dag_n\}_{n=0}^\infty$. Its states can be labeled as follows
\beq
\label{eq:ScalarBasis}
\ket{\psi} = 
\frac{1}{\mathcal{N}_{\ket{\psi}}} \prod_{k \geq 0} (a_k^\dagger)^{n_k} \ket{\Omega},
\qquad
\mca{N}_{\ket{\psi}} = \sqrt{{\prod}_{k \geq 0} n_k!}\,.
\eeq
The factor $\mathcal{N}_{\ket{\psi}}$ has been chosen to make such a state unit-normalized. A state of the form~\reef{eq:ScalarBasis} has energy
\beq 
H_0 \ket{\psi} = e_{\ket{\psi}} \ket{\psi},
\quad
e_{\ket{\psi}}=\sum_{k \geq 0} n_k(\Delta+k)~.
\eeq
In Hamiltonian truncation, we keep only states with energy $e \leq \La$ below a cutoff $\La$. Since the spectrum of $H_0$ is discrete, this means that $H$ becomes a finite matrix, denoted by $H(\La)$. If we let $\{\ket{\psi_i}\}$ be a basis of low-energy states, then the matrix elements of $H(\La)$ have the following form:
\beq
H(\La)_{ij} = \braket{\psi_i}{H_0}{\psi_j} + \bar{\la}_2 \braket{\psi_i}{V_2}{\psi_j} + \ldots =  e_{\ket{\psi_i}}\dd_{ij} + \bar{\la}_2 \braket{\psi_i}{V_2}{\psi_j} + \ldots
\eeq
Finally, the spectrum of $H$ can be approximated by explicitly diagonalizing $H(\La)$ for sufficiently large values of $\La$. We will explore this for both a $\phi^2$ and a $\phi^4$ deformation in the next sections.

\subsection{$\phi^2$ deformation}\label{sec:phi2}
In this section we study the $\phi^2$ deformation, meaning that in~\reef{eq:Smassdef} we allow for an arbitrary value of $\la = \la_2$, but all other couplings are set to zero. Since this theory is exactly solvable, it provides a way test the proposed diagonalization procedure in a controlled setting. To be precise, the spectrum of the theory with $\la \neq 0$ is that of a free theory with redefined mass $m^2 \to m^2 + 2\la$, or equivalently
\beq
\label{eq:DeltaRedefined}
\DD \to \DD(\bar{\lambda}) =  \half +  \sqrt{(\DD-\th)^2+ 2\bar{\lambda}}~.
\eeq
The spectrum of the theory with coupling $\la \neq 0$ consists of single-particle states with energy $\DD({\bar{\la}} )+ \mbb{N}_0$ --  $\mbb{N}_0$ being the set of non-negative integers -- two-particle states with energies $2 \DD({\bar{\la}})+ \mbb{N}_0$ (most of which are degenerate), and likewise there are $n$-particle states of energy $n \DD_{\bar{\la}} + \mbb{N}_0$ for any integer $n$. The energy $\DD({\bar{\la}})$ can be computed in perturbation theory; expanding~\reef{eq:DeltaRedefined} in a Taylor series around $\la = 0$, we find that Rayleigh-Schr\"{o}dinger perturbation theory converges if and only if
\beq
\label{eq:lambdaStar}
|\bar{\la}| <  \la_\star(\DD) = \half\!\left(\DD-\half\right)^2.
\eeq
For couplings larger than $\la_\star(\DD)$, the correct spectrum can only be computed nonperturbatively.

In order to set up the Hamiltonian truncation,
it will be convenient to express $V_2$ in terms of creation and annihilation operators
\beq
\label{eq:V2def}
  V_2 = \sum_{m,n = 0}^\infty A_{mn}(\DD) (a_m^\dag a_n^\dag + 2a_m^\dag a_n^\phd + a_m^\phd a_n^\phd)~,
\eeq
with coefficients
\beq
\label{eq:AmatP}
A_{mn}(\DD) = \int_{-\pi/2}^{\pi/2} \! \frac{dr}{\cos^2 r}\,  f_m(r) f_n(r) = \frac{\msc{V}_{mn}(\DD)}{2\DD - 1}~,
\eeq
where the matrix $\msc{V}_{mn}$ is computed in appendix \ref{subsec:constraintsEven}, and we report here the result for convenience:
\beq
\label{eq:VmatMain}
\msc{V}_{mn}(\DD) = \begin{cases} 0 & \text{if } m + n \text{ is odd}; \\
  \sqrt{(m+1)_{n-m}/(2\DD+m)_{n-m}} &  m \leq n,\; m+n \text{ even}; \\
  \msc{V}_{nm}(\DD) & m > n.
\end{cases}
\eeq
An algorithm that can be used to compute and diagonalize $H(\La)$ is described in great detail in section~\ref{app:algo}. In what follows, we will simply discuss the results of various numerical computations. In particular, we will check whether the spectrum of $H(\La)$ reproduces the exact spectrum predicted by Eq.~\reef{eq:DeltaRedefined}. 

Before turning to the numerical results, let us comment on the role of discrete symmetries. Both the $\phi^2$ and $\phi^4$ interactions are invariant under parity $\mtt{P}$ and the $\mathbb{Z}_2$ symmetry $\mca{Z} : \phi \to -\phi$. The individual creation operators $a_k^\dag$ have quantum numbers $(-1)^k$ under parity and $-1$ under $\mca{Z}$.\footnote{For the $V_2$ interaction, parity invariance follows from the fact that the coefficients $A_{mn}(\DD)$ from~\reef{eq:AmatP} vanish if $m+n$ is odd.}  Therefore, a Fock space state $\ket{\psi}= \ket{n_0,n_1,n_2,\ldots}$ has quantum numbers
\beq
\mtt{P} \ket{\psi} = (-1)^{\pi_\psi} \ket{\psi},
\quad
\mca{Z} \ket{\psi} = (-1)^{z_\psi} \ket{\psi},
\quad
\text{for}
\quad
\pi_\psi = \sum_{k \geq 0} k n_k,
\quad
z_\psi = \sum_{k \geq 0} n_k\,.
\eeq
Given these symmetries, the diagonalization of $H(\La)$ can be performed independently in the four different sectors of Hilbert space that contain states with quantum numbers $\mtt{P} = \pm 1$ and $\mca{Z} = \pm 1$. Notice that the Casimir energy is determined by the vacuum state which has quantum numbers $\mtt{P} = \mca{Z} = +1$, so this sector plays a special role. For fixed $\La$ and fixed couplings $\la_2,\ldots$, eigenvalues of $H(\La)$ corresponding to \emph{different} symmetry sectors are expected to cross. However, we stress that the physical energies are not quite the eigenvalues of $H(\La)$: the Casimir energy needs to be subtracted from the individual levels, in accordance to the discussion in section~\ref{sec:cutoff}.

\subsubsection{Cutoff effects}\label{sec:ph2conv}
\begin{figure}[htb]
     \centering
     \begin{subfigure}[htb]{0.45\textwidth}
         \centering
         \includegraphics[width=\textwidth]{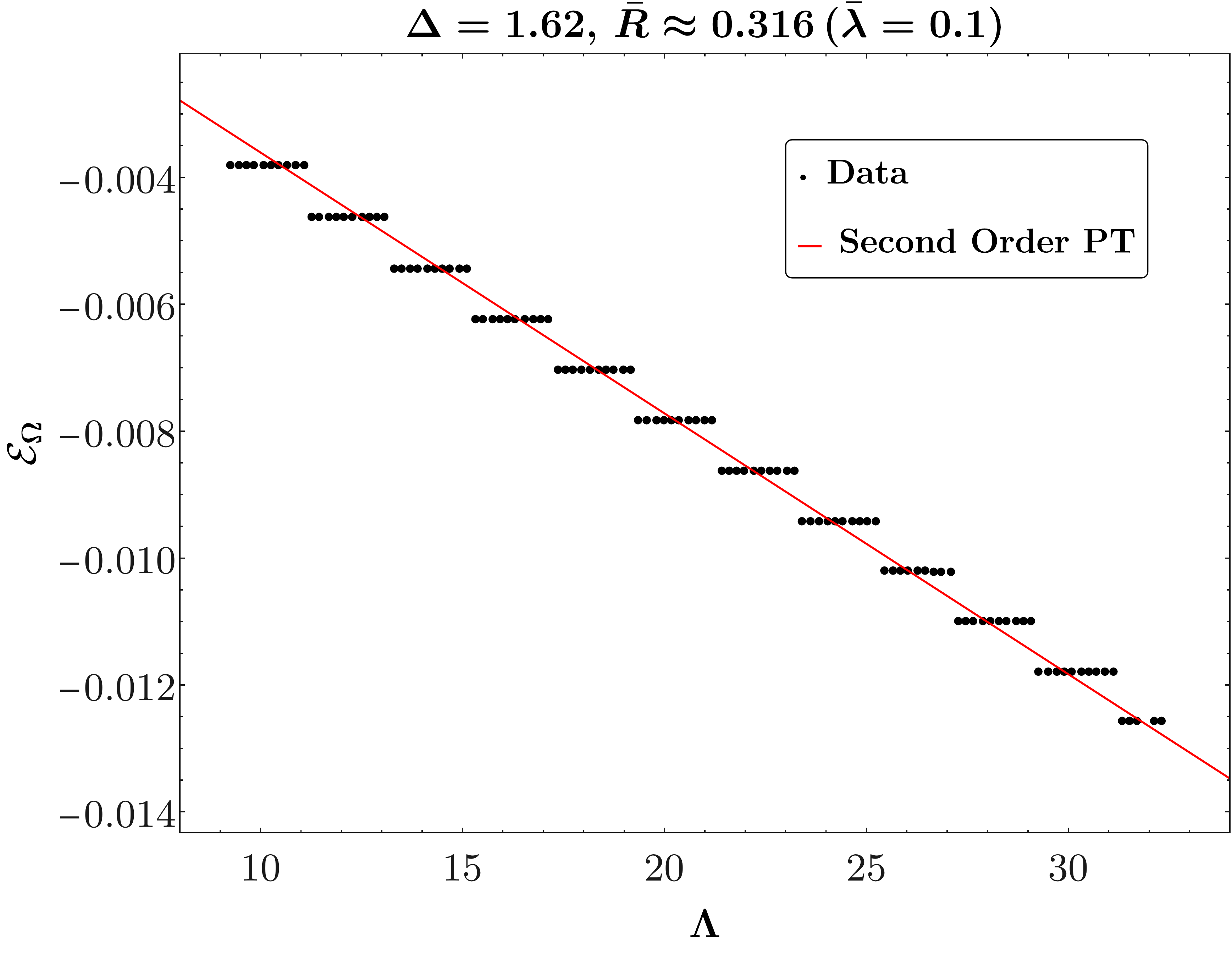}
     \end{subfigure}
     \hfill
     \begin{subfigure}[htb]{0.45\textwidth}
         \centering
         \includegraphics[width=\textwidth]{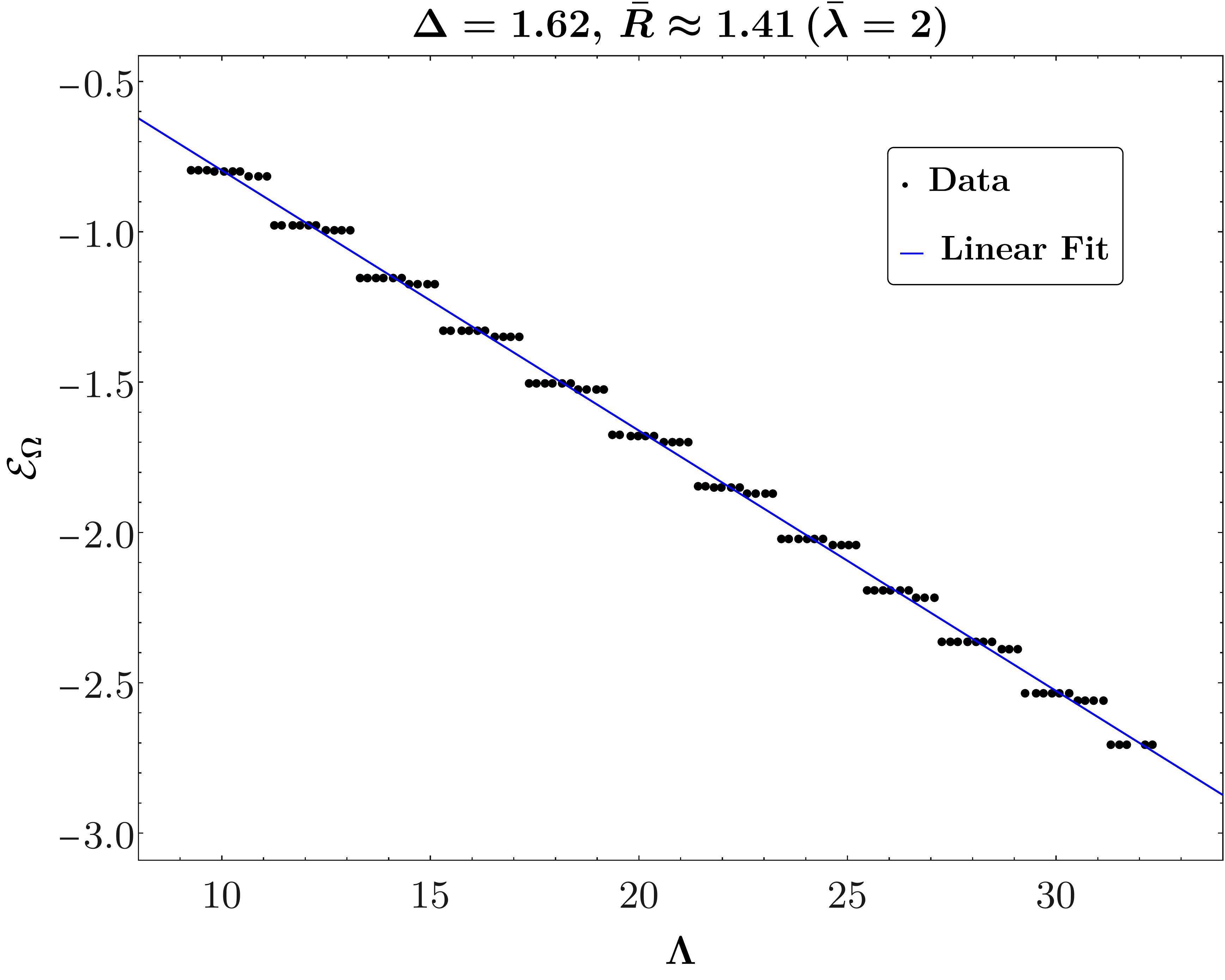}
     \end{subfigure}
             \caption{The vacuum energy linear growth in cutoff. The left plot shows the agreement of the data with second order perturbation theory. The right plot shows the linear growth beyond perturbation theory ($\bar{\lambda}=2$). $\bar{R}$ is the radius of AdS in units of the coupling, as defined in eq. \eqref{RbarDef}.}
                  \label{fig:VacuumEnergyphi2}
     \end{figure}

As mentioned in section~\ref{sec:exNew}, one expects that the vacuum energy $\mca{E}_{\Omega}(\La)$ grows linearly with the cutoff $\Lambda$. We can check this both perturbatively (by means of a Rayleigh-Schr\"{o}dinger computation at second order in $\la^2$) and non-perturbatively, using Hamiltonian truncation. Details for the perturbative computation are given in appendix~\ref{sec:ExplicitSecondOrder}. 
The result of both computations is shown in Figure~\ref{fig:VacuumEnergyphi2}. For concreteness, we take the single-particle energy in the free theory to be $\DD = 1.62$.\footnote{This corresponds to a mass $m^2 R^2 = \DD(\DD-1) \approx 1$, so a mass $\approx 1$ in units of the AdS radius.} For the nonperturbative plot on the right hand side we have set $\bar{\la} = 2$, beyond the radius of convergence of perturbation theory~\reef{eq:lambdaStar}. By eye, the linear growth of $\mca{E}_{}(\La) \limu{\La \to \infty} \La$ can easily be seen in both cases.

Next, we can study the energy shift of the first excited state $\ket{\chi} = a_0^\dag \ket{\Omega}$. From~\reef{eq:DeltaRedefined}, we see that
\beq
\label{eq:enDShift}
\DD_\chi(\la)  = \DD  + \frac{\bar{\la}}{\DD - \th} - c_\chi^{}  \bar{\la}^2 + \sO(\bar{\la}^3),
\qquad
c_\chi^{} = \frac{4}{(2\DD-1)^3}~.
\eeq
We can first of all reproduce this result analytically, using Rayleigh-Schr\"{o}dinger perturbation theory. This involves computing the difference $g_\chi(\tau)$ of the connected two-point correlators of $V_2$ inside the state $\ket{\chi}$ and the vacuum. For details, we refer to Appendix~\ref{app:RSPT}; the exact function $g_\chi(\tau)$ is spelled out in eq.~\reef{eq:GchiNiceResult}. Integrating this correlator over $\tau$ reproduces the exact result, as it should. 
At small $\tau$, the function $g_\chi(\tau)$ behaves as
\beq
\label{eq:H2an}
g_\chi(\tau) \; \limu{\tau \to 0} \; - \frac{2\pi}{(\DD-\th)\sin(2\pi \DD)} \; \tau^{2\DD-2} + \text{analytic} + \sO(\tau^{2\DD-1}).
\eeq
While a logarithmic singularity appears for integer $\Delta$, here we will assume for simplicity that $\DD$ is generic (and $\DD > 1$). The expression~\reef{eq:H2an} can be used to deduce the large-$\La$ behavior of Hamiltonian truncation. Indeed, the Laplace transform of $g_\chi(\tau)$ must behave as
\beq
\label{eq:Gchi}
\bDelta \rho_\chi (\alpha)
\limu{\a \to \infty} \frac{\Gamma(2\DD)}{(\DD-\th)^2}\; \a^{1-2\DD},
\quad
g_\chi(\tau) = \int_0^\infty\!d\a \, e^{-(\a-\DD)\tau} \, \bDelta \rho_\chi (\alpha)
\eeq
so the cutoff error can be estimated to be
\beq
\label{eq:chiError}
\bar{\la}^2 \int_{\La}^\infty\!\frac{d\a}{\a-\DD}\,\bDelta \rho_\chi (\alpha) \limu{\La \to \infty} \frac{\Gamma(2\DD-1)}{(\DD-\th)^2} \frac{ \bar{\la}^2}{\La^{2\DD-1}}.
\eeq
This convergence rate is in complete agreement with the discussion from section~\ref{sec:rate}, since the lowest-dimension boundary state that can be generated is the two-particle state  with dimension $2\DD$. 
We therefore predict that at finite cutoff $\La$, we measure an energy gap
\begin{equation}
\begin{split}
\bDelta\mca{E}_\chi &= e_\chi + \frac{\bar{\la}}{\DD-\th}  - \bar{\la}^2 \left( c_\chi -  \frac{\Gamma(2\DD-1)}{(\DD-\th)^2} \frac{1}{\La^{2\DD-1}} +\dots \right) + \sO(\bar{\la}^3) \\
&\approx 1.62 + 0.893 \bar{\la} - \bar{\la}^2 \left( 0.356 - \frac{0.898}{\La^{2.24}} +\dots \right)+ \sO(\bar{\la}^3) 
\end{split}
\label{eq:chiPrediction}
\end{equation}
plugging in $\DD = 1.62$ when passing from the first to the second line.

\begin{figure}[htb]
     \begin{subfigure}[htb]{0.45\textwidth}
         \centering
         \includegraphics[width=\textwidth]{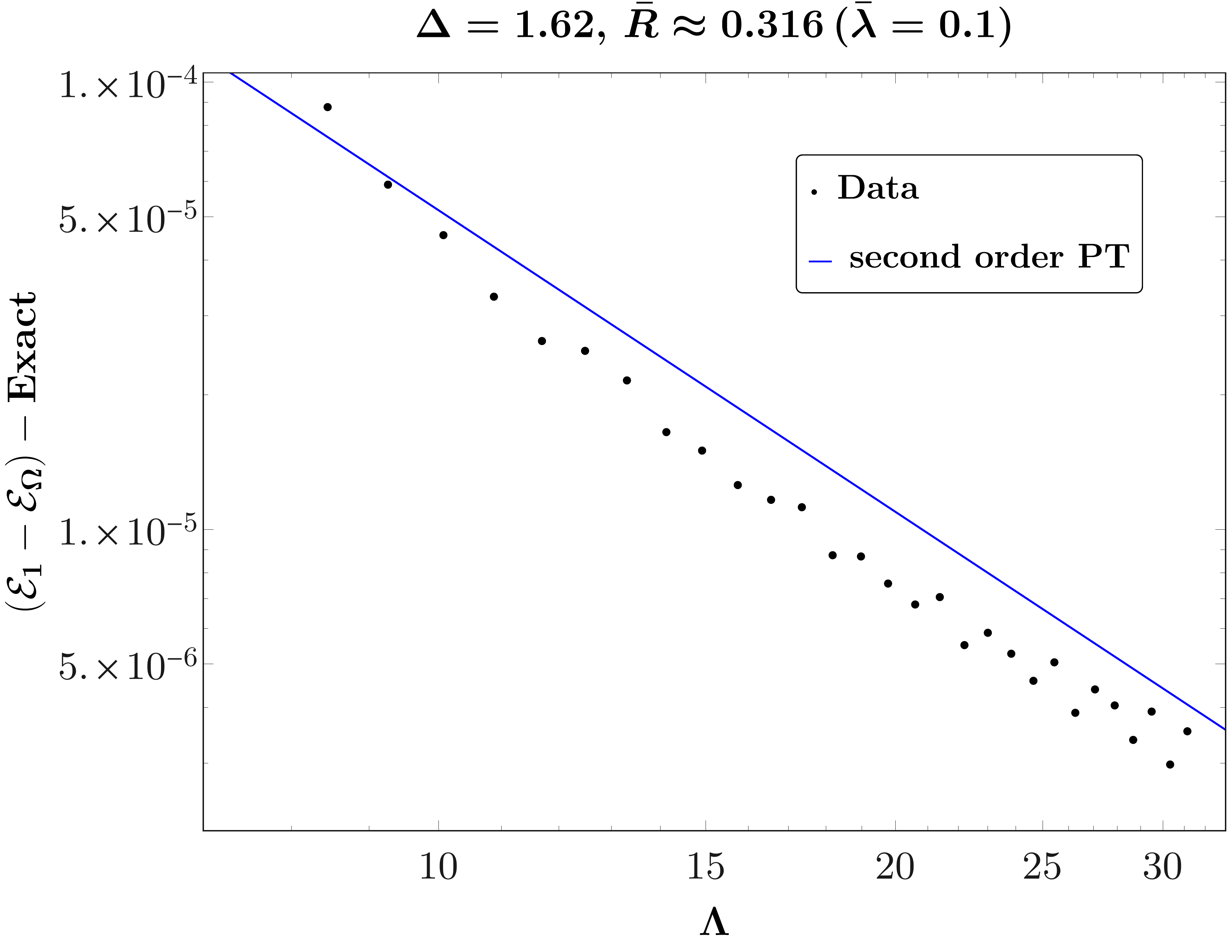}
     \end{subfigure}
     \hfill
     \begin{subfigure}[htb]{0.45\textwidth}
         \centering
         \includegraphics[width=\textwidth]{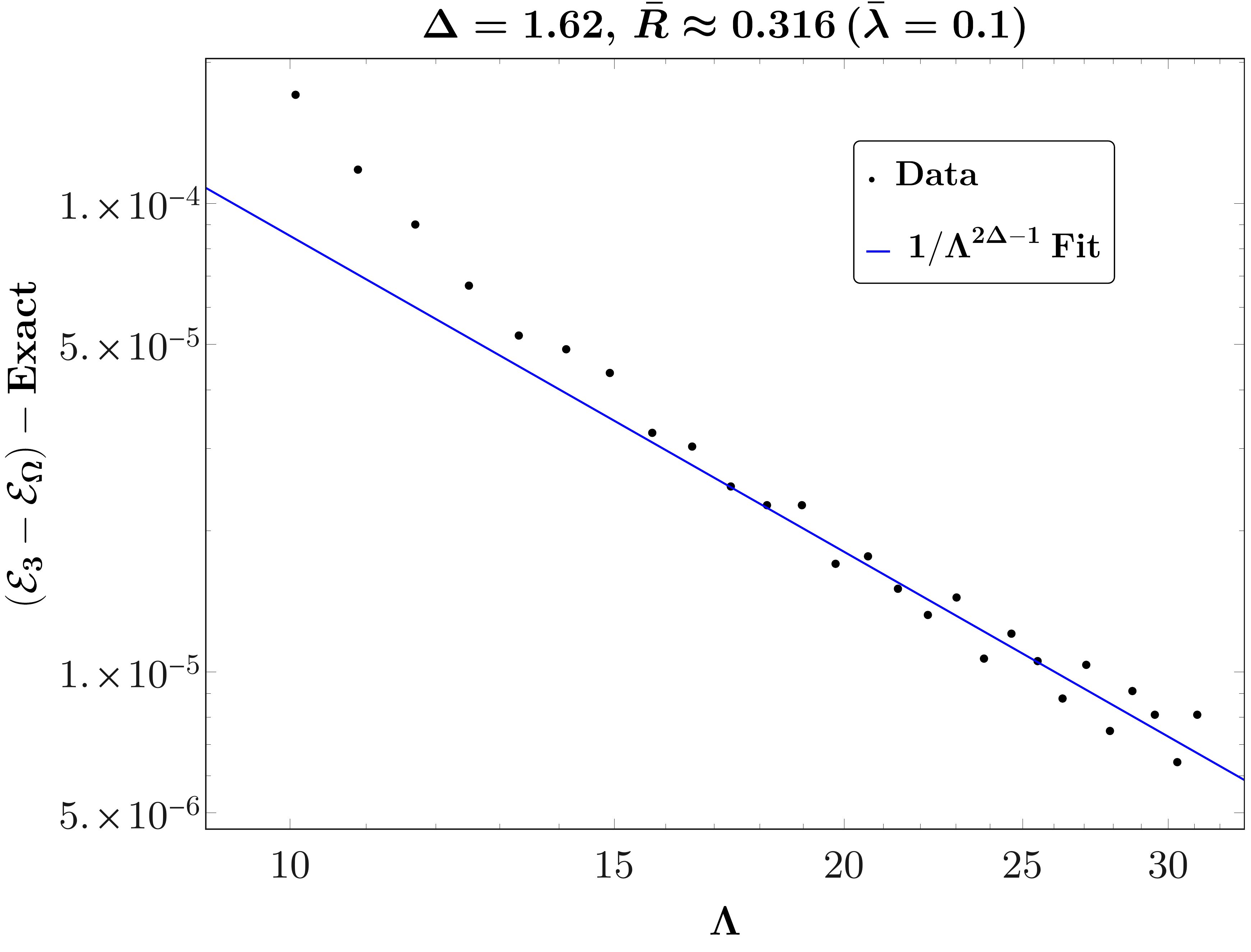}
    \end{subfigure}
              \caption{Log–log plot of Hamiltonian truncation deviation from the exact value as a function of the cutoff $\Lambda$ for coupling $\bar{\lambda}=0.1$. The left plot shows the convergence for the first excited state with the blue line equal to $-\bar{\lambda}^2 \frac{0.898}{\Lambda^{2.24}}$ extracted from eq.~\reef{eq:chiPrediction}. The right plot shows the convergence for the third excited state corresponding to a state with two particles at rest. The blue line is a single parameter fit  $\frac{a}{\Lambda^{2.24}}$.}
\label{fig:FirstExcitedPhi2}
\end{figure}

In the left plot of figure~\ref{fig:FirstExcitedPhi2}, we compare Eq.~\reef{eq:chiPrediction} to Hamiltonian truncation data, setting $\bar{\la} = 0.1$ such that terms of order $\bar{\la}^3$ and higher are subleading. We find excellent agreement between the numerical results and the analytical prediction~\reef{eq:chiPrediction}. This agreement relies crucially on the prescription~\reef{prescription}: without the correct subtraction of the Casimir energy, there would be a mismatch of order $\sO(1) \times \bar{\la}^2$.
The second excited state is the first $SL(2,\mbb{R})$ descendent of $\ket{\chi}$, namely $\ket{\chi'} = a_1^\dagger \ket{\Omega}$, which in the full theory has energy $\bDelta\mca{E}_{\chi'} = \bDelta\mca{E}_\chi + 1$: we will come back to this state in the next paragraph. After that, the third excitated state is $(a_0^\dagger)^2 \ket{\Omega}$, which describes two particles at rest. In the right plot of figure~\ref{fig:FirstExcitedPhi2}, we compare Hamiltonian truncation data for this state to the exact result $\bDelta\mca{E} = 2 \DD(\bar{\la})$. 

\begin{figure}[htb]
     \begin{subfigure}[b]{0.47\textwidth}
         \centering
         \includegraphics[width=\textwidth]{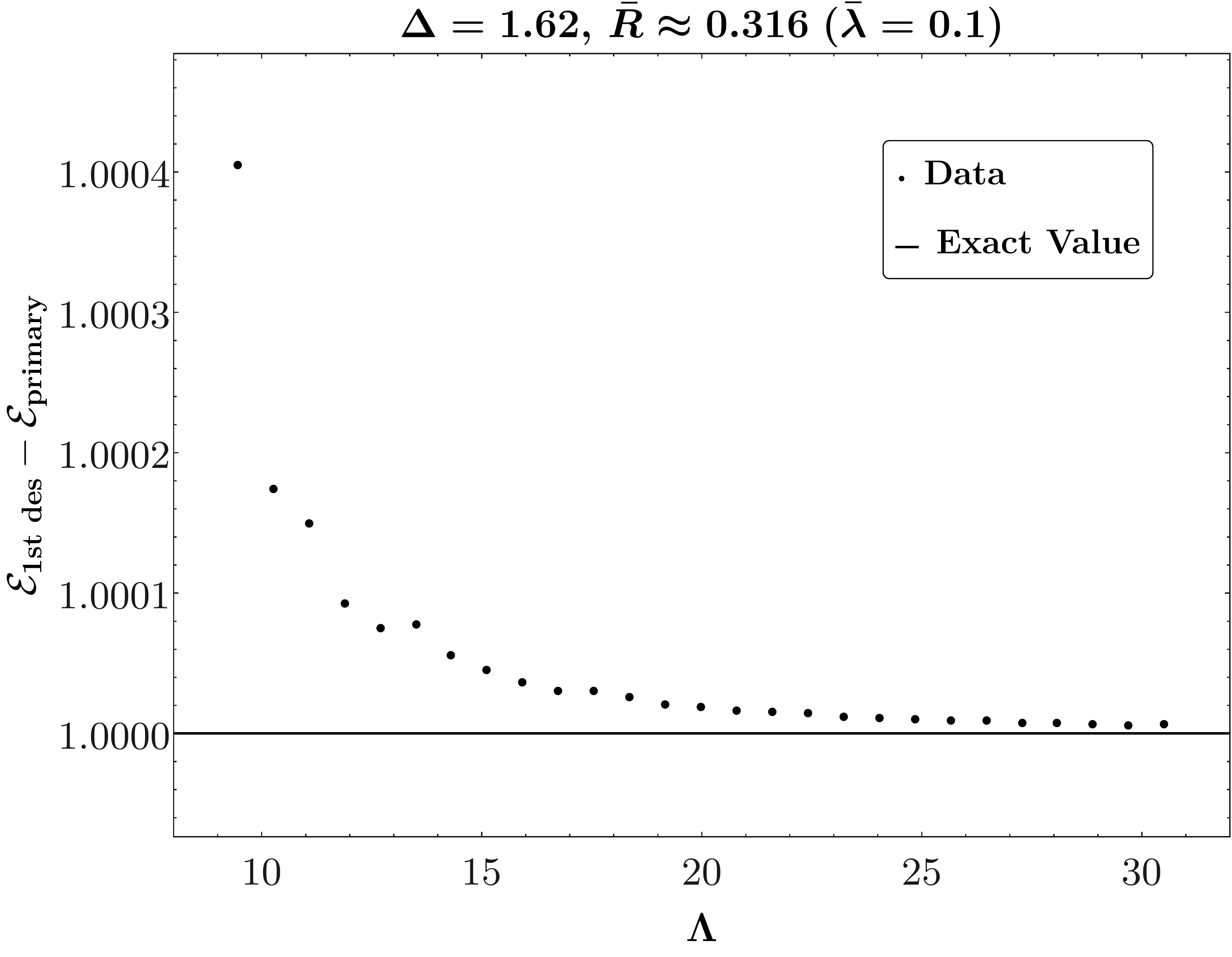}
     \end{subfigure}
     \hfill
     \begin{subfigure}[b]{0.45\textwidth}
         \centering
         \includegraphics[width=\textwidth]{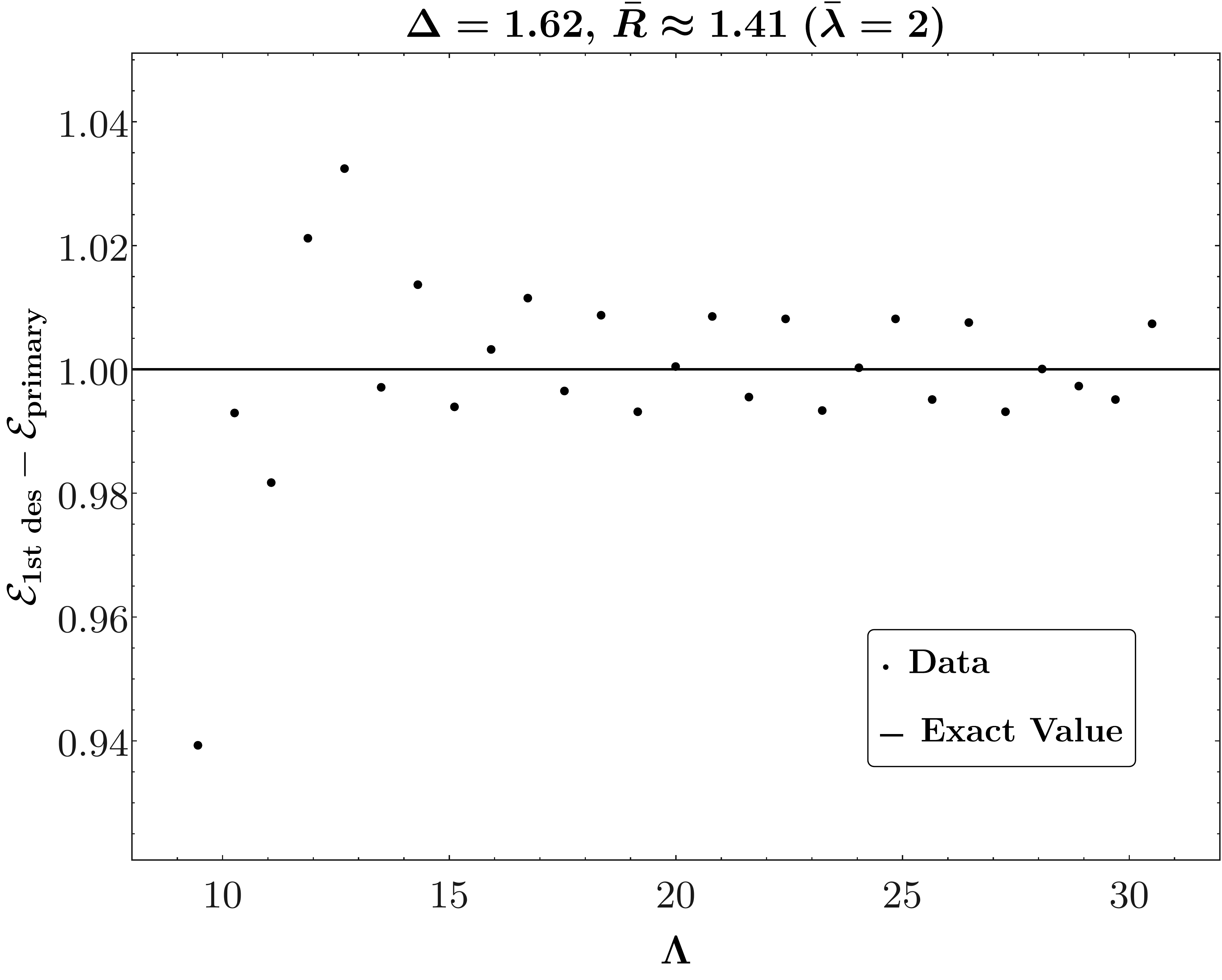}
    \end{subfigure}
              \caption{Difference of energies between the first primary state $\ket{\chi}$ and its first descendant, verifying that the $SL(2,\mbb{R})$ spacetime symmetry of AdS is restored in the continuum limit $\La \to \infty$.}
     \label{fig:DesMinusPrimaryPhi2}
\end{figure}

An additional test of the truncation procedure involves computing differences of energies between a primary state and its descendants. As a consequence of $SL(2,\mbb{R})$ invariance, such differences must be exactly integer. However, the truncation breaks the $SL(2,\mbb{R})$ symmetry, which can only be recovered in the continuum $\La \to \infty$. We expect that the energy of the second state in the spectrum $\ket{\chi'} = a_1^\dag \ket{\Omega}$ scales as
\beq
\label{eq:sl2rcorr}
\bDelta\mca{E}_{{\chi'}}(\La) - \bDelta\mca{E}_{{\chi}}(\La) \limu{\La \to \infty} 1 + \frac{a}{\La^b} + \ldots
\eeq
for some power $b > 0$ and some coefficient $a$. In figure~\ref{fig:DesMinusPrimaryPhi2}, we check this prediction numerically for different values of the coupling $\bar{\la}$.  Both for small ($\bar{\la} = 0.1$) and large ($\bar{\la} = 2$) values of the coupling, we observe that $SL(2,\mbb{R})$ is restored in the continuum; however, the plots show that this phenomenon is slower for the larger values of the coupling. In passing, let us mention that it would be interesting to predict the coefficients $a,\, b$ appearing in~\reef{eq:sl2rcorr}, by analyzing the $SL(2,\mbb{R})$ breaking directly. We have not studied this problem in more detail in the present work. 

\subsubsection{Spectrum}\label{sec:VaryingGPhi2}

\begin{figure}[!htb]
         \centering
         \includegraphics[width=\textwidth]{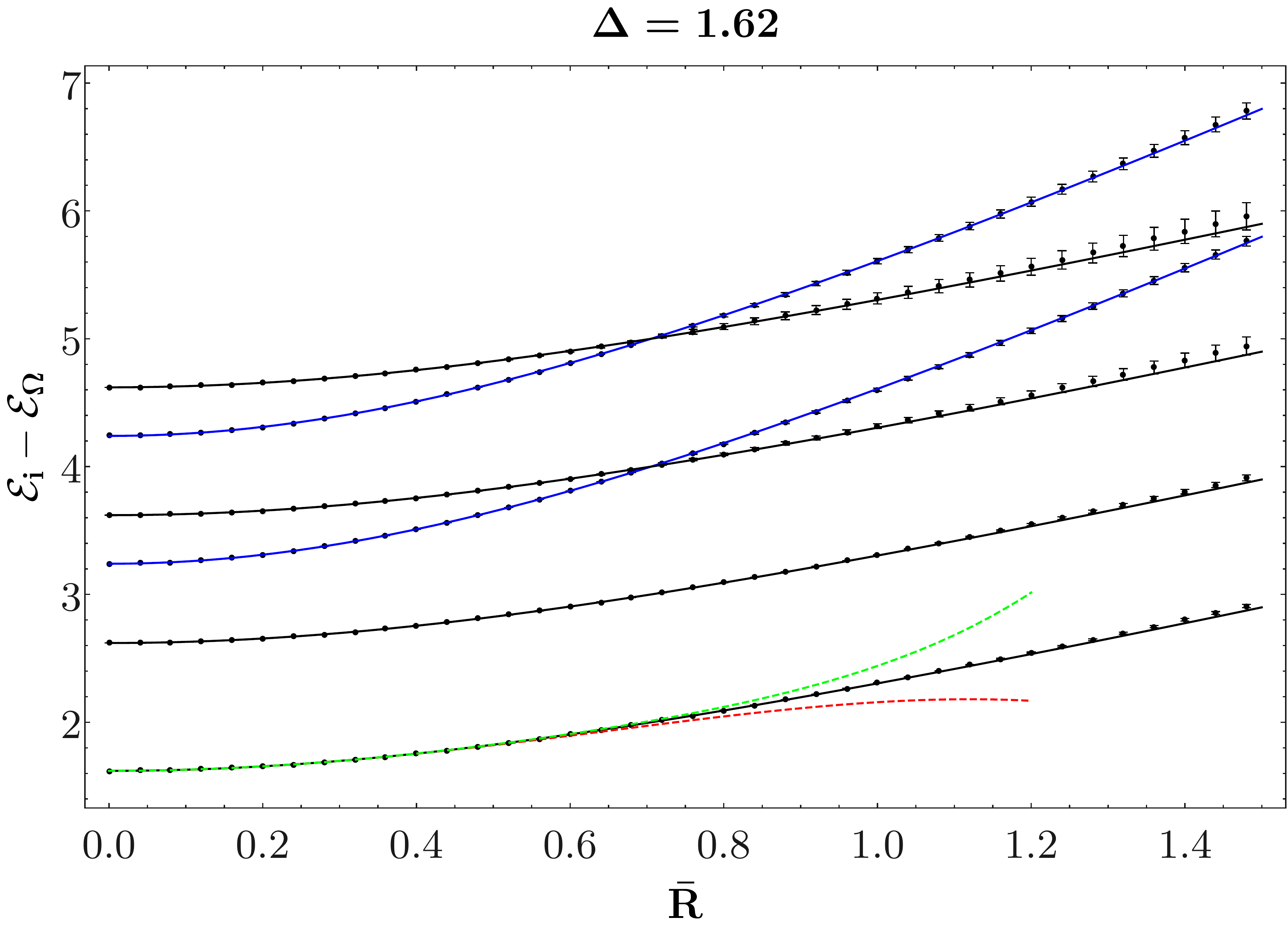}
         \caption{Spectrum of the first six excited energies in the $\phi^2$ theory with $\DD=1.62$. Dots with error bars: Hamiltonian truncation data. Black and blue solid curves represent the exact spectrum; the red and green dashed lines are exact results for the first level, truncated to second resp.\@ third order in perturbation theory. The black curves are the one-particle state $\ket{\chi}$ and its first three descendants; the blue curves are the two-particle state $(a_0^\dag)^2 \ket{\Omega}$ and its first descendant. } 
         \label{fig:spectrum}
\end{figure}

So far, we have checked in some examples that with the prescription~\reef{prescription}, Hamiltonian truncation in AdS agrees with exact results in the limit where the cutoff $\La$ is sent to infinity. At this point, we can be more systematic and compute the first six energy levels of the $\phi^2$ theory for a range of couplings, once more comparing the numerical data to exact results.  The results are shown in figure~\ref{fig:spectrum}. Dots correspond to Hamiltonian truncation data; the solid curves are the exact values. This plot probes couplings beyond the perturbation theory radius of convergence $\la_\star(\DD)$, which for $\DD = 1.62$ equals $\la_*(1.62) \approx 0.63$. To show the breakdown of perturbation theory, we have in addition plotted the energy of the state $\ket{\chi}$ computed up to order $\bar{\la}^2$ (shown in red) resp.\@ $\bar{\la}^3$ (green) in perturbation theory; it is clear that these perturbative curves deviate from the exact levels when $\bar{\la} \gtrsim \la_\star(\DD)$. In contrast, the Hamiltonian truncation data agrees with the exact data within error bars also for larger values of $\bar{\la}$. Finally, we want to mention that differences between energies are indeed approximately integer, in accordance with $SL(2,\mbb{R})$ symmetry.

In figure~\ref{fig:spectrum}, we notice crossings between one- and two-particle states, shown in black resp.\@ blue. These states have different quantum numbers: one-particle states have $\mca{Z} = -1$, whereas the blue curves have $\mca{Z} = 1$. Therefore, there is no level repulsion even at finite cutoff.

The data points in the above plot are extracted from ``raw'' Hamiltonian truncation data, computed at some finite cutoff $\La$, by extrapolating to $\La = \infty$. Let us explain in some detail the procedure that is used to go from unprocessed eigenvalues to these extrapolated data with error bars. For the $n$-th excited state $\bDelta\mca{E}_n$, we gather a sequence of energies $\{\bDelta\mca{E}_n(\La) \; | \; \La \leq \La_\text{max}\}$ by varying the cutoff $\La$, going up to some maximal cutoff $\La_\text{max}$.  We then fit this sequence to the function
\beq
\label{eq:fitForm}
\bDelta \mca{E}_n(\La|\La_\text{max}) \approx a_n(\La_\text{max}) + \frac{b_n(\La_\text{max})}{\La^{2\DD-1}}
\eeq
which gives an optimal value of $a_n(\La_\text{max})$, the energy in the limit $\La \to \infty$. There are two sources of error associated to this procedure:
\begin{itemize}
\item For a given cutoff $\La_\text{max}$, the truncation data don't exactly match the fit~\reef{eq:fitForm}. This is the usual $\chi^2$ or goodness-of-fit error, caused by the fact that we are only taking the leading cutoff error $1/\Lambda^{2\DD-1}$ into account.  The size of this error can be extracted from the $\mtt{ParameterTable}$ option of $\mtt{NonLinearModelFit}$ in $\mtt{Mathematica}$.
\item The restriction to cutoffs up to $\La_\text{max}$ is arbitrary: if we changed $\La_\text{max}$ by one or several units, the fit procedure would lead to a different extrapolated energy $a_n$. By computing $a_n$ for a range of maximal cutoffs  $\La_\text{max}$, we can obtain an estimate for the uncertainty associated to this phenomenon. 
\end{itemize}
The total error bar we assign to an energy $\mca{E}_n$ is the sum of these uncertainties.

\subsection{$\phi^4$ flow}\label{sec:phi4}

So far, we have discussed in detail Hamiltonian truncation result for the $\phi^2$ flow in AdS${}_2$. This flow is exactly solvable, so it provided a good laboratory to test the prescription~\reef{prescription}. It is of course more interesting to study flows that are \emph{not} exactly solvable, like the $\phi^4$ interaction. To be precise, we will consider Hamiltonians of the form~\reef{eq:LGham} with $\bar{\la}_4 \neq 0$. The two-dimensional $\phi^4$ flow has already been studied in detail using Hamiltonian truncation, but on $\mbb{R} \times S^1$ (see~\cite{Rychkov:2014eea,Rychkov:2015vap,Elias-Miro:2017xxf,Elias-Miro:2017tup}): here we will explore the same QFT, but defined in AdS${}_2$ instead of the cylinder.

Let us briefly discuss the physics to be expected as a function of the bare quartic coupling $\la_4$ and the bare mass $m^2$. On $\mbb{R}^2$, the theory displays spontaneous symmetry breaking (SSB): if the dimensionless ratio $m^2/\la_4$ is sufficiently large, the vacuum is $\mbb{Z}_2$-invariant, but below some critical value the field $\phi$ obtains a VEV and the $\phi \to -\phi$ symmetry is broken.\footnote{The critical value of $m^2/\la_4$ is of course scheme-dependent.} The transition between both phases is continuous and describes the $2d$ Ising universality class. All of these effects have previously been observed in Hamiltonian truncation on $\mbb{R} \times S^1$, which is Weyl-equivalent to $\mbb{R}^2$. Away from this critical value, the mass gap is finite.

What is the fate of the phase transition in AdS? At finite radius, the physics depends now on the two dimensionless couplings $m^2 R^2$ and $\bar{\lambda}_4$.\footnote{In this initial discussion, we absorb $\bar{\lambda}_2$ in $m^2 R^2$. The distinction will become relevant in the next paragraph, where $\bar{\lambda}_2$ will be chosen as a specific finite counterterm, to make the comparison to flat space easier.} As discussed in subsection \ref{sec:sbb}, SSB happens in AdS in a way analogous to flat space, up to boundary effects which might make the false vacuum stable. Therefore, in the limit where $m^2 R^2,\,\bar{\lambda}_4 \to +\infty$ with fixed ratio, the physics ought to be the same as in flat space. In this paper, we mainly focus on testing this hypothesis. However, it is worth asking about the general features of the phase diagram in the $(m^2 R^2,\bar{\lambda}_4)$ plane. In particular, does the line of phase separation extend all the way to $\bar{\lambda}_4=0$? Is the phase transition continuous? Is it described by the $2d$ Ising CFT in AdS? As we shall see below, it is not easy to tackle these questions with Hamiltonian truncation, so we will leave most of them open to future work. Nevertheless, we can at least gather some information from the weak coupling side, $\bar{\lambda}_4\ll 1$. In this limit, we can trust the semi-classical analysis of appendix \ref{generalbubble}. In particular, let us recall that the line $\bar{\lambda}_4=0$, $m^2 R^2>-1/4$, with $\mathbb{Z_2}$ preserving boundary conditions, belongs to the symmetric phase. We expect therefore the line of phase transition to obey the equation $m^2 R^2=-1/4+ O(\bar{\lambda}_4)$. This answers in the negative two of the above questions. At the phase transition, the expectation value of $\phi$ jumps from zero to a finite value: \emph{i.e.} the phase transition is discontinuous. Furthermore, as $\bar{\lambda}_4 \to 0$, the spectrum is continuously connected to the one of a free \emph{boson} with $\Delta=1/2$, while, as we shall recall in subsection \ref{subsec:ising}, the Ising CFT is described by a free \emph{fermion}.

\subsubsection{Fixing the $\phi^2$ counterterm}\label{sec:fixing}

We are most of all interested in the flat-space limit $R \to \infty$, keeping the bare couplings $\la_2$ and $\la_4$ fixed. Any ``RG trajectory'' (i.e.\@ a sequence of spectra for a range of radii $R$) therefore depends on the ratio $\la_2/\la_4$ and in addition on the UV scaling dimension $\DD$ of the single-particle state (or equivalently on its mass $m^2 = \DD(\DD-1)/R^2$). Note that the Hamiltonian is normal-ordered at fixed radius $R$. Therefore a trajectory of the form
\beq
\label{eq:naiveH}
H = H_0 + \sum_{k=2,4} \bar{\la}_k V_k,
\quad
\la_2,\, \la_4 \;\; \text{fixed}
\eeq
does \emph{not} describe a good RG trajectory: the normal-ordering counterterms implicit in~\reef{eq:naiveH} are not the same as those in flat space, as has been explained in~\cite{Rychkov:2014eea}. Instead, we should use a Hamiltonian of the form
\beq
\label{eq:slightlybetterH}
H = H_0 + \bar{\la} V_4 + 6 z(\DD) \bar{\la} V_2
\eeq
omitting terms proportional to the identity operator, since the Casimir energy is not observable in AdS. We can compute the coefficient $z(\DD)$ by adapting the derivation from reference~\cite{Rychkov:2014eea}. First, we notice that in flat space, the local operators $\phi^2$ and $\phi^4$ are normal-ordered as follows
\beq
\normord{\phi^2}_\DD=\phi^2 - Z_\DD,\quad \normord{\phi^4}_\DD=\phi^4 - 6 Z_\DD\phi^2 + 3Z^2_\DD
\eeq
where $Z_{\DD}$ formally denotes the contraction
\beq
Z_\DD = \lim_{Y \to X} \left[\phi(X) \phi(Y) - \NO{\phi(X) \phi(Y)}\right] = \lim_{Y \rightarrow X} \expec{\phi(X)\phi(Y)}_R
\eeq
keeping the dependence of the correlator on the AdS radius explicit. This expression is formal, since the limit $Y \to X$ diverges. However, the difference of $Z$ measured at two different radii is physical. For instance, we can compare the theory at a given $R$ to the same QFT quantized in the flat-space limit $R = \infty$. This leads to
\beq
z(\DD) \equiv  Z_\DD - Z_{\infty} = \frac{1}{2 \pi}\left(- \half\log{\Delta (\Delta -1)} + \psi(\Delta)\right)
\eeq
where $\psi(\cdot)$ is the digamma function. This completely fixes the Hamiltonian~\reef{eq:slightlybetterH}.

\subsubsection{Convergence estimates}
\label{subsec:phi4conv}

\begin{figure}[!htb]
\centering
\begin{subfigure}[htb]{0.45\textwidth}
\includegraphics[width=\textwidth]{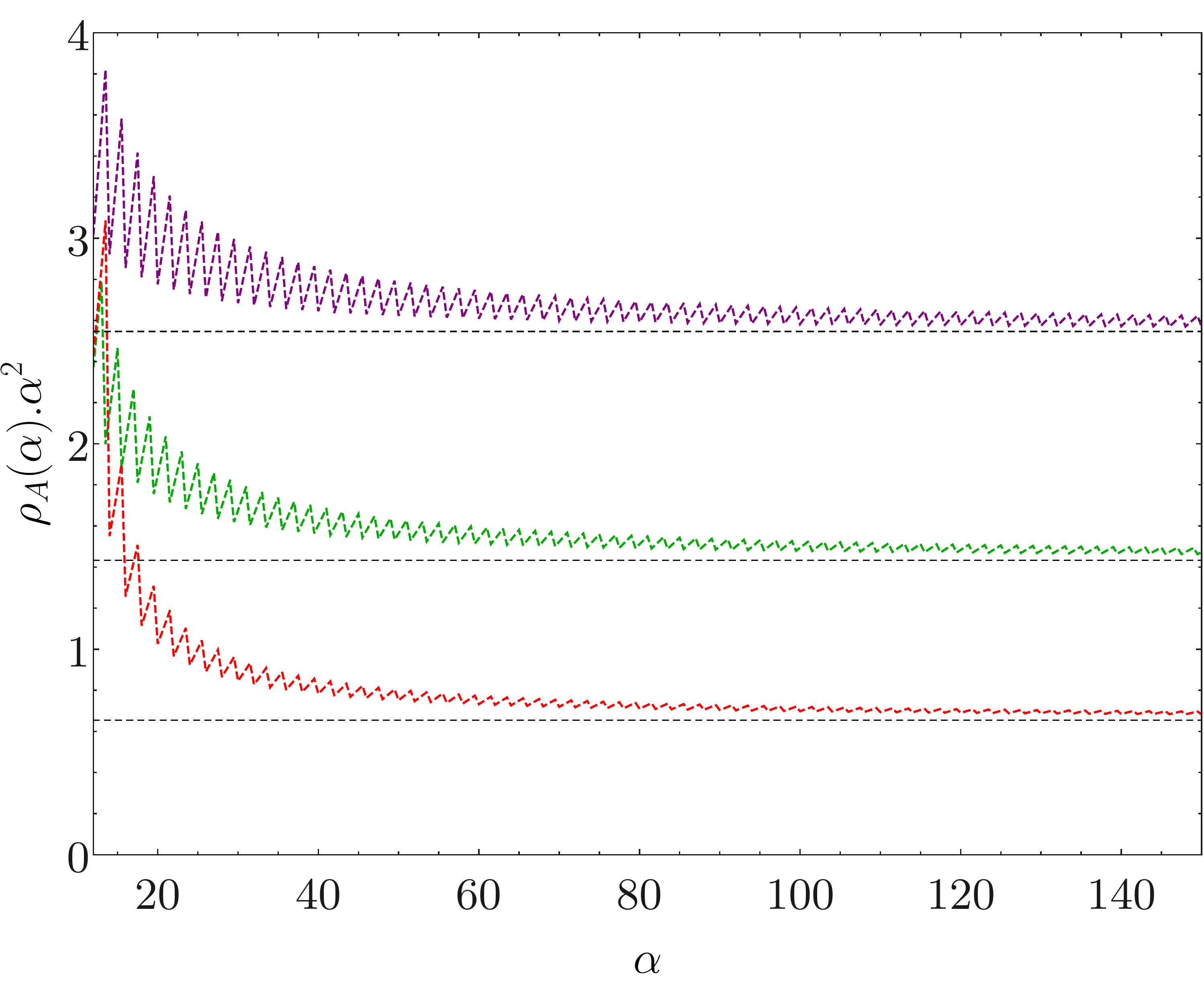}
\end{subfigure}
     \hfill
\begin{subfigure}[htb]{0.45\textwidth}
\includegraphics[width=\textwidth]{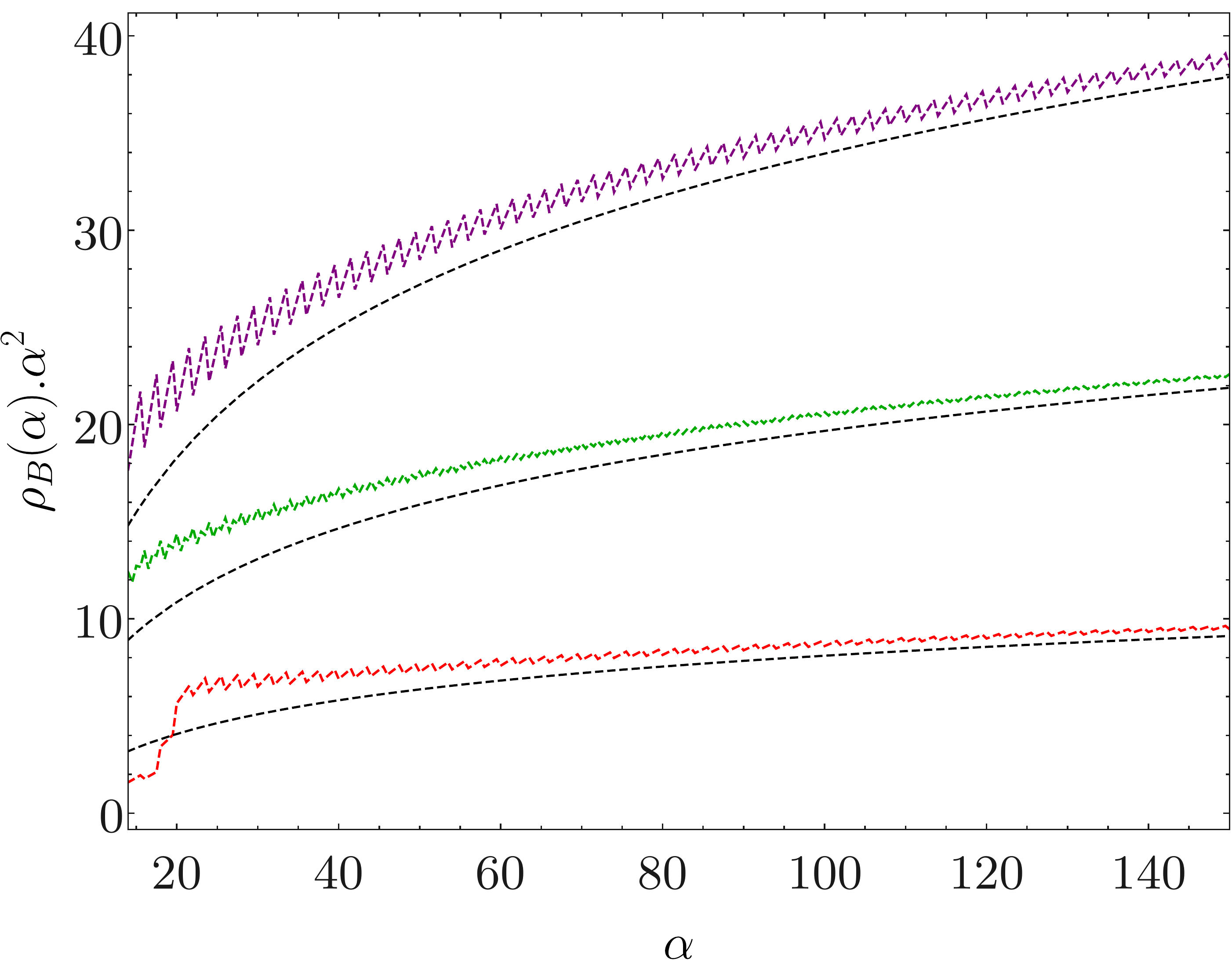}
\end{subfigure}
\caption{\label{fig:phi4convAn}Left: the spectral density $\rho_A(\a)$ multiplied by $\a^2$, for $\DD = 2$ (purple), $\DD=2.5$ (green)  and $\DD = 4$ (red), averaged over intervals $[\a-2,\a+2]$ (since the densities have delta function support). The horizontal dashed lines are the large-$\a$ predictions from formula~\reef{eq:rhoMM}. Right: the same plot for the spectral density $\rho_B(\a)$, plus comparison to the asymptotic formula~\reef{eq:rhoMM} (dashed).}
\end{figure}

Before we turn to numerical computations, let us first of all discuss the convergence rate, that is to say the error due to working at finite cutoff $\Lambda$. Although in the previous section we fixed a specific $\phi^2$ counterterm, let us for now consider the general Hamiltonian~\reef{eq:naiveH}
\beq
H = H_0 + \bar{\la}_2\, V_2 + \bar{\la}_4 \, V_4.
\eeq
For definiteness, let us focus on the leading energy shift of the first excited state $\ket{\chi} = a_0^\dagger \ket{0}$, although much of the discussion does not depend sensitively on the choice of state. In Rayleigh-Schr\"{o}dinger perturbation theory, we have
\beq
\bDelta\mca{E}_\chi(\bar{\la}_2,\bar{\la}_4) 
=  \DD + \frac{2}{2\DD-1}\,  \bar{\la}_2 - W \cdot \bar{\la}_2^2 - 2 A \cdot \bar{\la}_2 \bar{\la}_4 - B \cdot \bar{\la}_4^2 + \ldots
\eeq
subtracting the Casimir energy and omitting terms of cubic and higher order in perturbation theory. The coefficients $W$, $A$ and $B$ depend on $\DD$, and importantly they are sensitive to the cutoff $\La$, although they should have a finite limit as $\La \to \infty$. We have already encountered the $W$-term in section~\ref{sec:ph2conv}, and found that it evaluates to
\beq
W(\La) = \frac{4}{(2\DD-1)^3} + \sO\!\left(\frac{1}{\La^{2\DD-1}}\right).
\eeq
The coefficients $A$ and $B$ cannot be calculated in closed form, but they can be obtained as integrals over AdS:
\beq
A = \int_0^\infty\!d\tau\, \mca{A}(\tau)
\qaq
B = \int_0^\infty\!d\tau\, \mca{B}(\tau)
\eeq
where
\bsub
\label{eq:integralsToDo}
\begin{align}
\mca{A}(\tau) &= 24 \int_{-\pi/2}^{\pi/2}\!\frac{dr dr'}{(\cos r \cos r')^2} \, f_0(r')^2 \, G(\xi(\tau,r,r'))^2,
\\
\mca{B}(\tau) &= 192 \cosh(\DD \tau) \int_{-\pi/2}^{\pi/2}\!\frac{dr dr'}{(\cos r \cos r')^2} \, f_0(r)f_0(r') \, G(\xi(\tau,r,r'))^3.
\end{align}
\esub
The expressions were obtained by evaluating connected two-point correlation functions of $V_2$ and $V_4$ inside the state $\ket{\chi}$, and subtracting the same correlator inside the vacuum state. In particular, $G(\xi)$ was defined in eq. \eqref{eq:scalar2pt}, and the cross ratio $\xi$ of eq. \eqref{eq:crossratio} is evaluated at $\tau'=0$.
Let $\rho_A(\a)$ and $\rho_B(\a)$ be the spectral densities associated to $\mca{A}(\tau)$ resp.\@ $\mca{B}(\tau)$, that is to say
\beq
\mca{A}(\tau) = \int_0^\infty\!d\alpha\, \rho_A(\a) e^{-(\a-\DD)\tau}
\eeq
and likewise for $\mca{B}(\tau)$. We can likewise associate a spectral density $\rho_W(\a)$ to the diagram $W$ --- in fact, $\rho_W(\a)$ is exactly identical to $\bDelta \rho_\chi (\alpha)$ from~\reef{eq:Gchi}. The total cutoff error is then
\beq
- \int_\La^\infty\! \frac{d\a}{\a - \DD} \left[  \rho_W(\a) \cdot \bar{\la}_2^2 + 2 \rho_A(\a) \cdot \bar{\la}_2 \bar{\la}_4 + \rho_B(\a) \cdot  \bar{\la}_4^2 \right] + \text{subleading in pert.\@ theory}.
\eeq
If $\La$ is large, this error term is controlled by the asymptotic behavior of the densities $\rho_W(\a)$, $\rho_A(\a)$ and $\rho_B(\a)$. We have already seen that $\rho_W(\a) \limu{\a \to \infty} 1/\a^{2\DD-1}$. We will see that for $\DD > 3/2$, the error will be dominated by the other two terms. Indeed, we claim that
\beq
\label{eq:roughScaling}
\rho_A(\a) \limu{\a \to \infty} 1/\a^2
\qaq
\rho_{B}(\a) \limu{\a \to \infty} \ln(\a)/\a^2.
\eeq
Numerical evidence for this claim is presented in figure~\ref{fig:phi4convAn}. If this claim holds, then the total error term is of the order of
\beq
\label{eq:errorfor44}
\text{error term} = \left(2\vareps_A \cdot \bar{\la}_2 \bar{\la}_4 + \vareps_B \cdot \bar{\la}_4^2+\vareps_B' \cdot \bar{\la}_4^2\, \ln \La\right) \frac{1}{\La^2} + \ldots
\eeq
for some coefficients $\vareps_{A}$, $\vareps_B$ and $\vareps_B'$, omitting terms subleading in $\La$ and in perturbation theory. The leading large-$\a$ behavior of the diagrams $A$ and $B$ is due to \emph{bulk} effects, contrary to the diagram $W$ which was due to a contribution close to the boundary. Indeed, the same scaling has been observed on the cylinder $\mbb{R} \times S^1$~\cite{Rychkov:2014eea}.

To make~\reef{eq:errorfor44} precise, it makes sense to estimate the large-$\a$ asymptotics of $\rho_A$ and $\rho_B$ directly --- or equivalently, the small-$\tau$ asymptotics of $\mca{A}(\tau)$ and $\mca{B}(\tau)$. We can do so following the example of reference~\cite{Rychkov:2014eea}. In particular, we choose to study $d\mca{A}(\tau)/d\tau$ and $d\mca{B}(\tau)/d\tau$ instead of $\mca{A}(\tau)$ and $\mca{B}(\tau)$, and we replace the bulk-bulk propagator $G(\xi)$ by its Taylor series in the short-distance limit, to wit
\beq
\label{eq:taylorG}
G(\xi) \approx -\frac{\ln \xi +2 \ga_E +2 \psi(\DD)}{4\pi}
\eeq
dropping analytic terms that vanish as $\xi \to 0$. This yields
\bsub
\label{eq:rhoMM}
\begin{align}
\rho_A(\a) &\; \limu{\a \to \infty} \; \frac{24(\DD-1)}{\a^2 \pi (2\DD-3)(2\DD-1)},\\
\rho_B(\a) &\; \limu{\a \to \infty} \; \frac{144(\DD-1)}{\a^2 \pi^2(2\DD-3)(2\DD-1)}\left[ 2\ln \a - \frac{2}{\DD-1} + \psi(\DD-\tfrac{3}{2}) -3\psi(\DD-1)\right].
\end{align}
\esub
These formulas make precise the scaling predicted by~\reef{eq:roughScaling}. At the same time, we can compute the densities $\rho_{A,B}(\a)$ directly for finite values of $\alpha$, by expanding the integrals~\reef{eq:integralsToDo} around large $\tau$. In figure~\ref{fig:phi4convAn}, the asymptotic formulas are compared to these exact results for $\DD = 2,2.5$ and $4$, finding good agreement.

Let us make two comments about eqs.~\eqref{eq:rhoMM}. For one, both densities have a pole at $\DD = 3/2$. This is due to the truncation~\reef{eq:taylorG}, which leads to additional divergences close to the boundary of AdS --- yet the \emph{full} correlators $\mca{A}(\tau)$ and $\mca{B}(\tau)$ are integrable for dimensions $\DD > 1$, where eq.~\eqref{UVfiniteCond} holds. Second, the asymptotic density shown for $\rho_B(\a)$ can become negative for small values of $\a$; for instance for $\DD = 1.62$, the terms shown are only positive for $\a \gtrsim 43$. However, it can be shown that the full density $\rho_B(\a)$ is manifestly positive. This means that the asymptotic formula~\reef{eq:rhoMM} is not necessarily a good approximation to the actual density $\rho_B(\a)$ for values of $\a$ that can be realized in Hamiltonian truncation (say $\a \leq 25-40$), at least for small values of $\DD$. Nevertheless, figure~\ref{fig:phi4convAn} shows that for sufficiently large $\a$, the scaling behavior~\reef{eq:roughScaling} is observed.

\begin{figure}[!htb]
     \centering
     \begin{subfigure}[b]{0.45\textwidth}
         \centering
         \includegraphics[width=\textwidth]{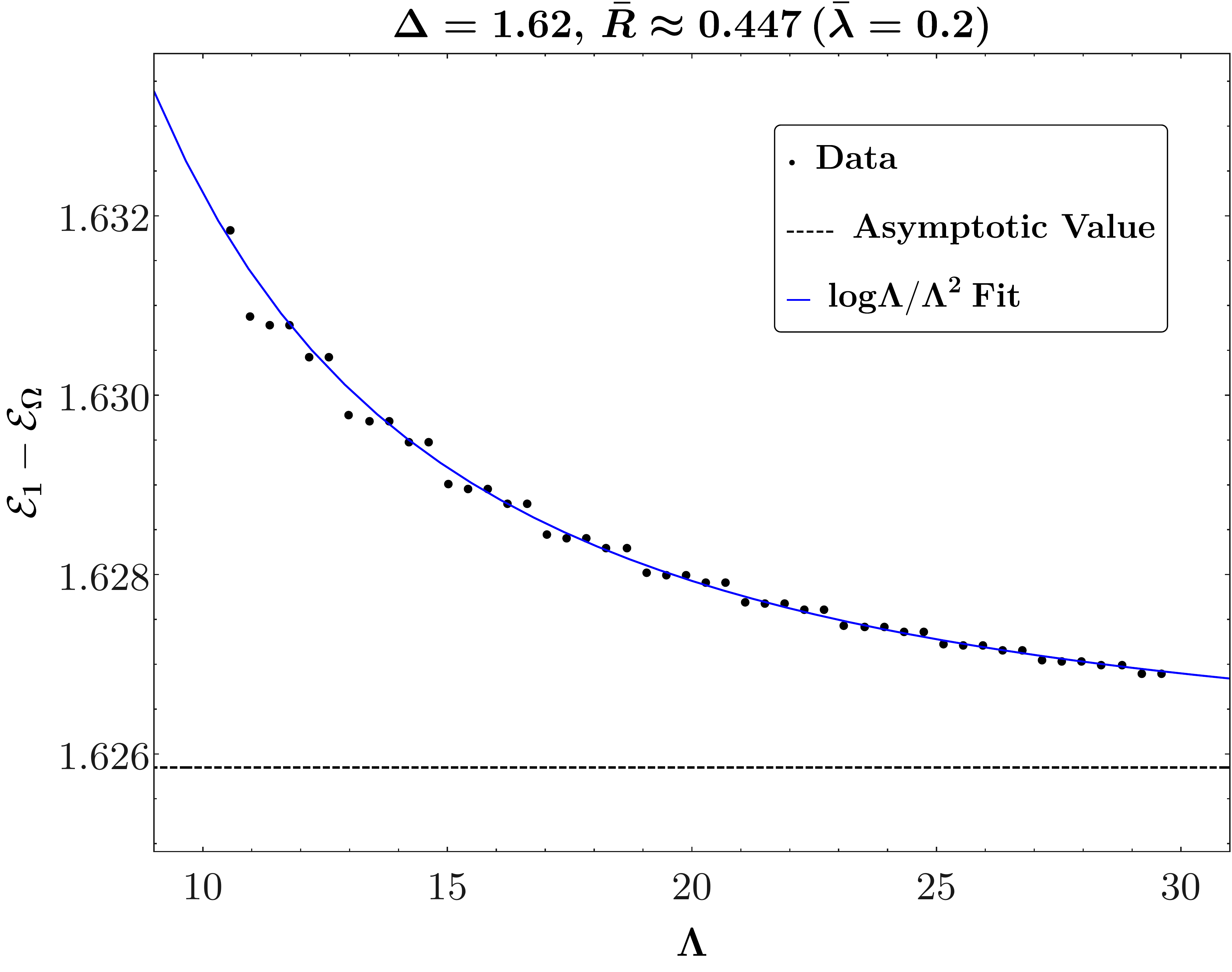}
         \caption{$\bar{\lambda}=0.2$}
         \label{fig:FirstExcitedPhi4g02}
     \end{subfigure}
     \hfill
     \begin{subfigure}[b]{0.45\textwidth}
         \centering
         \includegraphics[width=\textwidth]{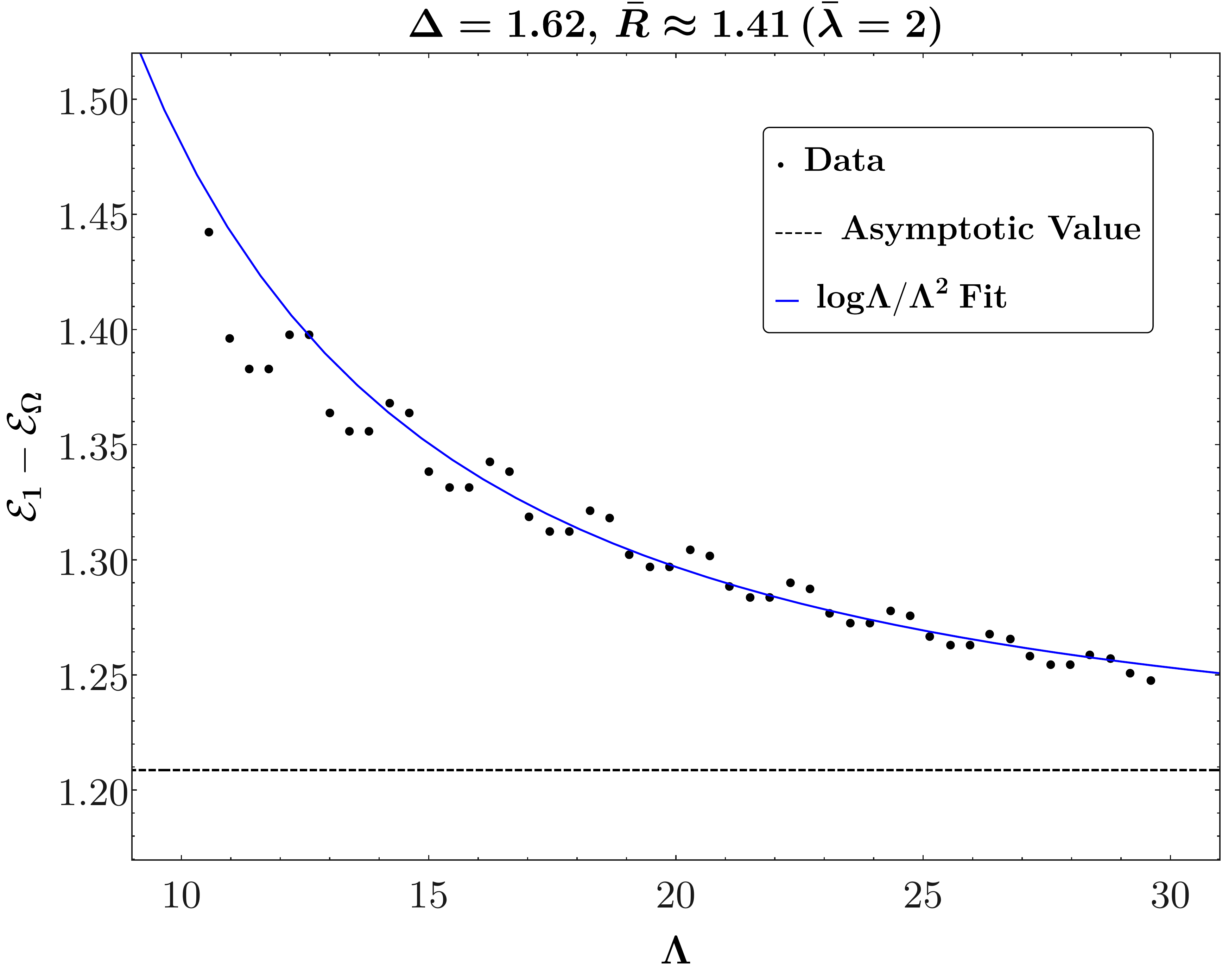}
         \caption{$\bar{\lambda}=2$}
         \label{fig:FirstExcitedPhi4g2}
     \end{subfigure}
              \caption{Hamiltonian truncation data for the energy shift of the state $\ket{\chi}$ due to the $\phi^4$ Hamiltonian~\reef{eq:slightlybetterH} with $\DD = 1.62$, for a range of cutoffs up to $\La = 30$. The left (right) plot corresponds to quartic coupling $\bar{\la} = 0.2$ ($\bar{\la} = 2$). Blue curve: fit of the form $u - v \ln(\La)/\La^2$, with the asymptotic value $u$ shown as a horizontal dotted line.}
              \label{fig:ConvergencePhi4}
\end{figure}

Let us finally compare this asymptotic prediction from second-order perturbation theory to Hamiltonian truncation data. It follows from the above discussion that the truncation error of the $\phi^4$ Hamiltonian should be of the order of $\ln(\La)/\La^2$, at least at second order in the coupling\footnote{We don't expect higher-order diagrams to be \emph{more} UV-sensitive.}. We can thus fix the $V_2$ counterterm according to section~\ref{sec:fixing} and diagonalize the Hamiltonian $H$ for finite values of the cutoff $\La$, computing the first excitated energy level non-perturbatively using the presription~\reef{prescription}. The results of this procedure are shown in figure~\ref{fig:ConvergencePhi4}. As expected, we observe that the truncation data converge to a constant in the limit $\La \to \infty$, and we see that the truncation error goes to zero as $\ln(\La)/\La^2$. 

\subsubsection{Evidence for a phase transition}

As we already discussed, we expect that the $\phi^4$ theory in AdS${}_2$ exhibits sponteanous symmetry breaking, just as in flat space. Already in finite volume, working on $\mbb{R} \times S^1$ where the $S^1$ has length $L$, truncation methods were used to study this transition~\cite{Rychkov:2014eea}. In that work, the bare mass $m^2$ was fixed, and for fixed but large $mL \gtrsim \sO(\text{few})$ the authors scanned over the quartic coupling $\la_4$. 
For $\la_4$ sufficiently small and positive bare mass, the system is in the unbroken phase.
In this phase, the energy  gap of the theory, closes continuously as $\la_4$ is increased to its critical value. 
For $\la_4$ larger than its critical value the system is in the broken phase. 

Here we can attempt to do the same thing: we can fix the bare mass $m^2$ and scan over the dimensionless coupling $\la/m^2$. Since we want to be close to the flat-space limit, we need to make sure that $R$ is large compared to $m^2$, or in other words that the UV scaling dimension $\DD$ is sufficiently large. To connect the notation used so far, this means that we fix $\DD \gtrsim \sO(\text{few})$ and scan over the dimensionless coupling $\bar{\la}/(m^2 R^2) = \bar{\la}/[\DD(\DD-1)] \rdef \msf{x}$. For any value of $\DD$ and $\msf{x}$, we can measure the mass gap. The resulting curves $\bDelta\mca{E}(\msf{x},\DD)$ will depend on $\DD$, but if there is any universal flat-space physics, they should have a well-defined limit as $\DD \to \infty$. The limiting curve $\bDelta\mca{E}(\msf{x},\infty)$ should exhibit a gap that closes in accordance with the critical exponents of the Ising universality class, and as a matter of principle one should be able to read off the Ising BCFT spectrum. In fact, we expect that the gap $\bDelta\mca{E}$ closes at the same critical coupling as in flat space.

\begin{figure}[!htb]
     \centering
         \includegraphics[width=\textwidth]{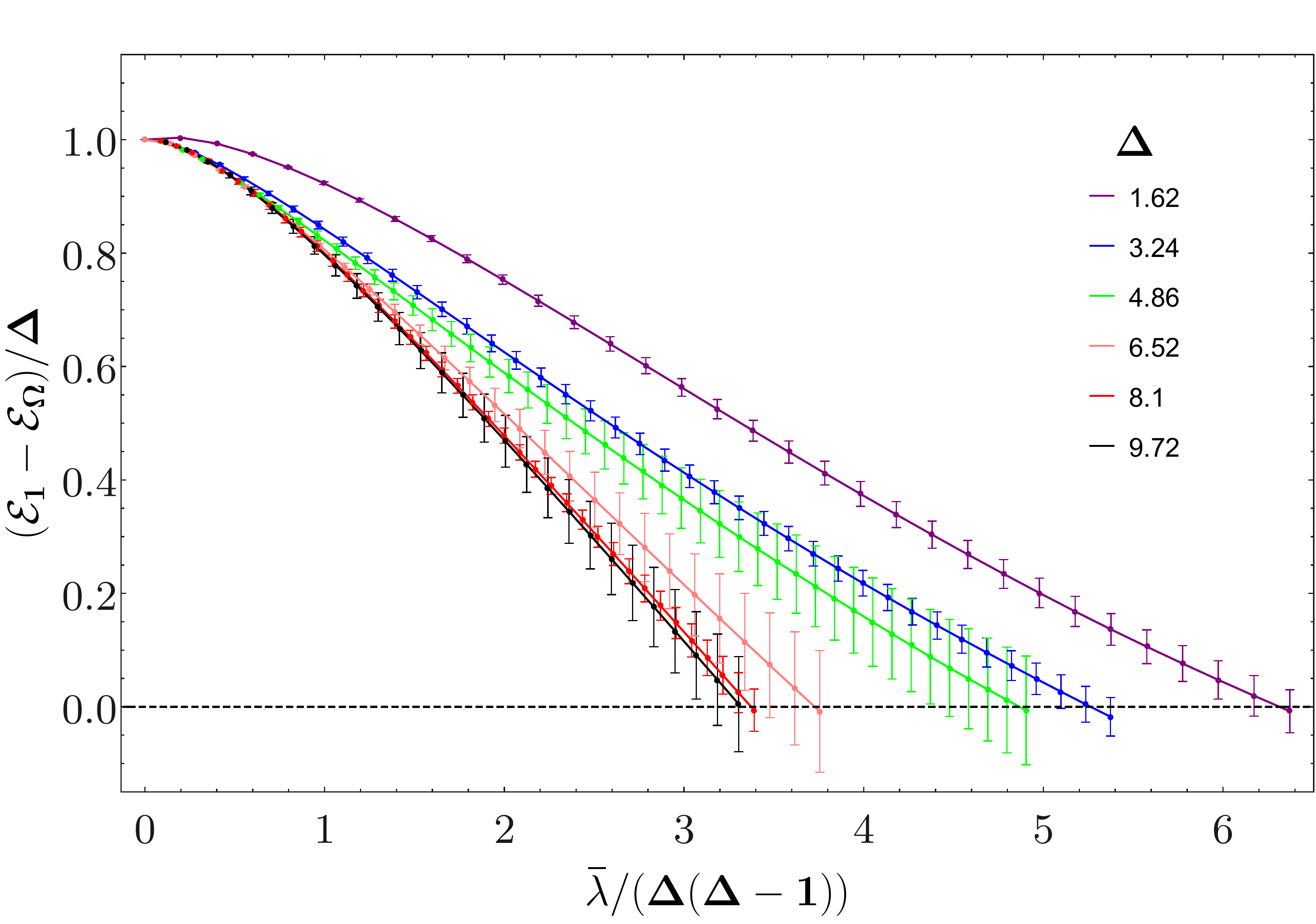}
         \caption{Mass gap of the $\phi^4$ Hamiltonian~\reef{eq:slightlybetterH} for a range of quartic couplings $\bar{\la}$ (x-axis), for a different set of UV scaling dimensions $\DD$ (different curves, see legend). }
         \label{fig:spectrumphi4}
\end{figure}

In figure~\ref{fig:spectrumphi4}, we plot precisely these curves $\bDelta\mca{E}(\msf{x},\DD)$. As always, we subtract the Casimir energy according to~\reef{prescription}, and we divide the mass gap by $\DD$ such that we can compare curves of different UV dimensions $\DD$. On the $x$-axis, we vary the dimensionless quartic coupling $\msf{x} = \bar{\la}/[\DD(\DD-1)]$. The plot provides evidence that the curves $\bDelta\mca{E}(\msf{x},\DD)$ indeed have a finite limit as $\DD \to \infty$, although we cannot increase $\DD$ further than $\DD \approx 10$ for computational reasons. Semi-quantitatively, the plot is consistent with a critical coupling of the order of $\la/m^2 \approx 3$. Numerically, the convergence rate decreases rapidly for large couplings: this can be seen for example by the large error bars appearing in figure~\ref{fig:spectrumphi4}. Therefore, analyzing the phase transion in more detail is not feasible in the current set-up, even by going to rather large cutoffs with $\sim5 \cdot 10^4$ states. 

The prescription~\reef{prescription} does not always lead to positive energies, contrary to differences of energies of a Hamiltonian, which are positive by construction. The value of e.g.\@ the first excited state $\mathcal{E}_1(\Lambda)-\mathcal{E}_\Omega(\Lambda-\DD)$ can indeed be negative, as seems to happen to points with large couplings in figure~\ref{fig:spectrumphi4}. This would indicate that the vacuum exchanges roles with an excited state (which has different quantum numbers!). However, we stress that for the data points in question, the error bars are very significant, and it is not excluded that the true energy in the limit $\La \to \infty$ is in fact slightly positive. For this reason, and to avoid clutter, we have not shown any points with a seemingly negative energy.
In appendix \ref{app:IHO}, we study an inverted harmonic oscillator with Hamiltonian truncation to illustrate how the states in the broken phase are expected to have a large overlap with unperturbed states close to the cutoff.

\section{Deformations of conformal field theories}
\label{sec:minimal}
%!TEX root = ../HTinAdS.tex
%%%%%%%%%%%%%%%%%%%%%%%%%%%%%%%%%%%%%%%%%%%%%

In this section, the unperturbed Hamiltonian corresponds to a conformal field theory in AdS. The deformation is triggered by a relevant operator. We shall focus, in particular, on the simplest two dimensional CFTs, \emph{i.e.} the Lee-Yang and the Ising models. Since AdS is conformally flat, a CFT placed in AdS is equivalent to a CFT in the presence of a flat boundary. More precisely, the equivalence is true if the boundary conditions preserve the Weyl invariance of the bulk. This setup is usually referred to as a boundary conformal field theory (BCFT) \cite{Cardy:1984bb}. Since the subgroup of the conformal group preserved by a flat boundary coincides with the isometries of AdS, any CFT placed in AdS with isometry preserving boundary conditions will be related to a BCFT by a Weyl transformation.\footnote{We are assuming that conformal symmetry automatically promotes to Weyl invariance.}

\vspace{0.01\textheight}
In this section we adopt the usual $2d$ CFT convention, reserving the name primaries for the Virasoro primaries. On the other hand, the $SL(2,\mathbb{R})$ primaries will be sometimes denoted as quasiprimaries.

\subsection{Virasoro minimal models and their deformations in AdS$_2$}
\label{sec:introBCFT}

We begin with a reminder of the basics of minimal models \cite{Belavin:1984vu} in the presence of a boundary, before moving on to AdS. A detailed exposition can be found in the original references \cite{Cardy:1984bb,Cardy:1986gw,Cardy:1989ir,Cardy:1991tv,Behrend:1999bn}, in chapter 11 of \cite{DiFrancesco:1997nk}, or in the dedicated book \cite{Recknagel:2013uja}.
A flat boundary preserves a single copy of the Virasoro algebra, whose generators $L_n$, $n \in \mbb{Z}$ obey the usual commutation relation
\beq
[L_m,L_n] = (m-n)L_{m+n} + \frac{c}{12} m(m^2-1) \dd_{m+n,0}~,
\label{VirGen}
\eeq
where $c$ is the central charge of the 2$d$ CFT.\footnote{A 2$d$ CFT admits a conformal boundary only if the left and right central charges coincide, $c=\bar{c}$ \cite{Billo:2016cpy}.} The $SL(2,\mbb{R})$ generators can be identified with Virasoro generators as follows:
\beq
H = L_0,
\quad
P = L_{-1},
\quad
K = L_{1}.
\label{sl2RgenL}
\eeq
The spectrum of boundary scaling operators, corresponding to the constant time slice Hilbert space  in AdS, is organized in representations of the Virasoro algebra. In the minimal models, there are a finite number of so called Verma modules, which have a Virasoro primary $\ket{\chi_i}$ as a lowest-weight state. The primary satisfies $L_n \ket{\chi_i} = 0$ for all $n \geq 1$.
The full Hilbert space is spanned by states of the form
\beq
L_{-n_1} \dotsm L_{-n_p}\ket{\chi_i}
\label{LchiStates}
\eeq
running over all Virasoro primaries $\ket{\chi_i}$ and all integers $n_1,\ldots,n_p \geq 1$. The $SL(2,\mbb{R})$ primaries are those states $\ket{\Psi}$ satisying $L_1 \ket{\Psi} = 0$. Under the parity transformation $\mtt{P}:\,r \mapsto -r$, the Virasoro generators transform as 
\beq
\mtt{P}L_n \mtt{P}^{-1} =(-1)^n L_n
\label{parityVir}
\eeq
which generalizes eq. \eqref{eq:parity}. 

Given a minimal model, the set of consistent boundary conditions is known \cite{Cardy:1989ir,Behrend:1999bn}. They are in one to one correspondence with the scalar Virasoro primary operators of the CFT. In the following, we will only consider diagonal minimal models, where all primaries are spinless. For instance, the Ising model has three Virasoro primaries: the identity, the spin field $\sigma$, the energy field $\epsilon$. Correspondingly, there are three conformal boundary conditions. The boundary spectrum supported by each boundary condition is determined as follows. Consider the fusion coefficients in the OPE
\beq
V_\phi \times V_\phi \sim n_{\phi \phi}{}^i V_i~,
\eeq
where $V_\phi$ and $V_i$ denote representations of the Virasoro algebra. Then, the boundary condition labeled by $\phi$ supports boundary operators falling in the representations such that $n_{\phi \phi}{}^i \neq 0$. For instance, the boundary spectrum of the $\sigma$ boundary condition in the Ising model comprises the identity and the $\epsilon$ module, whose Virasoro primaries have dimension $\Delta_0=0$ and $\Delta_\epsilon=1/2$ respectively. 

It is sometimes useful to map the theory to the exterior of the unit circle. In this setup, the boundary condition generates a state in radial quantization, the so called Cardy state, which allows for a decomposition in terms of local operators of the bulk CFT \cite{Cardy:1991tv}. Only the operators appearing in this decomposition acquire a one-point function in the presence of the boundary. For instance, the boundary state generated by the $\sigma$ boundary condition in the Ising model is
\beq
\ket{\sigma}_\textup{Cardy} = \dket{1} - \dket{\epsilon}~,
\eeq
where the Ishibashi states on the right hand side are a sum of left-right symmetric descendants of the Virasoro primary, whose specific form we will not need \cite{Ishibashi:1988kg}:
\beq
\dket{\phi}=\ket{\phi}+\frac{L_{-1}\bar{L}_{-1}}{\Delta_\phi}\ket{\phi} + \dots
\eeq

A special boundary condition exists in all the minimal models. It is the one labeled by the identity operator. It only supports the identity module, hence it does not allow for relevant deformations. In the generic minimal model, denoted by $\mca{M}(p,q)$ in \cite{DiFrancesco:1997nk}, the identity module has two null vectors, at level 1 and $(p-1)(q-1)$. In particular, the non trivial null states are found at level 4 in the Lee-Yang model and at level 6 in the Ising model. The leading boundary operator supported by the identity Cardy state is the displacement operator \cite{Billo:2016cpy}, \emph{i.e.} $L_{-2}\ket{\Omega}$,\footnote{Recall that $\ket{\Omega}$ is the vacuum in radial quantization around a point on the defect. This quantization scheme, which is the one relevant to the Hamiltonian truncation, should not be confused with the scheme that gives rise to the Cardy states, as explained above.} which has $\Delta=2$. This means that the rate of convergence of TCSA in AdS cannot be faster than $\sim 1/\Lambda$, as explained in subsection \ref{sec:rate}. 
 The identity boundary condition corresponds to the so called extraordinary transition in the corresponding statistical models, \emph{i.e.} it breaks all the global symmetries of the CFT. Indeed, the boundary OPE \eqref{bulkToBound} of any bulk operator cannot be empty, and therefore it must contain the identity module. Hence, all the bulk scalar operators acquire a one-point function in the presence of the boundary. Although we will comment about other boundary conditions in what follows, we shall be mainly interested in the one labeled by the identity. 

When mapping the CFT with a flat boundary to AdS$_2$, it is important to notice that the anomalous contribution to the stress tensor vanishes. Hence, the ground state energy vanishes.\footnote{This should be contrasted with the BCFT on a strip, which is a finite volume system and as such has a Casimir energy proportional to the central charge.} This is most easily checked in Poincaré coordinates. Using a pair of complex coordinates $z,\,\bar{z}$, the metric reads
\beq
ds^2_\textup{AdS} = \frac{R^2}{(\Im z)^2} \,dz d\bar{z}~.
\eeq
Under the Weyl map from flat space, the stress tensor is in fact unchanged:
\beq
ds^2_{\textup{AdS}} = e^{2\sigma} ds^2_\textup{flat}~,\quad 
\sigma = \log \frac{R}{\Im z}~, 
\qquad T^\textup{AdS}_{zz} = T^\textup{flat}_{zz}
+\frac{c}{12\pi}\left(\partial_z^2\sigma-(\partial_z \sigma)^2\right)
=T^\textup{flat}_{zz}~.
\eeq
It is then easy to check that the generators \eqref{VirGen}, obtained as modes of the stress tensor, provide conserved quantities in AdS as well. 

In what follows, we will trigger an RG flow in AdS by turning on a relevant Virasoro scalar primary $\mca{V}$. The relation between such operator in AdS and in flat space is
\beq
\mca{V}_\textup{AdS} = \left(\frac{\Im z}{R}\right)^{\Delta_\mca{V}} \mca{V}_\textup{flat}~.
\label{VtoV}
\eeq
One can then work out the action of the Virasoro generators on the AdS operator:
\beq
\left[L_n,\mca{V}_\textup{AdS} \right] =
\left[z^{n+1} \partial_z+\bar{z}^{n+1} \partial_{\bar{z}}+
\frac{\Delta_\mca{V}}{2} \left((n+1)(z^n+\bar{z}^n)-2\,\frac{z^{n+1}-\bar{z}^{n+1}}{z-\bar{z}}\right)\right] \mca{V}_\textup{AdS}~.
\label{VirOnV}
\eeq
As expected, the term proportional to $\Delta_\mca{V}$ vanishes for the $SL(2,\mathbb{R})$ generators \eqref{sl2RgenL}, which simply act as diffeomorphisms. Henceforth, we drop the subscript from $\mca{V}_\textup{AdS}$, and resume the convention set up in eq. \eqref{eq:Sin}. Eq. \eqref{VirOnV} allows to compute all the matrix elements of the perturbing operator $\mca{V}$ between states of the kind \eqref{LchiStates}, and hence the matrix elements of the potential $V$, once the three-point functions $\braket{\chi_i}{\mca{V}}{\chi_i}$ are known.\footnote{Of course, to obtain the potential one can alternatively compute the matrix elements of $\mca{V}_\textup{flat}$ in flat space, and multiply them by the Weyl factor in eq. \eqref{VtoV} before integrating on a constant time slice.} Since the potential preserves the isometries of AdS, not all the matrix elements between states belonging to the same $SL(2,\mathbb{R})$ families are in fact independent. Rather, there are recurrence relations between them, as is described in appendix \ref{app:constraints}.
Nevertheless these relations do not speed up the computation of the matrix elements significantly, as we discuss at the end of appendix \ref{app:algo}; instead, we compute all matrix elements individually in our algorithm.

\subsection{The Lee-Yang model}
\label{sec:leeyang}

The Lee-Yang CFT is the diagonal modular invariant of the simplest minimal model, denoted $\mca{M}(5,2)$ in \cite{DiFrancesco:1997nk}. This model was the theater where TCSA was first applied \cite{Yurov:1989yu}, and was studied with the same technique on a strip in \cite{Dorey:1997yg,Dorey:1999cj}. It has central charge
\beq
c=-\frac{22}{5}~,
\eeq
and it is therefore non unitary. It contains two scalar Virasoro primaries, the identity $\mathbb{1}$ and the field $\mca{V}$, with scaling dimensions
\beq
\Delta_\mathbb{1}=0~,\qquad \Delta_\mca{V}=-\frac{2}{5}~.
\eeq
Consequently, it allows for two conformal boundary conditions\footnote{In order for the coefficients of the boundary state to be real, it is necessary to normalize the vacuum in radial quantization around a point \emph{in the bulk} as $\brakket{\mathbb{1}}{\mathbb{1}}=-1$, see \emph{e.g.} \cite{Dorey:1999cj}.}:
\begin{subequations}
\begin{align}
\ket{\mathbb{1}}_\textup{Cardy} &=\left(\frac{\sqrt{5}+1}{2\sqrt{5}}\right)^{1/4} \dket{\mathbb{1}}+
\left(\frac{\sqrt{5}-1}{2\sqrt{5}}\right)^{1/4} \dket{\mca{V}}~,
\label{idCardyLY} \\
\ket{\mca{V}}_\textup{Cardy} &=-\left(\frac{\sqrt{5}-2}{\sqrt{5}}\right)^{1/4} \dket{\mathbb{1}}+
\left(\frac{\sqrt{5}+2}{\sqrt{5}}\right)^{1/4} \dket{\mca{V}}~.
\label{VCardyLY}
\end{align}
\end{subequations}
As mentioned, we shall be primarily interested in the identity Cardy state.
Eq. \eqref{idCardyLY} implies that the operator $\mca{V}$ acquires a one-point function in the presence of the boundary:
\beq
\brakket{\mca{V}}{\mathbb{1}}_\textup{Cardy}
=\left(\frac{\sqrt{5}-1}{2\sqrt{5}}\right)^{1/4}~.
\eeq
With the boundary state \eqref{idCardyLY}, the partition function on the disk is not normalized to one, rather
\beq
\brakket{\mathbb{1}}{\mathbb{1}}_\textup{Cardy}=\brakket{\Omega}{\Omega}=-\left(\frac{\sqrt{5}+1}{2\sqrt{5}}\right)^{1/4}~,
\eeq
where we emphasized that the same quantity gives the normalization of the vacuum in AdS. This normalization is irrelevant to the problem of finding the spectrum of the Hamiltonian, and we shall find it convenient to remove this factor from our basis of states, multiplying each one by $\brakket{\Omega}{\Omega}^{-1/2}$.\footnote{The fact that this normalization is purely imaginary is not an issue. We simply define the scalar product to be linear rather than sesquilinear, so that the vacuum is now unit normalized.} Then, the expectation value of $\mca{V}$ in AdS, with unit normalized vacuum, is
\beq
\expec{\mca{V}}=(2R)^{-\Delta_\mca{V}}\frac{\brakket{\mca{V}}{\mathbb{1}}_\textup{Cardy}}{\brakket{\Omega}{\Omega}}=-(2R)^{-\Delta_\mca{V}}\left(\frac{2}{1+\sqrt{5}}\right)^{1/2}
\equiv -R^{-\Delta_\mca{V}} a_\mca{V}~.
\eeq
For the matrix elements of the potential to be finite, the one-point function of $\mca{V}$ must be fine tuned in the Hamiltonian. We therefore perturb the conformal fixed point with the following coupling:
\beq
\label{MinimalV}
\bar{\lambda} V=\bar{\lambda}  \int_{-\pi/2}^{\pi/2} \frac{dr}{(\cos r)^2}\, 
\left(\frac{R^{\DD_\mca{V}}}{a_\mca{V}}\mca{V}(\tau=0,r)+1\right)~.
\eeq
The theory can be studied for both signs of the coupling. The expectation from the flow in flat space \cite{Yurov:1989yu} is that $\bar{\lambda}>0$ leads to a well defined flat space limit. This will be confirmed by the numerics.

The Hilbert space of the model at the conformal fixed point consists of the states \eqref{LchiStates}, where the primary $\ket{\chi}=\ket{\Omega}$ is created by the identity operator on the boundary. The first few states and their degeneracies can be found in the following table:
\begin{center}
\begin{tabular}{*{12}{c}}
\toprule
$\Delta$  & 0 & 1 & 2  & 3 & 4 & 5 & 6 & 7 & 8 & 9 & 10 \\
\midrule
states & 1 & 0 & 1 & 1 & 1 & 1 & 2 & 2 & 3 & 3 & 4  \\
$sl(2)$ primaries & 1 & & 1 & & & & 1 & & 1 & & 1  \\
\bottomrule
\end{tabular}
\end{center}

In the flat space limit, the Lee-Yang model flows to a massive theory, whose spectrum includes a single stable particle. Since the flat space flow is integrable, the mass is computable as a function of the scale set in the UV \cite{Cardy:1989fw,Zamolodchikov:1995xk}. Recalling that $\bar{\lambda}=\lambda R^{2-\Delta_\mca{V}}$, we approach the flat space limit by taking $R\to \infty$ after the rescaling $(\tau,r)\to (\tau,r)/R$. The flat space coupling is then found to be $\lambda/a_\mca{V}$. The mass of the stable particle is expressed in term of the coupling as follows \cite{Zamolodchikov:1995xk}:
\beq
m=\kappa \left(\frac{\lambda}{a_\mca{V}}\right)^{\frac{1}{2-\Delta_\mca{V}}}~, \qquad
\kappa=2^{19/12}\sqrt{\pi} \frac{\left(\Gamma(3/5)\Gamma(4/5)\right)^{5/12}}{5^{5/16}\Gamma(2/3)\Gamma(5/6)}=2.643\dots
\eeq
Plugging this relation in eq. \eqref{eq:midef}, we obtain a prediction for the slope of the energy levels of single particle states in the flat space limit:
\beq
\frac{\Delta_\textup{single particle}}{\bar{R}} \underset{\bar{R}\to\infty}{\longrightarrow} \frac{\kappa}{a_\mca{V}^{\frac{1}{2-\Delta_\mca{V}}}} = 2.603\dots
\label{LYslopeflat}
\eeq
Here we used the definition of the AdS radius in units of the coupling, which in this case reads
\beq
\bar{R}=R \lambda^{\frac{1}{2-\Delta_\mca{V}}}=R \lambda^{5/12}~.
\eeq
Eq. \eqref{LYslopeflat} is useful because it allows to estimate the finite size corrections in AdS. As we shall see, the non unitary nature of this model hinders our ability to reach large values of the radius within our computational limits.

\subsubsection{Cutoff effects}\label{sec:LYConv}

\begin{figure}[!h]
     \centering
     \begin{subfigure}{0.47\textwidth}
         \centering
         \includegraphics[width=\textwidth]{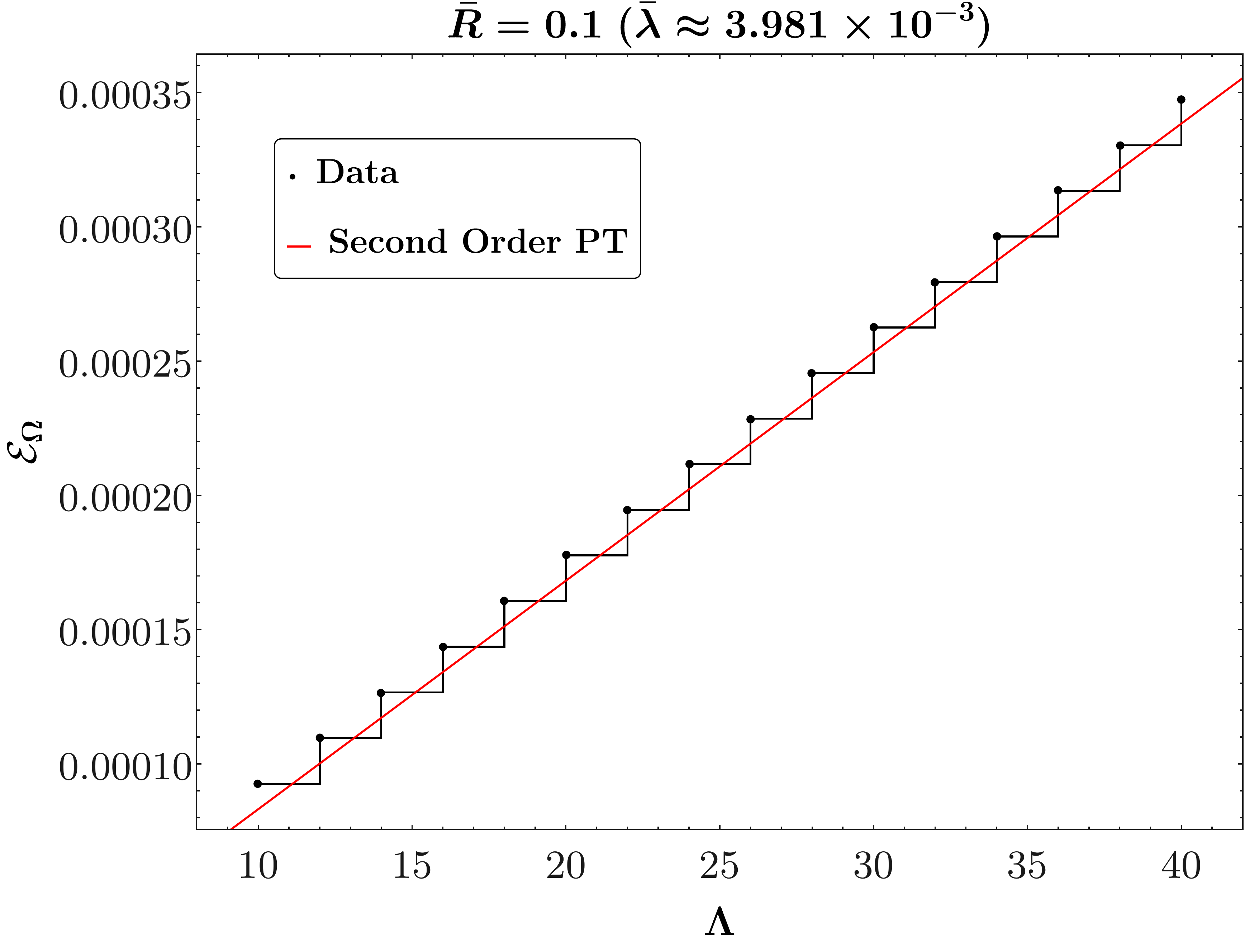}
     \end{subfigure}
     \hfill
     \begin{subfigure}{0.44\textwidth}
         \centering
         \includegraphics[width=\textwidth]{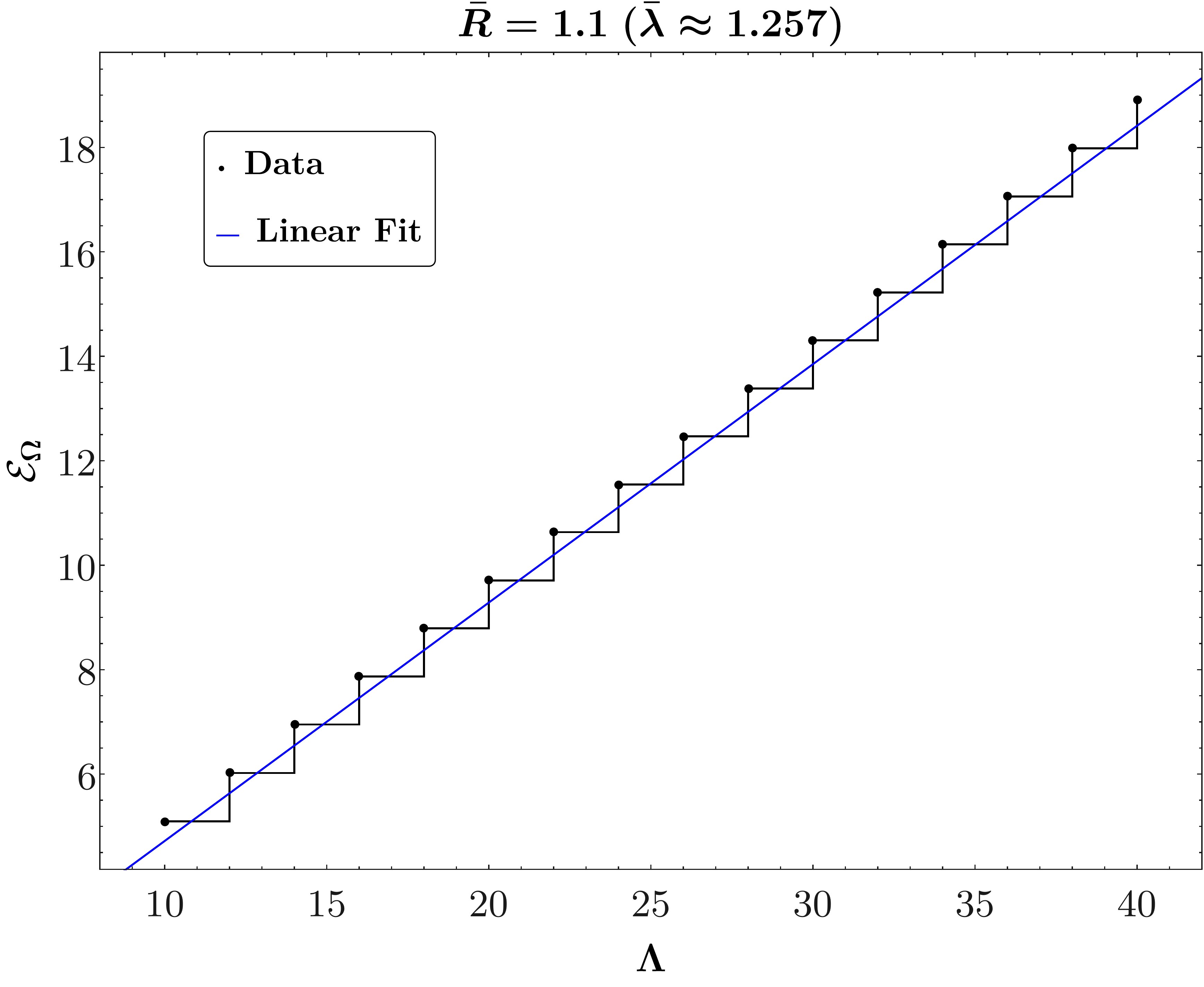}
     \end{subfigure}
\caption{Vacuum energy as a function of the cutoff in the Lee-Yang model. The black dots are the Hamiltonian truncation data: since the vacuum is parity even, the Hamiltonian needs to be diagonalized anew only when $\Lambda$ jumps by an even integer. In the left plot, the solid red line is eq. \eqref{slopeLY2nd}. In the right plot, the solid blue line is a fit to the data.}
     \label{fig:LYVacuumEnergy}
 \end{figure}

As with every QFT in AdS, the vacuum energy is not observable in the continuum limit: it diverges linearly with the cutoff $\Lambda$, as showcased in figure \ref{fig:LYVacuumEnergy}. However, we explained in section \ref{subsec:VacSecond} that the vacuum energy is computable in the Hamiltonian truncation framework. In particular, we can compare the function $\mca{E}_\Omega(\Lambda)$ at small $\bar{\lambda}$ with its value at second order in perturbation theory. We recall from section \ref{subsec:VacSecond} that the linear divergence is given by
\beq
\mca{E}_\Omega(\Lambda) \underset{\Lambda\to\infty}{\longrightarrow} c \left(\frac{\bar{\lambda}}{a_\mca{V}}\right)^2 \Lambda + O(\bar{\lambda}^3)~, \qquad
c=-8\int_0^\infty \!d\xi\, \textup{arcsinh} \left(2 \sqrt{\xi(\xi+1)}\right) F(\xi)~,
\label{lindivLY2nd}
\eeq
where $F(\xi)$ is the connected two-point function of the perturbation:
\beq
F(\xi) = R^{2\Delta_\mca{V}}\expec{\mca{V}(\tau,r)\mca{V}(0,0)}-a_\mca{V}^2~,
\eeq
with the cross ratio $\xi$ defined in eq. \eqref{eq:crossratio}. The correlation function is easily written as a sum of two Virasoro blocks, corresponding to the fusion rule $\mca{V}\times \mca{V}=\mathbb{1}+\mca{V}$, see \emph{e.g.} \cite{Dorey:1999cj}:
\beq
F(\xi) = -f_\mathbb{1}(\xi)+\frac{2}{1+\sqrt{5}}\frac{\Gamma(1/5)\Gamma(6/5)}{\Gamma(3/5)\Gamma(4/5)}f_\mca{V}(\xi)-a_\mca{V}^2~,
\eeq
with
\beq
f_\mathbb{1}(\xi) = \left(\frac{4\xi}{1+\xi}\right)^{2/5} {}_2 F_1 
\left(\frac{3}{5},\frac{4}{5},\frac{6}{5},\frac{\xi}{1+\xi}\right)~,\qquad
f_\mca{V}(\xi)=\left(\frac{16\xi}{1+\xi}\right)^{1/5} {}_2 F_1 
\left(\frac{2}{5},\frac{3}{5},\frac{4}{5},\frac{\xi}{1+\xi}\right)~.
\eeq
Performing the integration in eq. \eqref{lindivLY2nd} numerically, we obtain
\beq
c=0.5786\dots~, \qquad \mca{E}_\Omega(\Lambda) \underset{\Lambda\to\infty}{\longrightarrow} 0.5377\, \bar{\lambda}^2 \Lambda + O(\bar{\lambda}^3)~.
\label{slopeLY2nd}
\eeq
The left panel of figure \ref{fig:LYVacuumEnergy} shows that eq. \eqref{slopeLY2nd} does indeed fit the data accurately at weak coupling. In the right panel, we see that the behavior is still linear when the coupling is order one, as expected.
 
\begin{figure}[!h]
     \centering
     \begin{subfigure}{0.47\textwidth}
         \centering
         \includegraphics[width=\textwidth]{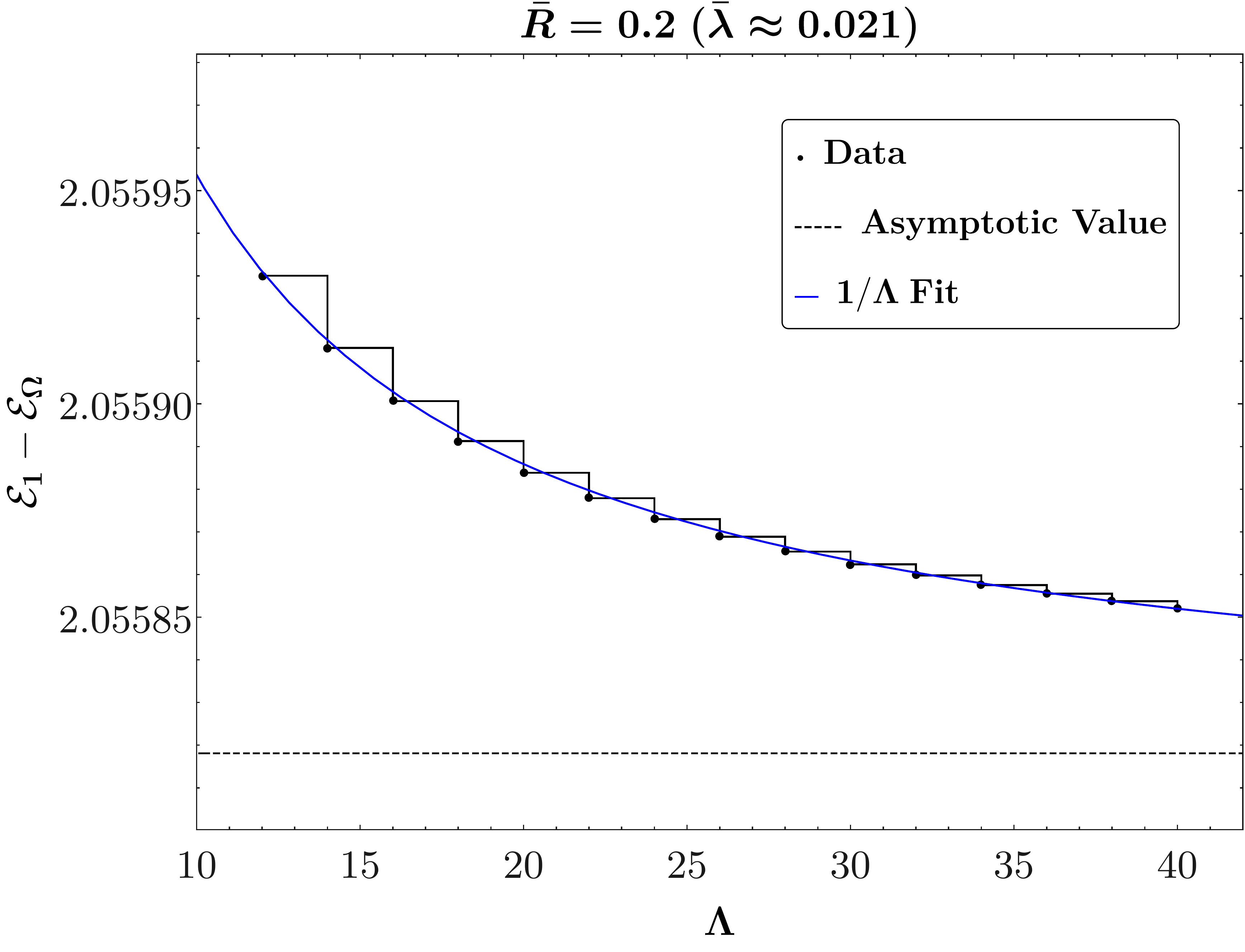}
     \end{subfigure}
     \hfill
     \begin{subfigure}{0.45\textwidth}
         \centering
         \includegraphics[width=\textwidth]{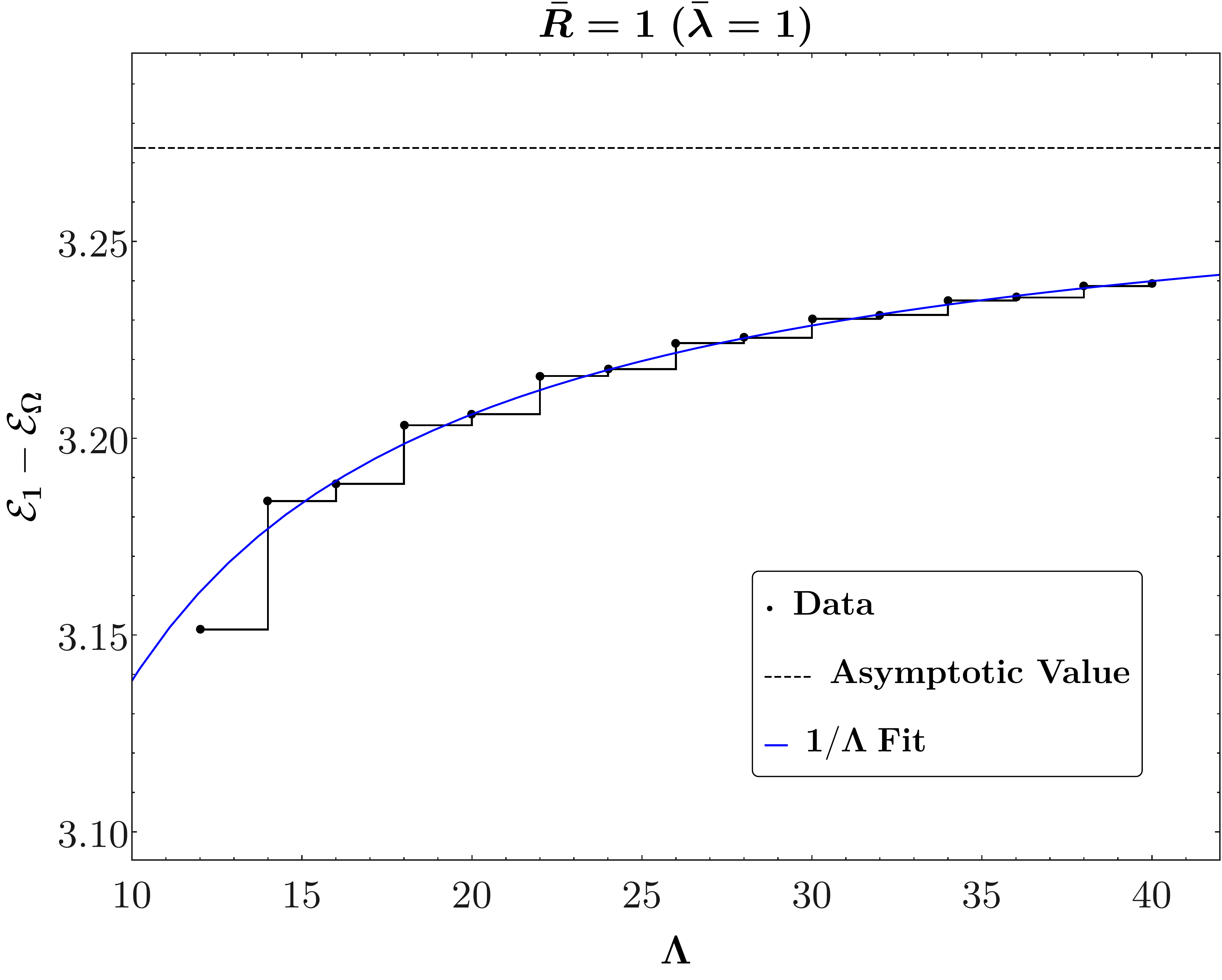}
     \end{subfigure}
               \caption{Energy of the first excited state in the Lee-Yang model, as a function of the cutoff at weak and strong coupling. Notice that the first excited state is parity even, hence again only points were $\Lambda$ is even need to be computed. The asymptotic value is obtained from the fit.}
              \label{fig:convergenceLY}
\end{figure}

Since the lightest operator in the spectrum is the displacement operator, with $\Delta=2$, the rate of convergence of the truncated spectrum to the exact values is $\Lambda^{-1}$ -- see section \ref{sec:rate}. This can be checked up to couplings of order one, as shown in figure \ref{fig:convergenceLY}.

\begin{figure}[!h]
     \centering
         \includegraphics[width=0.8\textwidth]{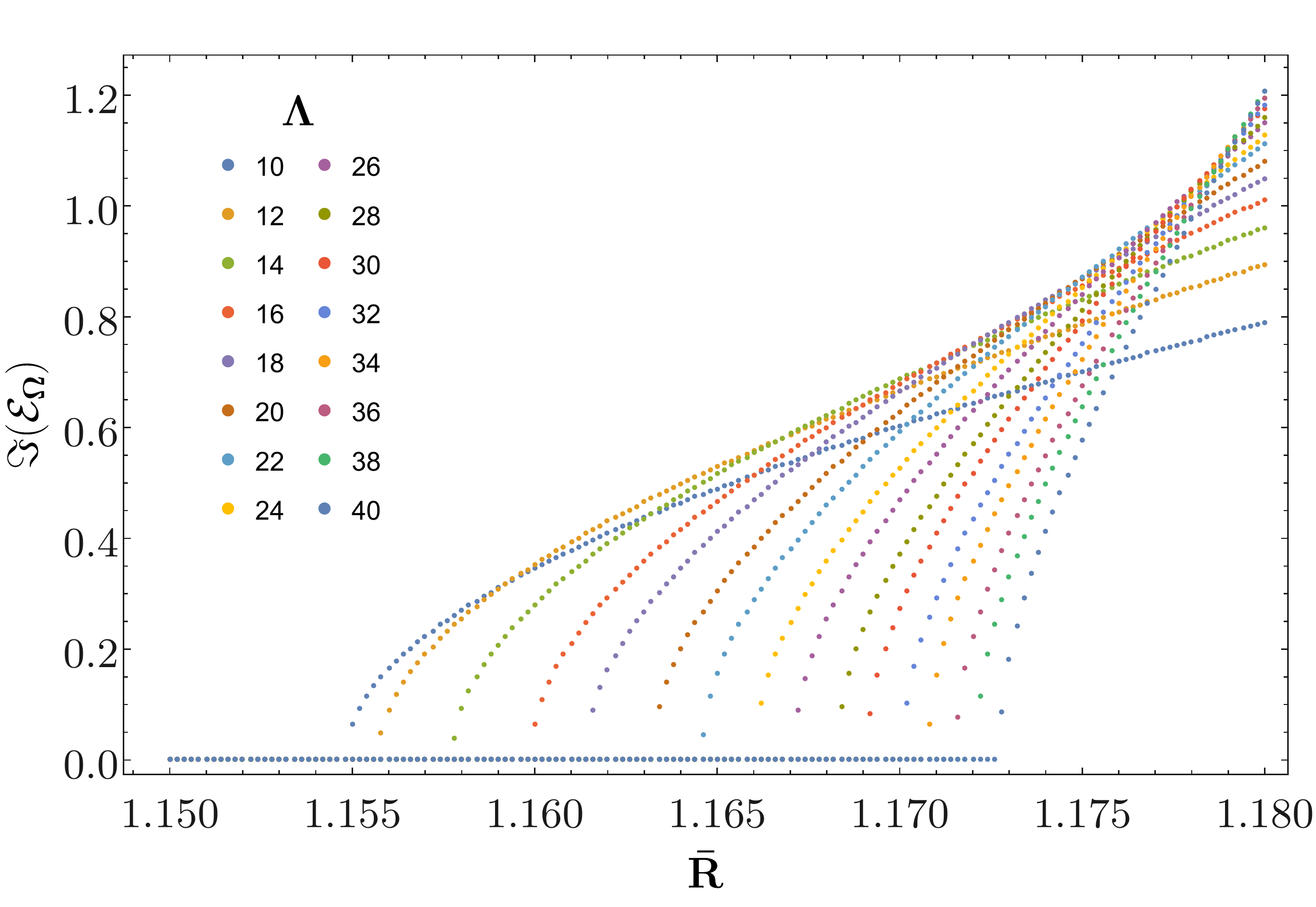}
         \caption{Imaginary part of the vacuum energy in the Lee-Yang model as a function of the radius of AdS, for different cutoffs.}
         \label{fig:LYImVacuum}
\end{figure}

 On the other hand, the Lee-Yang model presents a specific obstruction to the computation of the spectrum for large values of the coupling. Since the Hamiltonian is not Hermitian, its eigenvalues are not guaranteed to be real at any finite value of the cutoff. In the continuum limit, the spectrum is real in the flat space limit, and it is real at the conformal fixed point. Our working hypothesis is that the Lee-Yang model is well defined in the continuum in AdS for any value of the radius. However, for finite $\Lambda$ the eigenvalues do pick up an imaginary part when the radius is larger than a certain value $\bar{R}^*$. For the vacuum energy, this can be seen in figure \ref{fig:LYImVacuum}.
 
\begin{figure}[!h]
         \centering
         \includegraphics[width=0.55\textwidth]{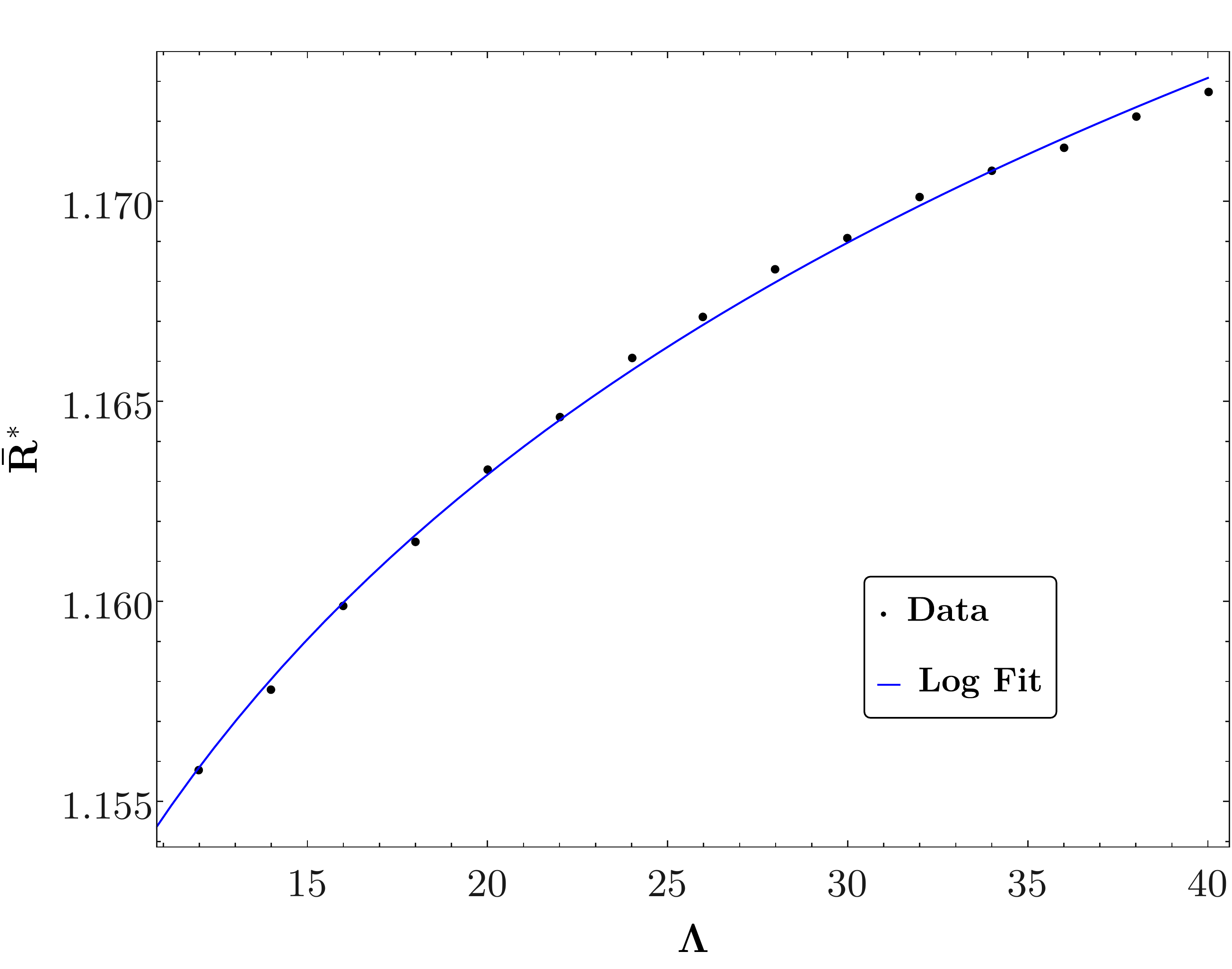}
         \caption{$\bar{R}^*$ is the value of the coupling beyond which the vacuum energy in the Lee-Yang model becomes complex. The logarithmic fit is phenomenological, and the best fitting function is $\bar{R}^*=1.12 + 0.01 \log(\Lambda)$. Notice that with the available range of values for $\Lambda$ it is hard to distinguish, for instance, between a logarithm and a power law with a sufficiently small exponent.}
         \label{fig:LYlambdastar}
\end{figure}

In order to provide evidence for the reality of the spectrum in AdS, we explored the dependence of $\bar{R}^*$ on the cutoff. Figure \ref{fig:LYlambdastar} shows that a logarithmic fit describes reasonably well the data in our possession. If this behavior is valid at asymptotically large values of $\Lambda$, the Lee-Yang model does indeed possess real energy levels in the continuum limit. However, the computational effort to reach this regime is severe: since the number of states grows exponentially with the cutoff, the spectrum cannot be real for a given value $\bar{R}$ unless we include a number of states of the order $\sim \exp(\alpha\exp (\beta\bar{R}))$, for some positive constants $\alpha$ and $\beta$.

\subsubsection{The spectrum}
\begin{figure}[!h]
         \centering
         \includegraphics[width=\textwidth]{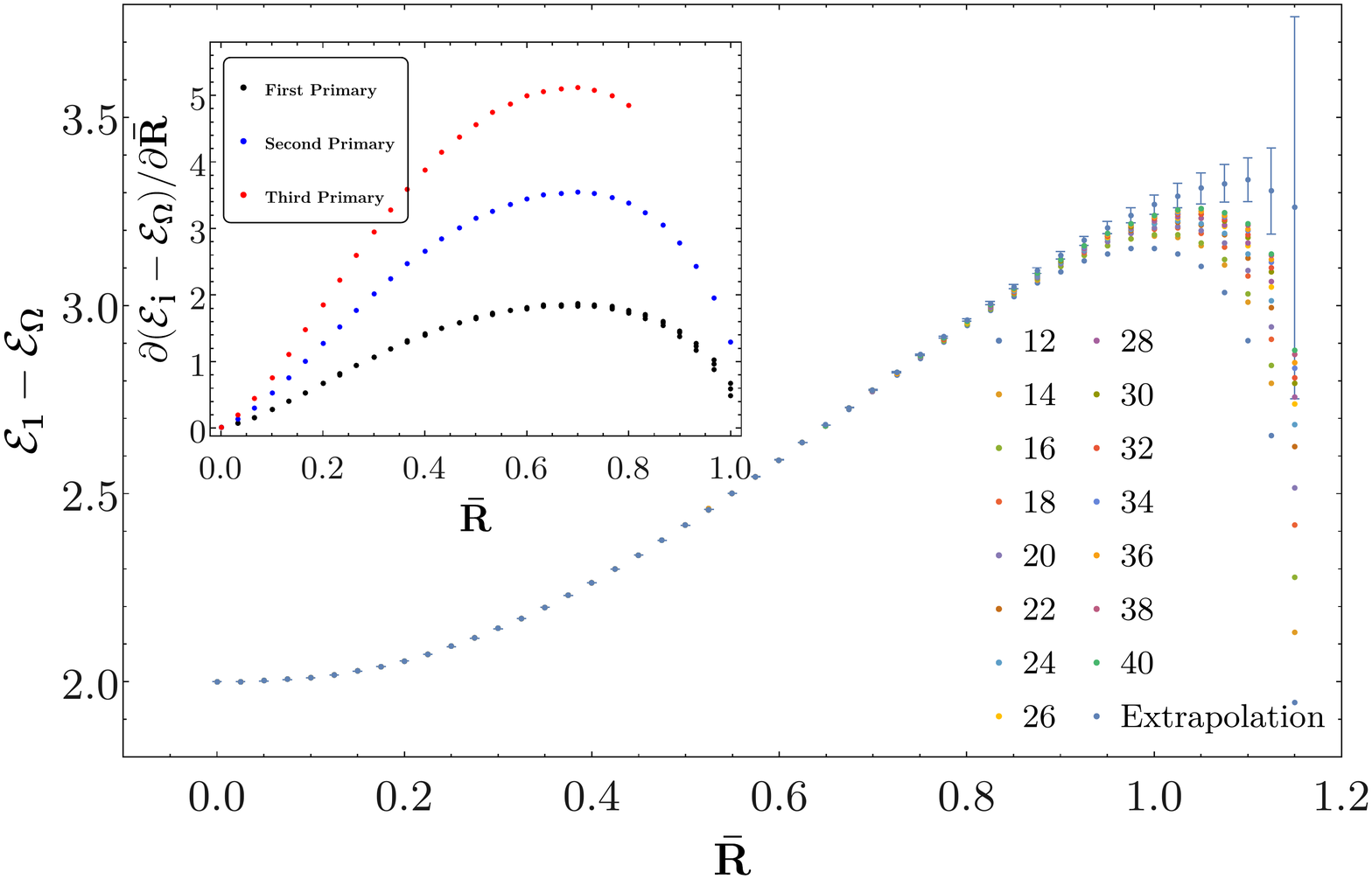}
         \caption{Main plot: energy of the first excited state in the Lee Yang model, as a function of the radius of AdS, for various choices of cutoff. The error bars are computed according to the procedure explained in subsection \ref{sec:VaryingGPhi2}. In the inset, the derivative of the gap is computed discretely, by measuring the gap at two nearby values of $\bar{R}$.} 
         \label{fig:LYExtp}
\end{figure}

Figure \ref{fig:LYExtp} shows the energy of the first excited state as a function of the radius of AdS. At small radius, the slope of the curve is fixed by perturbation theory. The first excited state starts its life as the displacement operator, with dimension $\Delta=2$, and deviates from that value as $\sim R^{2-\Delta_\mca{V}}=R^{12/5}$. On the other hand, at radii of order 1 in units of the coupling, the truncation effects become big, and our extrapolation to the continuum limit correspondingly poor. This is related to the appearance of complex energies described in the previous subsection. Since the Hamiltonian is real, complex eigenvalues can only arise in pairs, and the vacuum energy in particular can only become complex after colliding with the first excited state. This is clearly visible in figure \ref{fig:LYExtp}. Finally, the bulk of the plot contains genuine non perturbative data. The extrapolation to infinite cutoff is reliable, and the curve can be trusted. Although the energy of the first excited state is order one, we may be tempted to recognize a region where the dependence on the radius is approximately linear, see eq. \eqref{eq:midef}. To test whether this is the case, in the inset of figure \ref{fig:LYExtp} we plot the derivative of the energy of a few states with respect to the radius of AdS. The curves do not show a pronounced plateau, which casts doubts on the fact that flat space physics is already visible in the plot. A quantitative check in this direction is provided by comparing the maximal slope of the first excited state, about $\sim 1.85$, with the flat space prediction $\sim 2.60$ in eq. \eqref{LYslopeflat}. The inset also reports the slope of the energy levels which in the flat space limit we can tentatively associate to two and three particle states, with all particles at rest. The peaks in this case lie at about $\sim 3.5$ and $\sim 5.1$ respectively. The ratios of these slopes to the slope of the first excited state are 1.9 and 2.8 respectively, which are not too far from the value 2 and 3 expected in the flat space limit.\footnote{Notice that the deviation of the energies of the multiparticle states from the sum of their constituents is a measure of the strength of the (non gravitational) interactions, and they do not need to be of the same order as the deviation of the energy of the single particle state from its value in the flat space limit. For instance, in the case of a free scalar the binding energies are exactly zero for any value of the radius.} 

\begin{figure}[!h]
         \centering
         \includegraphics[width=\textwidth]{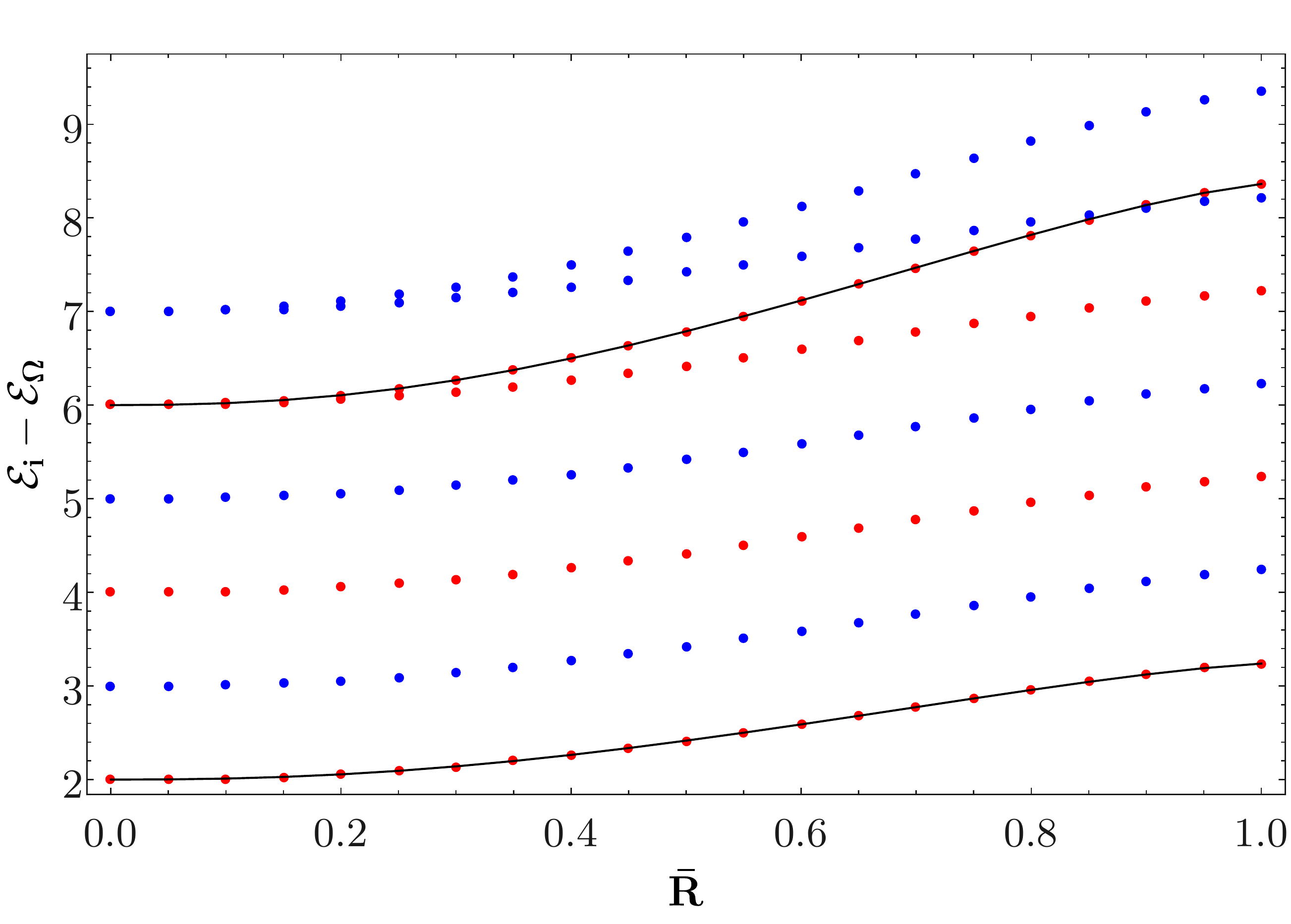}
         \caption{Spectrum of the Lee-Yang model. All the data points are computed with cutoff $\Lambda=40$. To avoid clutter, we did not plot the extrapolated data with their error bars, although the latter would be pretty small up to $\bar{R}\sim0.8$ as in figure \ref{fig:LYExtp}. As discussed in text, we cannot trust the data beyond $\bar{R}\sim 1$, due to large truncation effects.}
         \label{fig:LYSpectrum}
\end{figure}

Figure \ref{fig:LYSpectrum} shows the dependence of the energy of the first few excited states on the AdS radius. We highlighted in red the parity even and in blue the parity odd eigenstates of the Hamiltonian, recognizable from their CFT ancestors according to eq. \eqref{parityVir}. As it can be checked via the character of the vacuum module, there are two quasi-primaries in the Lee-Yang CFT in the range of scaling dimensions shown in figure \ref{fig:LYSpectrum}: the displacement operator at $\Delta=2$ and a new quasi-primary at $\Delta=6$. These are marked with a solid line in the figure. Correspondingly, all states up to level 5 are non-degenerate, while the states at level 6 and 7 are two-fold degenerate. The quasiprimary state starting at level 6 is on its way to become a two-particle state in the flat space limit. The connection between quasiprimaries in the BCFT and multiparticle states in the flat space limit has been explained in section \ref{sec:pheno}: since in the continuum the $SL(2,\mbb{R})$ structure forces smooth level crossing between a quasiprimary and the descendants of a different conformal family, we can track the quasiprimary state all the way from its origin in the BCFT to the flat space limit.

\begin{figure}[!h]
         \centering
         \includegraphics[width=\textwidth]{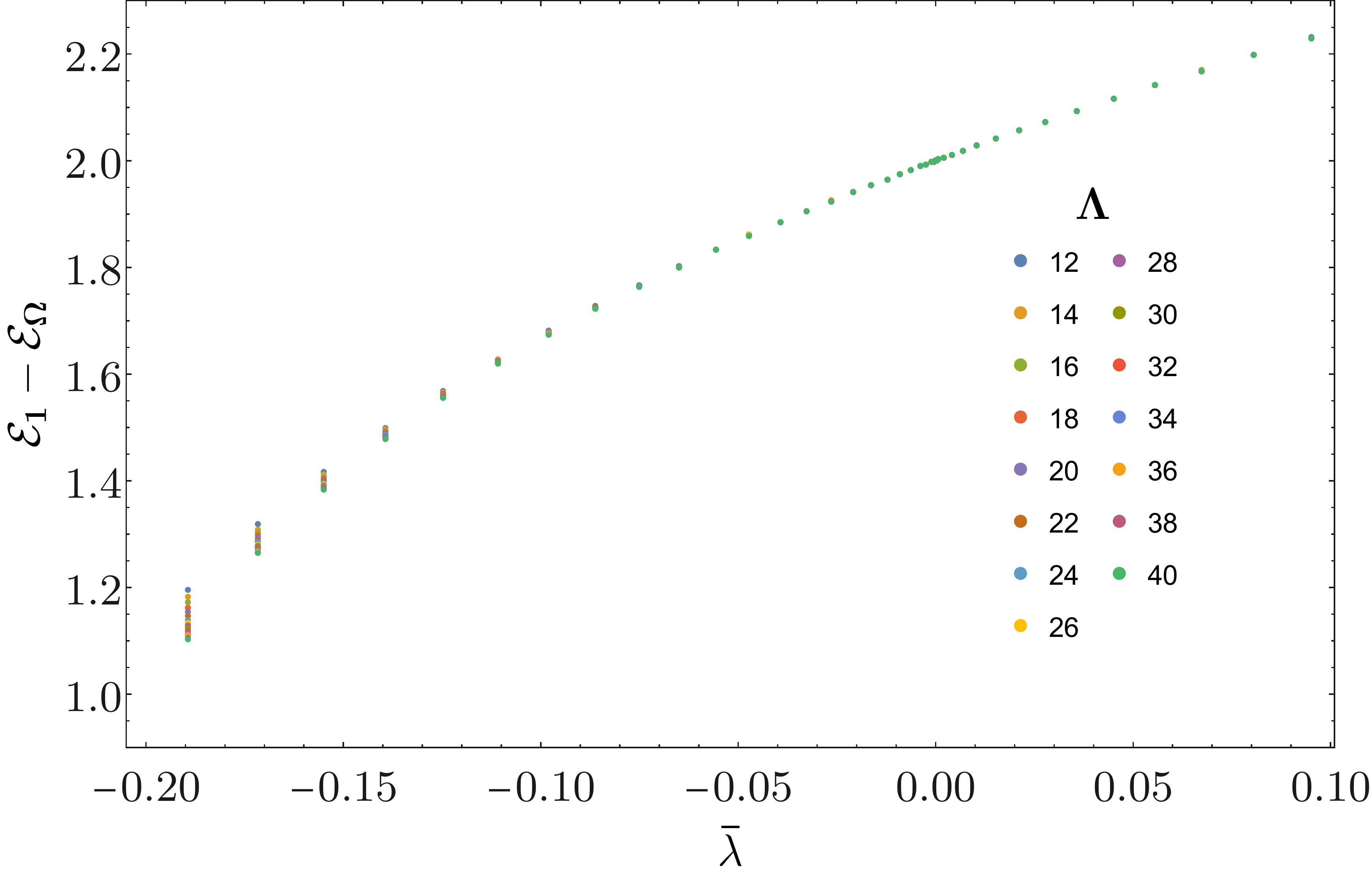}
         \caption{First excited state energy in the Lee-Yang model in a range of couplings around $\bar{\lambda}=0$. For the values of $\bar{\lambda}$ towards the left of the plot, the convergence with the cutoff $\Lambda$ is very slow, and does not allow to extrapolate the value $\bar{\lambda}_\textup{min}$ defined in the main text.} 
         \label{fig:LYNegRbar}
\end{figure}

Finally, figure \ref{fig:LYNegRbar} shows the gap as a function of $\bar{\lambda}$, for an interval extending to negative values. The energy of the first excited state is monotonic with $\bar{\lambda}$, and we expect that for a certain value $\bar{\lambda}_\textup{min}$ of the coupling a new boundary operator with dimension $\Delta=1$ emerges. We cannot reliably compute the spectrum for $\bar{\lambda}$ close to $\bar{\lambda}_\textup{min}$, since the truncation errors become large, as it is visible towards the left of figure \ref{fig:LYNegRbar}. Nevertheless, we can speculate that $\bar{\lambda}_\textup{min}$ is the minimal value of the coupling for which the spectrum in the continuum limit is real. Indeed, when the bulk perturbation turns on a marginal coupling, the latter cannot be fine tuned, and generically a fixed point exists only on one side of  $\bar{\lambda}_\textup{min}$. In fact, a natural scenario is that when $\bar{\lambda}=\bar{\lambda}_\textup{min}$ two fixed points collide and annihilate. In this case, the only other available fixed point is obtained by deforming the Lee-Yang CFT with the coupling \eqref{MinimalV}, but this time starting from the boundary condition captured by the Cardy state $\ket{\mca{V}}_\textup{Cardy}$ \eqref{VCardyLY}. This suggests a scenario analogous to the structure of boundary conditions of a free scalar with negative mass squared in AdS. We shall further discuss this hypothesis in section \ref{sec:disc}. For the moment, let us conclude by pointing out an obvious consequence of these considerations: the theory defined by deforming the Lee-Yang CFT with a coefficient $\lambda<0$ does not admit a flat space limit. On the contrary, it features a maximum radius $R_\textup{max} = (\bar{\lambda}_\textup{min}/\lambda)^{5/12}$.

\subsection{The Ising model}
\label{subsec:ising}

The Ising model is the diagonal minimal model $\mca{M}(4,3)$ with central charge
\beq
c=\frac{1}{2}~.
\eeq
Its field content consists of three Virasoro families:
\beq
\Delta_\mathbb{1}=0~, \qquad \Delta_\sigma=\frac{1}{8}~, \qquad
\Delta_\epsilon=1~.
\eeq
The spin operator $\sigma$ is odd under the $\mathbb{Z}_2$ symmetry of the model, while the  energy operator $\epsilon$ is even. The Ising model enjoys a Kramers-Wannier duality \cite{Kramers:1941kn} which flips the sign of the $\eps$ operator and replaces the $\sigma$ operator with its disorder counterpart $\mu$. 
The most general bulk RG emanating from the 2d Ising model which preserves the AdS isometries is given by the action
\beq
\label{eq:IsingFlow}
S = S_\text{Ising CFT} - \la_\epsilon\,  \int_{\mrm{AdS}}\!\sqrt{g}\,d^2x \; \epsilon(x) -  \la_\sigma\, \int_{\mrm{AdS}}\!\sqrt{g}\,d^2x \; \sigma(x) + \la_\mathbb{1}\, \int_{\mrm{AdS}}\!\sqrt{g}\,d^2x~,
\eeq
where $\la_\epsilon$ (resp.\@ $\la_\sigma$) has mass dimension $1$ (resp.\@ $15/8$) -- there are no other relevant scalar operators in the theory. The signs in front of $\la_\epsilon$ and $\la_\sigma$ are chosen for later convenience. As before, the cosmological constant $\la_\mathbb{1}$ must be tuned to make the theory finite. Physically, $\la_\epsilon$ and $\la_\sigma$ can be interpreted as a temperature $T-T_c$ resp.\@ a magnetic field. The action~\reef{eq:IsingFlow} is sometimes referred to as the \emph{Ising field theory}. 

In the following sections, we will study the action~\reef{eq:IsingFlow} in AdS${}_2$, picking a definite boundary state. Let us first discuss the phenomenology of~\reef{eq:IsingFlow} in flat space, where it has been studied in detail. Notice that the long-distance physics of~\reef{eq:IsingFlow} is controlled by the dimensionless coupling $\eta = \la_\epsilon/|\la_\sigma|^{8/15}$, where the absolute value is a consequence of the $\mbb{Z}_2$ symmetry. Due to Kramers-Wannier duality, $\eta$ parametrizes the projective line: the points $\eta = \pm \infty$ are identified.
There are two integrable points in the phase diagram: $\eta = 0$ ($\la_\eps = 0$), called the \emph{magnetic} deformation, and $\eta = \infty$ ($\la_\sigma = 0$), the \emph{thermal} deformation. The thermal deformation is identical to the theory of a Majorana fermion with mass $m = 2\pi \la_\epsilon$, provided that $\epsilon$ has the canonical CFT normalization $\expec{\eps(x)\eps(0)} = 1/|x|^2$. Surprisingly, the magnetic deformation can be exactly solved as well~\cite{Zamolodchikov:1989fp,Zamolodchikov:1989hfa}, and its particle content is known. For generic $\eta$, the theory in flat space or on the cylinder cannot be solved exactly, but a quantitative understanding of the spectrum has been obtained in the series of papers~\cite{Fonseca:2001dc,Fonseca:2006au,Zamolodchikov:2011wd,Zamolodchikov:2013ama, Gabai:2019ryw}, some of which use variants of Hamiltonian truncation on the cylinder $\mbb{R} \times S^1$. Finally, the thermal deformation of the Ising model in AdS${}_2$ was studied in Ref.~\cite{Doyon:2004fv}.

Let us now turn to the definition of the model~\reef{eq:IsingFlow} in AdS. At the conformal fixed point, as explained above, the relevant boundary conditions are labeled by the three primaries of the model:
\begin{subequations}
\label{CardyIsing}
\begin{align}
&\ket{\mathbb{1}}_\textup{Cardy} =   \frac{1}{2^{1/2}} \dket{\mathbb{1}}+\frac{1}{2^{1/2}} \dket{\eps}+\frac{1}{2^{1/4}} \dket{\sigma}~, \label{CardyIsingId} \\
&\ket{\eps}_\textup{Cardy} =  \frac{1}{2^{1/2}} \dket{\mathbb{1}}+\frac{1}{2^{1/2}} \dket{\eps}-\frac{1}{2^{1/4}} \dket{\sigma}~,   \\
&\ket{\sigma}_\textup{Cardy} =  \dket{\mathbb{1}}- \dket{\eps}~. \label{CardyIsingSi}
\end{align}
\end{subequations}
The boundary conditions labeled by $\mathbb{1}$ and $\eps$ are mapped to each other by a spin flip: they are fixed boundary conditions for the microscopic degrees of freedom. The spectrum supported on both boundary conditions only includes the identity module. The $\sigma$ Cardy state, which is invariant under $\mathbb{Z}_2$, corresponds to free boundary conditions. Its spectrum contains the identity module and the $\mathbb{Z}_2$ odd module labeled by $\Delta=1/2$, which can be identified with the family of states with an odd number of particles in the free fermion description. 

In the critical model, Kramers-Wannier duality can be implemented by a topological defect $D_\sigma$ \cite{Frohlich:2004ef}, whose action on the boundary states teaches us how the boundary conditions are mapped to each other:
\beq
D_\sigma \ket{\mathbb{1}}_\textup{Cardy} = D_\sigma \ket{\eps}_\textup{Cardy} = \ket{\sigma}_\textup{Cardy}~,
\qquad D_\sigma \ket{\sigma}_\textup{Cardy}= \ket{\mathbb{1}}_\textup{Cardy}+ \ket{\eps}_\textup{Cardy}~.
\label{KramersWannierIsing}
\eeq
As we shall discuss in a moment, the action of the duality is useful in understanding the RG flow triggered by deforming the theory in AdS by the $\epsilon$ operator.

We shall perform a Hamiltonian truncation study of the RG flows emanating from the critical Ising model with $\mathbb{1}$ boundary condition. Here are the first few states and their degeneracies:
\begin{center}
\begin{tabular}{*{12}{c}}
\toprule
$\Delta$  & 0 & 1 & 2  & 3 & 4 & 5 & 6 & 7 & 8 & 9 & 10 \\
\midrule
states & 1 & 0 & 1 & 1 & 2 & 2 & 3 & 3 & 5 & 5 & 7  \\
$sl(2)$ primaries & 1 & & 1 & & 1 & & 1 & & 2 & & 2  \\
\bottomrule
\end{tabular}
\end{center}
We shall comment along the way about the other boundary conditions. Once we normalize the vacuum state to one, the expectation values of the deforming operators are
\begin{align}
&\expec{\mca{\eps}}=(2R)^{-\Delta_\mca{\eps}}\frac{\brakket{\eps}{\mathbb{1}}_\textup{Cardy}}{\brakket{\Omega}{\Omega}}=\frac{1}{2R}~, \label{epsVev}\\
&\expec{\mca{\sigma}}=(2R)^{-\Delta_\mca{\sigma}}\frac{\brakket{\sigma}{\mathbb{1}}_\textup{Cardy}}{\brakket{\Omega}{\Omega}}=\frac{2^{1/4}}{(2R)^{1/8}}~. \label{sigmaVev}
\end{align}
These values determine $\lambda_\mathbb{1}$ in eq. \eqref{eq:IsingFlow}, as a function of $\lambda_\sigma$ and $\lambda_\eps$.

\subsection{The Ising model with thermal deformation}
\label{subsec:IsingT}

In this section, we will consider the purely thermal deformation, setting $\lambda_\sigma = 0$ in eq. \eqref{eq:IsingFlow}. Keeping into account eq. \eqref{epsVev}, the potential reads
\beq
\label{MinimalVEps}
\bar{\lambda} V=-\bar{\lambda}  \int_{-\pi/2}^{\pi/2} \frac{dr}{(\cos r)^2}\, 
\left(R\,\eps(\tau=0,r)-\frac{1}{2}\right)~,
\eeq
with $\bar{\lambda}= R \la_\eps$.
This flow is exactly solvable, since, as advertised, it is described by a free Majorana fermion in AdS, with mass $m = 2\pi \la_\eps$. In fact, there is hardly the need to perform any computation in order to describe the spectrum. Let us start from the critical theory. A free Majorana fermion on the upper half plane admits two boundary conditions, $\psi=\nu \bar{\psi}$, with $\nu=\pm 1$. The chiral flip $(\psi,\bar{\psi}) \to (-\psi,\bar{\psi})$, which corresponds to the Kramers-Wannier duality of the Ising model, flips the sign of the mass and swaps the boundary conditions. Hence, the theory in AdS is only allowed to depend on the product $m\nu$. Furthermore, leading order perturbation theory is exact, so the trajectories $\mathcal{E}(m R)$ must be linear. Finally, looking at the sign of $m$ which admits a flat space limit, we conclude that the slope of the first excited state is precisely the fermion mass: 
\beq
\mathcal{E}=m \nu R+\frac{1}{2}~, \qquad \nu = \pm 1~.
\label{FFspecPm}
\eeq
The constant term is just the scaling dimension of a free massless fermion. The rest of the spectrum is easily obtained by the usual Fock space construction. At the critical point, a fermion in mode $n$ has energy
\begin{equation}
e_f = \frac{1}{2}+n~,
\end{equation}
and following the Pauli exclusion principle, no pair of fermions share the same mode. Therefore, the energy of a state with $N$ fermions is
\begin{equation}
\mathcal{E}_N=N m \nu R+\frac{N}{2}+\sum_{i=1}^{i=N}n_i~, \qquad n_i\neq n_j
\label{FFspecN}
\end{equation}
where $n_i$ is a non negative integer: $n_i\in\mathbb{N}_0$. 

In fact, the argument above fixes eq. \eqref{FFspecPm} only up to a sign, but we can define the sign in front of $m$ in the Hamiltonian so as to obtain eq. \eqref{FFspecPm} as written. Of course, the sign might be fixed by computing the leading (and only) order in Rayleigh-Schr\"{o}dinger (RS) perturbation theory. Further details about the quantization of the free Majorana fermion in AdS$_2$ can be found in \cite{Doyon:2004fv}. Eq. \eqref{FFspecPm} determines the spectrum of the Ising model with any of the boundary states \eqref{CardyIsing}. Before comparing this prediction with the result of Hamiltonian truncation, let us briefly discuss the cutoff effects.

\subsubsection{Cutoff effects}

\begin{figure}[!h]
     \centering
     \begin{subfigure}{0.47\textwidth}
         \centering
         \includegraphics[width=\textwidth]{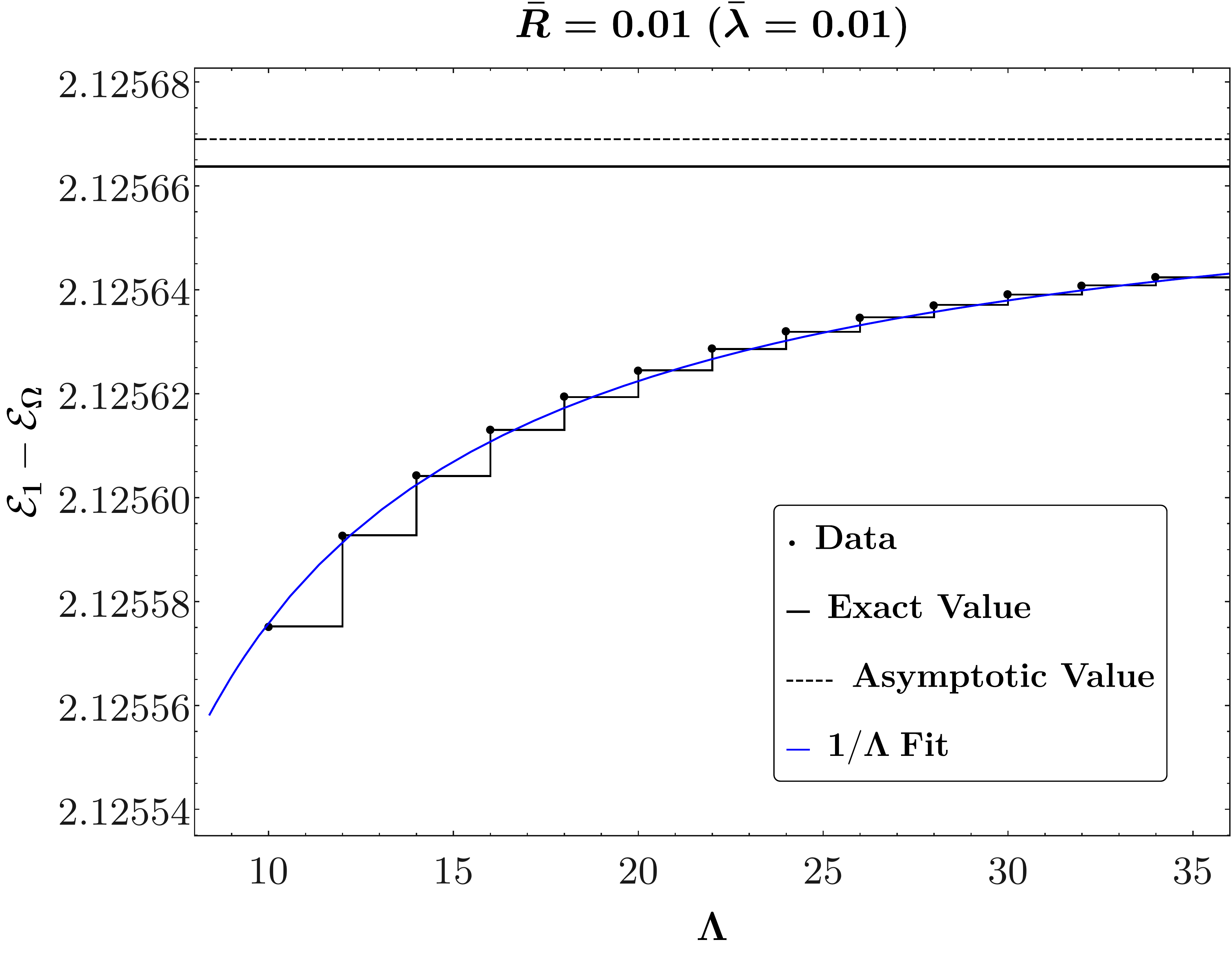}
     \end{subfigure}
     \hfill
     \begin{subfigure}{0.445\textwidth}
         \centering
         \includegraphics[width=\textwidth]{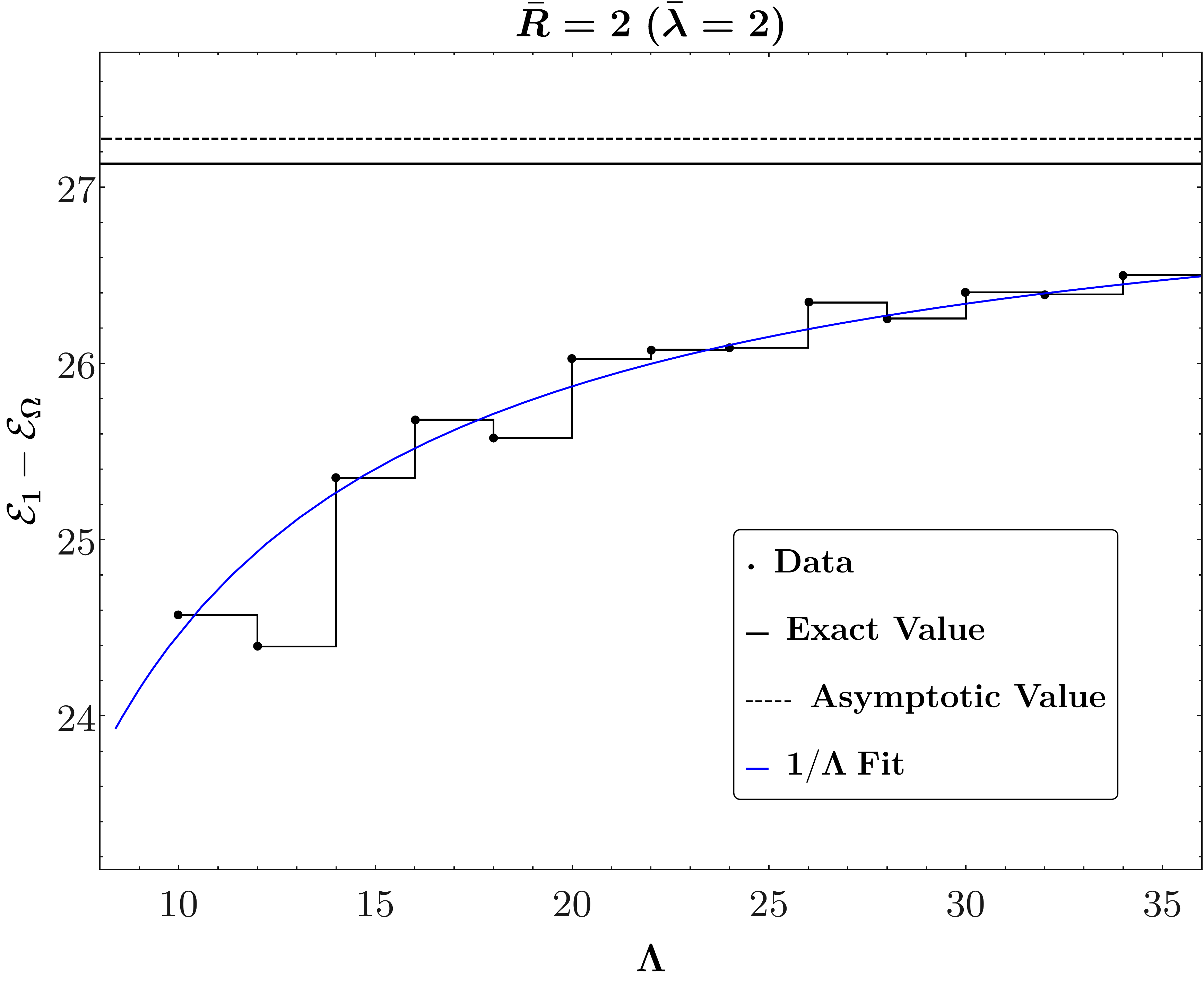}
     \end{subfigure}
     \caption{Energy of the first excited state in the Ising model deformed by $\eps$ versus the cutoff, for weak and strong couplings. The dashed lines are the asymptotic values extracted from the fit. The error bar can be found in figure~\ref{fig:TIsingFirstExcitedSpec}.}
     \label{fig:TIsingFirstExcited}
 \end{figure}

As for any minimal model with the boundary condition labeled by the identity, the leading operator at the conformal fixed point has $\Delta=2$. Hence, we expect the convergence rate to approach $\Lambda^{-1}$ for a large enough cutoff. This is indeed the case, as it can be seen in figure \ref{fig:TIsingFirstExcited}. The right panel, in particular, shows that in this case we can follow the first excited state up to a larger coupling than in our previous examples: $\bar{\lambda}=2$ corresponds to $R\, m=4\pi \simeq 12.6$, so the radius of AdS is more than ten times larger than the Compton wavelength of the fermion. On the other hand, it is worth mentioning a feature of the convergence rate at weak coupling. Following appendix \ref{sec:twopgen}, one can compute the coefficient of the $\Lambda^{-1}$ term -- \emph{i.e.} $c_i$ in eq. \eqref{eq:bdyerrors} -- at second order in perturbation theory. It turns out that $c_1 = O(\bar{\lambda}^3)$, $c_1$ being the coefficient associated to the convergence of the first excited state. By reproducing the left plot of figure \ref{fig:TIsingFirstExcited} at different values of $\bar{\lambda}$, it is possible to confirm this fact.

\subsubsection{Spectrum}
\label{subsec:IsingTspec}

\begin{figure}[!h]
         \centering
         \includegraphics[width=\textwidth]{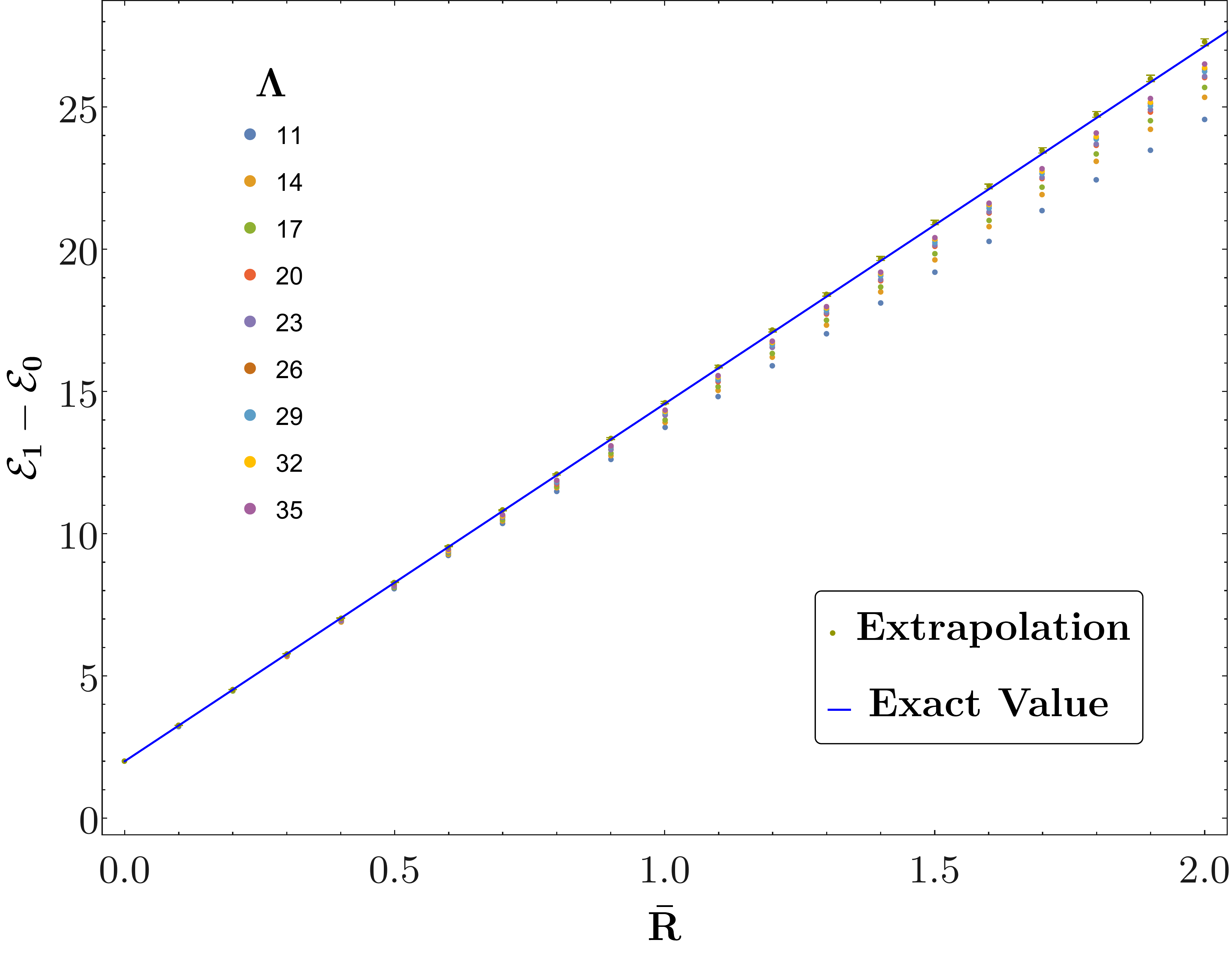}
         \caption{Energy of the first excited state in the Ising model with thermal deformation as a function of $\bar{R}$.}
         \label{fig:TIsingFirstExcitedSpec}
\end{figure}

\begin{figure}[!h]
         \centering
         \includegraphics[width=\textwidth]{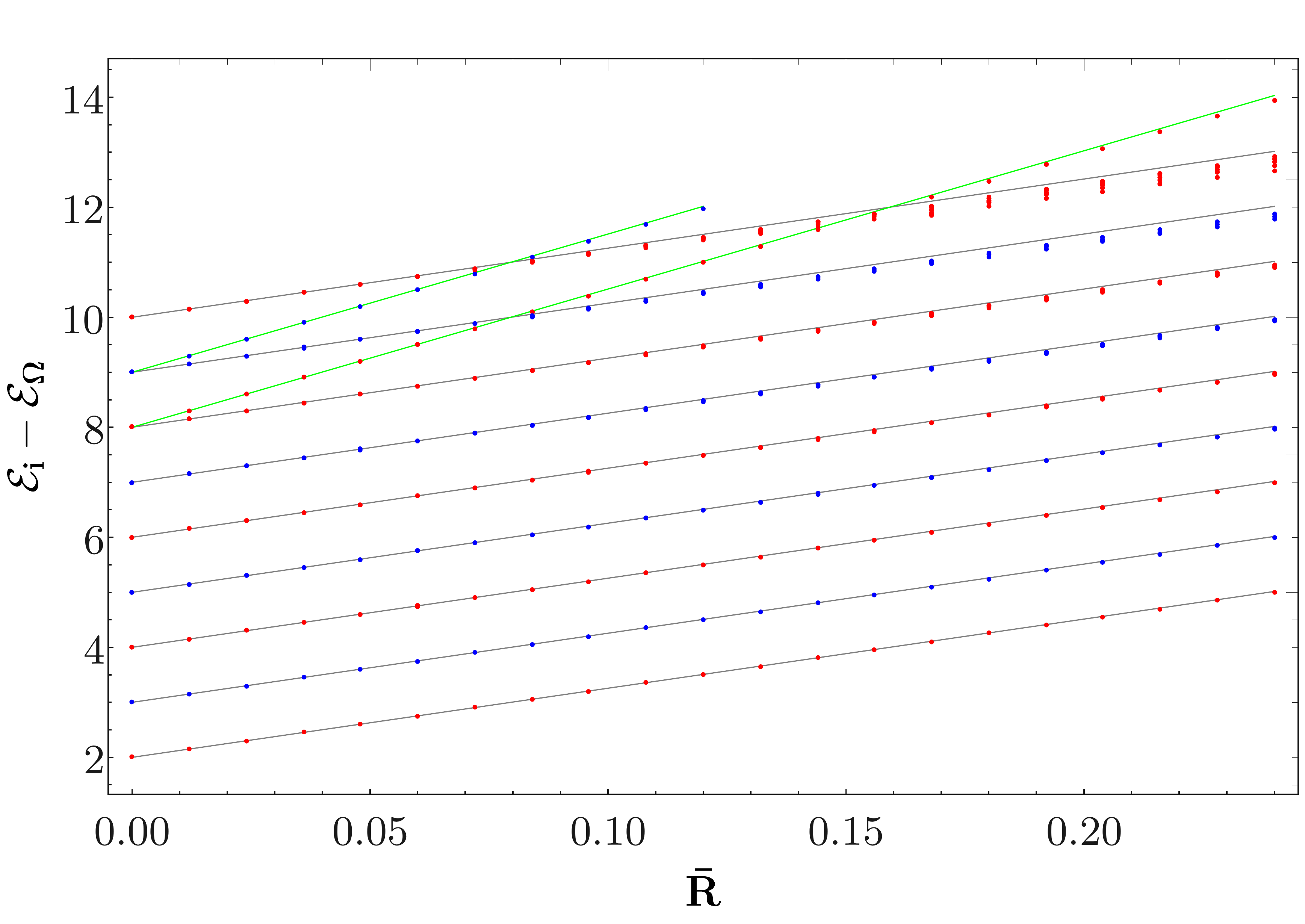}
         \caption{Spectrum of the Ising model with thermal deformation. The color distinguishes parity even (red) from parity odd (blue) eigenstates. The data is computed at $\Lambda=28$. Black and green lines mark the exact spectrum for the states with two and four fermions respectively.}
         \label{fig:TIsingSpectrum}
\end{figure}

Let us now associate the spectrum \eqref{FFspecPm} to the boundary states \eqref{CardyIsing} of the Ising model. Again, let us start from the UV fixed point. The $\mathbb{1}$ and $\eps$ boundary conditions only support the vacuum Virasoro character. Hence, their spectrum at the fixed point is integer spaced, and can be obtained from the Majorana fermion by projecting onto the states with even fermion number.\footnote{This projection is obtained by orbifolding the $\mathbb{Z}_2$ symmetry, the twisted sector being implemented by Ramond boundary conditions. In fact, the Ramond Ishibashi state for the Majorana fermion equals $\dket{\sigma}$ in eq. \eqref{CardyIsing} \cite{Bachas:2012bj}. Correspondingly, the two available boundary conditions for periodic (Neveu-Schwarz) fermions equal $\ket{\mathbb{1}}_\textup{Cardy}+\ket{\eps}_\textup{Cardy}$ and $\ket{\sigma}_\textup{Cardy}$ respectively.} 
The projection is obviously maintained by the deformation \eqref{MinimalVEps}. Therefore, the spectrum in AdS is obtained from eq. \eqref{FFspecN} with the same projection. In particular, at $\bar{\lambda}=0$ the first excited state is made of two fermions in modes $n=0$ and $n=1$, with energy $\mathcal{E}=2$. Similarly, the lowest energy state with four fermions appears at $\mathcal{E}=8$.
In order to identify the spectrum, we still need to fix the sign of $\nu$ in eq. \eqref{FFspecPm}. The simplest way to do it is to perform a leading order computation in RS perturbation theory. The result shows that, with the identification $m=2\pi \lambda_\eps$, $\nu=+1$ for the $\mathbb{1}$ boundary condition.

We are ready to compare with the results of Hamiltonian truncation. The gap is shown in figure \ref{fig:TIsingFirstExcitedSpec}, where the good convergence up to large values of the coupling is evident. Figure \ref{fig:TIsingSpectrum} gathers several low lying states, up to the first couple of four-particle states. Comparing with the table of quasiprimaries presented above, we see that, contrary to the other examples considered in this work, the second quasiprimary above the vacuum ($\Delta=4$ at $\bar{\lambda}=0$) has the same slope as the first excited state. This agrees with the usual expectation: while all the excited single particle states are $SL(2,\mathbb{R})$ descendants, the spectrum of two-particle states includes infinitely many quasiprimaries. Truncation errors are small, but visible: they begin to lift the degeneracies in the upper right corner of the plot. 

The level crossing among the first four-particle state and the two-particle states leaving from $\Delta=10$ deserves a special discussion. This is the first example where level crossing happens among states in the same parity sector. As we pointed out in subsection \ref{sec:pheno}, in the continuum limit the $SL(2,\mathbb{R})$ symmetry implies exact level crossing. On the other hand, at every finite truncation one expects level repulsion to take place. This creates a puzzle for the prescription \eqref{prescription}. What is the correct bare energy $e_i$ to be used in the formula?  The question is non trivial since our argument for the validity of eq. \eqref{prescription} relies on studying the theory with a spatial cutoff, which again breaks $SL(2,\mathbb{R})$. We find experimentally that the correct procedure is to assign the bare energy $e_i$ following the exact ($\Lambda=\infty$) lines. In other words, the red data points on the green line on the right of the crossing points are obtained from eq. \eqref{prescription} with $e_i=8$. Since level repulsion is forbidden in the continuum, the prescription is in principle well defined. However, for a finite cutoff $\Lambda$ the levels do not exactly cross, so the procedure becomes ambiguous in a region around the crossing point, whose size shrinks as the cutoff is increased.

\begin{figure}[!h]
         \centering
         \includegraphics[width=0.7\textwidth]{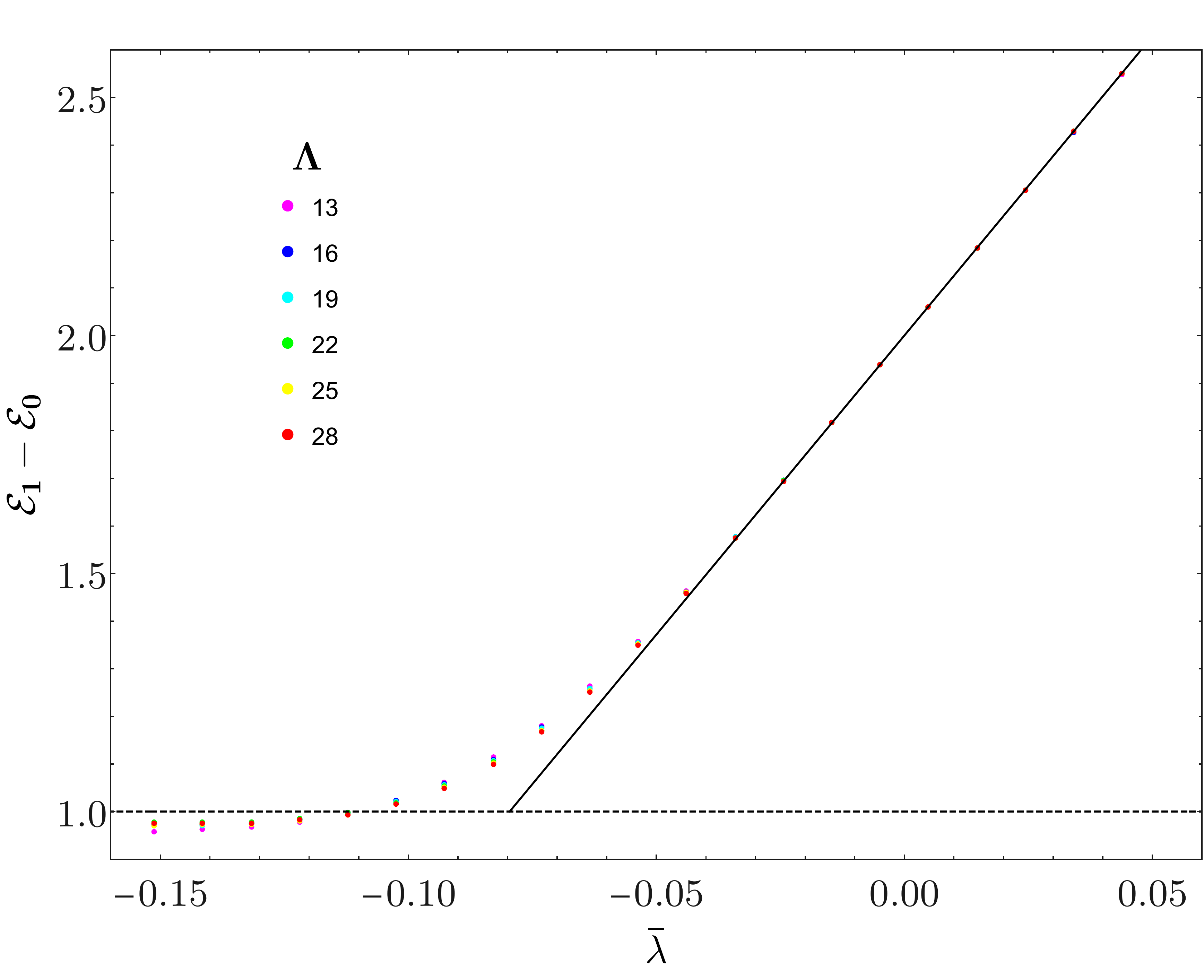}
         \caption{Energy of the first excited state in the Ising model with thermal deformation, in a range of couplings including $\bar{\lambda}<0$. The solid line is the exact solution. Notice that in this model $\bar{\lambda}=\bar{R}$, hence this plot is the continuation of the plot in figure \ref{fig:TIsingFirstExcitedSpec}. We use the label $\bar{\lambda}$ for consistency with the other examples.} 
         \label{fig:TIsingNeg}
\end{figure}

\begin{figure}[!h]
     \centering
     \begin{subfigure}{0.47\textwidth}
         \centering
         \includegraphics[width=\textwidth]{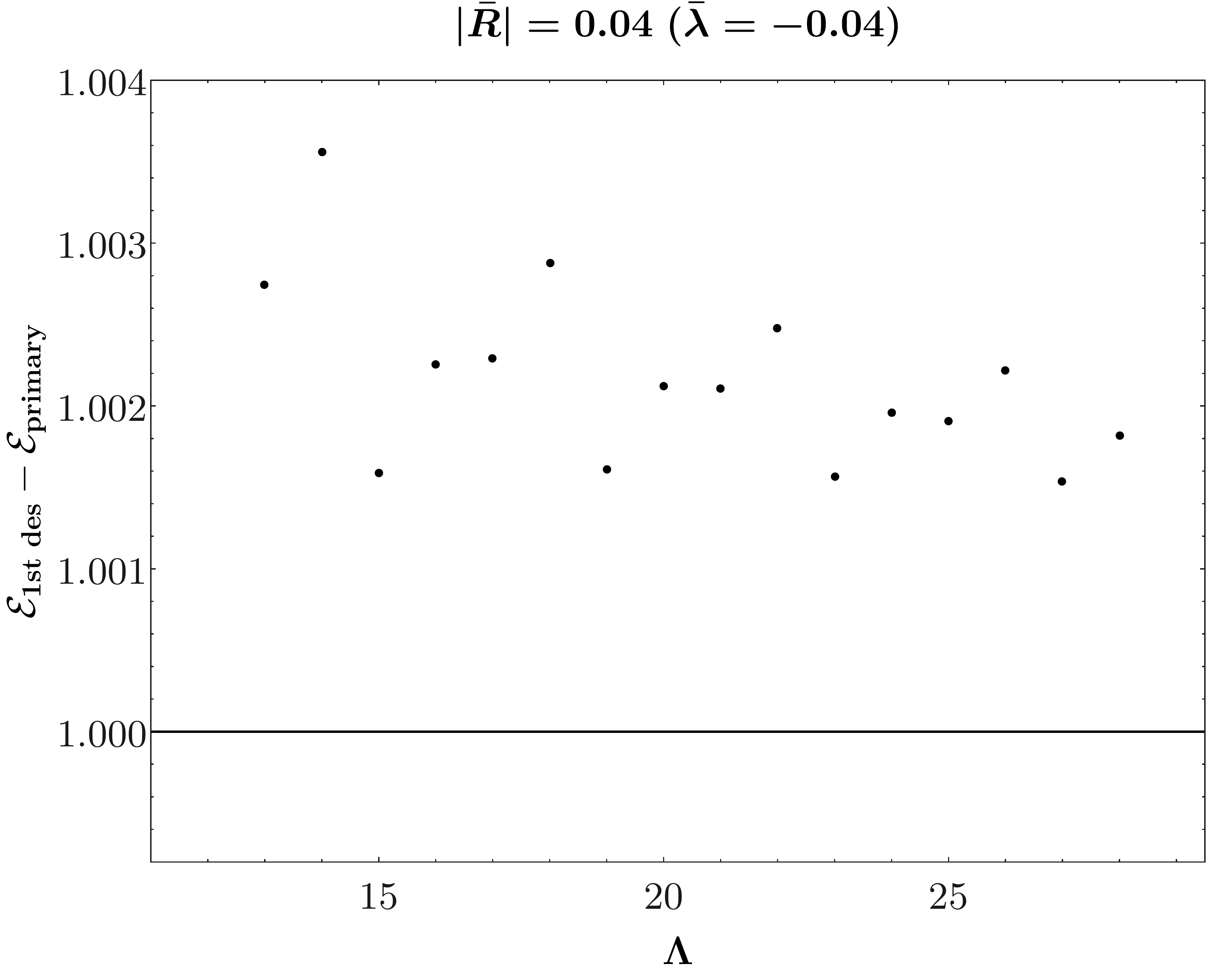}
     \end{subfigure}
     \hfill
     \begin{subfigure}{0.47\textwidth}
         \centering
         \includegraphics[width=\textwidth]{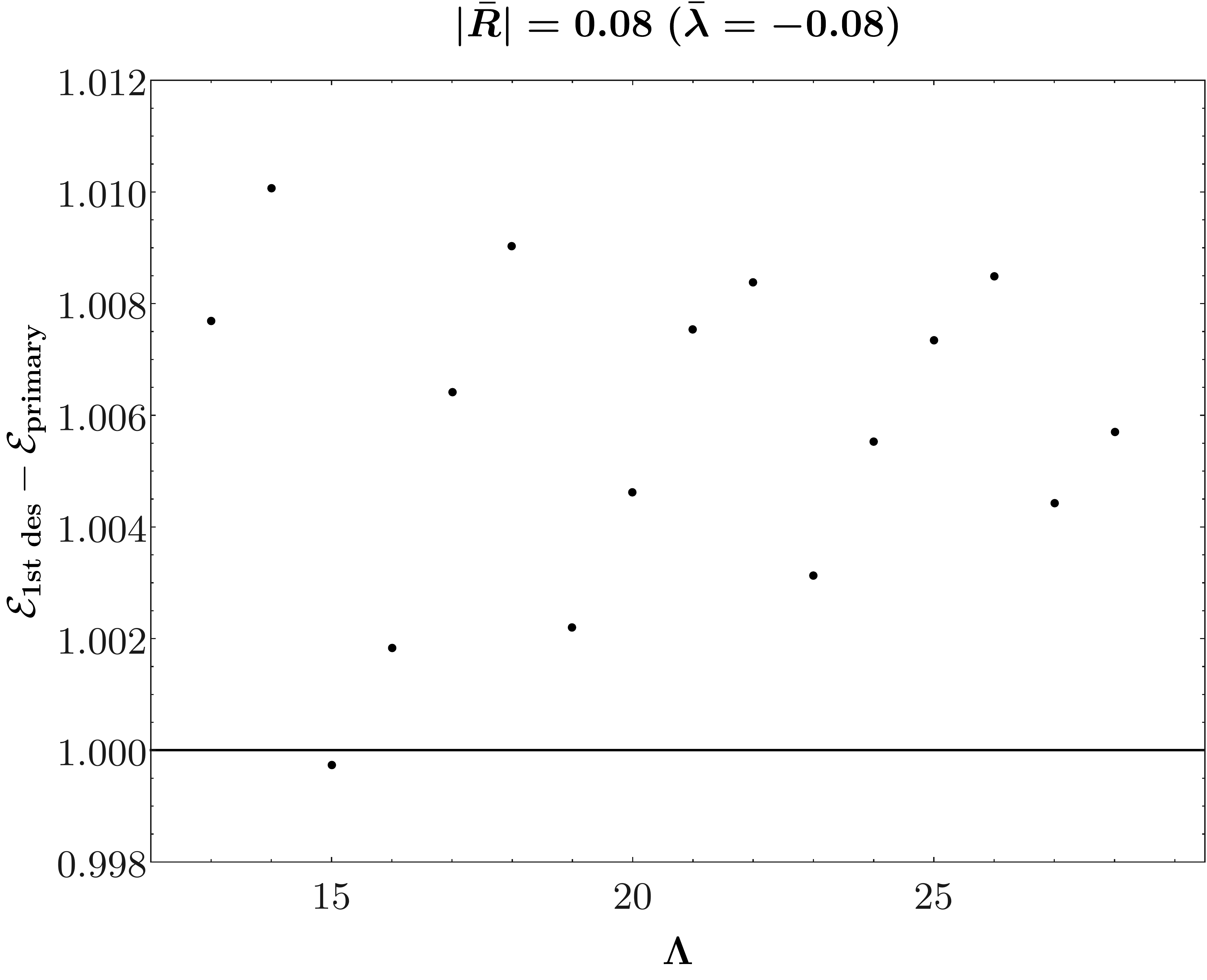}
     \end{subfigure}
     \caption{Gap between the first excited state and its leading descendant, in the Ising model with thermal deformation, for $\bar{\lambda}<0$. When the coupling is sufficiently negative, the restoration of conformal invariance is extremely slow, if at all present.}
     \label{fig:TIsingDesMinusPrimaryNeg}
\end{figure}

Let us now briefly discuss the $\sigma$ boundary condition, eq. \eqref{CardyIsingSi}. Since Kramers-Wannier duality maps $\mathbb{1}$ to $\sigma$ -- see eq. \eqref{KramersWannierIsing} -- the spectrum originating from the thermal deformation of the latter is obtained by choosing $\nu=-1$ in eq. \eqref{FFspecN}. Since the $\sigma$ boundary condition admits a state with $\Delta=1/2$, the gap is given by the single particle state in eq. \eqref{FFspecPm}, with slope $-2\pi$ as a function of $\bar{R}$. In general, the spectrum is composed of both even and odd fermion states. The gap as a function of the radius of AdS, for all the boundary conditions, is presented together in fig \ref{fig:specIsing}. 

Finally, let us again discuss what happens if we turn on the coupling with the \emph{wrong} sign, \emph{i.e.} the sign which does not allow to reach the flat space limit. We expored this question with Hamiltonian truncation, the result being shown in figure \ref{fig:TIsingNeg}. Like with the Lee-Yang model, our data become unreliable as the gap approaches $\Delta=1$. This is showcased not only by the discrepancy between the data and the exact solution, but also by the lack of $SL(2,\mathbb{R})$ invariance of the spectrum -- see figure \ref{fig:TIsingDesMinusPrimaryNeg}.
In the exact theory, the gap $\Delta=1$ corresponds to $\bar{\lambda}=-1/4\pi$. There, the free fermion Hamiltonian, with the boundary conditions we are considering, stops being well defined. Indeed, the integral of the Hamiltonian density diverges due to the behavior of the Fermi fields close to the boundary \cite{Doyon:2004fv}. On the other hand, the theory emanating from the $\sigma$ boundary condition has rising trajectories in the direction of negative $\bar{\lambda}$, and is well defined beyond the $\Delta=1$ point. Precisely at $\bar{\lambda}=-1/4\pi$, the spectra of the two theories, remarkably, match, including all the multiplicities, as one can check by plugging $mR=-1/2$ in eq. \eqref{FFspecN}, and recalling that when $\nu=1$ the Ising model is obtained by restricting to even $N$. This situation is coherent with both the free boson spectrum at the BF bound and the picture we sketched while discussing the Lee-Yang model. Again, we delay until section \ref{sec:disc} a unified discussion.

\subsection{The Ising model in a magnetic field}

We now move on to discuss the deformation triggered by $\lambda_\sigma$ in eq. \eqref{eq:IsingFlow}. More precisely, we perturb the fixed point Hamiltonian with
\beq
\bar{\lambda} V=-\bar{\lambda}  \int_{-\pi/2}^{\pi/2} \frac{dr}{(\cos r)^2}\, 
\left(\frac{R^{\Delta_\sigma}}{a_\sigma}\,\sigma(\tau=0,r)-1\right)~, \qquad a_\sigma=2^{1/8}~.
\eeq
Again, we consider the flow originating from the boundary condition labeled by the identity. The sign is chosen so that, deforming the fixed point Hamiltonian with $\bar{\lambda}>0$, the energies monotonically increase with $R$. Notice that, since the $\mathbb{Z}_2$ symmetry maps the $\mathbb{1}$ boundary condition into $\eps$, the sign of the magnetic field is meaningful. In particular, comparing with eq. \eqref{sigmaVev}, we see that $\bar{\lambda}>0$ corresponds to a magnetic field in the same direction in the bulk as on the boundary. We shall analyze both positive and negative values of $\bar{\lambda}$ below.

In the $R\to \infty$ limit, the particle content is known thanks to the integrability of the flow in flat space. The integrable flat space theory contains 8 particles, 3 of which have mass below the two-particle continuum \cite{Zamolodchikov:1989fp}. The mass ratio of the two lightest particles is given by
\beq
\label{m1m2Ising}
\frac{m_2}{m_1} = 2\cos \frac{\pi}{5} \approx 1.618\ .
\eeq
The Ising model with a magnetic field is not exactly solvable in AdS, and the results for the spectrum presented below are new.

\subsubsection{Cutoff effects}

\begin{figure}[!h]
     \centering
     \begin{subfigure}{0.47\textwidth}
         \centering
         \includegraphics[width=\textwidth]{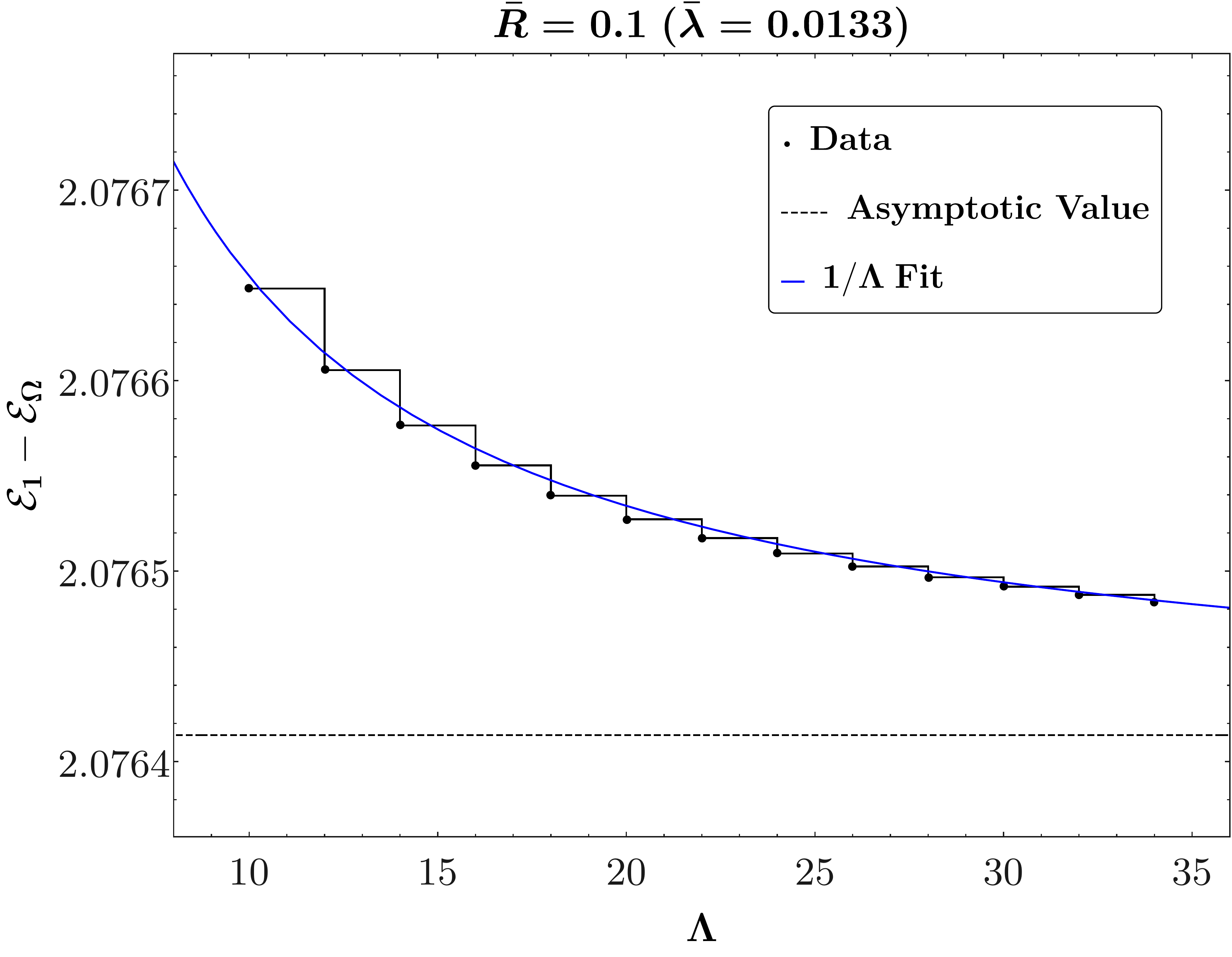}
     \end{subfigure}
     \hfill
     \begin{subfigure}{0.46\textwidth}
         \centering
         \includegraphics[width=\textwidth]{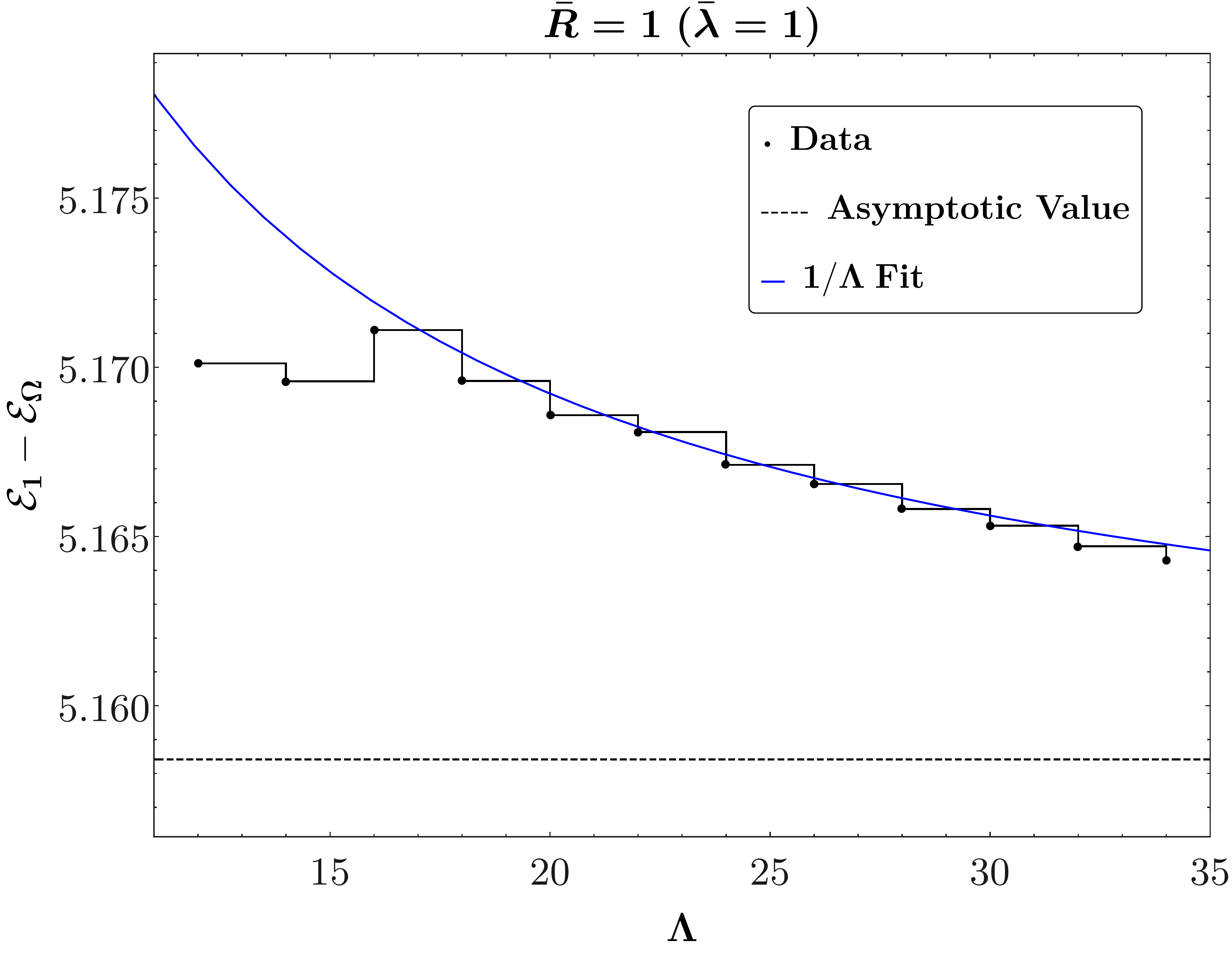}
     \end{subfigure}
     \caption{Energy of the first excited state in the Ising model deformed by $\sigma$, as a function of the cutoff, at weak and strong coupling.}
     \label{fig:MIsingFirstExcitedRbar}
 \end{figure}

\begin{figure}[!h]
     \centering
     \begin{subfigure}{0.45\textwidth}
         \centering
         \includegraphics[width=\textwidth]{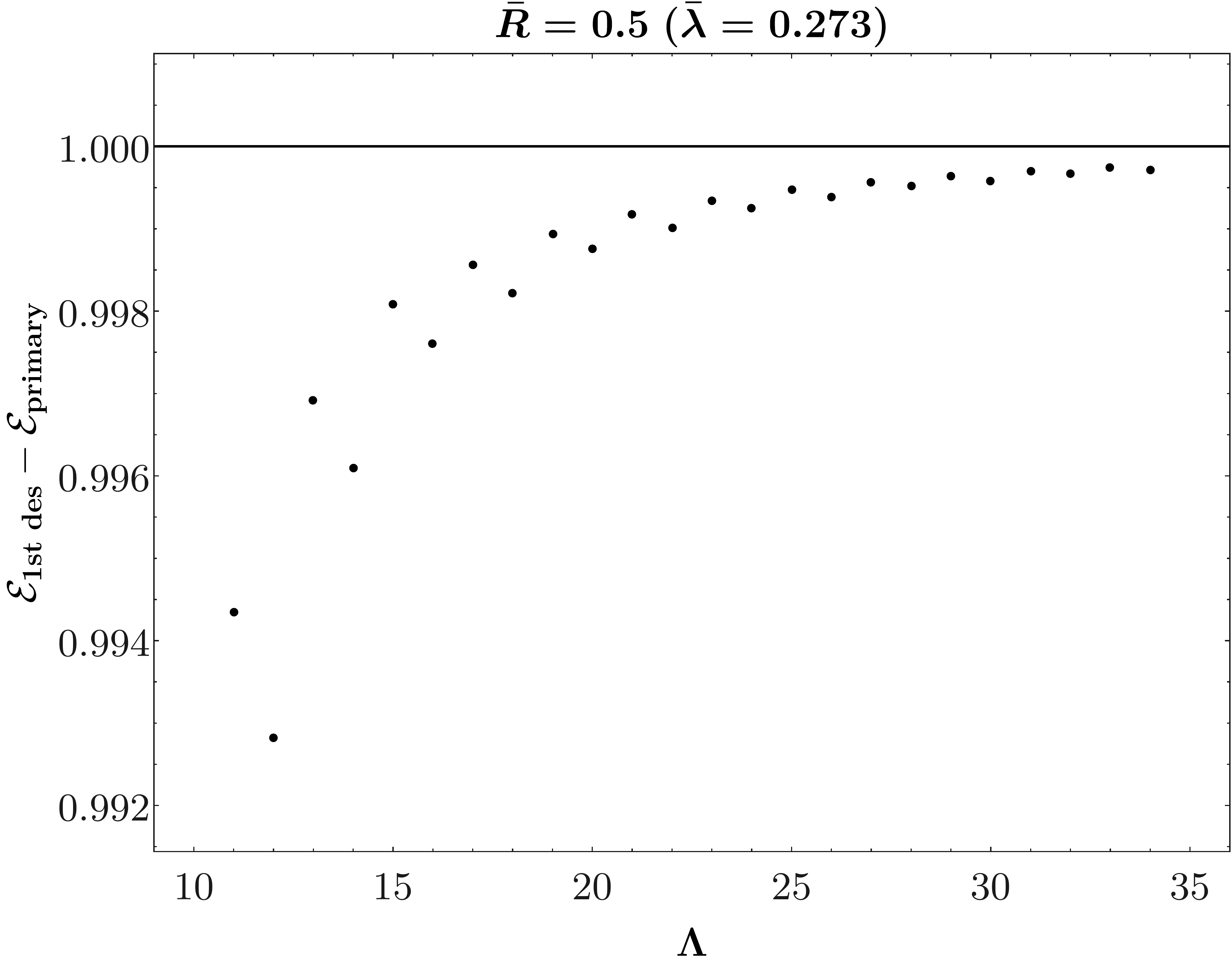}
     \end{subfigure}
     \hfill
     \begin{subfigure}{0.44\textwidth}
         \centering
         \includegraphics[width=\textwidth]{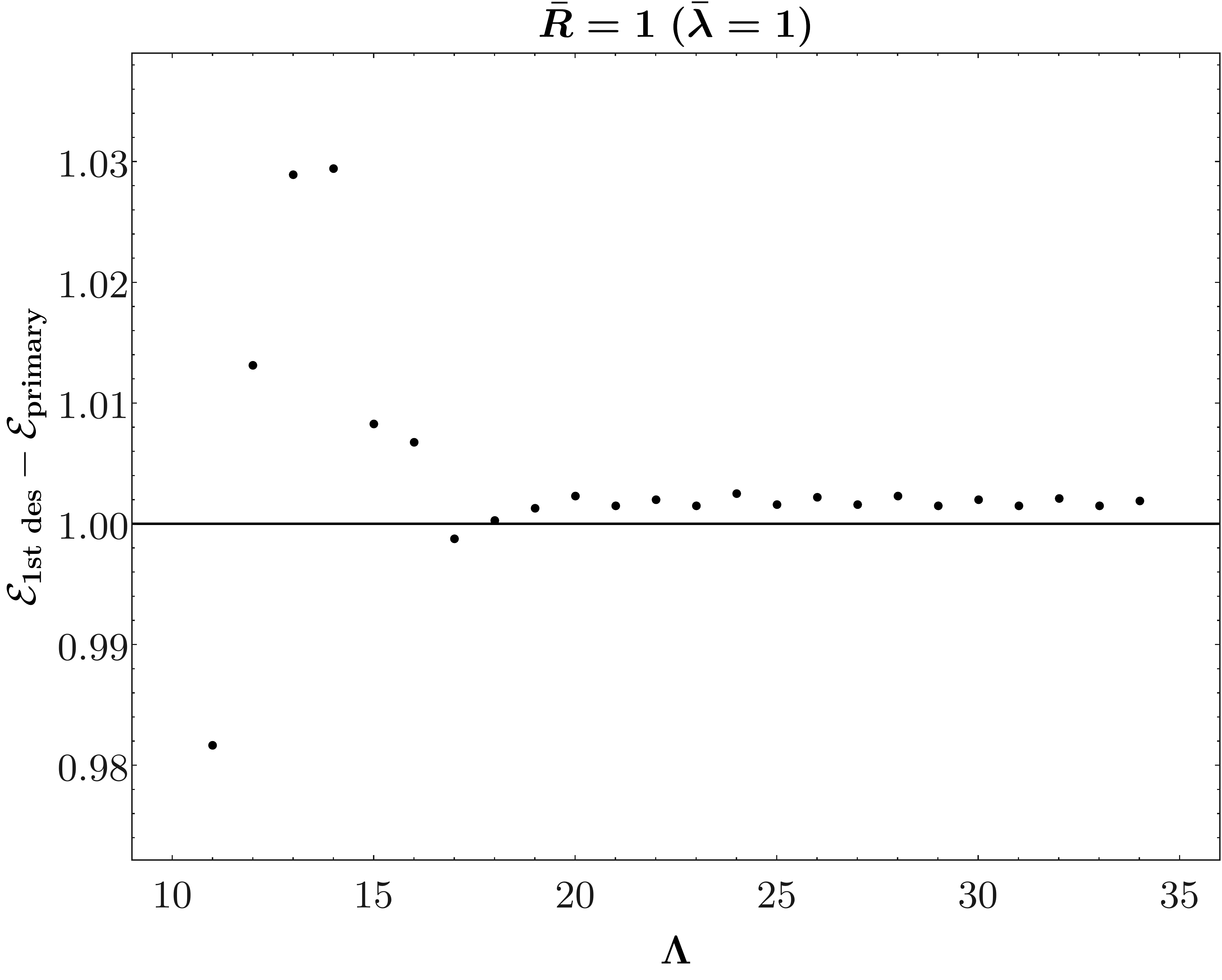}
     \end{subfigure}
     \caption{Energy gap between the first primary and its first descendant for two different couplings, in the Ising model with a magnetic field.}
     \label{fig:MIsingDesMinusPrimaryRbar}
 \end{figure}

The analysis of the cutoff effects proceeds in parallel with the previous examples. In figure \ref{fig:MIsingFirstExcitedRbar} the expected $1/\Lambda$ convergence rate is tested, for $\bar{\lambda}>0$, both at weak and at strong coupling. While the left plot gives confidence that the asymptotic region has been reached, the fit in the right plot is poorer. A similar situation emerges from the study of the gap between the first excited state and its first descendant. In the left panel of figure \ref{fig:MIsingDesMinusPrimaryRbar}, we observe a steady convergence to the unit gap expected in the continuum. In the right panel, the size of the violation of the $SL(2,\mathbb{R})$ symmetry rapidly drops, but then the convergence becomes slow. These are signs that around $\bar{\lambda}=1$ it may be worth  pushing the numerics further in the future, or improving the convergence rate along the lines of subsection \ref{sec:rate}. However, it is not easy to draw an indication of the size of the error. For instance, primaries and descendants may be affected by cutoff effects in a similar way, thus reducing the violation of conformal symmetry visible in figure \ref{fig:MIsingDesMinusPrimaryRbar}. On the other hand, the procedure explained in subsection \ref{sec:VaryingGPhi2}, which extracts error bars from the quality of the fits in figure \ref{fig:MIsingFirstExcitedRbar}, yields a small error: the error bars are smaller than the size of the points in figure \ref{fig:MIsingSpectrum}.

\begin{figure}[!h]
     \centering
     \begin{subfigure}{0.45\textwidth}
         \centering
         \includegraphics[width=\textwidth]{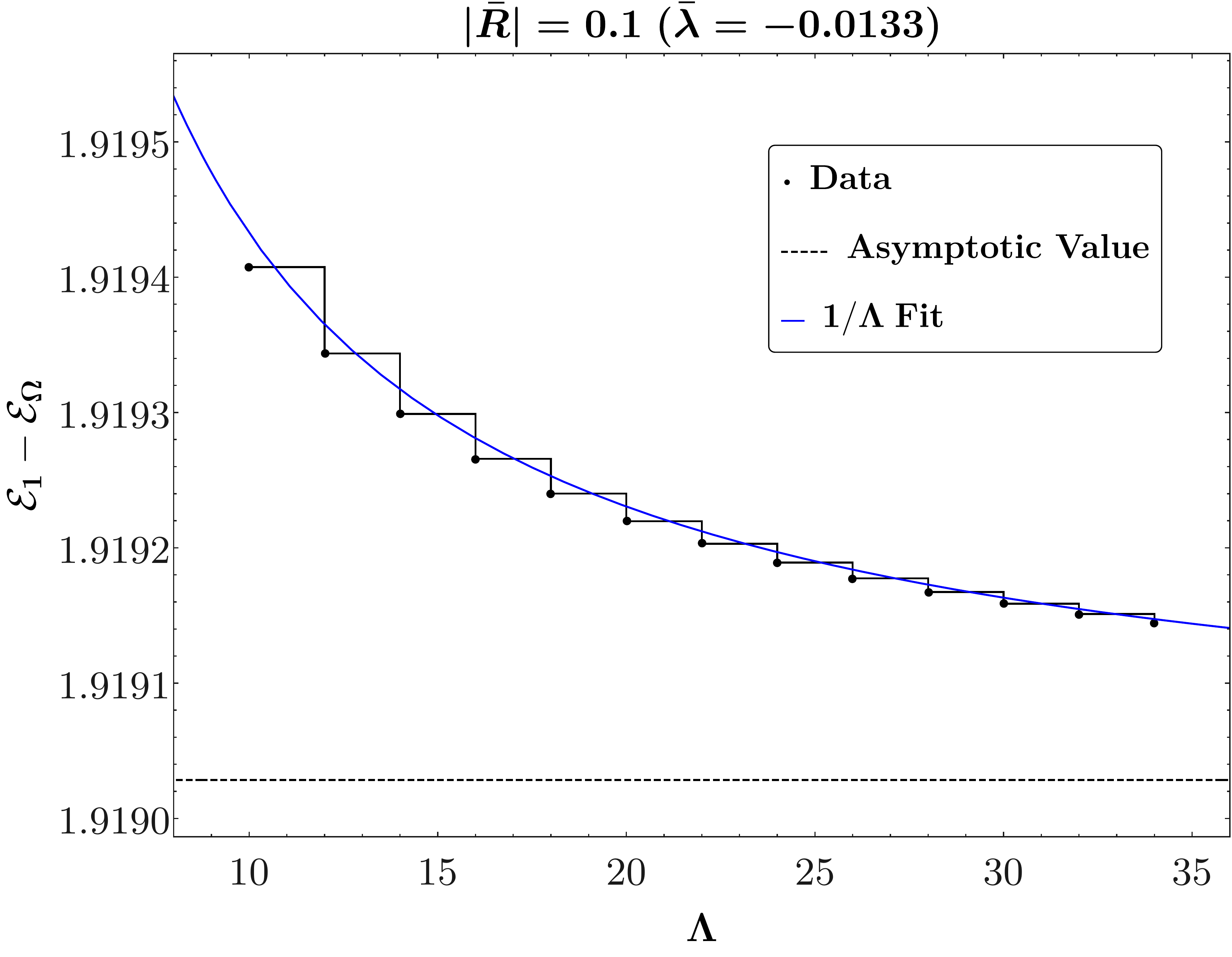}
     \end{subfigure}
     \hfill
     \begin{subfigure}{0.44\textwidth}
         \centering
         \includegraphics[width=\textwidth]{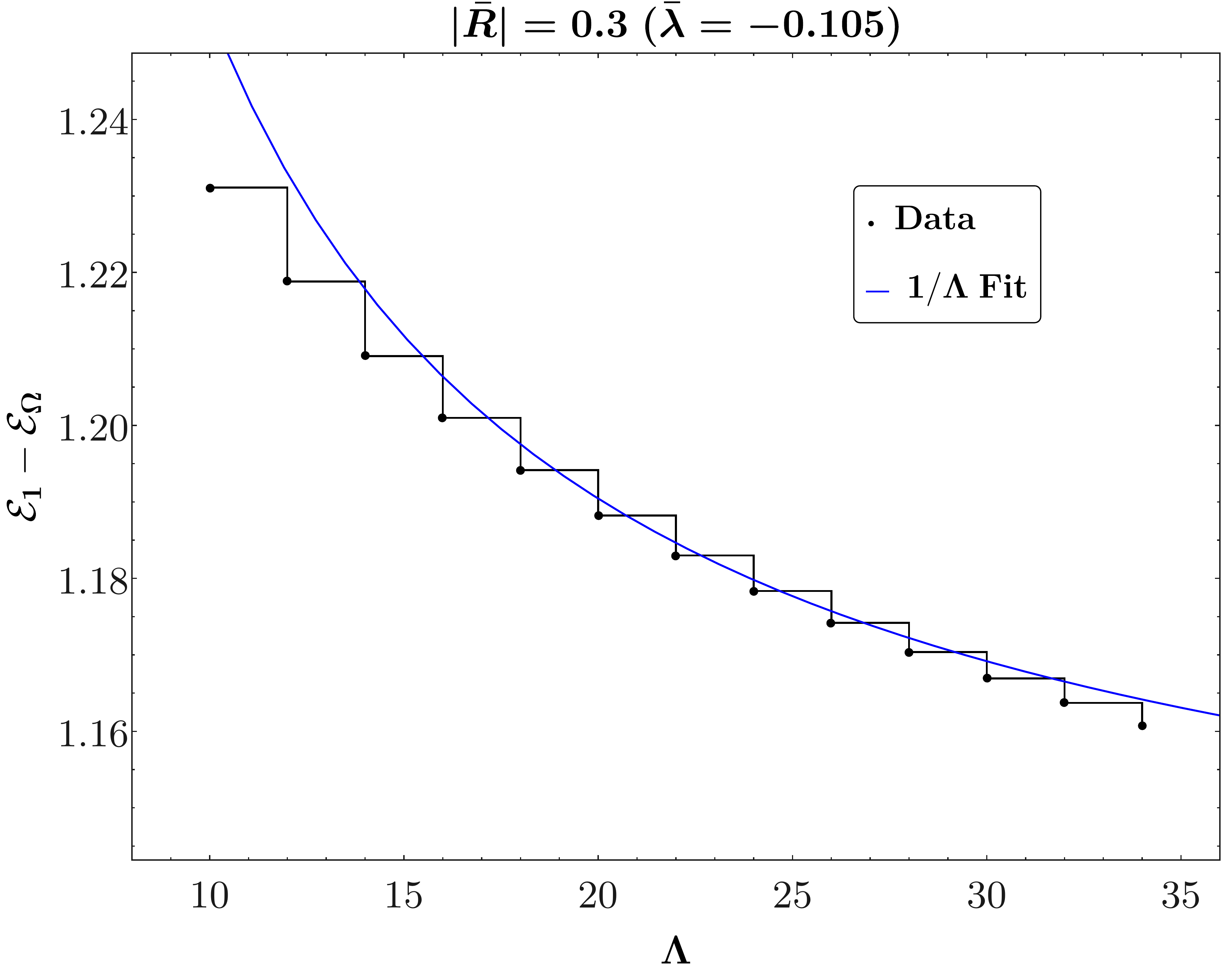}
     \end{subfigure}
     \caption{Energy of the first excited state in the Ising model, as a function of the cutoff, for two choices of negative magnetic field, $\bar{\lambda}<0$.}
     \label{fig:MNegIsingConvergence}
\end{figure}

\begin{figure}[!h]
     \centering
     \begin{subfigure}{0.46\textwidth}
         \centering
         \includegraphics[width=\textwidth]{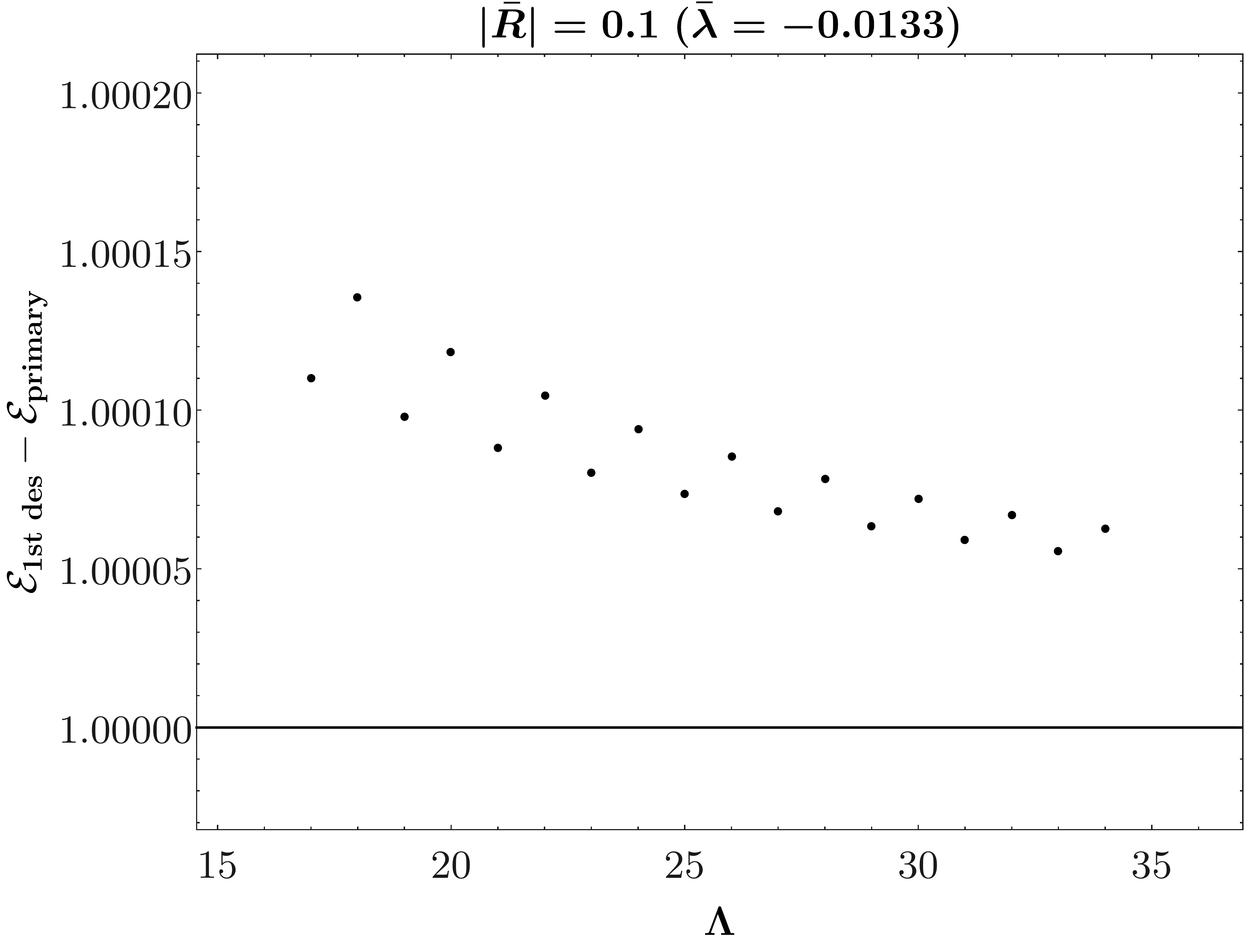}
     \end{subfigure}
     \hfill
     \begin{subfigure}{0.44\textwidth}
         \centering
         \includegraphics[width=\textwidth]{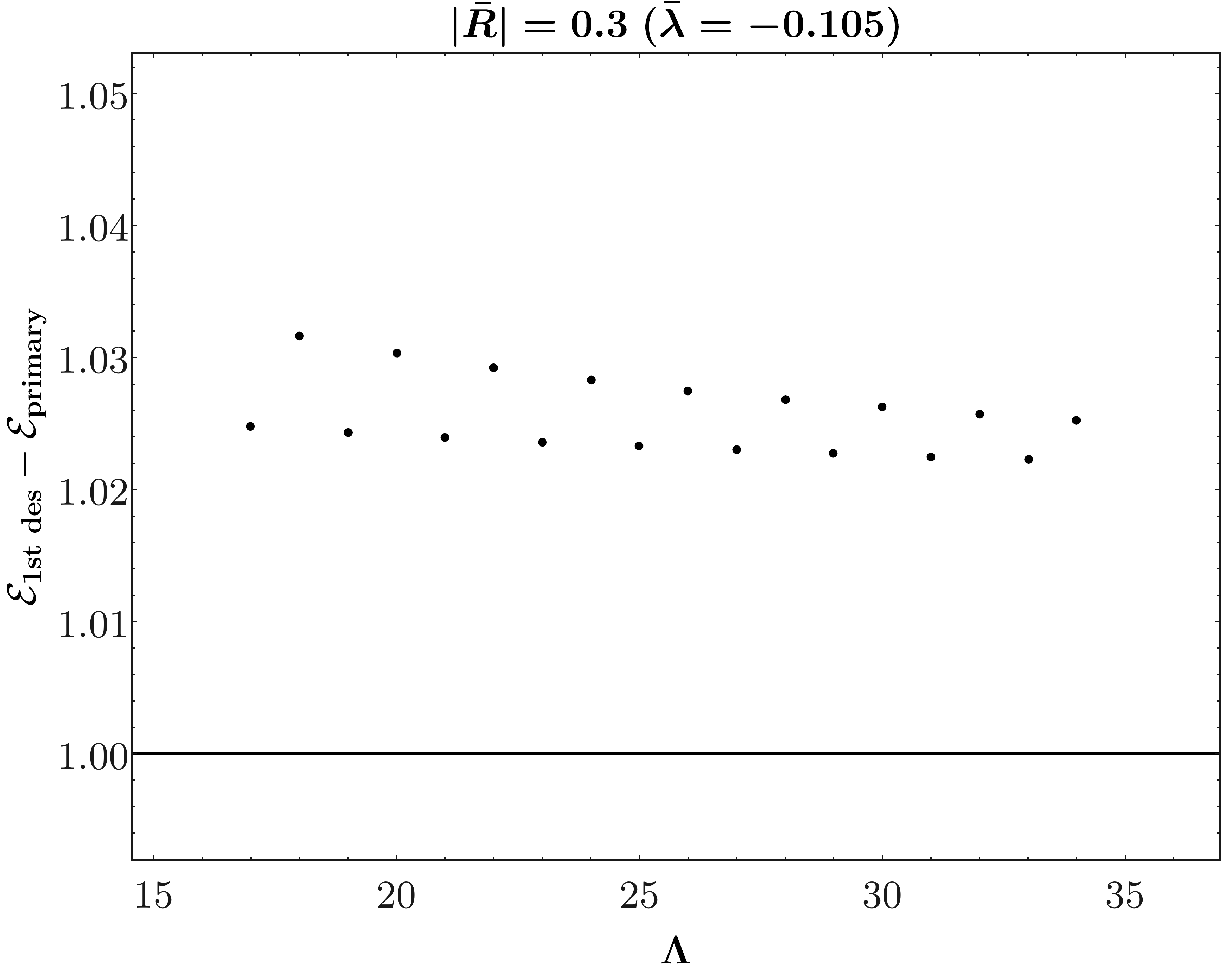}
     \end{subfigure}
     \caption{Energy gap between the first primary and its first descendant, as a function of the cutoff, for two choices of negative magnetic field, $\bar{\lambda}<0$.}
     \label{fig:MNegIsingDesMinusPimaryRbar}
\end{figure}

Finally, in figures \ref{fig:MNegIsingConvergence} and \ref{fig:MNegIsingDesMinusPimaryRbar}, we repeated the analysis with a choice of two negative values for $\bar{\lambda}$. As in the previous examples, if the sign of the coupling does not allow for a flat space limit, the convergence rapidly deteriorates. As shown in the right panel of both figures, at a coupling $\bar{\lambda} \simeq -0.1$ the spectrum is hardly converged in the range of cutoffs available to us. 
 
\subsubsection{Spectrum}

\begin{figure}[!h]
         \centering
         \includegraphics[width=\textwidth]{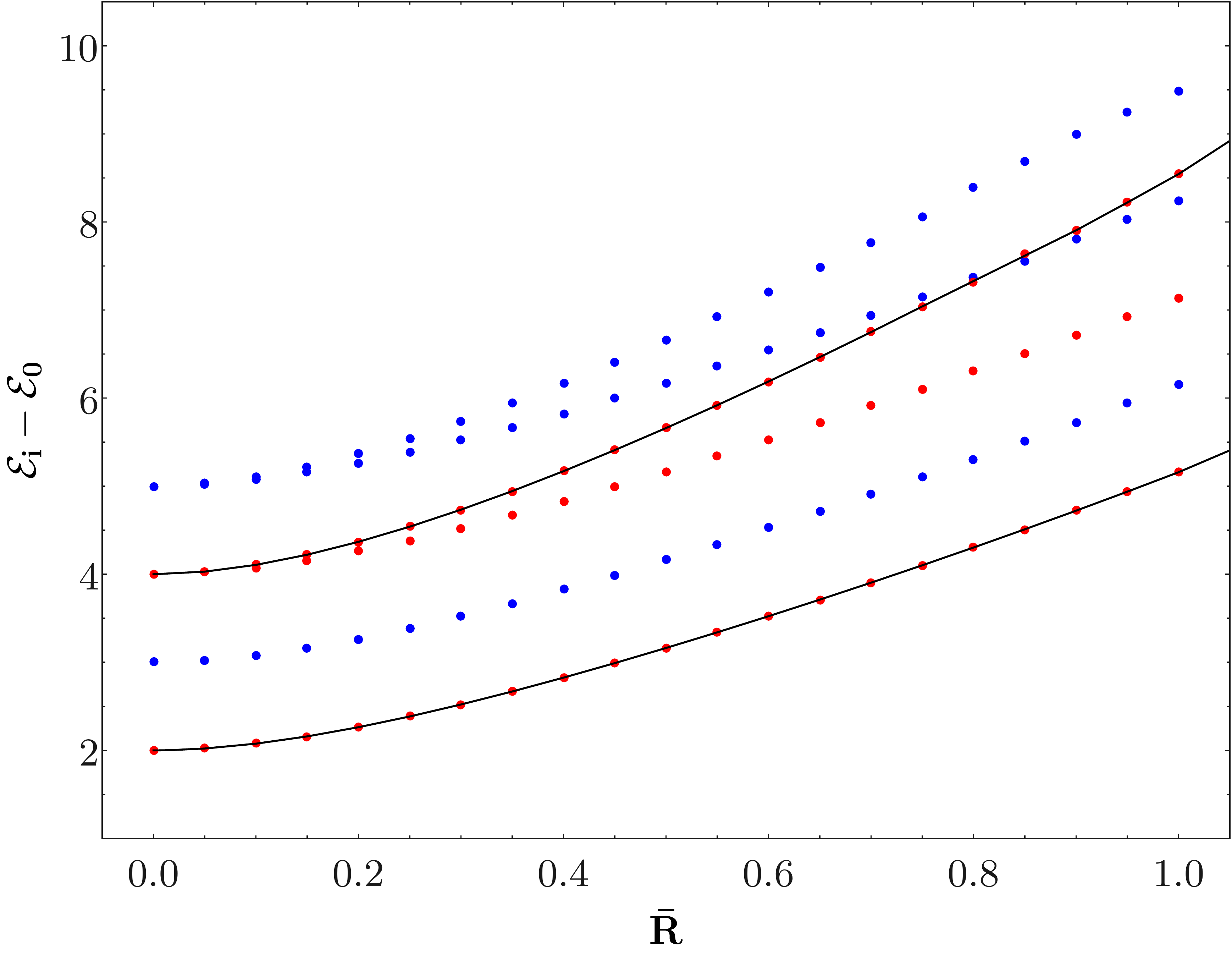}
         \caption{Spectrum of the Ising model in a positive magnetic field, $\bar{\lambda}>0$, as a function of the radius of AdS. The red and blue points are the extrapolated data, even and odd respectively under parity. The black lines highlight the first two quasiprimaries, and are only meant to guide the eye.}
         \label{fig:MIsingSpectrum}
\end{figure}

We are ready to present the results of the TCSA applied to this model. Figure \ref{fig:MIsingSpectrum} shows the growth of the energy gaps with $R$, when the magnetic field in the bulk and on the boundary are aligned ($\bar{\lambda}>0$). Two conformal families are visible, the quasiprimaries being highlighted by black lines. They are on their way to become the first two particles in the flat space spectrum with masses satisfying \eqref{m1m2Ising}. However, the effects of the curvature are still sizable at these values of $\bar{R}$: the ratio of the two slopes is at most $\sim 1.5$, to be compared with the exact value $1.618$.  

\begin{figure}[!h]
         \centering
         \includegraphics[width=\textwidth]{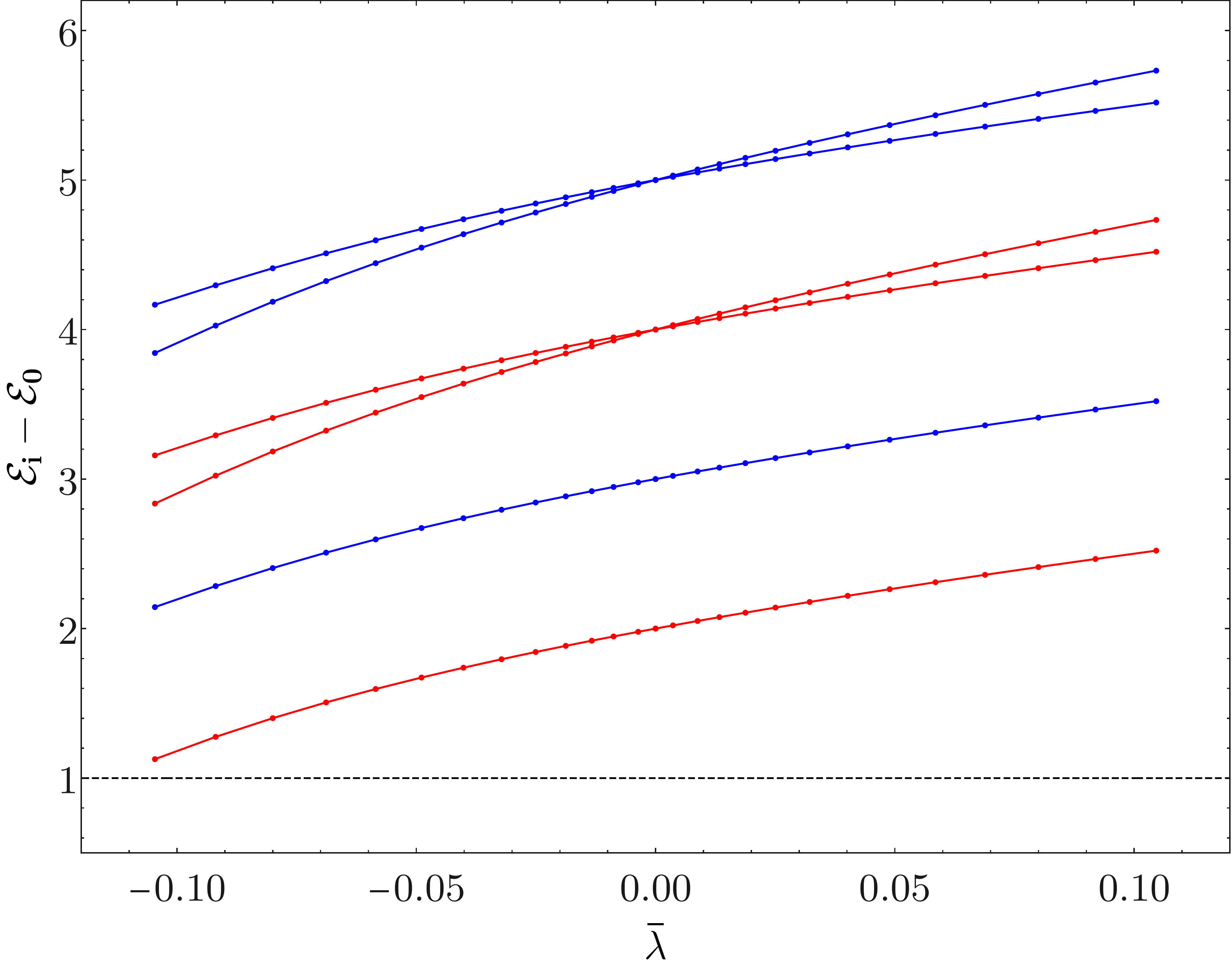}
         \caption{Spectrum of the Ising model in a magnetic field, as a function of the magnitude of the field. Error bars as measured from the uncertainty in the extrapolation are not shown: they would be small, throughout the plot, but as discussed in the text they probably underestimate the deviation from the continuum limit.}
         \label{fig:MNegIsingSpectrum}
\end{figure}

Figure \ref{fig:MNegIsingSpectrum} again shows the low lying spectrum, this time as a function of the intensity of the magnetic field $\bar{\lambda}$. For negative values of $\bar{\lambda}$, the first excited state approaches the threshold $\Delta=1$. As we pointed out in the previous examples, we expect the theory to become unstable when this happens. Correspondingly, the convergence of the truncation is poor for the leftmost points in the plot, and we cannot reliably draw conclusions about the value of $\bar{\lambda}$ where a new marginal operator arises, nor about the dependence $\Delta(\bar{\lambda})$ close to that point.

\section{Discussion}
\label{sec:disc}
%!TEX root = ../HTinAdS.tex
%%%%%%%%%%%%%%%%%%%%%%%%%%%%%%%%%%%%%%%%%%%%%

AdS spacetime provides a maximally symmetric IR regulator for QFT. 
Furthermore, one can place operators at the conformal boundary of AdS and their correlations functions obey all the conformal bootstrap axioms (except for the existence of a stress tensor).
This opens the possibility to study massive QFT with the non-perturbative conformal bootstrap methods \cite{Paulos:2016fap}.
There are two main difficulties to realize this idea in practice. The first is technical: the standard numerical conformal bootstrap methods converge poorly for large scaling dimension of the external operators, which corresponds to the regime of large AdS radius.
The second is conceptual: how to specify the particular solutions of the bootstrap equations that corresponds to a particular QFT in AdS?
One of our motivations to develop the present work was the need for more data to help address the second question.

In this work, we showed that QFT in AdS$_2$ can be studied non-perturbatively using Hamiltonian truncation. Our main target was the computation of the low-lying spectrum of the Hamiltonian conjugate to global time (or equivalently the scaling dimensions of boundary operators). We presented a concrete method to renormalize  divergences in Hamiltonian truncation and obtain finite physical results for the energy spectrum. In this way, this work opens a new non-perturbative window into QFT on AdS spacetime. 

\begin{figure}[!h]
         \centering
       \begin{overpic}[width=0.6\textwidth]{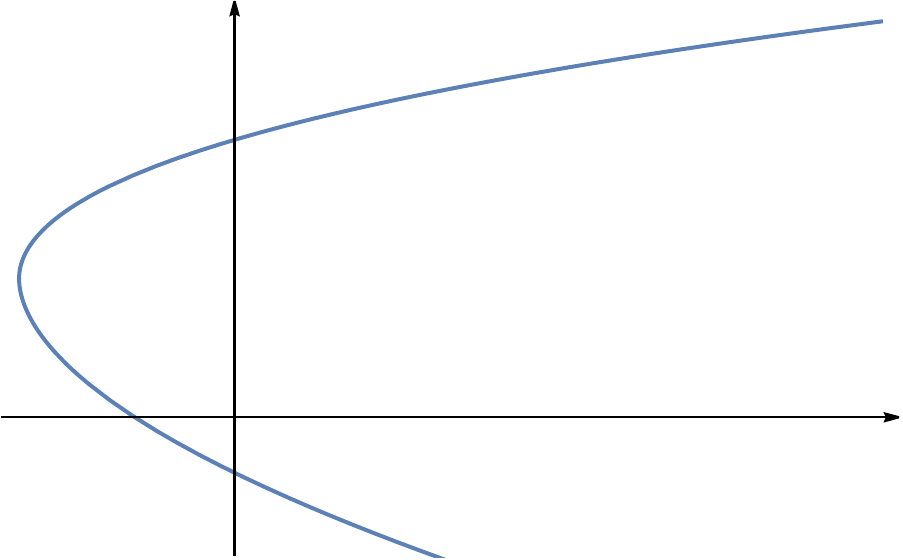}
       \put(21,60){$\Delta$}
       \put(100,11){$\bar{\lambda}$}
       \put(23,47){$2$}
       \put(27,43){\Large$\mathbb{1}$}
       \put (26,46.5) {\circle*{1}}
        \put(19.4,6){$-\frac{2}{5}$}
       \put(27,10){\Large$\mca{V}$}
        \put (26,9.5) {\circle*{1}}
       \multiput(2,30.5)(2,0){12}{\line(1,0){1}}
       \multiput(2,30.5)(0,-2){8}{\line(0,-1){1}}
       \put(23,31){$1$}
       \put(0,11){$\bar{\lambda}_\textup{min}$}
       \put(38,5){\vector(0,1){44}}
       \put(39,28){$\int_\textup{bry} \mca{V}_b$}
       \end{overpic}    
         \caption{         Sketch of the scaling dimension of the first boundary operator (above the identity) for the scaling Lee-Yang model in AdS$_2$. At $\lambda=0$, the bulk theory is conformal and the two black dots correspond to the two possible BCFTs (labeled by $\mathbb{1}$ and $\mca{V}$) for the Lee-Yang minimal model.
         We conjecture that the two curves merge (smoothly) at $\bar{\lambda} = \bar{\lambda}_\textup{min} \approx - 0.5$ and $\Delta =1$ (see  figure \ref{fig:LYNegRbar}). 
         We expect that $\Delta$ becomes complex for $ \bar{\lambda}< \bar{\lambda}_\textup{min}$ similarly to what happens for complex CFTs \cite{Gorbenko:2018ncu}.
         There is also a boundary RG flow generated by the relevant  boundary operator with $\Delta<1$ that goes from  the lower to the upper curve.         
         }
         \label{fig:specLY}
\end{figure}

\begin{figure}[!h]
         \centering
         \begin{subfigure}[]{0.46\textwidth}
         \begin{overpic}[width=\textwidth]{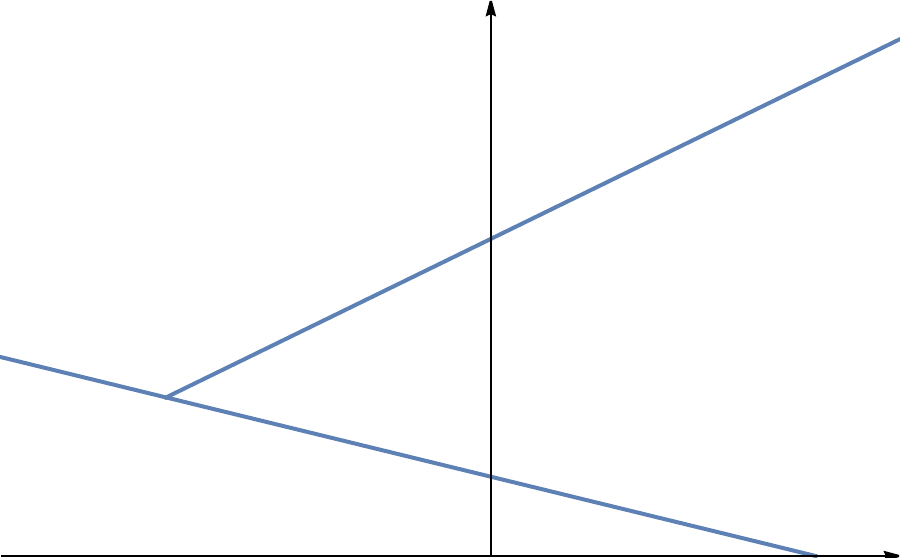}
       \put(49,58){$\Delta$}
       \put(97,-5){$\bar{\lambda}$}
       \put(51,36){$2$}
       \put(55,31){\Large$\mathbb{1}$,$\epsilon$}
         \put (54.4,35.5) {\circle*{1}}
        \put(51,4){$\frac{1}{2}$}
       \put(55,10){\Large$\sigma$}
        \put (54.4,9.2) {\circle*{1}}
       \multiput(19,17.9)(2,0){18}{\line(1,0){1}}
       \multiput(19,17.9)(0,-2){9}{\line(0,-1){1}}
       \put(51,18.5){$1$}
       \put(18,-5){$\bar{\lambda}_\textup{min}$}
       \put(75,5){\vector(0,1){39.5}}
       \put(77,25){$\int_\textup{bry} \sigma_b$}
       \end{overpic}
       \end{subfigure}
       \hspace{0.5cm}
         \begin{subfigure}{0.46\textwidth}
         \begin{overpic}[width=\textwidth]{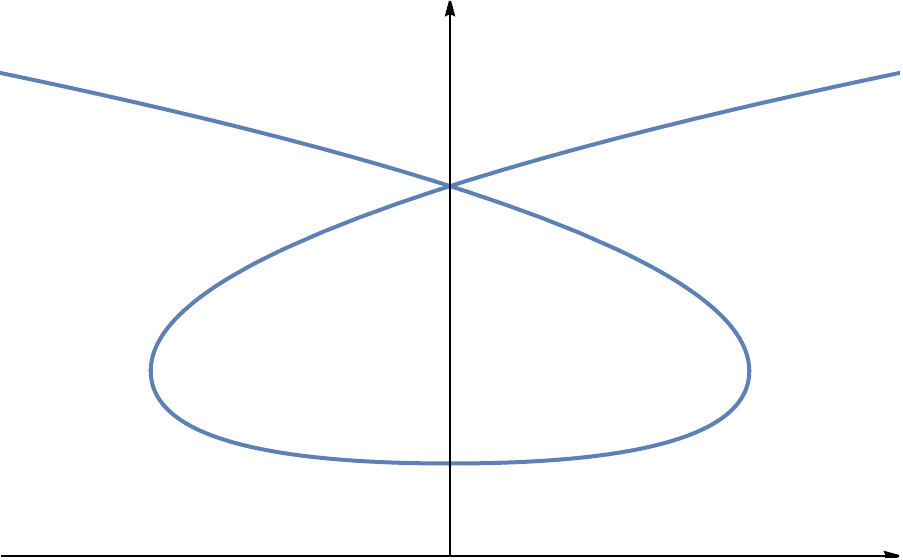}
       \put(44,58){$\Delta$}
       \put(97,-5){$\bar{\lambda}$}
       \put(46,36){$2$}
        \put (49.9,41.5) {\circle*{1}}
       \put(55,45){\Large$\mathbb{1}$}
       \put(40,45){\Large$\epsilon$}
        \put(46,5){$\frac{1}{2}$}
       \put(51,12){\Large$\sigma$}
        \put (49.9,10.5) {\circle*{1}}
       \multiput(16.8,20.9)(2,0){17}{\line(1,0){1}}
       \multiput(16.8,20.9)(0,-2){10}{\line(0,-1){1}}
       \put(46,21.5){$1$}
       \put(16,-5){$\bar{\lambda}_\textup{min}$}
       \put(62,11.8){\vector(0,1){24.2}}
       \put(63,25){$\int_\textup{bry} \sigma_b$}
       \end{overpic}
       \end{subfigure}
       \vspace{0.3cm}
         \caption{
         Sketch of the scaling dimension of the first boundary operator (above the identity) for the  Ising field theory  in AdS$_2$. 
   The black dots mark the three possible BCFTs (labeled by $\mathbb{1}$, $\epsilon$ and $\sigma$) for the Ising minimal model.
          On the left, we show the thermal deformation that is exactly solvable in terms of a massive free fermion \cite{Doyon:2004fv}. On the right, we show (our best guess for) the magnetic deformation. 
        The curve marked $\mathbb{1}$ was studied with Hamiltonian truncation (see figure \ref{fig:MNegIsingSpectrum}). The curve marked $\epsilon$ is obtained from this one by the $\mathbb{Z}_2$ (spin flip) symmetry of the Ising model. This $\mathbb{Z}_2$ symmetry implies that the right picture should be symmetric under the reflection $\bar{\lambda} \to -\bar{\lambda}$.
        We conjecture that the  curves merge (smoothly) at $\bar{\lambda} =\pm \bar{\lambda}_\textup{min} \approx \pm 0.1$ and $\Delta =1$.           There is also a boundary RG flow generated by the relevant  boundary operator with $\Delta<1$ that goes from  the lower to the upper curves. In the right plot, the flow is expected to land on the first stable fixed point. This is clearly visible from the perturbative beta function \eqref{betag} close to $\bar{\lambda}_\textup{min}$, which describes a one-parameter flow.    
         }
         \label{fig:specIsing}
\end{figure}

\begin{figure}[!h]
         \centering
         \begin{overpic}[width=0.6\textwidth]{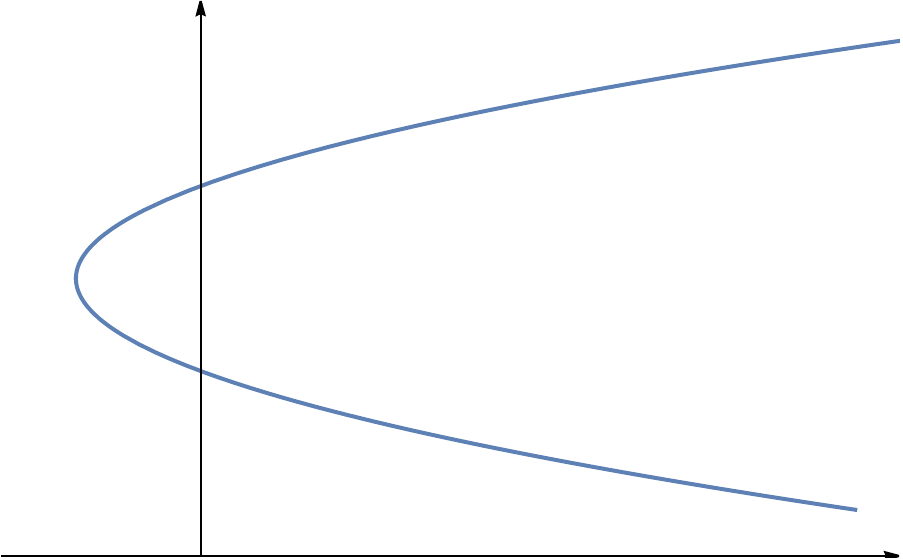}
       \put(17,59){$\Delta$}
       \put(97,-4){$\bar{\lambda}$}
       \put(13,42){$d+1$}
       \put(23,37){\Large$D$}
       \put (22.3,41.3) {\circle*{1}}
        \put(13,18){$d-1$}
       \put(23,14){\Large$N$}
        \put (22.3,20.8) {\circle*{1}}
       \multiput(8.1,31)(2,0){7}{\line(1,0){1}}
       \multiput(8.1,31)(0,-2){15}{\line(0,-1){1}}
       \put(19,32){$d$}
       \put(8,-4){$\bar{\lambda}_\textup{min}$}
       \put(65,11){\vector(0,1){39.5}}
       \put(66,28){$\int_\textup{bry} \phi^2$}
       \end{overpic}
       \vspace{0.3cm}
         \caption{ 
         Scaling dimension of the first $\mathbb{Z}_2$-even boundary operator (above the identity) for a free massive scalar field   in AdS$_{d+1}$, namely $\Delta= d\pm \sqrt{1+4\bar{\lambda}}$.
         The value $\bar{\lambda}=0$ corresponds to a conformally coupled scalar. The black dots mark the two possible BCFTs (Dirichlet or Neunmann boundary conditions). 
         We see that the 2 curves merge smoothly at $\bar{\lambda}=-\frac{1}{4}$ and $\Delta=d$.
         There is also a ($\mathbb{Z}_2$ preserving) boundary RG flow generated by the relevant  boundary operator ($\phi^2$) with $\Delta<d$ that goes from  the lower to the upper curve.                  }
         \label{fig:freescalarpic}
\end{figure}

Along the way, we gathered data about the spectrum at finite radius of a variety of two dimensional QFTs. A striking feature of all the models we studied is the existence of a minimal value of the coupling, $\lambda_\textup{min}$, beyond which the convergence of Hamiltonian truncation becomes poor. When the model is exactly solvable, \emph{i.e.} for the free boson and the Ising model with zero magnetic field, the minimal coupling can be precisely identified: it is the point where the spectra of the theories defined in the UV with different boundary conditions become identical. This leads us to the following speculation:

\emph{All conformally invariant boundary conditions for a given CFT are connected by  relevant deformations of this CFT in AdS.}
\\
\vspace{-.3cm}
\\
In figures \ref{fig:specLY} and  \ref{fig:specIsing} we depict how this may work for the spectrum of boundary operators in the Lee-Yang and the Ising field theories. This speculation is supported  by the numerical results in figures \ref{fig:LYNegRbar} and \ref{fig:MNegIsingSpectrum}.
In figure \ref{fig:freescalarpic}, we  show the analogous figure for the exactly solvable model of a free massive scalar in AdS$_{d+1}$. It is interesting to notice that the free massive fermion provides an additional, somewhat degenerate, instance of this phenomenon. Indeed, the spectra obtained by quantizing the theory with the two possible boundary conditions are reported in eq. \eqref{FFspecN}: rather than being connected by a flow in AdS, they precisely coincide at the conformal fixed point $\bar{\lambda}=0$.\footnote{Notice instead that they do not coincide at $\lambda_\textup{min}$ as defined in figure \ref{fig:specIsing}. As explained in subsection \ref{subsec:IsingT}, the spectrum of the Ising model with the $\ket{\mathbb{1}}_\textup{Cardy}$ boundary condition equals the free fermion spectrum, projected onto the even fermion number. It is this projected spectrum which merges with the full fermionic theory flowing from the $\ket{\sigma}_\textup{Cardy}$ boundary condition. Without the projection, the two spectra have different multiplicities at $\lambda=\lambda_\textup{min}$. Moreover, the $\ket{\mathbb{1}}_\textup{Cardy}$ boundary condition has an additional single fermion state with vanishing energy.}

Leaving aside the case of the free fermion, let us comment on the expected behavior close to the merging point of two curves. 
For concreteness, consider figure \ref{fig:specLY}. The theory emanating from the $\mca{V}$ boundary condition has a relevant operator, which generically has to be fine tuned on the boundary, in order to preserve the AdS isometries. Let us call $g$ the coupling of this operator. The fine tuning is possible as long as the operator is not marginal, so this leads us to identify $\bar{\lambda}_\textup{min}$ with the coupling where $\Delta=d=1$. If we perturb $\bar{\lambda}$ away from this value, the beta function has the form
\beq
\beta(g) = (\bar{\lambda} -\bar{\lambda}_\textup{min})  - g^2 +\dots
\label{betag}
\eeq
where we neglect higher order terms in the boundary coupling $g$ and in the bulk coupling $(\bar{\lambda} -\bar{\lambda}_\textup{min}) $.\footnote{A term proportional to $g (\bar{\lambda} -\bar{\lambda}_\textup{min})$ can be eliminated by shifting $g$.} We are also careless about the magnitude of the coefficients in eq. \eqref{betag}: they can be related to the appropriate bulk-to-boundary two-point function and boundary three-point function. On the other hand, the signs have been chosen to match the phenomenology in figure \ref{fig:specLY}. Indeed,
for $\bar{\lambda} >\bar{\lambda}_\textup{min}$ there are two fixed points with $g=g_\pm=\pm \sqrt{\bar{\lambda} -\bar{\lambda}_\textup{min}} $.
These fixed points correspond to the two curves merging at $\bar{\lambda} =\bar{\lambda}_\textup{min} $. 
In fact, the corresponding scaling dimension of the lightest boundary operator is given by
\beq
\Delta_\pm=d+\frac{d\beta}{dg}(g_\pm) = d\pm 2 \sqrt{\bar{\lambda} -\bar{\lambda}_\textup{min}}\,,
\eeq
 in agreement with our sketch. The theory at $\bar{\lambda} <\bar{\lambda}_\textup{min}$ should instead be described by a complex CFT \cite{Kaplan:2009kr,Gorbenko:2018ncu}. The same features present themselves in all of the figures in this section, except for the left panel in figure \ref{fig:specIsing}. In this case, the two curves cross rather than merging smoothly, and the lower curve is allowed to proceed to the left towards the flat space limit. The reason for this exception is simple: the relevant operator $\sigma_b$ is $\mathbb{Z_2}$ odd, while the bulk perturbation is $\mathbb{Z_2}$ even. No fine tuning is necessary in this case, hence no dangerous marginal operator is turned on at $\bar{\lambda}=\bar{\lambda}_\textup{min}$. Finally, this discussion suggests that turning on the relevant boundary operator allows to flow between the theories which merge at $\bar{\lambda}_\textup{min}$.
 
We look forward to learn about other studies that confirm or disprove the speculation above.

In this work, we also explored other aspects of QFT in curved space. In particular, we discussed spontaneous symmetry breaking in AdS, both in the classical theory, in subsection \ref{sec:sbb}, and via Hamiltonian truncation for $\phi^4$ in subsection \ref{sec:phi4}. Our considerations leave open a few interesting questions, related to the nature of symmetry breaking at finite AdS radius. Probably the most pressing one is: what is the signal on the boundary of the phase transition in the bulk? When analyzing the $\phi^4$ theory, we provided evidence for the fact that the bulk theory is not described by a CFT at the phase transition. This evidence was based at weak coupling on connecting the phase transition line to a free massive boson at the BF bound, whose stress tensor is not traceless. Hence, we should not be looking for Virasoro symmetry in the boundary spectrum. On the other hand, the appearance of the BF bound raises the tantalizing hypothesis that the (in)stability of the false vacuum shares some similarities with the previously discussed (in)stability of a theory under bulk deformations beyond a coupling $\bar{\lambda}_\textup{min}$, due to the flow generated by a marginal operator. It would be interesting to scrutinize this hypothesis further, perhaps via a higher order computation at weak coupling.

Moving on to the more technical aspects, a lot of effort was devoted in this paper to overcome the challenges presented by Hamiltonian truncation in infinite volume, leading up to the prescription \eqref{prescription}. We hope that the lessons learned here will be useful in studying other UV sensitive flows with the same framework. While the prescription has been tested in a variety of situations, we came short of rigorously proving its validity. The main obstacle is the lack of positivity of the spectral densities involved in the computation, whose large energy limit is therefore not easily tied to the short distance limit of the Laplace transform. Famously \cite{Pappadopulo:2012jk}, this problem would not be present if we wanted to estimate the large energy limit of the spectral density averaged over a smooth measure. This suggests that it may be worth applying to this problem the ideas of \cite{Rutter:2018aog}, where a smooth cutoff was used instead of the usual sharp truncation of the Hilbert space to a finite subspace. Another direction worth exploring is the possibility for eq. \eqref{prescription} to emerge from a renormalization procedure, where the high energy tail of the spectrum is integrated out systematically~\cite{Hogervorst:2014rta}. The difficulty is of course that the Hamiltonian in AdS does not possess a continuum limit to start from. One possible starting point is the theory with a spatial cutoff $\epsilon$, as in eq. \eqref{eq:regV}, which is well defined. An RG derivation of the prescription might be useful in elucidating the UV sensitivity of eq. \eqref{prescription}, which depends explicitly on the bare energies. For instance, this feature gives rise to an ambiguity along the flow, namely the identification of the correct shift in the neighborhood of a level crossing, as discussed in subsection \ref{subsec:IsingTspec}.

With the aim of obtaining high precision results in Hamiltonian truncation at larger values of the the AdS radius, it will be necessary to systematically improve the action along the lines explained in subsection \ref{sec:rate}. This was important in previous works~\cite{Rychkov:2014eea,Rychkov:2015vap,Elias-Miro:2017tup}, and it would be especially worth trying it when the UV boundary spectrum contains operators with low scaling dimension, like the in the case of the minimal models.

A more ambitious way of improving the procedure is related to the fact that, in this work, we did not take full advantage of the symmetries of AdS spacetime. The isometry group of AdS$_2$ is $SL(2,\mathbb{R})$ and states must fall into irreducible representations of this group. We  used this fact to test numerical convergence, but it would be interesting to exploit it to accelerate the algorithm. Perhaps, we could start by finding the subspace of the Hilbert space generated by primary operators by looking at the kernel of the symmetry generator $K$, and then diagonalize $H$ in this subspace.
We leave the exploration of this idea for the future.

One can also wonder if Hamiltonian truncation can be used to compute observables other than the spectrum. One motivation is the following. It is well known that one can extract scattering amplitudes from the  finite size behaviour of the energy levels on a compact space \cite{Luscher:1986pf}. However, in its standard form, this method only  gives access to the elastic region of the 2 to 2 scattering amplitude. One advantage of AdS is that one can extract scattering amplitudes at arbitrarily high energy from the flat space limit of the boundary correlation functions \cite{Gary:2009ae, Penedones:2010ue, Paulos:2016fap, Dubovsky:2017cnj, Hijano:2019qmi, Komatsu:2020sag}. It would be interesting to adapt Hamiltonian truncation to compute boundary correlation functions and test this idea in practice. 

Finally, a rewarding byproduct of our efforts was the formula \eqref{eq:muitoImportante}, which summarizes the cumbersome sums characterizing Rayleigh-Schr\" odinger perturbation theory in a compact and useful expression. We proved eq. \eqref{eq:muitoImportante} for non-degenerate perturbation theory. It would be nice to generalize the formula to the degenerate case.

\subsubsection*{Acknowledgements}

We thank Andrea Guerrieri, Jan-Joris Stauffer and Kiarash Naderi for collaboration at the early stage of this project. We also thank Nathan B. Agmon, Barak Gabai, Slava Rychkov, Balt van Rees, Matthew Walters and Xi Yin for useful discussions. The authors are supported by the Simons Foundation grant 488649 (Simons Collaboration on the Nonperturbative Bootstrap) and by the Swiss National Science Foundation through the project 200021-169132 and through the National Centre of Competence in Research SwissMAP. The computations in this paper were run on the EPFL SCITAS cluster.

\appendix

\section{Symmetry constraints on matrix elements}
\label{app:constraints}
%!TEX root = ../HTinAdS.tex
%%%%%%%%%%%%%%%%%%%%%%%%%%%%%%%%%%%%%%%%%%%%%

In this appendix, we consider constraints coming from AdS isometries (or boundary conformal symmetry) on matrix elements of the operator
\beq
V \ldef R^{\DD_\mca{V}} \int_{-\pi/2}^{\pi/2} \!\frac{dr}{(\cos r)^2}\; \mca{V}(\tau=0,r)
\eeq
where $\mca{V}$ is any local bulk operator. To this end, we fix two boundary primary states $\ket{\psi_i}$ and $\ket{\psi_j}$ of dimension $\DD_i$ and  $\DD_j$. We will use ${SL}(2,\mbb{R})$ symmetry to derive relations between different matrix elements between descendants, that is to say between the matrix elements
\beq
\mbf{V}\ud{j,m}{i,n} \ldef \braket{\psi_j,m}{V}{\psi_i,n}.
\label{VmatelApp}
\eeq
The argument will make use of a second operator $W$, defined as follows:
\beq
W \ldef R^{\DD_\mca{V}}  \int_{-\pi/2}^{\pi/2} \frac{dr}{(\cos r)^2}\, \sin r\, \mca{V}(0,r)
\eeq
along with its matrix elements which we denote as $\mbf{W}\ud{j,m}{i,n}$ as in eq.~\reef{VmatelApp}. Parts of our discussion will be similar in spirit to section\@ 2.1 of reference~\cite{Fitzpatrick:2010zm}, where relations between matrix elements in AdS${}_3$ were derived using ${SO}(2,2)$ group theory.

\subsection{Relations between matrix elements of descendants}
\label{sec:relrelrel}

Let us proceed with the derivation of the promised relations. We derive them simply by evaluating Ward identities. If $G$ is a generator of ${SL}(2,\mbb{R})$, then $G \ket{\psi_i,n}$ can be computed using eq.~\reef{eq:KPonstates}. To wit
\bsub
\begin{align}
H \ket{\psi_i,n} &= (\DD_i + n) \ket{\psi_i,n},\\
K \ket{\psi_i,n} &= \gamma(\DD_i,n) \ket{\psi_i,n-1},\\
P \ket{\psi_i,n} &= \gamma(\DD_i,n+1) \ket{\psi_i,n+1}
\end{align}
\esub
where $\ga(\DD,n) = \sqrt{n(2\DD+n-1)}$. At the same time, $G$ acts on the local operators $\mca{V}$ as a differential operator $\mca{D}_G$, as specified in Eq.~\reef{eq:genAct}.
For general $G$, we then  have the identity
\bsub
\begin{align}
\braket{\psi_j,m}{[G,\mca{V}(\tau,r)]}{\psi_i,n} &= \mca{D}_G \cdot \braket{\psi_j,m}{\mca{V}(\tau,r)}{\psi_i,n}\\
&= (\bra{\psi_j,m}G) \mca{V}(\tau,r) \ket{\psi_i,n} - \bra{\psi_j,m} \mca{V}(\tau,r) (G \ket{\psi_i,n}).
\end{align}
\esub
For $G = H$, this implies that
\beq
\label{eq:Drell}
\left(\frac{\pd}{\pd \tau} + \DD_{ij} + n-m \right)\!  \braket{\psi_j,m}{\mca{V}(\tau,r)}{\psi_i,n} = 0
\eeq
where $\DD_{ij} \ldef \DD_i - \DD_j$. This completely fixes the $\tau$-dependence of $\braket{\psi_j,m}{\mca{V}(\tau,r)}{\psi_i,n}$. Likewise, for $P$ we get
\begin{multline}
\label{eq:Prell}
\mca{D}_P \cdot \braket{\psi_j,m}{\mca{V}(\tau,r)}{\psi_i,n} = \ga(\DD_j,m) \braket{\psi_j,m-1}{\mca{V}(\tau,r)}{\psi_i,n}\\ - \ga(\DD_i,n+1)  \braket{\psi_j,m}{\mca{V}(\tau,r)}{\psi_i,n+1}
\end{multline}
and finally for $K$ we have
\begin{multline}
\label{eq:Krell}
\mca{D}_K \cdot \braket{\psi_j,m}{\mca{V}(\tau,r)}{\psi_i,n} = \ga(\DD_j,m+1) \braket{\psi_j,m+1}{\mca{V}(\tau,r)}{\psi_i,n}\\ - \ga(\DD_i,n)  \braket{\psi_j,m}{\mca{V}(\tau,r)}{\psi_i,n-1}.
\end{multline}
To proceed, we explicitly apply the operators $\mca{D}_{P}$ and $\mca{D}_K$, which are first order both in $\pd_\tau$ and $\pd_r$. The $\tau$-derivatives can be computed by means of~\reef{eq:Drell}. The LHS of~\reef{eq:Prell} then becomes
\beq
(\ldots) = \left[(-\DD_{ij} + m -n) \sin r + \cos r\, \frac{\pd}{\pd r} \right]\!\braket{\psi_j,m}{\mca{V}(0,r)}{\psi_i,n}
\eeq
setting $\tau = 0$ from now on. The $\sin r$ term can be integrated over $r$ yielding something proportional to the matrix element $\mbf{W}\ud{j;m}{i;n}$. The second term can be computed similarly after employing integration by parts:
\bsub
\begin{align}
\int_{-\pi/2}^{\pi/2} \frac{dr}{(\cos r)^2} \cos r\, \frac{\pd}{\pd r} \braket{\psi_j,m}{\mca{V}(0,r)}{\psi_i,n} &= - \int_{-\pi/2}^{\pi/2} \frac{dr}{(\cos r)^2}  \sin r\,  \braket{\psi_j,m}{\mca{V}(0,r)}{\psi_i,n}\\
&= - \mbf{W}\ud{j;m}{i;n}
\end{align}
\esub
discarding a boundary term that vanishes. Integrating the RHS of~\reef{eq:Prell} at $\tau = 0$ as well, we obtain the relation
\bsub
\label{eq:VWrels}
\beq
(-\DD_{ij} + m-n-1) \mbf{W}\ud{j;m}{i;n} = \gamma(\DD_j,m) \mbf{V}\ud{j;m-1}{i;n} - \gamma(\DD_i,n+1) \mbf{V}\ud{j;m}{i;n+1}.
\eeq
If we repeat this for $K$, we get in addition
\beq
(-\DD_{ij} + m-n+1) \mbf{W}\ud{j;m}{i;n} = \gamma(\DD_j,m+1) \mbf{V}\ud{j;m+1}{i;n} - \gamma(\DD_i,n) \mbf{V}\ud{j;m}{i;n-1}.
\eeq
\esub
Moreover, matrix elements with either $n=0$ or $m=0$ can be computed as definite integrals over
\beq
f_{ji}(r) \ldef  \braket{\psi_j}{\mca{V}(0,r)}{\psi_i}
\eeq
which only involves the primary states $\ket{\psi_{i,j}}$. To wit:
\bsub
\label{eq:VWedge}
\begin{align}
  \label{eq:Vexp}
 \mbf{V}\ud{j;m}{i;0} &= (-1)^m \sqrt{\frac{m!}{(2\DD_j)_m}}  \int_{-\pi/2}^{\pi/2}\!\frac{dr}{(\cos r)^2} \, C_m^{-\half(m-\DD_{ij} )}(\sin r)  f_{ji}(r),\\
 \mbf{W}\ud{j;m}{i;0} &=  (-1)^{m+1} (m+1) \sqrt{\frac{m!}{(2\DD_j)_m}}  \int_{-\pi/2}^{\pi/2}\!\frac{dr}{(\cos r)^2} \, \frac{C_{m+1}^{-\half(m-\DD_{ij}+1 )}(\sin r)}{m-\DD_{ij}+1} \, f_{ji}(r).
\end{align}
\esub
Here $C_n^{(\nu)}(\cdot)$ denotes a Gegenbauer polynomial. Matrix elements with $m=0$ but $n \neq 0$ can be computed similarly, exchanging $\DD_i \lra \DD_j$ in the above formulas. In order to derive~\reef{eq:VWedge}, let
\beq
\mca{F}_m(r) \ldef \braket{\psi_j,m}{\mca{V}(0,r)}{\psi_i}
\eeq
omitting labels for the external states $\ket{\psi_{i,j}}$, and in particular $\mca{F}_0(r) = f_{ji}(r)$. Setting $n=0$ in~\reef{eq:Krell}, we get the recurrence relation
\beq
 \mca{F}_{m+1}(r) = \frac{1}{\ga(\DD_j,m+1)} \left[(-\DD_{ij} + m) \sin r - \cos r \, \frac{\pd}{\pd r} \right]\!\mca{F}_m(r).
\eeq
After repeatedly integrating by parts and discarding (vanishing) boundary terms, the formulas~\reef{eq:VWedge} are recovered.

The interpretation of the above results depends on the precise value of $\DD_{ij} = \DD_i - \DD_j$. We will discuss the cases where $\DD_{ij}$ is fractional and integer separately:
\begin{itemize}
\item If $\DD_{ij}$ is non-integer, the relations~\reef{eq:VWrels} together with the ``initial conditions''~\reef{eq:VWedge} are sufficient to compute \emph{any} matrix element of $\mbf{V}$. This can be done simply by eliminating $\mbf{W}$ from the equations~\reef{eq:VWrels}.
\item 
On the other hand, if $\DD_{ij}$ is an integer, it is not always possible to eliminate ${\mbf{W}}$ from the above equations. Instead, we find \emph{new} relations between specific matrix elements:
\bsub
\begin{align}
  \ga(\DD_j,m+1)\mbf{W}\ud{j;m+1}{i;m-\DD_{ij}} &= \ga(\DD_i,m-\DD_{ij})\mbf{W}\ud{j;m}{i;m-\DD_{ij}-1} + \mbf{V}\ud{j;m}{i;m-\DD_{ij}},\\
  \ga(\DD_i,m+1) \mbf{W}\ud{j;m+\DD_{ij}}{i;m+1} &= \gamma(\DD_j,m+\dd_{ij}) \mbf{W}\ud{j;m+\DD_{ij}-1}{i;m} + \mbf{V}\ud{j;m+\DD_{ij}}{i;m}
\end{align}
\esub
which are once more derived by a judicious application of integration by parts. It turns out that these identities are again sufficient to reduce the computation of the matrix $\mbf{V}$ to its boundary.
\end{itemize}

If $\DD_{ij}$ is integer, certain selection rules apply. For concreteness, consider the case $\DD_i = \DD_j$. Then by setting $m=n+1$ in the first relation of~\reef{eq:VWrels}, we obtain the identity
\beq
0 = \ga(\DD_i,n+1) \mbf{V}\ud{j;n}{i;n} - \ga(\DD_i,n+1) \mbf{V}\ud{i;n+1}{i;n+1}
\eeq
which implies that all matrix elements $\mbf{V}\ud{j;n}{i;n}$ with $n=0,1,2,\ldots$ are identical. This result has a simple physical meaning. Indeed, if $\psi_j = \psi_i$, the diagonal of the matrix $\mbf{V}\ud{i;n}{i;n}$  is proportional to the tree-level energy shift of the states $\ket{\psi_i,n}$. But by conformal symmetry, all such energy shifts (or anomalous dimensions) must be identical. Consequently, it is necessary that for $\DD_i = \DD_j$, the diagonal of the matrix $\mbf{V}$ is constant. 

Next, consider the case where $\DD_{ij}$ is a non-zero integer, say $\DD_{ij} \rdef s$ with $s \in \{1,2,3,\ldots\}$. Then from~\reef{eq:Vexp} it follows that $\mbf{V}\ud{j;s}{i;0}$ vanishes --- this is simply a property of the Gegenbauer polynomials. But from Eq.~\reef{eq:VWrels} it follows that
\beq
\ga(\DD_j,n+s+1) \mbf{V}\ud{j;n+s+1}{i;n+1} = \ga(\DD_i,n+1) \mbf{V}\ud{j;n+s}{i;n}
\eeq
for any $n$. Consequently, \emph{all} matrix elements of the form $\mbf{V}\ud{j;n+s}{i;n}$ vanish! If $\DD_{ij}$ is a negative integer, a similar argument can be made.

Let us finally comment on parity. If $\ket{\psi_{i,j}}$ have the same parity and $\mca{V}$ respects parity, then $f_{ji}(r)$ is even under $r \mapsto -r$. But then using~\reef{eq:VWedge} and~\reef{eq:VWrels}, it can be shown that matrix elements of $\mbf{V}$ (resp.\@ $\mbf{W}$) vanish if $m+n$ is odd (resp.\@ even). The same result could have been derived directly since
\beq
\braket{\psi_i,m}{\mtt{P}V\mtt{P}^{-1}}{\psi_j,n} = (-1)^{m+n}\braket{\psi_i,m}{V}{\psi_j,n} = + \braket{\psi_i,m}{V}{\psi_j,n}
\eeq
(using the fact that $\mtt{P}V\mtt{P}^{-1} = +V$) which for odd $m+n$ is only possible if $\mbf{V}\ud{j;m}{i;n} = 0$. A similar argument can be applied to the case where $\ket{\psi_{i,j}}$ have opposite parity.

\subsection{Example: vacuum amplitudes}

As an application of the above formalism, let's consider the case where the in-state is the vacuum $\ket{\Omega}$. By ${SL}(2,\mbb{R})$ symmetry, the wavefunction $f_{i,\Omega}$ is completely fixed, up to an overall constant:
\beq
f_{i,\Omega}(r) = \braket{\psi_i}{\mca{V}(r)}{\Omega} =  \a_i (\cos r)^{\DD_i}.
\eeq
The coefficient $\a_i$ can be interpreted as the bulk-boundary OPE coefficient of $\mca{V} \to \Oo_i$ (up to a convention-dependent constant), see~\reef{bulkToBound}. A short computation shows that
\beq
\mbf{V}\ud{i;2n+1}{\Omega} = 0,
\quad
\mbf{V}\ud{i;2n}{\Omega} = \a_i (\DD_i+2n)B(\th \DD_i - \th,\tfrac{3}{2})\sqrt{\frac{(\th)_n(\DD_i)_n}{n!(\DD_i+\th)_n}}
\eeq
where $B$ is the beta function. As an application of this result, we note that
\beq
\lim_{n \to \infty} \frac{\mbf{V}\ud{i;2n}{\Omega}}{\sqrt{n} \, \mbf{V}\ud{i;0}{\Omega}} = \frac{2}{\DD \sqrt{B(\th,\DD)}}
\eeq
that is to say that the matrix elements $\mbf{V}\ud{i;2n}{\Omega}$ grow as $\sqrt{n}$ at large $n$. In particular, this result implies that the Casimir energy diverges at least linearly at second order in perturbation theory.

\subsection{Universality of matrix elements if $\DD_i - \DD_j$ is an even integer}
\label{subsec:constraintsEven}

If $\DD_{ij}$ is an even integer, that is to say $|\DD_i - \DD_j| = 2s$ for some integer $s$, and the in- and out-states $\ket{\psi_{i,j}}$ have the same parity, the matrix $\mbf{V}\ud{j;m}{i;n}$ is severely constrained. In fact, we will prove that it is fixed in terms of $s+1$ constants.

{\bf Theorem}: let $\ket{\psi_{i,j}}$ be two primaries with $\pi_i = \pi_j$ that satisfy $\DD_j = \DD_i \pm 2 s$ with $s \in \{0,1,2,\ldots\}$, and let $V$ be a parity-preserving bulk perturbation. Then the matrix ${\bf V}\ud{j;m}{i;n}$ 
is completely fixed by ${SL}(2,\mbb{R})$ symmetry in terms of the $s+1$ constants
\beq
c_a(\psi_j,\psi_i) \ldef \int_{-\pi/2}^{\pi/2}\!\frac{dr}{(\cos r)^2}\, (\sin r)^{2a} f_{ji}(r),
\quad
a = 0,1,\ldots,s.
\eeq
{\it Proof}: by parity, only matrix elements with even $m+n$ can be non-zero. Thus all non-vanishing matrix elements with $n=0$ are proportional to the following integral:
\[
\mbf{V}\ud{j;2m}{i;0} \; \propto \; \int_{-\pi/2}^{\pi/2}\!\frac{dr}{(\cos r)^2}\,  C_{2m}^{-m\pm s}(\sin r) f_{ji}(r),
\quad
m = 0,1,2,\ldots
\]
as an immediate consequence of~\reef{eq:VWedge}.  But it's easy to show that any such Gegenbauer polynomial is a polynomial of degree  $\leq s$ in the variable $z = (\sin r)^2$. Indeed, we have
\beq
\label{eq:functionsGalore}
C_{2m}^{-m + s}(\sin r) = \frac{(1-s)_m }{m!}\, {}_2F_1(-m,s,\th,z),
\quad
C_{2m}^{-m-s}(\sin r) = \frac{(1+s)_m}{m!}\, {}_2F_1(-s,-m,\th,z).
\eeq
The first of these functions vanishes if $m \geq s$, due to the prefactor $(1-s)_m$; for $m<s$ it's indeed a polynomial of degree $m$ in the variable $z$. 
Likewise, the second function in~\reef{eq:functionsGalore} is a polynomial in $z$ of degree $\text{min}(m,s)$. Consequently, any matrix element with $n=0$ can be expressed in terms of the parameters $c_a(\psi_j,\psi_i)$.

Now, the same argument can be applied to matrix elements with $m=0$. Finally, all other matrix elements, having $m,n \neq 0$ can be computed in terms of the previous ones using the relations described in subsection~\ref{sec:relrelrel}.
\qed

In particular, for $s=0$ (i.e.\@ whenever $\DD_i = \DD_j$) the matrix $\mbf{V}\ud{j;m}{i;n}$ is completely universal, meaning that it is fixed by symmetry up to an overall normalization. To be precise, we have the following corollary:

{\bf Corollary}: let $\ket{\psi_{i,j}}$ be two primaries with $\DD_i = \DD_j$ and $\pi_i = \pi_j$. Then for any parity-preserving bulk perturbation $V$, we have
\beq
\label{eq:eckhart}
\braket{\psi_j,m}{V}{\psi_i,n} = \msc{V}_{mn}(\DD_i) \, \braket{\psi_j}{V}{\psi_i}
\eeq
where $\msc{V}_{mn}(\DD_i)$ is a matrix that only depends on $\DD_i$, to wit:
\beq
\label{eq:Vmat}
\msc{V}_{mn}(\DD) = \begin{cases} 0 & \text{if } m + n \text{ is odd}; \\
  \sqrt{(m+1)_{n-m}/(2\DD+m)_{n-m}} &  m \leq n,\; m+n \text{ even}; \\
  \msc{V}_{nm}(\DD) & m > n.
\end{cases}
\eeq
{\it Note}: setting $m=n$, we see that the diagonal matrix elements are constant (since $\msc{V}_{nn}(\DD) = 1$). This is consistent with the discussion from the end of subsection~\ref{sec:relrelrel}.\\
{\it Proof}. Since any possible solution to the Ward identities is unique in this case, it suffices to check that $\msc{V}_{mn}(\DD_i)$ solves~\reef{eq:VWrels}. Indeed~\reef{eq:Vmat} is a valid solution, if we also have
\bsub
\beq
\braket{\psi_j,m}{W}{\psi_i,n} = \msc{W}_{mn}(\DD_i) \, \braket{\psi_j}{V}{\psi_i}
\eeq for
     \beq
     \msc{W}_{mn}(\DD)= \begin{cases} 0 & \text{if } m + n \text{ is even}; \\
   \sqrt{(m+1)_{n-m}/(2\DD+m)_{n-m}} &  m < n,\; m+n \text{ odd}; \\
   \msc{W}_{nm}(\DD) & m > n.
     \end{cases} 
     \eeq
     \esub
This exercise is left to the reader.\qed

\section{Rayleigh-Schr\"{o}dinger perturbation theory revisited}
\label{app:RSPT}
%!TEX root = ../HTinAdS.tex
%%%%%%%%%%%%%%%%%%%%%%%%%%%%%%%%%%%%%%%%%%%%%

\subsection{Rayleigh-Schr\"{o}dinger coefficients and connected correlators}
\label{app:QMPT}

In this appendix, we prove a relation between coefficients in Rayleigh-Schr\"{o}dinger perturbation theory and correlation functions. To be precise, we have in mind a Hamiltonian of the form $H = H_0 + \la V$, where $H_0$ has a non-degenerate discrete spectrum
\[
0 = e_0 < e_1 < e_2 < \ldots
\]
associated to normalized eigenstates $\ket{i}$, such that $\ket{0}$ is the $H_0$ vacuum state. In perturbation theory, the interacting energies can be expanded as
\beq
\mca{E}_i = e_i + \sum_{n=1}^\infty (-1)^{n+1} {\Red c_{i,n}}  \la^n
\eeq
for some constants ${\Red c_{i,n}}$ to be determined. Moreover, we can expand the eigenstates $\ket{ \Omega_j}$  of $H$ in terms of the eigenstates of $H_0$ with certain coefficients:
\beq
\label{eq:anjidef}
|\brakket{i}{\Omega_j}|^2 = \dd_{ij} +  \sum_{n = 2}^\infty (-1)^n {\Blue a_n(j \to i)} \, \la^n.
\eeq
Normalization of the states $\ket{\Omega_i}$ and $\ket{i}$ means that
\beq
\label{eq:overlapNorm}
{\Blue a_n(i \to i)} = - \sum_{j \neq i} {\Blue a_n(i \to j)} = - \sum_{j \neq i} {\Blue a_n(j \to i)}.
\eeq
From now on, we will use rule~\reef{eq:overlapNorm} to only consider coefficients with $j \neq i$.

It's the goal of Rayleigh-Schr\"{o}dinger perturbation theory to provide formulas for ${\Red c_{i,n}}$ and ${\Blue a_n(j \to i)}$. Some simple formulas are
\beq
\label{eq:lowFormulas}
{\Red c_{i,1}} = \braket{i}{V}{i},
\quad
{\Red c_{i,2}} = \sum_{j \neq i} \frac{|\braket{i}{V}{j}|^2}{e_j - e_i}
\qaq
{\Blue a_2(j \to i)} = \frac{|\braket{i}{V}{j}|^2}{(e_j - e_i)^2}.
\eeq
At higher orders in perturbation theory, similar formulas quickly become cumbersome.

We claim that, roughly speaking, the coefficients ${\Red c_{i,n}}$ can be expressed as an integral over the $n$-point correlation function of the operator $V$ inside the state $\ket{i}$. Considering the vacuum state at second order in perturbation theory, it is for instance well-known that
\beq
\label{eq:vacFormula}
{\Red c_{0,2}} = \int_0^\infty\!d\tau\, \braket{0}{V(\tau)V(0)}{0}_\text{conn}
\eeq
where the subscript $\braket{i}{\Oo_n \dotsm \Oo_1}{i}_\text{conn}$ means that contractions between the operators $\Oo_j$ must be subtracted in the state $\ket{i}$, e.g.\@
\bsub
\begin{align}
 \braket{i}{V(\tau)V(0)}{i}_\text{conn} &= \braket{i}{V(\tau)V(0)}{i} - \braket{i}{V}{i}^2 \rdef \mfr{g}_i(\tau)\\
\braket{i}{V(\tau_1 + \tau_{2})V(\tau_1)V(0)}{i}_\text{conn} &= \braket{i}{V(\tau_1 + \tau_{2})V(\tau_1)V(0)}{i} - \braket{i}{V}{i}^3 \nn\\
&\hspace{15mm} - \braket{i}{V}{i} \Big[ \mfr{g}_i(\tau_1) +  \mfr{g}_i(\tau_2) + \mfr{g}_i(\tau_1 + \tau_2) \Big]
\end{align}
\esub
and so forth. However, it's not possible to simply replace the in- and out-states in~\reef{eq:vacFormula} by $\ket{i}$ and $\bra{i}$ to reproduce the expression from Eq.~\reef{eq:lowFormulas} --- in fact, the resulting integral would diverge for a generic state $\ket{i}$. To proceed, we therefore define a spectral density $\rho_{i,n}$ as follows:
\beq
\label{eq:fSharpDef}
\hspace{-5mm} \boxed{
\braket{i}{V(\tau_1 + \ldots \tau_{n-1}) \dotsm V(\tau_1 + \tau_2)V(\tau_1)V(0)}{i}_\text{conn} = \prod_{\l=1}^{n-1} \int_{0}^\infty \!d\a_\l\, e^{-(\a_\l - e_i)\tau_\l} \, \rho_{i,n}(\a_1,\ldots,\a_{n-1})
}
\eeq
which has the following inverse:
\[
\label{eq:fSharpInv}
\rho_{i,n}(\vec{\a}) = \prod_{\l=1}^{n-1} \int_{\gamma-i\infty}^{\gamma + i \infty}\!\frac{d\tau_\l}{2\pi i} \, e^{(\a_\l-e_i) \tau_\l} \, \braket{i}{V(\tau_1 + \ldots \tau_{n-1}) \dotsm V(\tau_1 + \tau_2)V(\tau_1)V(0)}{i}_\text{conn}
\]
for $\gamma > 0$.\footnote{We use that on the right half plane $\Re(\tau_j) > 0$ the correlator is analytic.} For $n=1, 2,3$ we have for instance
\bsub
\begin{align}
\rho_{i,1} &=   \braket{i}{V}{i} \label{rho1states}\\
\rho_{i,2}(\a_1) &= \sum_{k \neq i} \dd(\a_1-e_k) \, |\braket{i}{V}{k}|^2 \label{rho2states}\\
  \rho_{i,3}(\a_1,\a_2) &= \sum_{k,\l \neq i} \dd(\a_1 - e_k)\dd(\a_2 - e_\l) \, \braket{i}{V}{\l}\big( \braket{\l}{V}{k}-\delta_{k\l}  \braket{i}{V}{i}
  \big)\braket{k}{V}{i}  \,. \label{eq:rho3n}
\end{align} 
\esub

We will see that the Rayleigh-Schr\"{o}dinger coefficients can be compactly expressed in terms of the densities $\rho_{i,n}$, to wit:
\beq
\label{eq:muitoImportante}
\boxed{
  {\Red c_{i,n}} = \prod_{\l=1}^{n-1} \int_0^\infty\!\frac{d\a_\l}{\a_\l - e_i} \, \rho_{i,n}(\a_1,\ldots,\a_{n-1}).
}
\eeq
For $n=1,2$, it is easy to check that~\reef{eq:muitoImportante} agrees with~\reef{eq:lowFormulas}. In the rest of this section, the proof of formula~\reef{eq:muitoImportante} will be explained.

First, let us comment on some general features of the $\rho_{i,n}(\vec{\a})$.\footnote{Here, we use the shorthand notation $\vec{\a} =(\a_1,\ldots,\a_{n-1})$.}   This object is generally a complex-valued distribution, in fact a sum (generally infinite) of delta functions. Also notice that $\rho$ only has support on $\a_\l \geq 0$ since there are no states with energy $e < 0$, and the distribution vanishes at $\a_\l = e_i$ since $\rho$ describes \emph{connected} correlation functions.
In fact, it will be useful in the proof of \eqref{eq:muitoImportante} to write $\rho$ as follows
\beq
\rho_{i,n}(\a_1,\ldots,\a_{n-1}) = \sum_{j_1\neq i} \dots
\sum_{j_{n-1}\neq i}  
\delta(\a_1-e_{j_1}) \dots \delta(\a_{n-1}-e_{j_{n-1}})U_{j_1,\dots,j_{n-1}}\,.
\label{rhogenform}
\eeq
 Moreover, since $V$ is a Hermitian operator it can be shown that
\beq
\label{eq:fS}
\overline{\rho_{i,n}(\a_1,\ldots,\a_k,\ldots,\a_{n-1})} = \rho_{i,n}(\a_{n-1},\ldots,\a_{n-k},\ldots,\a_1)
\eeq
reflecting a peculiarity from the configuration of the $\tau_j$ in the correlator~\reef{eq:fSharpDef}. For $n=2$ the above identity means that $\rho_{i,2}(\a) \in \mbb{R}$. We also remark that the behavior of the $\rho_{i,n}$ at infinity depends on the high-energy behavior of the quantum theory in question. For instance, if $\kappa \geq 1$ then
\beq
\label{eq:taub}
\mfr{g}_i(\tau) \; \limu{\tau \to 0}\;  \frac{c}{\tau^\kappa}
\quad
\lra
\quad
\rho_{i,2}(\a) \; \limu{\a \to \infty} \; \frac{c}{\Gamma(\kappa)} \, \a^{\kappa-1}
\eeq
appealing to a Tauberian theorem. In AdS${}_2$, this Tauberian formula applies with $\kappa = 2$, but in UV-finite quantum mechanics all correlators are integrable (having in particular $\kappa < 1$). In what follows it will be assumed that the integrals~\reef{eq:muitoImportante} are well-defined, without requiring any additional subtractions. 

The reader may worry about the seemingly divergent denominator $\prod_{\l=1}^{n-1} 1/(\a_\l - e_i)$ in~\reef{eq:muitoImportante}. However, recall that $\rho_{i,n}$ vanishes on some domain $|\a_\l - e_j| \leq \eps$ due to the connectedness of the correlator it describes, and consequently the integral~\reef{eq:muitoImportante} is well-defined. If necessary, one can therefore replace all of the integrals by their principal value
\beq
\label{eq:pv}
\prod_{\l=1}^{n-1} \int_{\mbb{R}} \frac{d\a_\l}{\a_\l - e_i} \, \rho_{i,n}(\vec{\a}) \to \lim_{\eps \to 0^+} \,  \prod_{\l = 1}^{n-1} \int_{\mbb{R} \setminus [e_i - \eps,e_i + \eps]}  \frac{d\a_\l}{\a_\l - e_i} \, \rho_{i,n}(\vec{\a})
\eeq
without changing the result.

\subsubsection*{Proof of~\reef{eq:muitoImportante}}

Let us now describe a proof of the identity~\reef{eq:muitoImportante}. The proof will rely on a systematic evaluation of the Dyson operator
\beq
\label{eq:DysonOp}
U(b,a) \ldef \mrm{T}\exp\!\left[-\la \int_a^b\!d\tau\,  V(\tau)\right]\!.
\eeq
inside matrix elements. The operator $U$ has the defining property
\beq
\label{eq:HandH0}
e^{-H(b-a)} = e^{-H_0b} U(b,a) e^{H_0 a}
\eeq
which can easily be shown (the identity~\reef{eq:HandH0} is a tautology for $b=a$, and the two sides obey the same differential equation with respect to $b$). 

For concreteness, let us first consider the vacuum state $\ket{0} = \ket{\text{vac}}$, for which the proof of~\reef{eq:muitoImportante} is easiest. There we have for any $T > 0$ the exact identity
\beq
\label{eq:vacPr1}
\braket{\text{vac}}{U(T,0)}{\text{vac}} = \sum_j\braket{\text{vac}}{e^{H_0 T} e^{-H T}}{\Omega_j}\brakket{\Omega_j}{\text{vac}}  = \sum_{i=0}^\infty \exp(-\mca{E}_i T) |\brakket{\text{vac}}{\Omega_j}|^2
\eeq
where we have inserted a complete basis of states $\ket{\Omega_j}$ and used Eq.~\reef{eq:HandH0} to rewrite $U$.  Taking the limit $T \to \infty$, we have\footnote{Notice that $\braket{\text{vac}}{U(T,0)}{\text{vac}} > 0$ so the logarithm is well-defined.}
\beq
\label{eq:vac222}
\mca{E}_0(\la) = - \lim_{T \to \infty} \frac{1}{T} \, \ln \, \braket{\text{vac}}{U(T,0)}{\text{vac}}.
\eeq
The object on the RHS is precisely the generator of connected correlation functions inside the vacuum state. At $n$-th order in perturbation theory, we read off that
\bsub
\begin{align}
\label{KLtype}
  c_{\text{vac},n} &= \prod_{\l=1}^{n-1} \int_0^\infty\!d{\tau_\l}\, \braket{\text{vac}}{V(\tau_1 + \ldots \tau_{n-1}) \dotsm V(\tau_1 + \tau_2)V(\tau_1)V(0)}{\text{vac}}_\text{conn} \\
  &=  \prod_{\l=1}^{n-1} \int_0^\infty\!\frac{\!d{\a_\l}}{{\a_\l}} \, \rho_{\text{vac},n}(\a_1,\ldots,\a_{n-1})
\end{align}
\esub
as claimed. Formula \eqref{KLtype} for the vacuum energy is not new --- see a disussion below equation~\reef{cinDensity} --- and a formal version of~\reef{eq:muitoImportante} appeared previously without proof in~\cite{Klassen:1991ze}.\footnote{The formula in question involves connected correlators that are integrated over all of spacetime. Beyond short-distance divergences, such correlators are also IR-divergent due to intermediate states of energy \emph{below} the energy $e_i$ of the external state $\ket{i}$. Our formula~\reef{eq:muitoImportante} takes such divergences into account and does not need any additional renormalization.}

For excited states the proof of~\reef{eq:muitoImportante} goes along the same lines, but it is more complicated. To proceed, we insert the operator $\exp(\mca{E}_i T)\exp(-HT)$ inside the unperturbed state $\ket{i}$. This yields
\begin{align}
e^{\dd \mca{E}_i T} \braket{i}{U(T,0)}{i} &= \sum_{j=0}^\infty   |\brakket{i}{\Omega_j}|^2 \, e^{-(\mca{E}_j -\mca{E}_i) T}  \notag\\
&=  1 - \sum_{j \neq i}^\infty   |\brakket{i}{\Omega_j}|^2 \left[1 - e^{-(\dd \mca{E}_j - \dd \mca{E}_i) T} e^{-(e_j - e_i)T}\right]
\label{eq:master}
\end{align}
writing $\dd \mca{E}_i = \mca{E}_i - e_i = \sO(\la)$. 
We have used the identity
\beq
|\brakket{i}{\Omega_i}|^2 + \sum_{j \neq i} |\brakket{i}{\Omega_j}|^2 = 1
\eeq
to obtain the above expression. We will show that formula \eqref{eq:muitoImportante} follows from matching the terms linear in $T$ on both sides of \eqref{eq:master}.

The {\bf LHS} of~\reef{eq:master} 
 can be recast as
\begin{align}
\ln(\msf{LHS}) &=  \dd \mca{E}_i(\la)  T +  \sum_{n=1}^\infty (-1)^n \la^n \mca{X}_{i,n}(T),\\
\mca{X}_{i,n}(T)& \ldef \int_0^T\!d\tau_1 \dotsm \int_{\tau_{n-1}}^T\!d\tau_n\,  \braket{i}{V(\tau_n) \dotsm V(\tau_1)}{i}_\text{conn}. 
\label{eq:Xndef}
\end{align}
Using the spectral representation \eqref{eq:fSharpDef} we can write \eqref{eq:Xndef} as follows
\beq 
\mca{X}_{i,n}(T) =   \int_0^\infty\! d \a_1 \dots d\a_{n-1} \,\mca{W}_n( \a_1-e_i,\dots,\a_{n-1}-e_i|T) \, \rho_{i,n} (\vec{\a})
\label{XnfromW}
\eeq
where \footnote{This follows from \eqref{eq:Xndef} changing integration variables  to $t_i=\tau_{i+1}-\tau_i$ with $i=0,1,\dots, n-1$ and $\tau_0=0$.}
\beq
\mca{W}_n(\vec{\a}|T) = \int_0^T dt_0 
\int_0^{T-t_0} dt_1\,e^{-\a_1 t_1} 
\int_0^{T-t_0-t_1} dt_2\,e^{-\a_2 t_2} \dots
\int_0^{T-t_0  \dots - t_{n-2}} dt_{n-1}\,e^{-\a_{n-1} t_{n-1}}\,.
\label{Wdef}
\eeq
Equivalently, we can define $\mca{W}$ by
\beq
\mca{W}_n(\vec{\a}|T) = \int_0^T dt_0 
\mca{Q}_{n-1}(\a_1,\dots,\a_{n-1}|T-t_0) 
\eeq
with $\mca{Q}$ defined recursively
\beq
\mca{Q}_i(\a_{1},\dots,\a_i |T) = \int_0^{T} dt\,e^{-\a_1 t} \mca{Q}_{i-1}(\a_{2},\dots,\a_i |T-t)\,,\qquad\qquad\mca{Q}_0=1\,.
\label{Qrec}
\eeq 
It is not hard to see that this recursion relation leads to
\beq
\mca{Q}_i(\a_{1},\dots,\a_i |T) = 
\frac{1}{\a_1 \a_2 \dots \a_i} -  \sum_{\l = 1}^{i} \frac{ e^{-\a_\l T}}{\a_\l } \prod_{j=1\atop j \neq \l}^i \frac{1}{\a_j - \a_\l}\,.
\label{Qexp}
\eeq 
This can be easily checked by plugging  \eqref{Qexp} in \eqref{Qrec} and using the identity
\beq
0 = 
\frac{1}{\a_1 \a_2 \dots \a_i} -  \sum_{\l = 1}^{i} \frac{1}{\a_\l } \prod_{j=1\atop j \neq \l}^i \frac{1}{\a_j - \a_\l}\,,
\label{rationalIdemtity}
\eeq 
which follows from the fact   this  rational function of $\alpha_1$   vanishes as $\alpha_1 \to \infty$ and does not have any poles (the reader may easily check that the apparent poles have vanishing residue).

We can now use \eqref{Qexp} in \eqref{Wdef} to obtain
\beq
\mca{W}_n(\vec{\a}|T) = \frac{T}{\a_1 \a_2 \dots \a_i} - \mca{K}_n(\vec{\a}|T)
\label{WintermsofK}
\eeq
where
\beq
\label{eq:Kndef}
\mca{K}_n(\vec{\a}|T) \ldef \sum_{\l = 1}^{n-1} \frac{1-e^{-\a_\l T}}{\a_\l^2} \prod_{j=1\atop j \neq \l}^{n-1} \frac{1}{\a_j - \a_\l}.
\eeq
This allows us to write \eqref{XnfromW}, for $n \geq 2$, as follows:
\beq
\label{eq:Xexpo}
\mca{X}_{i,n}(T) = T \int_0^\infty\!\frac{d\vec{\a}}{\vec{\a}-e_i} \, \rho_{i,n}(\vec{\a}) -  \int_0^\infty\!d\vec{\a} \; \mca{K}_n(\vec{\a}-e_i|T)  \rho_{i,n}(\vec{\a})\,.
\eeq
Here and in what follows we use a vector notation for spectral integrals, for instance
\beq
\int_0^\infty\!\frac{d\vec{\a}}{\vec{\a}-e_i} f(\vec{\a}) = \prod_{\l=1}^{n-1} \int_0^\infty\!\frac{d\a_\l}{\a_\l - e_i} f(\a_1, \ldots, \a_{n-1}).
\eeq

We will now show that the only term that grows linearly with $T$ in \eqref{eq:Xexpo} is the first term. More precisely, we will show that the second term in \eqref{eq:Xexpo} evaluates to something of the form
\beq
\text{const.} + \sum_{\beta\neq 0} P_\beta(T) e^{-\beta T}\,,
\label{notlinearT}
\eeq
for some values of $\beta\neq 0$ and  $P_\beta(T) $ some polynomial of $T$.
For $n=1$ we simply have 
\beq 
\mca{X}_{i,1}(T)  = T  \rho_{i,1}= T \braket{i}{V}{i}.
\eeq
For $n=2$ the second integral is straightforward to do:
\beq
\int_0^\infty\!d\a \, \mca{K}_2(\a - e_i;T) \rho_{i,2}(\a) = \sum_{j \neq i} |\braket{i}{V}{j}|^2 \mca{K}_2(e_j - e_i|T).
\eeq
However, for $n \geq 3$ we have to worry about the denominators $\a_j - \a_{\l}$. 
It is clear from the definition \eqref{Wdef} that $\mca{W}_n(\a_1,\dots,\a_{n-1}|T)$ is completely regular when some of its arguments $\a_\ell$  coincide. 
Therefore, from \eqref{WintermsofK} we conclude  the same must be true of $\mca{K}_n$ although this is not obvious from the explicit formula \eqref{eq:Kndef}.

For instance for $n=3$ we have
\beq
\lim_{\a' \to \a} \mca{K}_3(\a,\a'|T) = \frac{2}{\a^3} - \frac{2+\a T}{\a^3} \exp(-\a T) \rdef \mca{K}^\sharp_3(\a|T).
\eeq
Therefore, for $n=3$ the second term in~\eqref{eq:Xexpo} evaluates to
\begin{multline}
  \hspace{-5mm} \int_0^\infty\!d\vec{\a} \; \mca{K}_3(\vec{\a}-e_i|T)  \rho_{i,3}(\vec{\a}) = \sum_{j,k \neq i \; \wedge \; j \neq k} \braket{i}{V}{k}\braket{k}{V}{j}\braket{j}{V}{i}\,  \mca{K}_3(e_j-e_i,e_k-e_i|T) \\
  + \sum_{j \neq i} (\braket{j}{V}{j} - \braket{i}{V}{i}) |\braket{i}{V}{j}|^2 \, \mca{K}_3^\sharp(e_j - e_i|T),
\end{multline}
which is of the general form \eqref{notlinearT}.

For general $n$, we can use \eqref{rhogenform} to write the second term in \eqref{eq:Xexpo} as follows
\beq 
\int_0^\infty\!d\vec{\a} \; \mca{K}_n(\vec{\a}-e_i|T)  \rho_{i,n}(\vec{\a})
= \sum_{j_1\neq i} \dots
\sum_{j_{n-1}\neq i}  
 U_{j_1,\dots,j_{n-1}} \mca{K}_n(e_{j_1}-e_i, \dots,e_{j_{n-1}}-e_i |T)\,.
 \label{sum2ndterm}
\eeq
Clearly the arguments of $\mca{K}_n$ are always different from zero.
Therefore, from \eqref{eq:Kndef} we conclude that this is of the general form
 \eqref{notlinearT}.
 The only subtlety is that for some terms in the sum \eqref{sum2ndterm} the arguments of $\mca{K}_n$ are exactly equal. In this case, one cannot use \eqref{eq:Kndef} due to the presence of vanishing denominators.
However, as explained above $\mca{K}_n$ always has a finite limit when several of its arguments coincide.
Moreover,  we can write 
\beq
\mca{K}_n(\vec{\a}|T) = g_0(\vec{\a})
 - \sum_{\l = 1}^{n-1} e^{-\a_\l T} g_\l(\vec{\a})\,,
\eeq
with $g_\l(\vec{\a})$ some rational functions that can be read off from \eqref{eq:Kndef}.
If $\a_1$ and $\a_2$ coincide then
\begin{align}
\mca{K}_n(\a_1,\a_1,\a_3,\dots |T) &=   g_0(\vec{\a})
 - \sum_{\l = 3}^{n-1} e^{-\a_\l T} g_\l(\vec{\a}) -
e^{-\a_1 T} \tilde{g}_1(\vec{\a}|T)\,,
\end{align}
where $g_0$ and $g_{\l\ge 3}$ have finite limits when $\a_2 \to \a_1$ and 
\begin{align}
\tilde{g}_1(\vec{\a}|T)&=  \lim_{\epsilon \to 0} \left [ g_1(\a_1,\a_1+\epsilon,\a_3,\dots)
+e^{-\epsilon T} g_2(\a_1,\a_1+\epsilon,\a_3,\dots) \right]\\
&= T \frac{1}{\a_1^2} \prod_{j=3}^{n-1}\frac{1}{\a_j-\a_1} - \frac{\partial }{\partial \a_1} \left[\frac{1}{\a_1^2} \prod_{j=3}^{n-1}\frac{1}{\a_j-\a_1}\right]
\end{align}
Due to the collision  $\a_2 \to \a_1$ the coefficient of the exponential $e^{-\a_1 T}$ became a linear function of $T$.
This is a general phenomena. When several $\a$'s coincide the coefficient of the associated exponential $e^{-\a_\l T}$ becomes a polynomial in $T$.\footnote{For example,
\beq
\lim_{\a',\a'' \to \a} \mca{K}_4(\a,\a',\a''|T)=\frac{3}{\a^4} + \frac{6+4\a T+\a^2 T^2}{2\a^4} \, \exp(-\a T).
\eeq
}
However, as long as every $\a_\l \neq 0$ the exponential never disappears and therefore \eqref{sum2ndterm} is indeed of the form  \eqref{notlinearT}.

The {\bf RHS} of~\reef{eq:master} can also be expanded to 
\beq
\label{eq:RHSRHS}
\ln(\msf{RHS}) = \sum_{n=2}^\infty (-1)^n \la^n \mca{Y}_{i,n}(T)
\eeq
for some functions $\mca{Y}_{i,n}(T)$:
\bsub
\begin{align}
  \mca{Y}_{i,2}(T) &= - \sum_{j \neq i} {\Blue a_2(j \to i)} \left[1-e^{-(e_j - e_i)T}\right] \\
  \mca{Y}_{i,3}(T) &= - \sum_{j \neq i} {\Blue a_3(j \to i)} \left[1-e^{-(e_j - e_i)T}\right] - T \sum_{j \neq i} ({\Red c_{j,1}} - {\Red c_{i,1}}) {\Blue a_2(j \to i)} e^{-(e_j - e_i)T}
\end{align}
\esub
and likewise for higher $n$. The coefficients  ${\Blue a_n}$ were defined in~\reef{eq:anjidef}. 

We can now directly compare~\reef{eq:Xndef} and~\reef{eq:RHSRHS} to obtain the desired formula, Eq.~\reef{eq:muitoImportante}. For $n=2$, we obtain the relation
\beq
T \left[ {\Red c_{i,2}}  -  \int_0^\infty\!\frac{d{\a}}{\a-e_i}  \, \rho_{i,2}(\a)\right] =   \int_0^\infty\!d\a \; \frac{1-e^{-(\a-e_i)T}}{(\a-e_i)^2}  \rho_{i,2}({\a}) -  \sum_{j \neq i} {\Blue a_2(j \to i)} \left[1-e^{-(e_j - e_i)T}\right].
\eeq
Since this equation must hold for \emph{any} $T > 0$, we recover the $n=2$ case of~\reef{eq:muitoImportante} along with the standard result
\beq
{\Blue a_2(j \to i)} = \frac{|\braket{i}{V}{j}|^2}{(e_i - e_j)^2}.
\eeq
The same strategy can be used to prove~\reef{eq:muitoImportante} for higher $n$. The logic is always the same, namely to match all terms of order $\la^n$ appearing in Eqs.~\reef{eq:Xndef} and~\reef{eq:RHSRHS}. The LHS has two purely linear pieces in $T$, namely
\beq
(-1)^{n+1} {\Red c_{i,n}} T + (-1)^n T \int_0^\infty\!\frac{d\vec{\a}}{\vec{\a}-e_i} \rho_{i,n}(\vec{\a}) + \ldots.
\eeq
which perfectly cancel, provided that~\reef{eq:muitoImportante} holds. The remaining term on the LHS comes from the kernel $\mca{K}_n$, and crucially it never contains a term growing linearly with $T$ at large $T$: at most it contains a constant and polynomials in $T$ multiplied by (growing or decaying) exponentials. This remainder term can be matched against the RHS --- which does not contain any term growing linearly with $T$  --- and this procedure gives expressions for the coefficients ${\Blue a_n(j \to i)}$.

As mentioned at the beginning, the spectrum of the unperturbed Hamiltonian was taken to be non-degenerate. Let us briefly discuss how the proof breaks down in case $H_0$ is degenerate. Suppose for instance that there are two vacua, that is to say $g \geq 2$ states $\ket{\text{vac}_\a}$ with $e_\a = 0$. Equation~\reef{eq:vacPr1} still holds when $\ket{\text{vac}}$ is replaced by any of the $\ket{\text{vac}_\a}$, or in fact by any linear combination $\ket{v} = c^\a \ket{\text{vac}_\a}$. Equation~\reef{eq:vac222} also holds, provided that the true vacuum $\ket{\Omega}$ at finite coupling has non-zero overlap with the starting state $\ket{v}$. However, it is no longer true that the coefficients $c_{\text{vac},n}$ are described by equation~\reef{KLtype}, and indeed such integrated correlators generically diverge. In other words, the $T\to \infty$ limit and the perturbative expansion do not commute.

\subsection{Feynman diagrams}\label{sec:feynman}

\begin{figure}[htb]
\centering
  \includegraphics[scale=1.5]{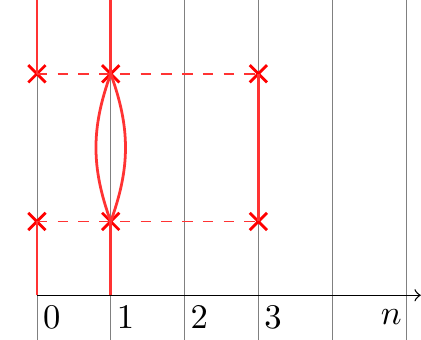}
  \caption{An example Feynman diagram arising at second order in $\phi^3$ perturbation theory. This diagram involves $\ket{0,1}$ resp.\@ $\bra{0,1}$ as in- and out-states and $\ket{1,1,3}$ as an intermediate state.}
\label{fig:feyn0}
\end{figure}

This subsection is dedicated to a detailed explanation of the diagrammatic representation of RS perturbation theory, introduced in subsection \ref{sec:exNew}. The reader may refer to figure \ref{fig:feyn0} to fix ideas, although similar pictures can be drawn for a generic interaction term.
The diagram is to be read from bottom to top, with time flowing upwards. The horizontal axis corresponds to the mode numbers, \emph{e.g.}\@ the state $\ket{1,2,2,4}$ would be represented by single lines at $n=1,4$ and a double line at $n=2$. Polynomial vertices can either lower or raise occupation numbers. The diagram in figure~\ref{fig:feyn0} would occur at second order in perturbation theory for a theory with potential $V_3 = \int \NO{\phi^3}$~. In the time domain it can be read as
\beq
\braket{0,1}{V_3(\tau_2)}{1,1,3} \braket{1,1,3}{V_3(\tau_1)}{0,1}
\eeq
or when computing energies it would contribute an amount
\beq
\dd \mca{E}_{0,1} \supset \bar{\la}^2 \frac{|\braket{1,1,3}{V_3}{0,1}|^2}{E_{0,1} - E_{1,1,3}} = - \frac{72\bar{\la}^2}{\DD+4} \left| \int_{-\pi/2}^{\pi/2}\!\frac{dr}{(\cos r)^2}\, f_0(r) f_1(r) f_3(r) \right|^2
\eeq
to the energy of the state $\ket{0,1}$. The functions $f_n(r)$ are defined in eq. \eqref{eq:fnDef}. Vice versa, by ``cutting open'' diagrams like figure~\ref{fig:feyn0} horizontally, one easily recovers the intermediate state and its energy. 

\subsection{Explicit calculations at second order in perturbation theory}\label{sec:ExplicitSecondOrder}
This subsection is a complement to subsection \ref{sec:exNew}. There, after writing the potential $V$ in terms of ladder operators and introducing a truncation cutoff $\Lambda$, we expand the energy of the vacuum $\ket{\Omega}$ and of the excited state $\ket{\chi} \equiv a_0^\dagger\ket{\Omega}$ up to second order in $\bar{\lambda}$ using Rayleigh-Schrödinger perturbation theory:
\bsub
\label{eq:RSPerturb}
\begin{align}
\mca{E}_{\Omega}&= e_{\Omega} + \bar{\lambda} \langle \Omega|V|\Omega\rangle - \bar{\lambda} ^2 \sum_{i=\text{2 p.s.}} \frac{|\braket{\Omega}{V}{i}|^2}{e_i-e_\Omega}  + \sO(\bar{\la}^3)~, \\
\mca{E}_{\chi}&= e_{\chi} + \bar{\lambda}  \langle \chi|V|\chi\rangle - \bar{\lambda} ^2 \sum_{\substack{i=\text{1 p.s.,}\\ \text{3 p.s.}}} \frac{|\braket{\chi}{V}{i}|^2}{e_i-e_\chi}  + \sO(\bar{\la}^3)~.
\end{align}
\esub
in which ``$n$ p.s.'' stands for $n$-particle states and we used the fact that $\NO{\phi^2}$ acting on an $n$-particle state produces $n-2, n$ and $(n+2)$-particle states. 
Substituting the operator $V$ in the sums above in terms of $A_{mn}$ using eq. \ref{eq:AmatP} and summing the states up to cutoff $\Lambda$ gives

\bsub
\label{eq:SecondOrderEnergies}
\begin{align}
\mca{E}_{\Omega}(\Lambda)&= - 4\bar{\lambda}^2 \sum_{m,n=0}^{m+n+2\DD\leq\Lambda} \frac{1}{N(m,n)} \frac{A^2_{mn}(\DD)}{2\DD+m+n}~, \\
\mca{E}_{\chi}(\Lambda)&= \DD + 2 \bar{\lambda} A_{00} - 4\bar{\lambda}^2 \sum_{m=1}^{m+\DD\leq\Lambda} \frac{A^2_{0m}(\DD)}{m}  - 4\bar{\lambda}^2 \sum_{m,n=0}^{m+n+3\DD\leq\Lambda} \frac{1}{N(m,n,0)} \frac{(A_{mn}(\DD)+\delta_{0m}A_{0n}+\delta_{0n}A_{0m})^2}{2\DD+m+n}~.
\end{align}
\esub
where $N(\ell,m,\ldots,n)$ is the norm of the state $a_{\ell}^\dagger a_{m}^\dagger \dotsm a_{n}^\dagger \ket{\Omega}$. In particular,
\begin{align}
N(m,n)&=1+\delta_{m,n}~,\\
N(m,n,0) & =1+\delta_{m,n}~,\  \ \  m,n\neq0~, \\
N(m,0,0) &=2~, \qquad \qquad m\neq 0~, \\
N(0,0,0) &=6~.
\end{align}
In the large $\Lambda$ limit, the two double sums in eq. \eqref{eq:SecondOrderEnergies} have the same linear asymptotic behaviour. To see this, let us define 
\beq
\label{eq:sndef}
\sigma_N(\DD) \ldef \sum_{m+n = 2N} \frac{|\braket{\Omega}{V_2}{m,n}|^2}{2\DD+m+n}~.
\eeq
such that
\beq
\mca{E}_\Omega(\La) \limu{\La \to \infty} - \bar{\la}^2 \sum_{N=0}^{\half \La} \sigma_N(\DD) + \sO(\la^3)~.
\eeq
Eq.~\ref{eq:sndef} is a sum of $\sO(N)$ terms that can be expressed in terms of the explicit coefficients $\msc{V}_{mn}(\DD)$ from eq.~\ref{eq:VmatMain} . At large $N$ we can replace this sum by an integral 
\beq
  \lim_{N \to \infty} \sigma_N(\DD) = \sigma_\infty(\DD) = \frac{1}{(2\DD-1)^2} \int_0^1 dx\, \left(\frac{x}{2-x}\right)^{2\DD-1}~,
  \eeq
  which can be expressed in terms of special functions if desired. 
We then have
\bsub
\label{eq:LinearGrowthOfVac}
\begin{align}
\mca{E}_\Omega(\La) \limu{\La \to \infty} - \bar{\la}^2 \sigma_\infty(\DD) \La  + \sO(\la^3,\La^0)~,\\
\mca{E}_\chi(\La) \limu{\La \to \infty} - \bar{\la}^2 \sigma_\infty(\DD) \La  + \sO(\la^3,\La^0)~.
\end{align}
\esub

As emphasized in section \ref{sec:cutoff}, the naive difference $ \mca{E}_{\chi}(\Lambda)-\mca{E}_\Omega(\Lambda)$ does not reproduce the exact result \eqref{eq:DcorrMain}. Let us show instead that the prescription \eqref{prescription} does the job. We begin by computing
\beq
\mca{E}_{\chi}(\Lambda)-\mca{E}_{\Omega}(\Lambda-\DD)= \DD + 2 \bar{\lambda} A_{00} - 4\bar{\lambda}^2 \sum_{m=1}^{m+\DD\leq\Lambda} \frac{A^2_{0m}(\DD)}{m} - 4\bar{\lambda}^2 \sum_{m=0}^{m+3\DD\leq\Lambda} \frac{A^2_{0m}(\DD)}{m+2\DD}~.
\label{EpDiffLambda}
\eeq
In the limit $\Lambda \rightarrow \infty$, the above sums can be written in terms of hypergeometric functions: 
\begin{align}
  \label{eq:KamranSum}
&\sum_{m=1}^{\infty} \frac{A^2_{0m}(\DD)}{m} =\frac{2}{\DD(2\DD-1)^2}~\, _{3}F_{2}\!\left({{\half,1,1}~\atop~{\half+\DD,1+\DD}}\,\Bigg|\,1\right)~, \\
&\sum_{m=0}^{\infty} \frac{A^2_{0m}(\DD)}{m+2\DD} = \frac{2 }{\DD (2\DD+1) (2\DD-1)^2}\, _{3}F_{2}\!\left({{\frac{3}{2},1,1}~\atop~{\frac{3}{2}+\DD,1+\DD}}\,\Bigg|\,1\right)~.
\end{align}
These formulas furnish  a completely explicit expression for the energy shift~\reef{EpDiffLambda} in the limit $\La \to \infty$, which agrees with the expected result~\reef{eq:enDShift}.\footnote{This check requires verifying an identity involving a sum of two hypergeometric ${}_3F_2(1)$ functions. We have checked this identity numerically for many values of $\Delta$, but we do not have an analytical proof for any $\Delta$.}

Similarly, we have checked that the second-order energy shift of the states $a_n^\dagger \ket{\Omega}$ for small values of $n$ is correctly reproduced with our prescription. In addition, we also checked that the \emph{third}-order energy shift of the state $\ket{\chi}$ is correctly reproduced.

Alternatively, we can reproduce the second-order energy shift of the state $\ket{\chi}$ by computing the two-point connected correlator of $V_2(\tau)$ inside the state $\ket{i}$. Computing such correlators is an efficient way to extract spectral densities more generally, that is to say for other states or for energy shifts at higher orders in perturbation theory. Let us therefore briefly spell out this computation. The correlator we need to compute is given by
\beq
\label{eq:Gchidef}
g_\chi(\tau) \ldef \braket{\chi}{V_2(\tau)V_2(0)}{\chi}_\text{conn} -  \braket{\Omega}{V_2(\tau)V_2(0)}{\Omega}_\text{conn}.
\eeq
We will see that $g_\chi$ can be expressed in terms of the special function
\beq
\mfr{g}_\DD(z) \ldef \sum_{n=0}^\infty \msc{V}_{0,2n}(\DD)^2 z^n = {}_3F_2\!\left[{{\th,\, 1,\, 1}~\atop~{\DD,\, \DD+\th}}\, \Bigg| \, z\right].
\eeq
To derive this result, we recall that $\ket{\chi} = a_0^\dag \ket{\Omega}$ and use the identities
\begin{align*}
    [a_k,\,\NO{\phi^n(\tau,r)}] &= n \cdot e^{(\DD+k)\tau} \, f_k(r) \; \NO{\phi^{n-1}(\tau,r)}\,,\\
    [\NO{\phi^n(\tau,r)},a_k^\dagger] &= n \cdot e^{-(\DD+k)\tau} \, f_k(r) \; \NO{\phi^{n-1}(\tau,r)}
\end{align*}
for $k=0$ and $n=2$. Now, equation~\reef{eq:Gchidef} can be restated as
\beq
\label{eq:gxxx}
g_\chi(\tau) = \braket{\Omega}{[a_0,V_2(\tau)][V_2(0),a_0^\dag]}{\Omega} +  \braket{\Omega}{[V_2(\tau),a_0^\dag][a_0,V_2(0)]}{\Omega} - \braket{\chi}{V_2}{\chi}^2
\eeq
using the fact that $\braket{\Omega}{V_2}{\Omega}$ vanishes. The first term in~\reef{eq:gxxx} works out to
\begin{align}
  4 e^{\DD \tau} \int[dr][dr']\, f_0(r) f_0(r') G(\tau,r|0,r') &=   4 \sum_{n=0}^\infty e^{-2n\tau} A_{0,2n}(\DD)^2 \nn \\
  &= \frac{4}{(2\DD-1)^2} \, \mfr{g}_\DD(e^{-2\tau})
\end{align}
where $G = \expec{\phi \phi}$ is the scalar propagator and 
using the shorthand notation
\beq
\int[dr] \ldef \int_{-\pi/2}^{\pi/2}\!\frac{dr}{(\cos r)^2}.
\eeq
Here we have used that all \emph{odd} matrix elements $A_{0,2n+1}$ vanish, and that $A_{m,n}$ is proportional to $\msc{V}_{m,n}$. The second term in~\reef{eq:gxxx} works out to
\beq
  4 e^{-\DD \tau} \int[dr][dr']\, f_0(r) f_0(r') G(\tau,r|0,r') =  \frac{4}{(2\DD-1)^2} \, e^{-2\DD \tau} \, \mfr{g}_\DD(e^{-2\tau}).
\eeq
Finally,
\beq
\braket{\chi}{V_2}{\chi}^2 = \frac{4}{(2\DD-1)^2}
\eeq
and thus
\beq
\label{eq:GchiNiceResult}
g_\chi(\tau) = \frac{4}{(2\DD-1)^2} \left[ (1+e^{-2\DD \tau})\, \mfr{g}_\DD(e^{-2\tau})-1 \right].
\eeq
Indeed we recover (by testing for various values of $\DD$) that
\beq
\int_0^\infty\!d\tau\, g_\chi(\tau) = \frac{4}{(2\DD-1)^3}.
\eeq

\section{Short time singularities in the correlators of $V$}
\label{app:smalltau}
%!TEX root = ../HTinAdS.tex
%%%%%%%%%%%%%%%%%%%%%%%%%%%%%%%%%%%%%%%%%%%%%

In this appendix, we study the singularities of the following class of correlators,
\beq
\braket{i}{V_\epsilon(\tau_1 + \ldots \tau_{n-1}) \dotsm V_\epsilon(\tau_1 + \tau_2)V_\epsilon(\tau_1)V_\epsilon(0)}{i}_\text{conn}~,
\label{connected_corr_ep}
\eeq
when a subset of the $\tau_i$ goes to zero. The subscript $\epsilon$ denotes the regularized interaction, as defined in eq. \eqref{eq:regV}. 

In this appendix, it will be convenient to redefine the spatial coordinate as
\beq
\theta = r+\frac{\pi}{2}~,
\eeq
so that $\theta\in [0,\pi]$. Hence,
\beq
V_\epsilon(\tau) = R^{\DD_\mca{V}} \int_{\epsilon}^{\pi-\epsilon}\! \frac{d\theta}{(\sin \theta)^2}\, \mca{V}(\tau,\theta).
\eeq
Furthermore, we set $R=1$ from now on.

\subsection{The vacuum two-point function}
\label{sec:twoptvac}

Let us begin with the vacuum two-point function of the potential:
\beq
\braket{\Omega}{V_\epsilon(\tau)V_\epsilon(0)}{\Omega}
= \int_\eps^{\pi-\eps}\!\!
\frac{d\theta_1 d\theta_2}{\sin^2 \theta_1 \sin^2 \theta_2}\braket{\Omega}{\mca{V}(\tau,\theta_1)\mca{V}(0,\theta_2)}{\Omega}~.
\label{VVvacuum}
\eeq
The matrix elements in eq. \eqref{VVvacuum} are computed by a Euclidean correlator as long as $\tau>0$, which we assume throughout this subsection. Notice that, since the vacuum expectation value of $V_\eps$ vanishes by definition, the full correlator coincides with its connected part. It will be convenient to keep in mind that the two-point function of $\mca{V}$ depends on the position of the insertions only through their geodesic distance, or equivalently through the cross ratio $\xi$ defined in eq. \eqref{eq:crossratio}, which in the notation of this section reads
\beq
\xi =
 \frac{\cosh \tau-\cos (\theta_1-\theta_2)}{2 \sin \theta_1\sin \theta_2}~.
 \label{xitheta}
\eeq 
Hence, we also denote
\beq
\braket{\Omega}{\mca{V}(\tau,\theta_1)\mca{V}(0,\theta_2)}{\Omega}=f_\Omega(\xi)~.
\label{fvac}
\eeq
As discussed in subsection \ref{subsec:VacSecond}, the large energy limit of the Laplace transform of this correlator controls the cutoff dependence of the vacuum energy at second order in perturbation theory. 

We start by discussing the $\eps=0$ case. 
There are two possible sources of non-analyticities as $\tau \to 0$. First, the two-point function of the perturbing operator $\mca{V}$ has a short distance singularity fixed by the OPE at the UV fixed point:\footnote{Logarithms arise when the UV fixed point is a free theory, like in section \ref{sec:massive}. Also, notice that the subtraction of the vacuum expectation value of $\mca{V}$ has no effect on the leading singularity.}
\beq
f_\Omega(\xi) \sim (4\xi)^{-\Delta_\mca{V}}~,\qquad \xi\to 0~.
\label{OPEidentity}
\eeq 
In order to isolate the contribution of this short distance limit to eq. \eqref{VVvacuum}, we approximate $f_\Omega$ as in eq. \eqref{OPEidentity} and we rescale the difference $(\theta_1-\theta_2) \to \tau (\theta_1-\theta_2)$, keeping $\theta_1+\theta_2$ fixed. After sending $\tau \to 0$ and integrating, we obtain
\beq
\braket{\Omega}{V(\tau)V(0)}{\Omega} \sim \frac{\pi\, \Gamma\left(\Delta_\mca{V}-\frac{3}{2}\right)\Gamma\left(\Delta_\mca{V}-\frac{1}{2}\right)}{\Gamma (\Delta_\mca{V})\Gamma(\Delta_\mca{V}-1)}\,\tau^{-2\Delta_\mca{V}+1}+\dots\qquad \tau\to 0~.
\label{VacSmallTauBulk}
\eeq
It is interesting to notice that the exponent is a property of the short distance physics, and accordingly the integral over $(\theta_1-\theta_2)$ reduces to an integral over a flat space two-point function. It converges as long as $\Delta_\mca{V}>1/2$, otherwise an IR divergence makes the approximation \eqref{OPEidentity} inadequate. On the other hand, the integral over $(\theta_1+\theta_2)$ is sensitive to the AdS geometry, and requires the stronger condition $\Delta_\mca{V}>3/2$. As we shall see momentarily, precisely when this condition is not satisfied the leading small $\tau$ singularity is \emph{not} given by eq. \eqref{VacSmallTauBulk}.

Indeed, a second source of non-analyticity arises from the region where both insertions of $\mca{V}$ reach the same point on the boundary. In this case, the important contribution comes from  the region where $\theta_1 \sim \theta_2 \sim \tau$, and from its image under parity $(\pi-\theta_1) \sim (\pi-\theta_2) \sim \tau$. It is simple to guess that this is the case by considering the cross ratio $\xi$ in eq. \eqref{xitheta}.
When $\theta_i$ and $\tau$ all vanish at the same rate, $\xi$ attains a finite limit. After performing the change of variables $\theta_i = z_i \tau$ in eq. \eqref{VVvacuum} -- still at $\eps=0$ -- and multiplying by 2 to keep into account the contribution from the other AdS boundary, the leading contribution as $\tau \to 0$ is
\beq
\braket{\Omega}{V(\tau)V(0)}{\Omega}
\sim \frac{2}{\tau^2} \int_0^\infty \frac{dz_1dz_2}{z_1^2z_2^2} f_\Omega\left(\frac{1-z^2_{12}}{4z_1z_2}\right)+\dots
\label{VacSmallTauBoundary}
\eeq
Let us briefly discuss the convergence of this integral. As at least one of the $z_i$'s goes to zero, the boundary OPE can be used to factorize the double integral into a product. We conclude that the integral converges at small $z_i$ if the boundary OPE of $\mca{V}$ does not contain operators with scaling dimension $\Delta\leq 1$. This is precisely the condition \eqref{eq:as} which ensures finiteness of the matrix elements of $V$, and is therefore satisfied by assumption. At the other end of the integration region the integrand is suppressed, unless $z_{12}$ stays finite. In the latter case, $f_\Omega$ is dominated by the bulk OPE limit as in eq. \eqref{OPEidentity}.
Hence, the integral in \eqref{VacSmallTauBoundary} converges as long as $\Delta_\mca{V}<3/2$~. In the opposite case, the leading small $\tau$ singularity comes instead from the bulk OPE limit considered above, see eq. \eqref{VacSmallTauBulk}. 

Notice that, contrary to eq. \eqref{VacSmallTauBulk}, the theory dependence of the singularity in eq. \eqref{VacSmallTauBoundary} is through the full two-point function $f_\Omega$. Although we could still perform one of the two integrals to simplify the coefficient, we shall later obtain precisely this result as a by-product of a more general analysis -- see eqs. (\ref{VV_cor_kernel},\ref{VV_smalltau}).

\begin{figure}[t]
\centering
\begin{tikzpicture}[scale=1.4]
\draw [thick]  (-4,0) -- (4,0);
\draw [dashed] (3,0) arc [start angle=0, end angle=180, radius=3];
\draw [dashed] (2.5,0) arc [start angle=0, end angle=180, radius=2.5];
\draw (0,0) -- (2.31,0.97); 
\filldraw (2.31,0.97) circle [radius=0.05];
\node at (2.5,1.1) {$\mca{V}$};
\draw (0,0) -- (1.5,2.6); 
\filldraw (1.5,2.6) circle [radius=0.05];
\node at (1.7,2.7) {$\mca{V}$};
\draw [dashed] (1.4,0) arc [start angle=0, end angle=22.5, radius=1.4]; 
\node at (1.6,0.25) {$\theta_2$};
\draw [dashed] (0.7,0) arc [start angle=0, end angle=60, radius=0.7]; 
\node at (0.75,0.65) {$\theta_1$};
\draw (1.5,3.01) circle [radius=1.53];
\filldraw [red] (0.06,2.5) circle [radius=0.05];
\filldraw [red] (1.96,1.56) circle [radius=0.05];
\node at (2.5,-0.2) {1};
\node at (3,-0.2) {$e^\tau$};
\end{tikzpicture}
\caption{The dashed semi-circles are constant global time surfaces. The solid circle is a surface of fixed geodesic distance from the $\mca{V}$ insertion at angle $\theta_1$. The $\mca{V}$ insertion at angle $\theta_2$ is generic. The $\delta$ function in eq. \eqref{VVkernel} has support at the location of the two red points.}
\label{fig:circles}
\end{figure}
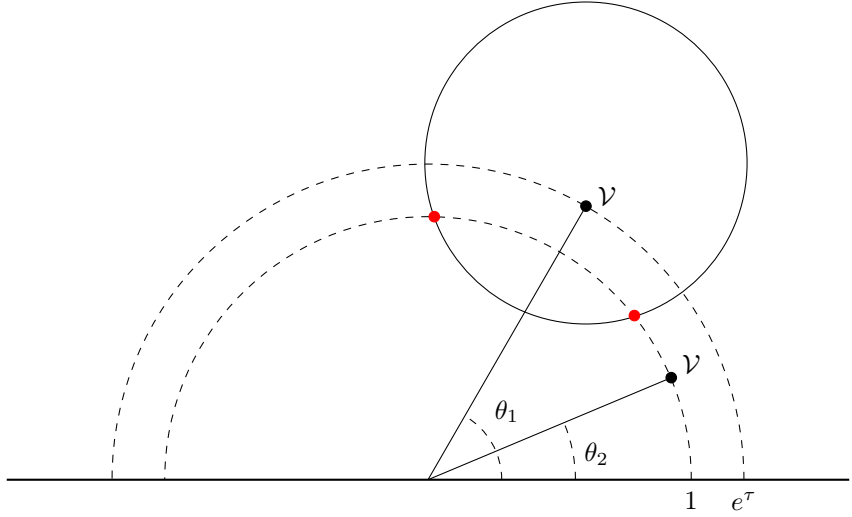

Although the previous analysis is sufficient to isolate the leading small $\tau$ singularities, for completeness we now develop a more systematic formalism, which yields the matrix element in eq. \eqref{VVvacuum}, with $\eps=0$, as a single integral over the correlator \eqref{fvac}. Since $f_\Omega$ only depends on the coordinates through the cross ratio $\xi$, it is convenient to separate the kinematical from the theory dependent data as follows:
\beq
\braket{\Omega}{V(\tau)V(0)}{\Omega}=\int_{\xi_\text{min}(\tau)}^\infty\! d\xi\, K(\xi|\tau) f_\Omega(\xi)~, \qquad  \xi_\text{min}(\tau) = \th(\cosh \tau-1)~.
\label{VV_cor_kernel}
\eeq
 The kernel $K$ is defined as
\beq
K(\xi|\tau) \ldef\int_{0}^{\pi}\!\frac{d\theta_1 \, d\theta_2}{\sin^2 \theta_1 \sin^2 \theta_2}\,\dd(\xi(\tau,\theta_1,\theta_2) - \xi), \qquad \xi(\tau,\theta_1,\theta_2) =\frac{\cosh\tau-\cos(\theta_1-\theta_2)}{2\sin\theta_1 \sin\theta_2}~.
\label{VVkernel}
\eeq
The kernel \eqref{VVkernel} can be expressed in terms of an elliptic integral for generic $\tau>0$. We will first integrate in $\theta_2$.
In Poincaré coordinates, the curve of constant geodesic distance from a point $x_1=(z,\bar{z})$, $z=\exp (\tau+\ii \theta)$, is a circle:
\begin{multline}
\xi = \frac{(z-z')(\zb-\zb')}{4\, \Im z\, \Im z'}~, \quad 
\iff \quad    (\Re z'-\Re z)^2+(\Im z'-c)^2= r^2, \\
c= \Im z (1+2\xi)~,\ \ 
r=2 \Im z \sqrt{\xi(\xi+1)}~.
\label{const_xi_circles}
\end{multline}
Therefore, the $\delta$-function in eq. \eqref{VVkernel} has support at most in two points, at fixed $\theta_1$, see fig. \ref{fig:circles}.
We can then evaluate the integral in $d\theta_2$ by changing variables, being careful about the sign of the Jacobian $\partial \xi(\tau,\theta_1,\theta_2)/\partial \theta_2$. Referring again to fig. \ref{fig:circles}, the geodesic distance of the marked point at angle $\theta_2$ from the point at angle $\theta_1$ is decreasing while the former enters the solid circle, and increasing while it exits it. This fixes the sign of the Jacobian. We obtain the following expression for the kernel:
\beq
K(\xi|\tau) = \frac{4 (1+2\xi)\cosh \tau}{\sqrt{\xi(\xi+1)}}  \int_{\theta_\textup{min}(\tau)}^{\pi/2} \! d\theta_1\,
\frac{1}{\left(\sin^2\theta_1+4 \xi(1+\xi)\sin^2\theta_\textup{min}(\tau)\right) \sqrt{\sin^2\theta_1-\sin^2\theta_\textup{min}(\tau) }}  ~,
\label{VVkernel_int}
\eeq
where
\beq
\qquad \theta_\textup{min}(\tau)= \arcsin \frac{\sinh \tau}{2 \sqrt{\xi(1+\xi)}}~,
\eeq
We used the reflection symmetry $\theta \to \pi-\theta$ to restrict the integration region to $\theta_1<\pi/2$. At the minimal angle $\theta_\textup{min}$ the solid circle of fig. \ref{fig:circles} becomes tangent to the semi-circle of radius $1$.
As promised, eq. \eqref{VVkernel_int} expresses the kernel in terms of the incomplete elliptic integral of the third kind:
\beq
K(\xi|\tau)= \frac{4 (1+2\xi)}{\sqrt{\xi(\xi+1)}\cosh \tau} 
\frac{1}{\sqrt{1-\sin^2 \theta_\textup{min}(\tau)}} \,
\Pi\left(\frac{1}{\cosh^2\tau},\frac{\pi}{2}- \theta_\textup{min}(\tau),\frac{1}{1-\sin^2 \theta_\textup{min}(\tau)}  \right)~.
\eeq
The full matrix element at fixed $\tau$ may now be obtained by integrating the kernel as in eq. \eqref{VV_cor_kernel}. Alternatively, one can recover the small $\tau$ limits in eqs. (\ref{VacSmallTauBulk},\ref{VacSmallTauBoundary}) as limits of the kernel itself. 

In particular, the bulk channel singularity is obtained by rescaling $\xi \to \tau^2 \xi$. In the $\tau \to 0$ limit, this reduces the kernel to a complete elliptic integral of the second kind:
\beq
K(\tau^2 \xi|\tau) \sim \frac{16}{\tau} \sqrt{\xi} \,E\left(1-\frac{1}{4\xi}\right)~,
\qquad \tau\to0.
\eeq
This approximation to the kernel can then be used in eq. \eqref{VV_cor_kernel} together with eq. \eqref{OPEidentity}, to recover eq. \eqref{VacSmallTauBulk}.

The kernel $K(\xi|\tau)$ is more useful in deriving the coefficient of the boundary channel small $\tau$ singularity, eq. \eqref{VacSmallTauBoundary}. Indeed, in this case the cross ratio is held fixed, and the rescaling $\theta\to\tau\,\theta$ in the integral form \eqref{VVkernel_int} yields a simple algebraic result:
\beq
K(\xi|\tau) \sim \frac{8}{\tau^2}\, \textup{arcsinh} \left(2 \sqrt{\xi(\xi+1)}\right)~, \qquad \tau \to 0~.
\label{VV_smalltau}
\eeq
Replacing eq. \eqref{VV_smalltau} in eq. \eqref{VV_cor_kernel}, we obtain the coefficient of the small $\tau$ singularity in eq. \eqref{VacSmallTauBoundary} as a single integral over the cross ratio, as promised.

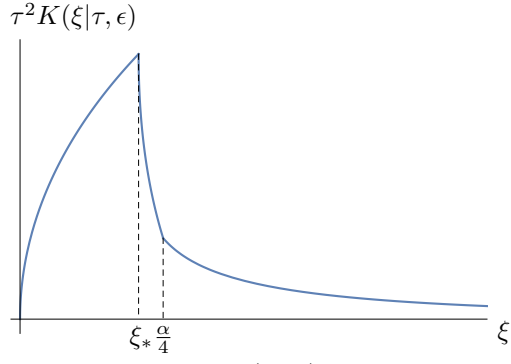
\begin{figure}[t]
\centering
\begin{overpic}[scale=0.5]{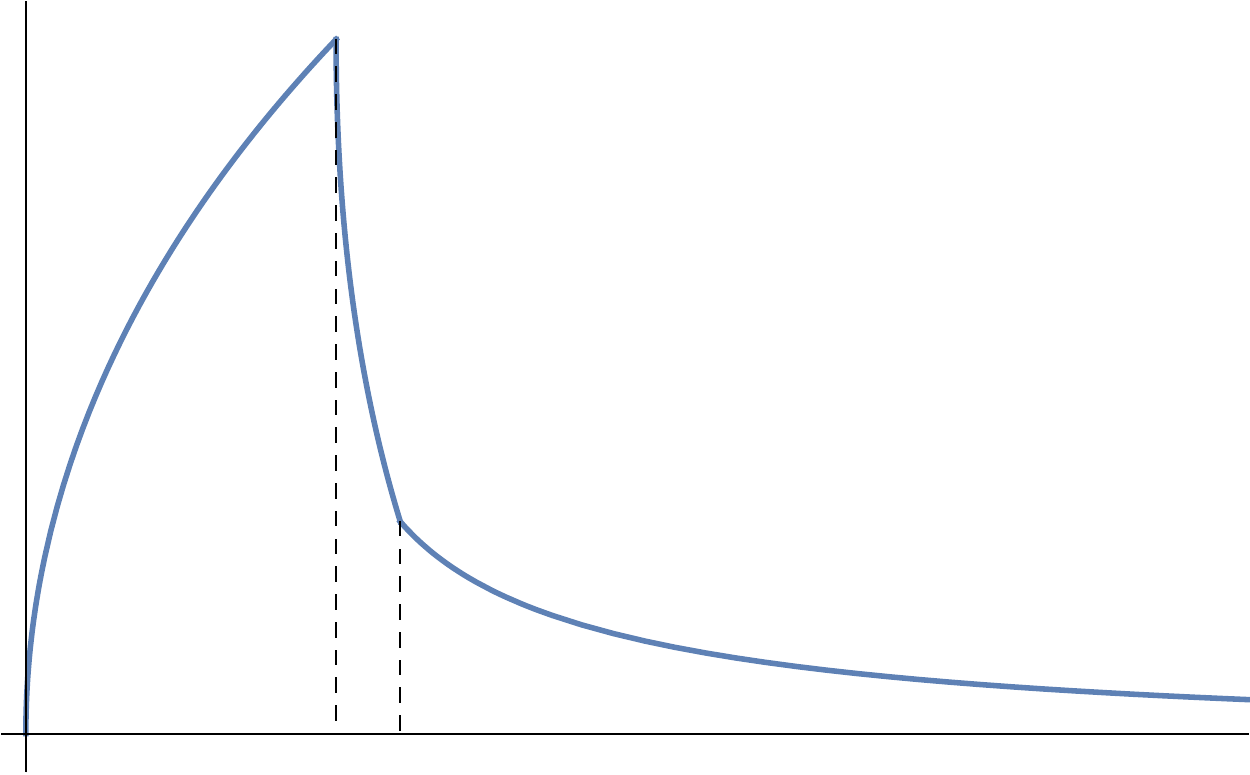}
\put(102,-1){$\xi$}
\put(0,65){$\tau^2 K(\xi|\tau,\eps)$}
\put(25,-2){$\xi_*$}
\put(30,-2){$\frac{\a}{4}$}
\end{overpic}
\caption{Plot of eq. \eqref{Kpiece} for a choice of $\a$.}
\label{fig:Keptau}
\end{figure}

To conclude this subsection, let us describe how the introduction of a non vanishing $\epsilon$ cutoff changes the kernel \eqref{VVkernel}. Referring to fig. \ref{fig:circles}, now the integration region of the potential only extends on the wedge defined by $\theta \in [\epsilon, \pi-\epsilon]$. Depending on the values of $\xi,\,\tau$ and $\epsilon$, one of three cases happens: none, one or both of the intersections marked in red lie inside this wedge. Hence, the kernel is a piecewise continuous function. We shall focus on the limit $\epsilon,\,\tau \to 0$, while keeping $\eps/\tau$ fixed. It is useful to define the following constants:
\beq
\alpha = \frac{\tau^2}{\eps^2}~, \qquad \a_*=4 \xi (\xi+1)~, \qquad 
\xi_*=\frac{1}{2}\left(\sqrt{\a+1}-1\right)~.
\eeq 
If we perform the change of variable $\theta \to \eps \theta$, it is not difficult to find the following result:
\beq
K(\xi|\tau,\eps) \sim \frac{1}{\tau^2}
\left\lbrace
\begin{array}{ll}
8\, \textup{arcsinh} \sqrt{\a_*}  & \xi<\xi_*~, \\
-4  \log \left[ \frac{\sqrt{\a_*+1}-\sqrt{\a_*}}{\sqrt{\a_*+1}+\sqrt{\a_*}}
\left(\frac{\sqrt{\a_*+1}+\sqrt{\a_*-\a}}{\sqrt{\a_*+1}-\sqrt{\a_*-\a}}\right)^2 \right] & \xi_*<\xi<\a/4~, \\
-2 \log\left[(\a+1)\left(\frac{(\sqrt{\a_*+1}-\sqrt{\a_*})(\sqrt{\a_*+1}+\sqrt{\a_*-\a})}{\sqrt{\a_*+1}+\sqrt{\a_*}}\right)^2\frac{\sqrt{\a_*+1}+\sqrt{\a_*-\a}}{\sqrt{\a_*+1}-\sqrt{\a_*-\a}}\right]
& \xi>\a/4~.
\end{array}
\right.
\label{Kpiece}
\eeq
We plot this function in fig \ref{fig:Keptau}. As expected, the contribution to the two-point function \eqref{VVvacuum} of points whose cross ratio is larger than the value $\xi_*$ is suppressed.

\subsection{The two-point function in a generic state}
\label{sec:twopgen}

The aim of this subsection is to prove eq. \eqref{gepbound}. As discussed in subsection \ref{subsec:second}, this result is instrumental in deriving a bound on the UV behavior of the spectral density $\bDelta \rho^\eps_{i,2}(\alpha)$. The object of interest is the following difference of correlation functions:
\beq
g_\eps(\tau)=\braket{i}{V_\eps(\tau)V_\eps(0)}{i}_\text{conn} - \braket{\Omega}{V_\eps(\tau)V_\eps(0)}{\Omega}_\text{conn}~.
\label{gepApp}
\eeq
The operators in the expectation values are ordered as written. Hence, only for $\tau>0$ is $g_\eps(\tau)$ computed by a Euclidean correlator. The left hand side of eq. \eqref{gepApp} is defined by analytic continuation everywhere else. We shall make the simplifying assumption that the leading non-analytic behavior in $g_\epsilon(\tau)$ as $\tau \to 0$ can be detected by approaching the limit from real and positive $\tau$. This is true in particular if $\tau=0$ is not an essential singularity.\footnote{For instance, logarithmic singularities are allowed as well.}

As discussed in the previous subsection, there are two sources of non analyticities at small $\tau$: the bulk $\mca{V}\times \mca{V}$ OPE, and the simultaneous fusion of the same operators with the boundary of AdS. Let us first focus on the latter, more complicated scenario. Clearly, as long as $\epsilon$ is finite, this singularity cannot arise. In other words, assuming for a moment that the bulk OPE is sufficiently soft, $g_\eps(\tau)$ is bounded by a constant as $\tau \to 0$. However, the bound is lost as we also take $\eps \to 0$. What we need is a different bound, which persist as the spatial cutoff is removed. 

Let us begin by recalling that the connected two-point function with respect to any state $\ket{i}$ can be written as
\beq
\braket{i}{\mca{V}(\tau,\theta_1)\mca{V}(0,\theta_2)}{i}_\text{conn} = 
\braket{i}{\mca{V}(\tau,\theta_1)\mca{V}(0,\theta_2)}{i}-\braket{i}{\mca{V}(\tau,\theta_1)}{i}\braket{i}{\mca{V}(0,\theta_2)}{i}~.
\label{Ctwopconn}
\eeq
Although the connected correlator is invariant under a shift of $\mca{V}$ by any c-number, recall that our definition of $\mca{V}$ includes the subtraction of the cosmological constant, when needed to make the matrix elements of the potential well defined. In this way, the second addend on the right hand side of eq. \eqref{Ctwopconn} is $\tau$ independent and, upon integration, gives a finite contribution to $g_\eps$ also in the $\eps\to0$ limit. Since we are interested in the non-analytic part, we shall drop it from now on.
Hence, we replace eq. \eqref{gepApp} by the following:
\begin{multline}
G_\epsilon(\tau)=\int_\eps^{\pi-\eps}\!\!
\frac{d\theta_1 d\theta_2}{\sin^2 \theta_1 \sin^2 \theta_2}f(\tau,\theta_1,\theta_2)~,\\
f(\tau,\theta_1,\theta_2)=
\braket{i}{\mca{V}(\tau,\theta_1)\mca{V}(0,\theta_2)}{i}-
\braket{\Omega}{\mca{V}(\tau,\theta_1)\mca{V}(0,\theta_2)}{\Omega}~.
\label{gepComplete}
\end{multline}
We immediately obtain an $\epsilon$-independent bound as follows:
\beq
|G_\epsilon(\tau)| \leq \int_0^{\pi}\!\!
\frac{d\theta_1 d\theta_2}{\sin^2 \theta_1 \sin^2 \theta_2}|f(\tau,\theta_1,\theta_2)|~.
\label{gepNoep}
\eeq
Our aim will be now to bound the right hand side as $\tau \to0$.
Let us make some simplifying assumptions, which will be reconsidered at the end:
\begin{itemize}
\item[1] the state $\ket{i}$ is created by a primary operator $\bop_i$,
\item[2] the boundary spectrum contains at least one operator with $\Delta \leq 2$, and the three-point function $\expec{O_\Delta \mca{V}\mca{V}}$ is non zero,
\item[3] the bulk OPE $\mca{V}\times \mca{V}$ does not contain operators of dimension smaller than $2\Delta_\mca{V}$, apart from the identity.
\end{itemize} 
Since we are interested in the limit where both $\mca{V}$'s go to the boundary simultaneously, it is convenient to introduce the $\rho$-coordinates depicted in fig. \ref{fig:VVboundary}. These are related to the $(\tau,\theta)$ coordinates as follows:
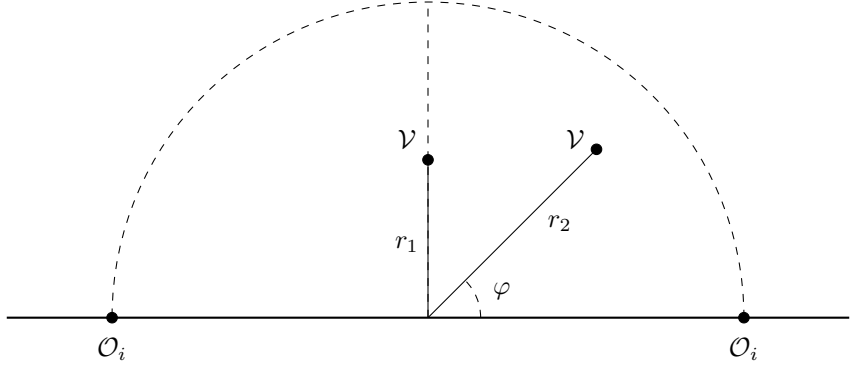
\begin{figure}[t]
\centering
\begin{tikzpicture}[scale=1.4]
\draw [thick]  (-4,0) -- (4,0);
\draw [dashed] (3,0) arc [start angle=0, end angle=180, radius=3];
\draw [dashed] (0,0) -- (0,3);
\draw (0,0) -- (0,1.5); 
\node at (-0.2,0.7) {$r_1$};
\filldraw (0,1.5) circle [radius=0.05];
\node at (-0.2,1.7) {$\mca{V}$};
\draw (0,0) -- (1.6,1.6); 
\node at (1.25,0.9) {$r_2$};
\filldraw (1.6,1.6) circle [radius=0.05];
\node at (1.4,1.7) {$\mca{V}$};
\draw [dashed] (0.5,0) arc [start angle=0, end angle=45, radius=0.5]; 
\node at (0.7,0.25) {$\varphi$};
\filldraw (-3,0) circle [radius=0.05];
\filldraw (3,0) circle [radius=0.05];
\node at (-3,-0.3) {$\mca{O}_i$};
\node at (3,-0.3) {$\mca{O}_i$};
\end{tikzpicture}
\caption{The configuration which defines the $\rho$-coordinates. The largest dashed semicircle has unit radius. Notice that an isometry can be used to place one of the bulk operators perpendicularly above the middle point between the boundary operators, without displacing the latter. Furthermore, using an inversion one can set $r_1<1$.}
\label{fig:VVboundary}
\end{figure}
\beq
r_1 =\tan \frac{\theta_1}{2}~, \quad
r_2 =\left(\frac{1+e^{2\tau}-2 e^\tau \cos \theta_2}{1+e^{2\tau}+2 e^\tau \cos \theta_2}\right)^{1/2}~, \quad
\cos \varphi = \frac{1-e^{-2\tau}}{\sqrt{1+e^{-4\tau}-2e^{-2\tau}\cos 2\theta_2}}~.
\label{tautor}
\eeq
Notice that the inversion symmetry $(r_1,r_2,\cos\varphi) \to (1/r_1,1/r_2,\cos\varphi)$ is mapped to the symmetry $(\theta_1,\theta_2)\to (\pi-\theta_1, \pi-\theta_2)$. We used it to restrict $\theta_1$ to lie in the interval $[0,\pi/2]$, and correspondingly $r_1\in [0,1]$. On the other hand, the symmetry under swapping $x_1 \leftrightarrow x_2$ is not explicit in eq. \eqref{tautor}. The inverse map is
\beq
\theta_1=2 \arctan r_1~, \quad
\cos \theta_2 = 
\frac{1-r_2^2}{\sqrt{1+(2-4 \cos^2\varphi)r_2^2+r_2^4}}~,
\quad
\tau=-\frac{1}{2}\log \frac{1-2 r_2 \cos\varphi +r_2^2}{1+2r_2 \cos\varphi+r_2^2}~.
\label{rtotau}
\eeq
Although the integration domain in eq. \eqref{gepNoep} covers a region where $r_2>1$, the singularity at small $\tau$ arise from the limits $\theta_1,\,\theta_2\to 0$ or $\pi$. Up to the inversion symmetry, both of the limits are mapped to $r_1,\, r_2 \to 0$. Hence, we can focus on the region where $r_1,\, r_2<r_0<1$. It is convenient to trade $(\theta_1,\theta_2)$ for $r_1$ and $r_2$ in the integral \eqref{gepNoep}, while keeping the $\tau$ dependence explicit. Therefore, we use
\beq
\theta_1= 2\arctan r_1~, \quad
\theta_2 = \arccos \left(\frac{1-r_2^2}{1+r_2^2}\cosh \tau\right)~, \quad
\cos \varphi =\frac{1+r_2^2}{2 r_2} \tanh \tau~,
\eeq
to obtain
\begin{multline}
|G_\eps(\tau)|\leq \frac{\cosh \tau}{2 \cosh^6 \frac{\tau}{2}}
\int_{0}^{r_0}\!dr_1
\int_{\tanh (\tau/2)}^{r_0}\!\!  dr_2\,
\frac{1+r_1^2}{r_1^2}\frac{r_2 (1+r_2^2)}{\left[\left(r_2^2-\tanh^2\frac{\tau}{2}\right)\left(1-r_2^2\tanh^2\frac{\tau}{2}\right)\right]^{3/2}}
|f(r_1,r_2,\varphi(r_2,\tau))| \\
+ \textup{terms finite at } \tau =0~,
\label{gepChange}
\end{multline}
In eq. \eqref{gepChange}, we slightly abused notation in the argument of $f$. Notice also that we inserted a factor 2 accounting for the inversion symmetry. We can further simplify the integrand by bounding with constants all the parts which are regular, at $\tau=0$, throughout the region of integration. We thus obtain
\beq
|G_\eps(\tau)|\leq A(r_0,\tau) G(t)~, \quad t=\tanh \frac{\tau}{2}~,
\label{gepAG}
\eeq
\beq
G(t)= \int_{0}^{r_0}\!dr_1
\int_{t}^{r_0}\!\!  dr_2\,
\frac{r_2}{r_1^2\left(r_2^2-t^2\right)^{3/2}}
|f(r_1,r_2,\varphi(r_2,t))|~,\qquad \cos\varphi = \frac{(1+r_2^2)t}{(1+t^2)r_2}~.
\label{gt}
\eeq
The function $A$ in eq. \eqref{gepAG} has a finite value at $\tau=0$, and since $G(t)$ turns out to diverge at small $t$, $A$ can be chosen large enough to also account for the dropped finite terms in eq. \eqref{gepChange}. 

Let us discuss the dangerous limits in the region of integration of eq. \eqref{gt}. Due to assumption 3 and the subtraction of the vacuum correlator in eq. \eqref{gepComplete}, the bulk OPE is non singular, hence the limit $r_1 \to r_2$ as $t\to0$ will not be discussed. The singularities at the lower end of the integration range for $r_1$ and $r_2$ are associated to the infinite volume of AdS. At finite $t$, the condition \eqref{eq:as} guarantees the integrability of the function in the region where either $r_1$ or $(r_2-t)$ are small: notice in particular that both limits are boundary OPE limits, since $\varphi=0$ when $r_2=t$. On the other hand, the singularity of the integrand is enhanced at $t=0$, and the boundary OPE of each of the $\mca{V}$'s does not fix the behavior of $f$ when $r_1 \sim r_2 \to 0$. 
The usefulness of the $\rho$-coordinates is precisely that they allow to write a power expansion for the correlator centered around $r_1,\ r_2= 0$ with a finite radius of convergence. Indeed, the quantization on the unit half-circle dashed in fig. \ref{fig:VVboundary} yields an expansion in term of boundary operators. Dubbing $r_M=\max(r_1,r_2)$, $r_m=\min(r_1,r_2)$, we find\footnote{The operators $\mca{O}_i$ are normalized so that $\braket{\Omega}{\mca{O}_i(-1)\mca{O}_i(1)}{\Omega}=1.$}
\beq
f(r_1,r_2,\varphi)=\sum_{\DD>0} r_M^\DD\,\braket{\Omega}{\mca{O}_i(-1)\mca{O}_i(1)}{\DD}\braket{\DD}{\mca{V}(1,\varphi_M)\mca{V}(r_m/r_M,\varphi_m)}{\Omega}~,\qquad 
(\varphi_M,\varphi_m) \leftrightarrow \left(\varphi,\frac{\pi}{2}\right)~.
\label{VVbOPE}
\eeq
Again, we are abusing notation with respect to, \emph{e.g.}, eq. \eqref{gepComplete}: the arguments of $\mca{V}$ are $(r,\varphi)$ from now on. The assignment of $(\varphi_M,\varphi_m) $ depends on $r_M$. Here, the states $\ket{\DD}$ form a complete set of eigenstates of the unperturbed Hamiltonian, but notice that the identity is excluded from the sum, due to the subtraction of the vacuum correlator in eq. \eqref{gepComplete}. This subtraction differentiates the energy gaps from the eigenvalues of the Hamiltonian themselves, and is ultimately what makes the former finite. 

We first focus on the finitely many terms in eq. \eqref{VVbOPE} with $\DD\leq2$, which are present by the assumption 2 above. To this end, we separate them using the triangular inequality:
\begin{multline}
|f(r_1,r_2,\varphi)|=\left|\sum_{\DD\leq 2} c_{i,\Delta} \textup{block}_\Delta
+f_R(r_1,r_2,\varphi)\right|
\leq \sum_{\DD\leq 2} |c_{i,\Delta} \textup{block}_\Delta |+|f_R(r_1,r_2,\varphi)|~, \\
 c_{i,\Delta} = \braket{\Omega}{O_i(-1)O_i(1)}{\DD}~,
\label{ffRdecomp}
\end{multline}
where $\textup{block}_\Delta$ is a shorthand notation for each addend in eq. \eqref{VVbOPE} and $f_R$ is defined by the equation itself.
We dub $G_{\DD}$ the contribution of each block with $\DD\leq2$ to eq. \eqref{gt}. It is convenient to extract the singularity as $r_1\sim r_2 \sim t \to 0$ via the change of variables $r_i = z_i t$, which yields
\begin{multline}
G_{\DD}(t)=t^{\DD-2} \int_{0}^{\frac{r_0}{t}}dz_1 \int_{1}^{\frac{r_0}{t}} dz_2 \frac{z_2}{z_1^2\left(z_2^2-1\right)^{3/2}} z_M^\DD
\braket{\DD}{\mca{V}(1,\varphi_M)\mca{V}(z_m/z_M,\varphi_m)}{\Omega}~, \\
(\varphi_M,\varphi_m) \leftrightarrow \left(\arccos\left(\frac{1+(z_2 t)^2}{z_2(1+t^2)}\right),\frac{\pi}{2}\right)~.
\label{Gtz}
\end{multline}
Let us now discuss the behavior of the double integral as $t \to 0$. In this limit, when $1<z_2<r_0/t$, $\cos \varphi \to 1/z_2$ uniformly. Hence, the only $t$ dependence is left in the upper limits of integration.
But firstly, there are the infinite volume singularities of the measure, at $z_1=0$ and $z_2=1$. Close to this point, $z_M=z_2\sim 1$ and we can study the following simpler integral:
\beq
\int_{0} \frac{dz_1}{z_1^2} \int_{1} \frac{ dz_2}{(z_2^2-1)^{3/2}}
|\braket{\DD}{\mca{V}(1,\varphi(z_2))\mca{V}(z_1,\pi/2)}{\Omega}|~, \qquad 
\sin\varphi(z_2)\sim \sqrt{2}(z_2-1)^{1/2}~.
\eeq 
The $z_1,\,(z_2-1)\to 0$ limits are controlled by the boundary OPEs of the two insertions respectively. Hence, the integrals factorize and we get
\beq
\int_{0} \frac{dz_1}{z_1^2} z_1^{\DD_0}\int_{1} \frac{ dz_2}{(z_2^2-1)^{3/2}} (z_2-1)^{\DD_0/2}~.
\eeq 
Here, $\DD_0$ is the lightest operator in the boundary OPE of $\mca{V}$. As advertised, well-definiteness of the matrix elements of the potential $V(\tau)$ -- \emph{i.e.}  eq. \eqref{eq:as}  -- requires $\DD_0>1$. Therefore the integrals converge in this region. We also need to show convergence in the region $z_i \sim r_0/t \to \infty$. Here, we can use a different approximation to the full integral. Defining $z_1=\nu \cos\omega$ and $z_2=\nu \sin \omega$, the relevant integral is
\beq
\int^{\frac{1}{t}}d\nu \nu^{\DD-3}\int_0^{\frac{\pi}{2}}d\omega \frac{1}{\sin^2\omega\cos^2\omega}\max(\sin\omega,\cos\omega)^\DD |\braket{\DD}{\mca{V}\left(1,\frac{\pi}{2}\right)\mca{V}\left(\min(\tan \omega, \cot \omega),\frac{\pi}{2}\right)}{\Omega}|~.
\label{largenu}
\eeq
The limits $\omega \to 0,\pi/2$ are identical and are again OPE limits for the correlator. For instance, at small $\omega$ the integrand behaves as $\omega^{\DD_0-2}$, and so it converges. The integral over $\nu$ converges as $t\to 0$ if $\DD<2$. On the contrary, if $\DD=2$ this region gives the dominating logarithmic contribution to the bound of the correlator at small $\tau$.\footnote{We would like to remark that this bound does not need to be saturated: there is no fundamental reason why the correlator on the right hand side of eq. \eqref{largenu} cannot vanish in this specific configuration. This happens for instance in the Ising model with a thermal deformation -- see subsection \ref{subsec:IsingT} -- as it is easy to check directly. While this implies the absence of a $\log\tau$ for any external state at second order in perturbation theory, the connected correlator \eqref{gepApp} can be computed exactly at $\eps=0$ at least for $\ket{i}=L_{-2}\ket{\Omega}$: $g_0(\tau)=8\pi^2 \cosh 2\tau$ in this case. Not only there is no $\log\tau$, $g_0$ is in fact analytic at $\tau=0$. The corresponding spectral density has a finite number of states, whose contribution to the energy gap cancels exactly, as expected for the theory of a free fermion.}

We now need to go back and estimate the contribution to $G(t)$ of all the states with $\DD>2$ in the decomposition \eqref{VVbOPE}. Using eq. \eqref{ffRdecomp}, this amounts in replacing $f \to f_R$ in eq. \eqref{gt}. In this case, we can simply set $t=0$ in the right hand side of the equation,\footnote{One might object that $\cos\varphi(r_2,t)$ has a non uniform limit as $t\to 0$. However, $f_R$ depends on $\varphi$ only through the positions $(r_2\cos \varphi,r_2\sin \varphi)$, which approaches $(0,r_2)$ uniformly.} and study the following integral:
\beq
\int_{0}^{r_0}\!\frac{dr_1 dr_2}{r_1^2r_2^2}\,
|f_R(r_1,r_2,\pi/2)|~.
\eeq
Now, $f_R$ is a finite linear combination of two-point functions of $\mca{V}$ in the presence of various boundary operators, and as such, it has the same OPE bounds already discussed as $r_1 \to 0$ at fixed $r_2$ or vice versa. Hence, the integral converges there. On the other hand, when $r_1 \sim r_2 \sim \nu \to 0$, the subtractions make the behavior of the integrand softer: $f_R \sim \nu^{\Delta}$ with $\Delta>2$. To show rigorously that this suffices for the integral to converge, let us go to polar coordinates in the $(r_1,r_2)$ plane:
\beq
\int_0^{r_0} \frac{d\nu}{\nu^3} \int_0^{\frac{\pi}{2}}d\omega
\frac{1}{\cos^2\omega\sin^2\omega} |f_R(\nu,\omega)|~.
\eeq
At any fixed value of $\nu$, the angular integral converges. Since $f_R$ can be expressed as an absolutely convergent series there, using eq. \eqref{VVbOPE}, the series can be integrated term by term, and we obtain
\beq
\int_0^{r_0} \frac{d\nu}{\nu^3} \sum_{\Delta>3} \kappa_\Delta \nu^\Delta~.
\eeq
While the $\kappa_\Delta$ are theory dependent, the previous equation shows that the small $\nu$ behavior of the integrand is not enhanced by the integration over $\omega$. This is sufficient to show that the integral converges. This concludes the proof that eq. \eqref{Gtz} provides the leading singularity as $\tau\to0$. As we stated at the beginning of the subsection, we work under the assumption that the strength of the singularity is independent of the phase of $\tau$. Hence, although our manipulations are only valid for $\tau>0$, we extend the conclusion to a neighborhood of the origin:
\beq
|G_\epsilon(\tau)|<C(\tau_0) 
\left\lbrace
\begin{array}{ll}
|\tau|^{\Delta_*-2}~, & \Delta_*<2 \\
\log |\tau| ~,  & \Delta_*=2~,
\end{array}\right.
\quad |\tau|<\tau_0~.
\label{finalBound}
\eeq
Here, $\Delta_*$ is the lowest dimensional operator in the OPE \eqref{VVbOPE}.   

We would like now to comment on the assumptions 1 - 3 above, in reversed order. If the bulk OPE contains singularities above the identity, they can be added to the analysis. In this paper, the bulk OPE is relevant only for the $\phi^4$ deformation of a massive theory considered in subsection \ref{sec:phi4}, so we analyze this case separately in subsection \ref{subsec:phi4conv}. Here, we limit ourselves to the simple scenario where the unperturbed Hamiltonian is a CFT, and furthermore we assume the gap in the boundary spectrum is large enough to avoid any enhancement of the bulk OPE singularity by boundary effects. As we will see in a moment, a sufficient condition for this to happen is that the lowest boundary operator has dimension $\Delta>3$. The limit where the two operators collide is controlled by the OPE in the UV theory, schematically:
\beq
\mca{V} \times \mca{V} = 1+ \tilde{\mca{V}}~.
\label{bulkOPE}
\eeq
The contribution of the identity cancels in the difference in eq. \eqref{gepComplete}, so we concentrate on the first non-trivial operator. Its contribution to eq. \eqref{gepNoep} can be computed by rescaling $(\theta_1-\theta_2) \to \tau (\theta_1-\theta_2)$, as we did in subsection \ref{sec:twoptvac}. In the small $\tau$ limit, we obtain
\beq
G_\eps(\tau) \sim \tau^{1+\Delta_{\tilde{\mca{V}}}-2\Delta_{\mca{V}}} 
\int_\eps^{\pi-\eps}\!\frac{d\theta}{\sin^4\theta}
\braket{i}{\tilde{\mca{V}}(0,\theta)}{i} \int_{-\infty}^\infty d\theta_-
(1+\theta^2_-)^{\Delta_{\tilde{\mca{V}}}/2-\Delta_{\mca{V}}}~.
\label{GepsBulk}
\eeq
The integral over $\theta_-=(\theta_1-\theta_2)/\tau$ is dominated by the UV and is the same as in flat space. The integral over $\theta=(\theta_1+\theta_2)/2$ is sensitive to the AdS geometry. The former converges only when $\Delta_{\tilde{\mca{V}}}/2-\Delta_{\mca{V}}$ is large enough. However, this IR feature is spurious: one can for instance take derivatives of $G_\eps(\tau)$, improving convergence without modifying the non analytic parts of the result. On the other hand, the integral over $\theta$ converges when the leading boundary operator coupling to $\tilde{\mca{V}}$ has dimension $\Delta>3$. If this is not the case, the crude approximations in eqs. (\ref{bulkOPE},\ref{GepsBulk}) are inadequate: both bulk and boundary effects must be taken into account to compute the small $\tau$ singularity. We shall not treat this more general case here.

The remaining assumptions are easier to relax. The role of assumption 2 is to make the $\tau \to 0$ limit singular, so that any finite contribution to $G_\eps(\tau)$ can be disregarded. When this is not the case, one may consider derivatives of $G_\eps(\tau)$ instead. Clearly, the leading non analytic behavior can be made singular in this way. What needs to be proven, is that said non analytic behavior can also be traced to the region covered by eq. \eqref{Gtz}, and so still leads to an equation analogous to \eqref{finalBound}. Technically, this should come about because of the dependence of $r_2$ from $\tau$ in eq. \eqref{tautor}, which turns derivatives wrt $\tau$ into derivatives wrt $r_M$ in eq. \eqref{VVbOPE}, in the region where $r_M=r_2$. We have not done this analysis in detail, but these considerations prompt us to claim that eq. \eqref{finalBound} can be extended to $\Delta_*>2$, simply by taking enough derivatives of both sides so that the right hand side becomes singular. This is confirmed, for instance, in the case of the $\phi^2$ deformation in subsection \ref{sec:ph2conv}.

Finally, the purpose of assumption 1 is to make the transformation properties of the matrix elements involved under AdS isometries simple. This is used in eq. \eqref{VVbOPE} to write $f(r_1,r_2,\varphi)$ in terms of the scaling operator $O_i$ evaluated at special positions. Had the state $\ket{i}$ been created by a descendant in the origin, a more complicated linear combination of descendants would have appeared in eq. \eqref{VVbOPE}. However, this only contributes a numerical prefactor to the final result, and does not affect the $\tau$ dependence. Hence, the result \eqref{finalBound} immediately extends to generic states in the Hilbert space. This concludes our analysis.

\subsection{Free scalar example}
\label{app:periodicity}

In this example, we consider the potential 
\beq
V(\tau)=\int_{-\pi/2}^{\pi/2}\frac{dr}{\cos^2 r} \NO{\phi^2(\tau,r)}~.
\eeq
Let us concentrate to the the second-order contribution to the energy of the state $\ket{\chi} = a_0^\dag \ket{\text{vac}}$. The relevant object is
\beq
g_\chi(\tau) = \braket{\chi}{V(\tau)V(0)}{\chi}_\text{conn} - \braket{\Omega}{V(\tau)V(0)}{\Omega}_\text{conn}~.
\eeq
This difference was computed exactly in appendix \ref{sec:ExplicitSecondOrder}, with the result 
\beq
g_\chi(\tau) = \frac{4}{(2\DD-1)^2} \left[(1+e^{-2\DD\tau})\mfr{g}_\DD(e^{-2\tau}) - 1\right]~,
\label{gchiScalarC}
\eeq
where
\beq
\mfr{g}_\DD(z) \ldef \sum_{n=0}^\infty \msc{V}_{0,2n}(\DD)^2 z^n = {}_3F_2\!\left[{{\th,\, 1,\, 1}~\atop~{\DD,\, \DD+\th}}\, \Bigg| \, z\right].
\label{g3F2}
\eeq
We can use the result to explore in particular the behavior of $g_\chi(\tau)$ for \emph{imaginary} $\tau$, which escapes the analysis of the previous subsections. But first of all, at small $\tau$ we have
\beq
g_\chi(\tau) \limu{\tau \to 0} \text{analytic} - \frac{4\pi}{(2\DD-1)\sin(2\pi \DD)}\, \tau^{2\DD-2} + \ldots
\label{gchiSmallTauC}
\eeq
which is consistent with eq. \eqref{finalBound}, since the leading operator in the boundary OPE of $\phi^2$ has dimension $\Delta_*=2\Delta$.

\begin{figure}[htb]
\centering
\includegraphics[scale=0.42]{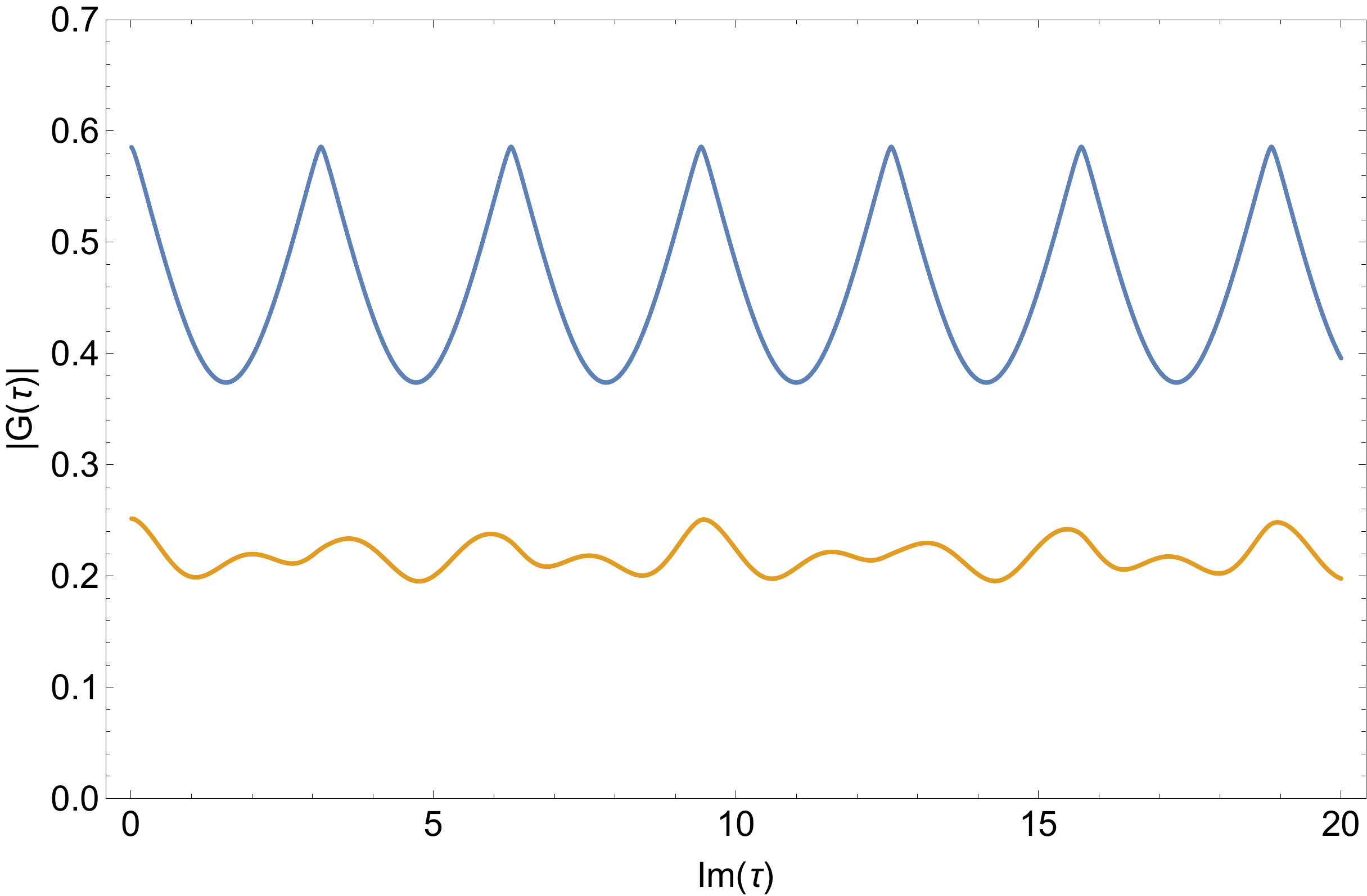}
\caption{Plot of the correlator $g_\chi(\tau)$ for imaginary $\tau$, for both $\DD = 2$ (blue) and $\DD = \sqrt{7}$ (orange). Notice the periodicity of the blue curve (rational $\DD$), whereas the orange curve (irrational $\DD$) does not have any obvious features.}
\label{fig:recur}
\end{figure}

On the other hand, the behavior of $g_\chi(\tau)$ for imaginary $\tau$ depends on $\DD$. The function $\mfr{g}_\DD(e^{-2\tau})$ is invariant under $\tau \to \tau + \pi i$. Now if $\DD$ is \emph{rational}, say of the form $\DD = p/q$, the exponential $e^{2\DD \tau} = e^{2p/q \cdot \tau}$ appearing in $g_\chi(\tau) $ is invariant under $\tau \to \tau + q\cdot \pi i$. Consequently, on the imaginary axis $g_\chi$ is exactly periodic, with period $q\cdot \pi$. The same behavior occurs in the case of minimal models in the identity module: there, all correlators are exactly invariant under $\tau \to \tau + \pi i$. In contrast, if $\DD$ is irrational, the function $g_\chi(\tau)$ cannot be exactly periodic. In Fig.~\ref{fig:recur} we  plot $|G(\tau = \ii t)|$ for a range of $t$, both for rational and irrational $\DD$ (namely $\DD = 2$ and $\DD = \sqrt{7}$). 
Nevertheless, in both cases we can check that there are no non-analyticities stronger than the one in eq. \eqref{gchiSmallTauC} for imaginary $\tau$. This fact simply follows from eqs. \eqref{gchiScalarC} and \eqref{g3F2}. Indeed, the only source of singularities in $g_\chi(\ii t)$ is the hypergeometric function, and, as stated above, $\mfr{g}_\DD(e^{-2\ii t})$ is periodic in $t$ with period $\pi$. Then, the only non-analyticity along the path is the branch point at $t=\pi n$, $n \in \mathbb{Z}$. We conclude that the analytic structure of $g_\chi(\tau)$ is of the form presented in figure \ref{fig:invLapCont}, with power law monodromies along the imaginary axis, all with the same exponent.

\section{Algorithms}
\label{app:algo}
%!TEX root = ../HTinAdS.tex
%%%%%%%%%%%%%%%%%%%%%%%%%%%%%%%%%%%%%%%%%%%%%

In this section, we present some of the algorithms we used in the numerics. The calculations consist of three parts:
\begin{itemize}
\item finding the basis of states,
\item constructing the Hamiltonian,
\item diagonalizing the Hamiltonian.
\end{itemize}
Depending on the theory we study and the algorithm we use, each of the above can be the bottleneck in runtime or memory.

\subsection{Scalar field}\label{sec:ScalarAlg}
As discussed in section \ref{sec:massive}, we construct the Hamiltonian in the truncated Fock space of the free theory. The basis at cutoff $\Lambda$ is made of states\footnote{The labeling of the states introduced here differs from the one used in the rest of the paper: see \emph{e.g.} eq.~\eqref{RSpta}.}
\beq
\label{eq:ScalarBasisD}
\ket{\psi}= \ket{n_0,n_1,n_2,...,n_{i_{\text{max}}}}\equiv \frac{1}{\mathcal{N}_{\ket{\psi}}} (a_{i_{\text{max}}}^\dagger)^{n_{i_{\text{max}}}}  \, ... \, (a_2^\dagger)^{n_2}(a_1^\dagger)^{n_1}(a_0^\dagger)^{n_0}\ket{\Omega}~,
\eeq
with energies $E_{\ket{\psi}}=\sum_{i=0}^{i=i_{\text{max}}} n_i(\Delta+i)\leq\Lambda$. In particular, $i_{\text{max}}=\floor{\Lambda-\Delta}$ is the highest mode possible for a one-particle state. The factor $\mathcal{N}_{\ket{\psi}}>0$ makes the state unit-normalized. Finding the set of states in the free scalar theory is very fast and requires little memory, compared to the other parts of the code. In table \ref{tab:NStatesScalar}, we present the number of states for a few different cutoffs. 
\begin{table}[ht]
\caption{Number of states at cutoff $\Lambda$ for $\Delta=1.62$ in scalar theory}
\begin{tabular}{ |p{1.45cm}||p{1.45cm}|p{1.45cm}|p{1.45cm}|p{1.45cm}| p{1.45cm}|p{1.45cm}|p{1.45cm}|}
 \hline
$\Lambda$&10&15&20&25&30&35&40\\
 \hline
 $\#\text{states}$&52&237&886&2904&8613&23620&60783\\
 \hline
\end{tabular}
\label{tab:NStatesScalar}
\end{table}

To build the Hamiltonian, we construct the $V$ matrix once for each desired $\Delta$ and then save it. The Hamiltonian can then quickly be calculated using $H=H_0+\bar{\lambda}V$.

One method to build the $V$ matrix is to directly compute $\bra{\psi_j}V\ket{\psi_i}$ by commuting the ladder operators.\footnote{Another way is to act with the $V$ operator on the state $\ket{\psi_i}$ and expand it in our basis: $V\ket{\psi_i} = \sum_j V^j_i \ket{\psi_j}$. Typically, the method we used takes up more memory, whereas this method requires longer runtimes.} A generic matrix element $V_{ij}$ is given by
\begin{align*}
\label{eq:V2def}
  (V_2)_{ij} &= \bra{\psi_j} \sum_{m,n = 0}^\infty A_{mn}(\DD) (a_m^\dag a_n^\dag + 2a_m^\dag a_n^\phd + a_m^\phd a_n^\phd) \ket{\psi_i}~,\\
    (V_4)_{ij} &= \bra{\psi_j} \sum_{m,n,p,q = 0}^\infty A_{mnpq}(\DD) (a_m^\dag a_n^\dag a_p^\dag a_q^\dag + 4 a_m^\dag a_n^\dag a_p^\dag a_q + 6 a_m^\dag a_n^\dag a_p a_q+ 4 a_m^\dag a_n a_p a_q  +  a_m a_n a_p a_q) \ket{\psi_i}~,
\end{align*}
where the states $\psi_i$ are defined as in eq. \reef{eq:ScalarBasisD}. Here $A_{mn}$ can be expressed in closed form in eq. \reef{eq:AmatP}, and if desired the $A_{mnpq}$ can be computed analytically via a recursion relation~\cite{TK}. However, we calculated each of the $A_{mnpq}$ numerically and independently, since the runtime of this operation is negligible compared to other parts of the code. 
Let us emphasize a few points regarding the calculation of $V$:
\begin{itemize}
\item Both the $\phi^2$ and $\phi^4$ deformations preserve parity and $\mathbb{Z}_2$ symmetry. Therefore, we divide the Fock space into four sectors according to the eigenvalue under these symmetries.
\item The matrix $A_{mn}$ is symmetric. So, the sum above can be reduced to $m\geq n$. This decreases the runtime by almost a factor of 2. Similarly, the runtime for the $\phi^4$ deformation decreases with a factor of $\sim 4!$, as we can reduce the sum $\sum_{m,n,p,q=0}$ to $\sum_{m \geq n\geq p\geq q\geq0}$. 
\item The potential is symmetric: $V_{ji}=V_{ij}$. So, it is sufficient to calculate the upper triangular part.
\item Most of the elements of the potential $V_{ij}$ vanish: $V$ is a \textit{sparse matrix}. This is because $\normord{\phi^{2}}$, acting on a state with $p$ particles, only produces states with $p$ and $p\pm 2$ particles. Of course, taking this into account drastically reduces the number of matrix elements which it is necessary to compute and store. 
\end{itemize}
The last step is the diagonalization of the Hamiltonian. As we mentioned, we diagonalize the four sectors with different parity and $\mathbb{Z}_2$ symmetry separately. In this work, we used \textit{Spectra}, a  \textit{C\texttt{++}}  library, which uses the Arnoldi/Lanczos method to find the smallest eigenvalues of a sparse matrix. 

Let us give the reader an idea of the runtime of a typical code. In the case of the $\phi^2$ deformation, for $\Delta=1.62$ and $\Lambda=34.13$ (which corresponds to $19939$ states), it takes $\sim44$ CPU-seconds to compute the matrix $V$ and $\sim5$ CPU-seconds to diagonalize the full Hamiltonian, for a single coupling $\bar{\lambda}$.

\subsection{Minimal models}
In the case of a minimal model, as discussed in section \ref{sec:minimal}, the Hilbert space is composed by states of the form:
\beq
\ket{\psi}=L_{-n_p} \dotsm L_{-n_1}\ket{\Omega}~,
\label{LchiStatesApp}
\eeq
where $n_1, \ldots, n_p \geq 1$. In fact, it is sufficient to choose, say, $1 \leq n_1\leq \ldots \leq n_p $ in order to span the whole Hilbert space. We dub this the \emph{ascending} basis. We also define the \emph{level} of the state \eqref{LchiStatesApp} as the sum $\sum_{i} n_i$. The level is the eigenvalue of the state \eqref{LchiStatesApp} under the application of the unperturbed Hamiltonian $L_0$.

The states \eqref{LchiStatesApp} form in general a reducible representation of the Virasoro algebra. Indeed, the identity module of a minimal model $\mca{M}(p,q)$ has two null primary states at level 1 and $(p-1)(q-1)$. The set of physical states is obtained by quotienting out the submodules composed of all the Virasoro descendants of these two states.
The null state at level 1 is $L_{-1}\ket{\Omega}$, for every minimal model. The other primary null state, for the Lee-Yang model and the Ising model respectively, is
\begin{subequations}
\label{nullLYIsing}
\begin{align}
\ket{\text{null}}_{\text{LY}}&= \left(L_{-4}-\frac{5}{3}L_{-2}^2\right)\ket{\Omega}~,\\
\ket{\text{null}}_{\text{Ising}}&= \left(L_{-6}+\frac{22}{9}L_{-4}L_{-2}-\frac{31}{36}L_{-3}^2-\frac{16}{27}L_{-2}^3\right)\ket{\Omega}~.
\end{align}
\end{subequations}
Let us emphasize that null states constitute the large majority of the states \eqref{LchiStatesApp}: for instance, in the Lee-Yang model, at level 30 there are 5604 states of the kind \eqref{LchiStatesApp}, and only 568 physical states.
Quotienting out the submodule generated by $L_{-1}\ket{\Omega}$ is trivial in the ascending basis: we just take $n_1>1$. On the contrary, in order to quotient out the descendants of \reef{nullLYIsing} we need to write them in the ascending basis, which requires performing a large number of commutators. Rather than performing the commutators directly, we employed a more efficient method, which we briefly describe with an example. Consider the state 
\beq
L_{-3}L_{-2}^2L_{-4}\ket{\Omega}~.
\label{exampleTrick}
\eeq
It is not hard to understand which states of the ascending basis are required to decompose it. We first consider all distinct subsets of the indices of the Virasoro generators which appear in eq.~\eqref{exampleTrick} already in ascending order, \emph{i.e.} excluding $-4$:\footnote{When decomposing more complicated states, in general not all the subsets contribute to the result. The condition for a subset to be relevant is immediate and easy to implement in the code.}
\beq
\big\{(-3),(-2),(-3,-2),(-2,-2),(-3,-2,-2) \big\}~.
\label{subsetsTrick}
\eeq
Then, we sum the elements of each subset to $-4$, and we apply the resulting Virasoro generator on the vacuum, together with the complement of the generators in the same subset. Finally, we add the obvious $L_{-4}L_{-3}L_{-2}^2\ket{\Omega}$ to the basis:
\begin{multline}
L_{-3}L_{-2}^2L_{-4}\ket{\Omega} \\
\in \textup{Span}\left\{L_{-7}L_{-2}^2\ket{\Omega},L_{-6}L_{-3}L_{-2}\ket{\Omega},L_{-9}L_{-2}\ket{\Omega},L_{-8}L_{-3}\ket{\Omega},L_{-11}\ket{\Omega},L_{-4}L_{-3}L_{-2}^2\ket{\Omega}\right\}~.
\end{multline}
The coefficient of each state in the decomposition can be easily computed knowing the multiplicities of the subsets \eqref{subsetsTrick} and the Virasoro algebra \eqref{VirGen}. The reason why this method is faster than the blind application of the Virasoro algebra follows precisely from the multiplicities of the subsets. When considering a descendant built by applying the same generator many times, the distinct subsets are far less than the total number. Correspondingly, when applying the commutation relations, a large number of identical sub-operations are performed. The method just described allows to avoid this redundancy.

Since $L_0$ is the unperturbed Hamiltonian, truncating the spectrum at cutoff $\Lambda$ translates into keeping physical states of level $p\leq \Lambda$. We collect in the table below the dimension of the truncated Hilbert space for a few values of the cutoff:

\resizebox{0.9\textwidth}{!}{
\begin{tabular}{ |p{2cm}||p{1cm}|p{1cm}|p{1cm}|p{1cm}|p{1cm}| p{1.cm}|p{1.cm}|p{1cm}|}
 \hline
$\Lambda$&5&10&15&20&25&30&35&40\\
 \hline
 $\specialcell{\#\text{states of}\\ \text{Lee-Yang}}$&5&19&52&126&276&568&1103&2059\\
 \hline
  $\specialcell{\#\text{states of}\\ \text{Ising}}$&7&30&95&260&632&1426&3019&6099\\
 \hline
\end{tabular}}

Contrary to the case of free theory, the basis of states obtained with the procedure above is not orthogonal. Hence, before proceeding to the computation of the potential $V$, we need to evaluate the Gram matrix $G$, \emph{i.e.} the matrix of scalar products in the basis of physical states. This essentially reduces to the computation of overlaps of the kind
\beq
\label{eq:GramEl}
\langle{\psi_j}|{\psi_i}\rangle = \bra{\Omega}L_{j_1} \dotsm L_{j_q}L_{-i_p} \dotsm L_{-i_1}\ket{\Omega}~,
\eeq
where the indices obey $i_p \geq \ldots \geq i_1 > 1 $ and $j_p \geq \ldots \geq j_1 > 1$. Since $L_0$ is Hermitian, and the states \eqref{LchiStatesApp} are eigenvectors of $L_0$ with eigenvalues equal to their level, the matrix element \eqref{eq:GramEl} vanishes unless $\sum_{m}j_m-\sum_{n}i_n=0$. Then, the overlap is simply computed by using the Virasoro algebra to commute the generator with a larger index towards the right, and recalling that the vacuum is annihilated by all the modes $L_n$, $n>-2$. 

The potential $V$ involves the integral over the local operator $\mca{V}$, see \emph{e.g.} eq. \eqref{MinimalV}.
Therefore, to compute its matrix elements we need the overlaps
\beq\label{SandVir}
\braket{\psi_j}{\mca{V}}{\psi_i}= \bra{\Omega}L_{j_1} \dotsm L_{j_q} \, {\mca{V}}\, L_{-i_p} \dotsm L_{-i_1}\ket{\Omega}~.
\eeq
Notice that this vanishes unless $\ket{\psi_i}$ and $\ket{\psi_j}$ share the same parity. These overlaps can be computed by applying the Virasoro algebra, together with the action of the generators on $\mca{V}$, eq. \reef{VirOnV},
\beq
\left[L_n,\mca{V} \right] =
\left[z^{n+1} \partial_z+\bar{z}^{n+1} \partial_{\bar{z}}+
\frac{\Delta_\mca{V}}{2} \left((n+1)(z^n+\bar{z}^n)-2\,\frac{z^{n+1}-\bar{z}^{n+1}}{z-\bar{z}}\right)\right] \mca{V}~ \equiv \hat{\mca{D}}_n({z,\bar{z}})\mca{V}~,
\label{eq:commuteD}
\eeq
and finally using the fact that the expectation value of $\mca{V}$ in the vacuum is a constant. More specifically, the strategy goes as follows. We first commute $\mca{V}$ all the way to the left vacuum in eq. \eqref{SandVir}, and we reorder the result, such that in 
\beq
 \bra{\Omega}{\mca{V}}\, L_{-i_{p'}} \dotsm L_{-i_1}\ket{\Omega}~
\eeq
there are only negative Virasoro modes. Then we commute the generators again to the left of $\mca{V}$. In the end, we find a sum over a set of differential operators acting on a constant $a_{\mca{V}}$:
 \beq
 \braket{\psi_j}{\mca{V}}{\psi_i} = \sum_l c_l \prod_{m_l}  \hat{\mca{D}}_{m_l}(z,\bar{z}) \; a_{\mca{V}} = \sum_l c_l \mathfrak{D}_l(z,\bar{z})\; a_{\mca{V}}~.
 \label{eq:DD}
 \eeq
In order to compute the integral of eq. \eqref{SandVir} over AdS, it is useful to describe the objects $\mathfrak{D}_l(z,\bar{z})$ in some detail. Since, albeit not explicitly, the operators \eqref{eq:commuteD} are polynomials in $z$ and $\bar{z}$,  it follows that $\mathfrak{D}_l(z,\bar{z})$ are polynomials as well. Furthermore, they are homogeneous, and their degree $d$ equals the sum of the indices $m_l$ in eq. \eqref{eq:DD}. This sum is independent of $l$, since it equals the difference of the levels of the states $\ket{\psi_j}$ and $\ket{\psi_i}$:
 \beq
 d=\sum m_l=(j_1+j_2+ \ldots +j_q) - (i_1+i_2+ \ldots +i_p)~.
 \eeq
Finally, $\mathfrak{D}_l(z,\bar{z})$ is symmetric under $z\leftrightarrow\bar{z}$. Now, $\mathfrak{D}_l(z,\bar{z})$ can be easily evaluated on the $\tau=0$ slice in global coordinates, which corresponds to $z=1/\bar{z}=\ii\,e^{\ii r}$. The previous constraints then translate in the following equality:
 \beq
 \mathfrak{D}_l=\sum_{k=0}^{d/2} \alpha_{l,k} \left(z^{2k} + z^{-2k}\right)
 =\sum_{k=0}^{d/2} \alpha_{l,k} \left(e^{\ii k(2r+\pi)}+e^{-\ii k(2r+\pi)}\right) = 
 \sum_{k=0}^{d/2} 2\alpha_{l,k} \cos(2kr+k\pi)~.
 \label{Dltrig}
 \eeq 
To find $V$, we now need to integrate these trigonometric functions against the AdS measure.
 It is simple to do this analytically, keeping in mind that we know that all the matrix elements are finite. Hence, the constant terms $\alpha_{l,0}$ in eq. \eqref{Dltrig} must all combine to yield integrals of the following form:
\begin{equation}
\int^{\frac{\pi}{2}}_{-\frac{\pi}{2}} \frac{\cos(2kr+k\pi)-1}{\cos^2(r)} dr = - 2k \pi ~,
\label{trigIntegrals}
\end{equation}
where $k \in \mathbb{N}$. 
Hence, the matrix elements of the potential are rational factors times $\pi$. 
Of course, computing $V$ once for the highest interesting cutoff yields the Hamiltonian for any value of the coupling $\lambda$.
Like in the free scalar case, both $V$ and the Gram matrix are symmetric, so we only compute their upper triangular part. Furthermore, $V$ is parity preserving, so $H$ is block diagonal.

The last step is to diagonalize the Hamiltonian. Due to the presence of a non-trivial Gram matrix $G$, we need to solve the following generalized eigenvalue problem:
\beq
H.v = E\; G.v~,
\eeq
Empirically, we find that a higher precision in numerics is required to diagonalize the Hamiltonian with respect to the scalar case. In fact, this step is the bottleneck in the code runtime. Notice that, since the potential couples states with arbitrarily different unperturbed energy, when $\bar{\lambda}$ is order one the off-diagonal matrix elements of the Hamiltonian are large, contrary to the example discussed in subsection \ref{sec:ScalarAlg}.

We wrote the code for the minimal models in Mathematica. To give the reader an idea of the runtime, it takes $\sim 100$ core-hours to find $V$ at cutoff $\Lambda =40$ (with 2059 states) in the Lee-Yang model, whereas it takes $\sim 400$ core-hours to find $V$ at cutoff $\Lambda =35$ (with 3019 states) in the Ising model with thermal deformation. The diagonalization also typically takes a core-hour for a single matrix. In order to produce a plot like the one in figure~\ref{fig:TIsingFirstExcitedSpec}, we need to diagonalize several Hamiltonians,  corresponding to different couplings $\bar{\lambda}$ and truncation cutoffs $\Lambda$.

In appendix~\ref{app:constraints}, relations between matrix elements of the operator $V$ were derived. As a matter of principle, these recursion relations make it possible to compute many matrix elements indirectly --- once a small number of matrix elements is computed using the algorithm discussed above, the rest of the matrix can be ``filled in'' using equations~\reef{eq:VWrels}. In our setup, we have instead chosen to compute all matrix elements directly. Indeed, computing matrix elements indirectly requires organizing the Hilbert space of the theory in terms of $SL(2,\mbb{R})$ primaries and descendants, which is doable but adds additional complexity to the algorithm. Moreover, for moderate values of the cutoff $\La$, a sizeable fraction of the Hilbert space consists of $SL(2,\mbb{R})$ primaries, which in turns means that many matrix elements of $V$ still need to be computed explicitly.

\section{Spontaneous symmetry breaking in AdS}
\label{app:negative}
%!TEX root = ../HTinAdS.tex
%%%%%%%%%%%%%%%%%%%%%%%%%%%%%%%%%%%%%%%%%%%%%

This appendix contain computations that have to do with symmetry breaking in AdS and different boundary conditions.

\subsection{All or nothing}
\label{app:SSB}

Consider the following Euclidean action for a scalar field in AdS
\beq
S= \int d^{d+1}x \sqrt{g} \left[ \frac{1}{2} (\partial \phi)^2 + V(\phi) \right]
\eeq
where $V(0)=V'(0)=0$.  The global minimum of $V(\phi)$ is attained at $\phi=\phi_t \neq 0$.
We use the AdS metric
\beq
\label{AdSrhometric}
ds^2 = R^2 \left[ d\rho^2 + \sinh^2 \rho \, d\Omega_d^2 \right] \,
\eeq
and impose boundary conditions
\beq
\label{phibc}
\lim_{\rho \to \infty} \phi(\rho) e^{\frac{d}{2} \rho} =0\ ,
\eeq
so that the action gets a finite contribution from $\rho \to \infty$. 
We claim that there are only two possibilities:
\begin{enumerate}
\item[{\bf 1.}] Within field configurations satisfying the boundary condition \eqref{phibc}, the global minimum of the action is zero and it is attained by the constant solution $\phi=0$.
\item[{\bf 2.}]  The action is not bounded from below and its value can always be decreased by setting $\phi=\phi_t$ in a bigger region of AdS.
\end{enumerate}

In \ref{thinwall} we will use the thin wall approximation to get some intuition for the claim above. Then, 
we will present a more general argument in \ref{generalbubble}.

\subsubsection{Thin wall approximation} \label{thinwall}

In the thin wall approximation \cite{Coleman:1977py}, the action is given in terms of the volume of the region in the true vacuum and the area of the walls separating it from the false vacuum $\phi=0$, 
\beq
S = V(\phi_t) \,{\rm volume} + T\, {\rm area}\ ,
\eeq
where $T$ is the tension of the walls.
To minimize the action, it is preferable to use spherical bubbles because these have the minimal area for a given volume. 
For this reason we consider a single spherical bubble of geodesic radius $\rho$, centred at $\rho=0$ in the coordinates \eqref{AdSrhometric}.
This gives
\beq
S(\rho) = T R^d \,{\rm Vol}(\Omega_d) \left[ \sinh^d \rho -b \int_0^\rho dt \sinh^d t
\right]\ ,\qquad\qquad
b\equiv \frac{|V(\phi_t)|R}{T}\ .
\eeq
Case {\bf 1.} above corresponds to $b\le d$ and the action is positive for bubbles of any size.
This follows from the fact that $S(\rho)$ is positive for very small $\rho$ and that its derivative, $\frac{dS}{d\rho} \propto d \cosh \rho - b \sinh \rho$, is always positive if $b \le d$.
Case {\bf 2.} corresponds to $b> d$ and very large bubbles decrease the action without bound.
Indeed, in this case $S(\rho) \sim e^{d\rho} \left(1- \frac{b}{d}\right)<0$ for large bubbles.

\subsubsection{General spherical bubbles } \label{generalbubble}
Let us now consider the case of a general potential $V(\phi)$ and drop the thin wall approximation.
We will analyse a single spherical bubble because breaking of spherical symmetry should increase the action (from the kinetic term).

It is convenient to trade the  boundary condition \eqref{phibc} for $\phi(\rho_\star)=0$ for some fixed $\rho_\star$. The action for the finite region $\rho<\rho_\star$ is clearly bounded from below if $V(\phi)$ is bounded from below. We will determine its absolute minimum and then study its dependence with $\rho_\star$. This class of field configurations is sufficient to show that the action can be unbounded from below in the limit $\rho_\star \to \infty$.

\begin{figure}[!htb]
\begin{center}
\includegraphics[scale=0.7]{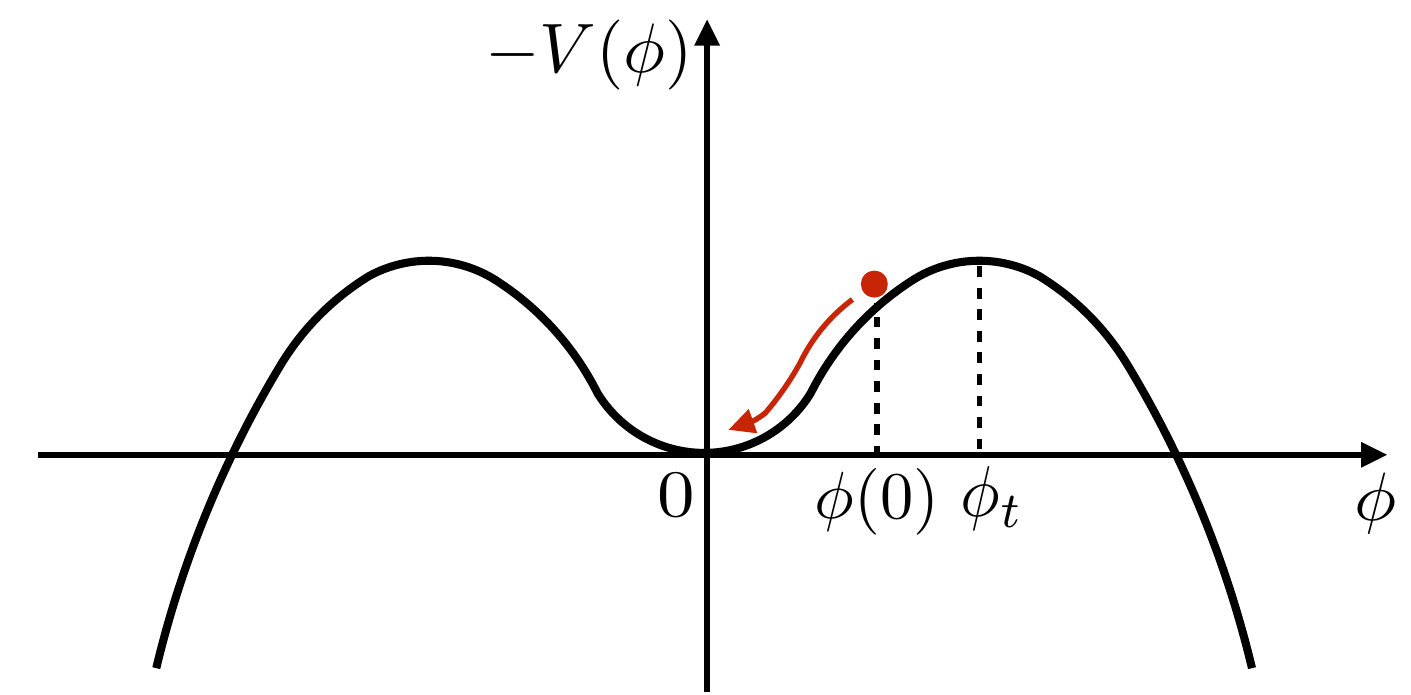}
\end{center}
\vspace{-4mm}
\caption{{ The equation of motion \eqref{ELeqs} can be interpreted as a particle moving in the potential $-V(\phi)$ with time-dependent   friction. 
 We are interested in trajectories that start from rest at $\rho=0$ and reach $\phi=0$ at the time $\rho=\rho_\star$.
Apart from the trivial trajectory $\phi(\rho)=0$, there may be other classical trajectories with lower action.
This depends on the potential. However, if such a trajectory exist for some $\rho_\star$ then it will also exist for larger $\rho_\star$. Indeed, it is sufficient to start with a larger $\phi(0)$, and $\phi(0) \to \phi_t$ when $\rho_\star \to \infty$.
If $V''(0)< -\frac{d^2}{4}$ then there is always another trajectory because the late time motion around $\phi=0$ is oscillatory. 
This corresponds to the Breitenlohner-Freedman criteria for the instability of the trivial vacuum around $\phi=0$. 
}}
\label{fig:invertedV}
\end{figure}

We are interested in the field profile $\phi(\rho)$ that obeys the boundary condition  $\phi(\rho_\star)=0$ and  minimizes the action
\beq
\label{Sspherical}
S = R^{d+1} \,{\rm Vol}(\Omega_d) \int_0^{\rho_\star} d\rho  \sinh^d \rho \left[
\frac{1}{2} \phi'(\rho)^2 +V(\phi(\rho))
\right]\ .
\eeq
The corresponding Euler-Lagrange equations,
\beq
\label{ELeqs}
\phi''(\rho) = V'(\phi(\rho)) - \frac{d}{\tanh \rho} \phi'(\rho)\ , 
\eeq
describe a particle with position $\phi$ moving in the potential $-V(\phi)$ with a friction force depending on the time $\rho$ (see figure \ref{fig:invertedV}). 
This equation of motion allows for the following behaviour when $\rho \to 0$: either  $\phi'(\rho)\sim \rho $ or $\phi$ diverges.
We exclude the second possibility because it would give an infinite contribution to the action \eqref{Sspherical}.\footnote{If the potential goes like  $V \sim \phi^{n}$ for large $\phi$ then the divergent solution has $\phi'(\rho)\sim \rho^{-\gamma} $ when $\rho \to 0$ with 
$\gamma = {\rm min} \left(d, \frac{n}{n-2} \right)$.  One can check that this leads to a divergent integral in \eqref{Sspherical}.
}
Therefore we can use the boundary condition $\phi'(0)=0$, which means we are interested in trajectories that start from rest.
It is clear that $\phi(\rho)=0$ is a solution but it may not be the only one.
The particle motion interpretation of \eqref{ELeqs}, depicted in figure \ref{fig:invertedV}, makes it clear that if there is a non-trivial solution that arrives at $\phi=0$ for some finite time $\rho_\star$ then a similar solution will also exist for an arbitrary large  $\rho_\star$.
Recall that we are interested in the limit of  large  $\rho_\star$ that allows for arbitrary large bubbles of true vacuum in AdS.
Using standard methods, we can compute\footnote{Recall  the classical result $\frac{\partial}{\partial t_f} S(q_i,t_i ; q_f, t_f )= - E$, relating the  action $S$ of the classical path from $q(t_i)=q_i$ to $q(t_f)=q_f$  to its energy $E$.}
\beq 
\frac{1}{R^{d+1} \,{\rm Vol}(\Omega_d)  \sinh ^d \rho_\star }\frac{d}{d\rho_\star}  S_{\rm min} = - \frac{1}{2}   \phi'(\rho_\star)^2 \ .
\eeq
Moreover, if a non-trivial solution exist then $ \phi'(\rho_\star)^2$ tends to a positive constant as $\rho_\star \to \infty$, which implies that $S_{\rm min} \to -\infty$ in this limit.\footnote{Indeed, suppose that a solution exists which starts at rest at $\phi(0)=\phi_0<\phi_t$ and arrives at $\phi(\rho_*)=0$. Then the solution with $\phi(0) \to \phi_t$ passes by $\phi_0$ with a non-vanishing velocity, and so arrives at $\phi=0$ with a larger velocity than the original solution.} This corresponds to case {\bf 2.} above. If the potential is such that for any $\rho_\star$ the minimum of \eqref{Sspherical} is zero, then we fall in case {\bf 1.}

It is convenient to analyse the asymptotic behaviour of the equation of motion \eqref{ELeqs}  for large $\rho$ and around $\phi=0$, where it reduces to a linear ODE:
\beq 
\phi''(\rho) + d \, \phi'(\rho)- V''(0)\phi(\rho) =0 \ .
\eeq
The general solution is $\phi(\rho) = c_+ e^{-\alpha_+ \rho} +c_- e^{-\alpha_- \rho}$ with
\beq
\alpha_\pm = \frac{d}{2} \pm \sqrt{ \frac{d^2}{4} + V''(0)} \ .
\eeq
We conclude that for $V''(0)< -\frac{d^2}{4}$ the late-time motion is oscillatory and therefore the particle reaches $\phi=0$ in finite time. This means that the vacuum $\phi=0$ is unstable if $V''(0)$ is below the Breitenlohner-Freedman bound \cite{Breitenlohner:1982jf}.
If $V''(0)>-\frac{d^2}{4}$ then the late-time motion is repulsive or over-damped and we cannot decide if we fall in case {\bf 1.} or {\bf 2.} with an asymptotic analysis.
A purely quadratic potential with $V''(0) >-\frac{d^2}{4}$  corresponds to case {\bf 1.}, as it can be checked explicitly.

\section{Inverted harmonic oscillator}
\label{app:IHO}

In this appendix we will consider the fact of the harmonic oscillator with a \emph{negative} mass, and in particular how such a system would be analyzed using Hamiltonian truncation methods. To be precise, we consider the following Hamiltonian:
\beq
H = a^\dag a + g \left[(a^\dag)^2 + a^2 + 2a^\dag a \right]
\eeq
where $[a,a^\dag] = 1$. Physically, this is nothing but a harmonic oscillator with unit frequency, perturbed by the normal-ordered operator $X^2$ with an (at least for now) arbitrary real coupling $g$.  The spectrum of $H$ can easily be obtained using a Bogoliubov transformation, using the Ansatz
\beq
\label{eq:bDef}
b = \cosh(\theta) a + \sinh(\theta) a^\dag
\eeq
for some $\theta \in \mbb{R}$ to be determined. By construction, the annihilation operator $b$ from~\reef{eq:bDef} and its partner $b^\dag$ obey the canonical commutation relation $[b,b^\dag] = 1$ for any $\theta$. Setting
\beq
\label{eq:Hrr}
\theta = \frac{1}{4} \ln(1+4g)
\eeq
we find that
\beq
\label{eq:Hrecast}
H = \sqrt{1+4g} \, b^\dag b + \mca{E}_\Omega,
\quad
\mca{E}_\Omega =  \half \sqrt{1+4g} - \half - g \leq 0.
\eeq
In other words, $H$ is a harmonic oscillator with frequency $\omega = \sqrt{1+4g}$ in disguise, up to a finite shift in the ground state energy.

The above reasoning applies to positive $g$, and in fact it's easy to see that the same formulas apply to slightly negative values of the coupling constant. By inspection of~\reef{eq:Hrr} and~\reef{eq:Hrecast} we see that the Bogoliubov transformation~\reef{eq:bDef} works for any $g \geq g_\star \ldef - 1/4$. However, the theory ceases to exist for $g < g_\star$ --- for instance, the energy $\sqrt{1+4g}$ has a branch cut starting at $g= g_\star $.

\begin{figure}[htb]
\begin{center}
\includegraphics[scale=0.32]{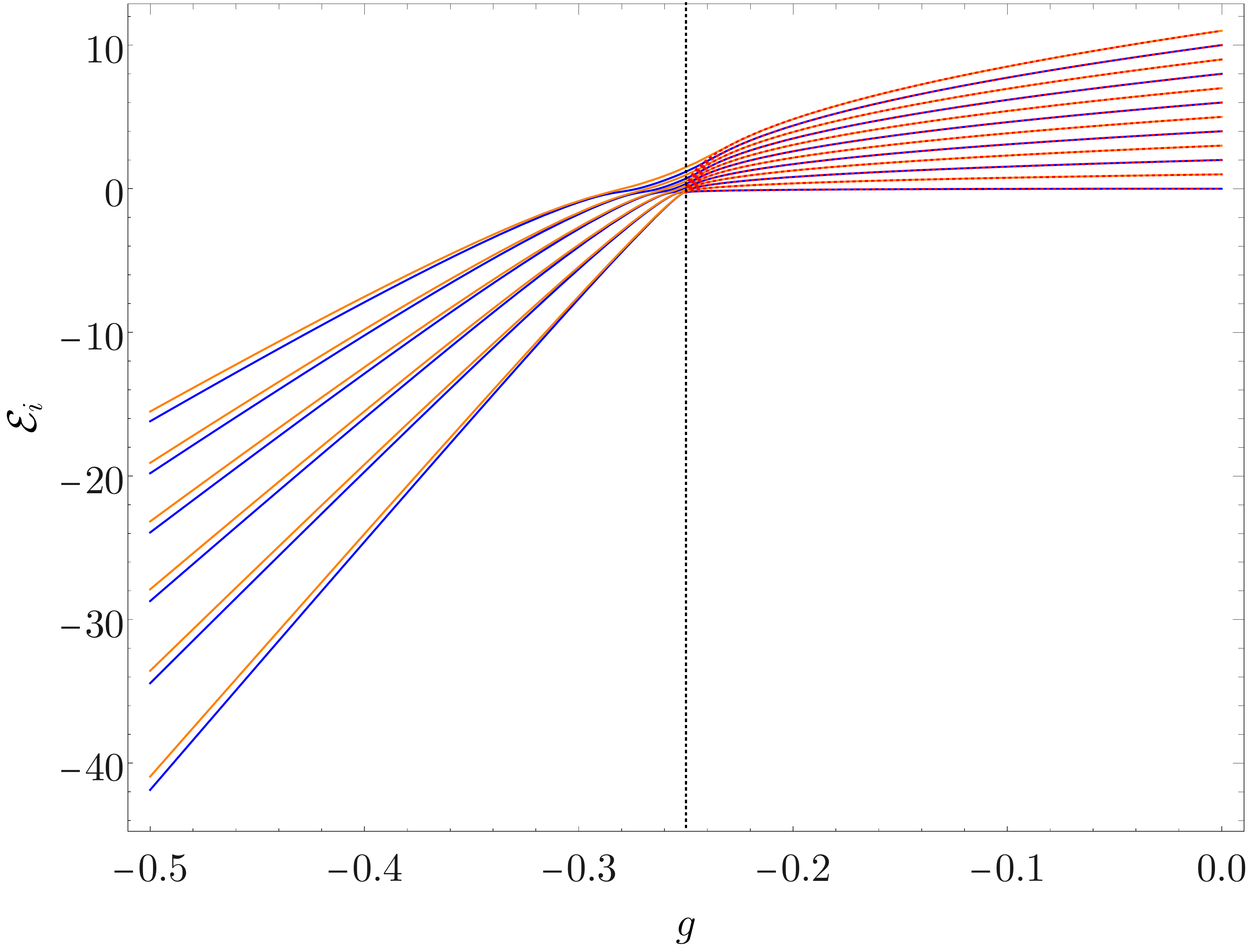}
\end{center}
\vspace{-4mm}
\caption{{Spectrum of the cutoff Hamiltonian $H_N(g)$ for $N=50$ for $g \in [-1/2,0]$. The blue (resp.\@ orange) curves have even (odd) $\mbb{Z}_2$ parity. The dotted red lines for $g \geq g_\star$ show the exact spectrum of the theory; the vertical black dotted line indicates $g_\star$. }}
\label{fig:bogoPlot}
\end{figure}

At the same time, we can analyze $H$ using truncation methods. To do so we work in Fock space, in a basis of the first $N$ eigenstates of the $g=0$ Hamiltonian, to wit
\beq
\ket{k} \ldef \frac{1}{\sqrt{k!}}\, (a^\dag)^k \ket{\text{vac}}
\quad
\text{for}
\quad
k = 0,\ldots,N.
\eeq
The Hamiltonian then becomes a Hermitian matrix $H_N(g)$ of size $(N+1) \times (N+1)$, which is diagonalizable for \emph{any} real value of $g$. Note that $H_N(g)$ does not mix states with even and odd mode number $k$, since the theory is invariant under the $\mbb{Z}_2$ symmetry $(a,a^\dag) \mapsto (-a,-a^\dag)$. In figure \ref{fig:bogoPlot} we plot the first 10 energy levels of $H_N(g)$ for a range of negative $g$, working at cutoff $N=50$. For $g \geq g_\star $ Hamiltonian truncation is in excellent agreement with the analytic formula~\reef{eq:Hrecast}. Yet for $g < g_\star $, we observe that states with even and odd $\mbb{Z}_2$ parity become near-degenerate, emulating spontaneous symmetry breaking in quantum field theory.

Although the theory is unphysical for $g<g_\star$, we can nevertheless study the eigenstates and eigenvalues of $H_N(g)$ numerically. When Hamiltonian truncation converges, the ground state $\ket{\Omega}$ should not be sensitive to the cutoff, $N$ in this case. This means that the overlap of $\ket{\Omega}$ with Fock space states $\ket{k}$ having $k \sim N$ ought to be negligeable when $N$ is large. To make this quantitative, we can introduce the integrated probability density
\beq
\mbb{P}_k^{(N)} = \sum_{j = 0}^k | \brakket{\Omega}{j}|^2
\eeq
which grows with $k$ and obeys $\mbb{P}_N^{(N)} = 1$ (since $\ket{\Omega}$ is normalized). Below in figure \ref{fig:bogoOverlap} we display  $\mbb{P}_k^{(N)}$ for various values of $g$, setting $N = 200$. For $g > g_\star$ we see that $\ket{\Omega}$ has little overlap with states close to the cutoff, as expected (because Hamiltonian truncation converges very rapidly in quantum mechanics). However, this behavior changes dramatically as we lower $g$ to $g_\star$. Already for $g = g_\star - 10^{-3}$, the ground state $\ket{\Omega}$ consists mostly of high-energy states, and as we lower $g$ this effect becomes even more important. Plots of $\mbb{P}_k^{(N)}$ are shown in figure \ref{fig:bogoOverlap}.
\begin{figure}[htb]
\begin{center}
\hspace{5mm} \includegraphics[scale=0.32]{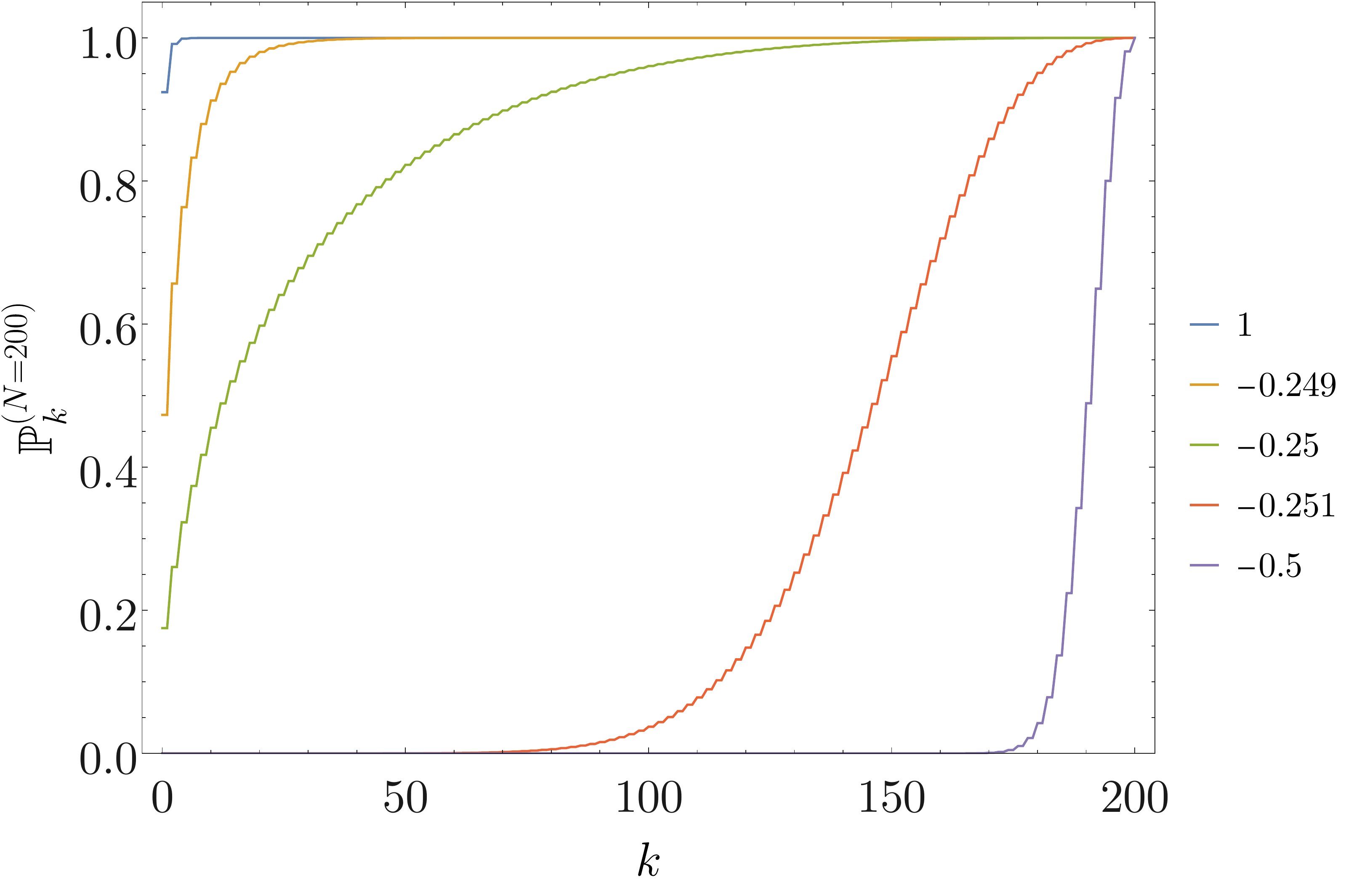}
\end{center}
\vspace{-4mm}
\caption{{ Probability $\mbb{P}_k^{(N)}$ to find the truncation ground state $\ket{\Omega}$ in the first $k$ Fock space states, working at cutoff $N = 200$. For $g=1$ the ground state has almost no overlap with high-energy states. As $g \searrow g_\star$, the overlap with states close to the cutoff grows dramatically, and for values of $g$ below $g_\star$, the ground state is very UV-sensitive.}}
\label{fig:bogoOverlap}
\end{figure}
\begin{figure}[htb]
\begin{center}
\includegraphics[scale=0.32]{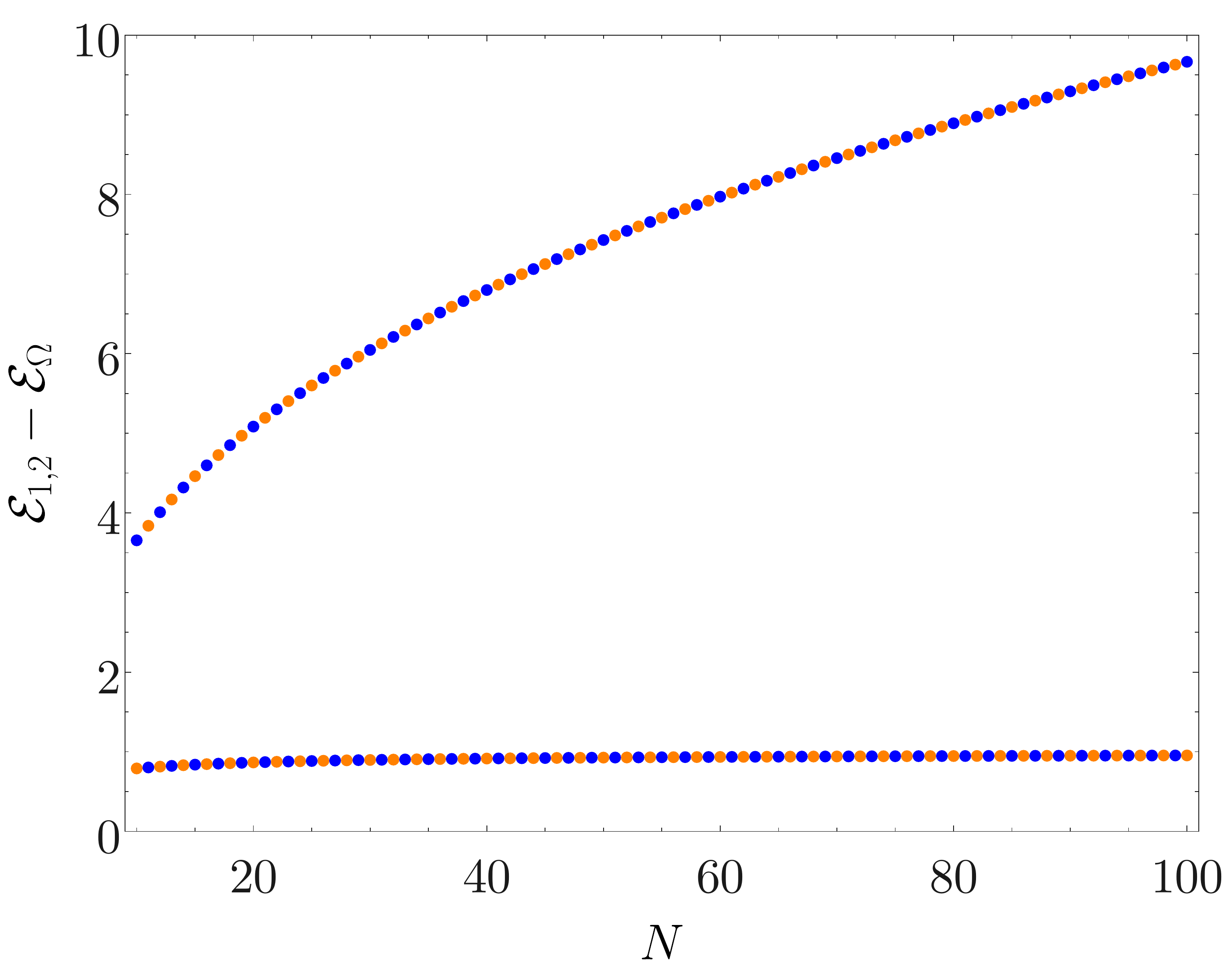}
\end{center}
\vspace{-4mm}
\caption{{ The first two energies $\mathcal{E}_1$ and $\mathcal{E}_2$ (after subtracting the Casimir energy $\mathcal{E}_\Omega$) of the Hamiltonian $H_N(g)$ for $g=-1/2$, scanning over a range of cutoffs $N$. Blue (resp.\@ orange) points label $\mbb{Z}_2$ even (odd) states. In particular, it is clear that $\mathcal{E}_2 - \mathcal{E}_\Omega$ does not have a finite limit as $N \to \infty$, and that the parity of the $k$-th eigenstate of $H$ depends on the sign $(-1)^N$.}}
\label{fig:bogoLevels}
\end{figure}
It is clear that for $g < g_\star$, at least some states in the spectrum of $H_N$ are extremely UV-sensitive; indeed, the spectrum $\mca{E}_\Omega, \, \mca{E}_1, \ldots$ for such values of $g$ depends strongly on the cutoff $N$, as shown in figure \ref{fig:bogoLevels}.\footnote{From the plot, it appears that $\mca{E}_1 - \mca{E}_\Omega$ has a finite limit as $N \to \infty$, but $\mca{E}_2 - \mca{E}_\Omega$ does not converge as the cutoff is removed.} Surprisingly, the parity of the $k$-th excited state goes as $(-1)^{k+N}$, that is to say that it alternates with $N$.

{\fns 
\bibliographystyle{utphys}
\bibliography{biblio}

\providecommand{\href}[2]{#2}\begingroup\raggedright\begin{thebibliography}{10}

\bibitem{Wilson:1971dc}
K.~G. Wilson and M.~E. Fisher, ``{Critical exponents in 3.99 dimensions},''
  \href{http://dx.doi.org/10.1103/PhysRevLett.28.240}{{\em Phys. Rev. Lett.}
  {\bfseries 28} (1972) 240--243}.

\bibitem{tHooft:1973alw}
G.~'t~Hooft, ``{A Planar Diagram Theory for Strong Interactions},''
  \href{http://dx.doi.org/10.1016/0550-3213(74)90154-0}{{\em Nucl. Phys. B}
  {\bfseries 72} (1974) 461}.

\bibitem{Tilloy:2021yre}
A.~Tilloy, ``{Relativistic continuous matrix product states for quantum fields
  without cutoff},'' \href{http://arxiv.org/abs/2102.07741}{{\ttfamily
  arXiv:2102.07741 [quant-ph]}}.

\bibitem{Tilloy:2021hhb}
A.~Tilloy, ``{Variational method in relativistic quantum field theory without
  cutoff},'' \href{http://arxiv.org/abs/2102.07733}{{\ttfamily arXiv:2102.07733
  [quant-ph]}}.

\bibitem{Yurov:1989yu}
V.~Yurov and A.~Zamolodchikov, ``{Truncated Conformal Space Approach to scaling
  Lee-Yang model},'' \href{http://dx.doi.org/10.1142/S0217751X9000218X}{{\em
  Int. J. Mod. Phys. A} {\bfseries 5} (1990) 3221--3246}.

\bibitem{James:2017cpc}
A.~J.~A. James, R.~M. Konik, P.~Lecheminant, N.~J. Robinson, and A.~M. Tsvelik,
  ``{Non-perturbative methodologies for low-dimensional strongly-correlated
  systems: From non-abelian bosonization to truncated spectrum methods},''
  \href{http://dx.doi.org/10.1088/1361-6633/aa91ea}{{\em Rept. Prog. Phys.}
  {\bfseries 81} no.~4, (2018) 046002},
  \href{http://arxiv.org/abs/1703.08421}{{\ttfamily arXiv:1703.08421
  [cond-mat.str-el]}}.

\bibitem{Paulos:2016fap}
M.~F. Paulos, J.~Penedones, J.~Toledo, B.~C. van Rees, and P.~Vieira, ``{The
  S-matrix bootstrap. Part I: QFT in AdS},''
  \href{http://dx.doi.org/10.1007/JHEP11(2017)133}{{\em JHEP} {\bfseries 11}
  (2017) 133}, \href{http://arxiv.org/abs/1607.06109}{{\ttfamily
  arXiv:1607.06109 [hep-th]}}.

\bibitem{Fitzpatrick:2010zm}
A.~L. Fitzpatrick, E.~Katz, D.~Poland, and D.~Simmons-Duffin, ``{Effective
  Conformal Theory and the Flat-Space Limit of AdS},''
  \href{http://dx.doi.org/10.1007/JHEP07(2011)023}{{\em JHEP} {\bfseries 07}
  (2011) 023},
\href{http://arxiv.org/abs/1007.2412}{{\ttfamily arXiv:1007.2412 [hep-th]}}.
%%CITATION = ARXIV:1007.2412;%%.

\bibitem{Brower:2019kyh}
R.~C. Brower, C.~V. Cogburn, A.~L. Fitzpatrick, D.~Howarth, and C.-I. Tan,
  ``{Lattice Setup for Quantum Field Theory in AdS$_2$},''
  \href{http://arxiv.org/abs/1912.07606}{{\ttfamily arXiv:1912.07606
  [hep-th]}}.

\bibitem{Katz:2016hxp}
E.~Katz, Z.~U. Khandker, and M.~T. Walters, ``{A Conformal Truncation Framework
  for Infinite-Volume Dynamics},''
  \href{http://dx.doi.org/10.1007/JHEP07(2016)140}{{\em JHEP} {\bfseries 07}
  (2016) 140}, \href{http://arxiv.org/abs/1604.01766}{{\ttfamily
  arXiv:1604.01766 [hep-th]}}.

\bibitem{Anand:2020gnn}
N.~Anand, A.~L. Fitzpatrick, E.~Katz, Z.~U. Khandker, M.~T. Walters, and
  Y.~Xin, ``{Introduction to Lightcone Conformal Truncation: QFT Dynamics from
  CFT Data},'' \href{http://arxiv.org/abs/2005.13544}{{\ttfamily
  arXiv:2005.13544 [hep-th]}}.

\bibitem{Anand:2020qnp}
N.~Anand, E.~Katz, Z.~U. Khandker, and M.~T. Walters, ``{Nonperturbative
  dynamics of (2+1)d $\phi^4$-theory from Hamiltonian truncation},''
  \href{http://arxiv.org/abs/2010.09730}{{\ttfamily arXiv:2010.09730
  [hep-th]}}.

\bibitem{KaplanBottom}
J.~Kaplan, ``{Lectures on AdS/CFT from the Bottom Up},'' 2016.
\newblock
  \url{https://sites.krieger.jhu.edu/jared-kaplan/files/2016/05/AdSCFTCourseNotesCurrentPublic.pdf}.

\bibitem{Penedones:2016voo}
J.~Penedones, \href{http://dx.doi.org/10.1142/9789813149441_0002}{``{TASI
  lectures on AdS/CFT},''} in {\em {Proceedings, Theoretical Advanced Study
  Institute in Elementary Particle Physics: New Frontiers in Fields and Strings
  (TASI 2015): Boulder, CO, USA, June 1-26, 2015}}, pp.~75--136.
\newblock 2017.
\newblock
\href{http://arxiv.org/abs/1608.04948}{{\ttfamily arXiv:1608.04948 [hep-th]}}.
\newblock
%%CITATION = ARXIV:1608.04948;%%.

\bibitem{Aharony:2010ay}
O.~Aharony, D.~Marolf, and M.~Rangamani, ``{Conformal field theories in anti-de
  Sitter space},'' \href{http://dx.doi.org/10.1007/JHEP02(2011)041}{{\em JHEP}
  {\bfseries 02} (2011) 041}, \href{http://arxiv.org/abs/1011.6144}{{\ttfamily
  arXiv:1011.6144 [hep-th]}}.

\bibitem{Dorey:1997yg}
P.~Dorey, A.~Pocklington, R.~Tateo, and G.~Watts, ``{TBA and TCSA with
  boundaries and excited states},''
  \href{http://dx.doi.org/10.1016/S0550-3213(98)00339-3}{{\em Nucl. Phys. B}
  {\bfseries 525} (1998) 641--663},
  \href{http://arxiv.org/abs/hep-th/9712197}{{\ttfamily arXiv:hep-th/9712197}}.

\bibitem{Hogervorst:2014rta}
M.~Hogervorst, S.~Rychkov, and B.~C. van Rees, ``{Truncated conformal space
  approach in $d$ dimensions: A cheap alternative to lattice field theory?},''
  \href{http://dx.doi.org/10.1103/PhysRevD.91.025005}{{\em Phys. Rev. D}
  {\bfseries 91} (2015) 025005},
  \href{http://arxiv.org/abs/1409.1581}{{\ttfamily arXiv:1409.1581 [hep-th]}}.

\bibitem{Carmi:2018qzm}
D.~Carmi, L.~Di~Pietro, and S.~Komatsu, ``{A Study of Quantum Field Theories in
  AdS at Finite Coupling},''
  \href{http://dx.doi.org/10.1007/JHEP01(2019)200}{{\em JHEP} {\bfseries 01}
  (2019) 200},
\href{http://arxiv.org/abs/1810.04185}{{\ttfamily arXiv:1810.04185 [hep-th]}}.
%%CITATION = ARXIV:1810.04185;%%.

\bibitem{Breitenlohner:1982jf}
P.~Breitenlohner and D.~Z. Freedman, ``{Stability in Gauged Extended
  Supergravity},'' \href{http://dx.doi.org/10.1016/0003-4916(82)90116-6}{{\em
  Annals Phys.} {\bfseries 144} (1982) 249}.

\bibitem{EliasMiro:2020uvk}
J.~Elias-Mir\'o and E.~Hardy, ``{Exploring Hamiltonian Truncation in
  $\bf{d=2+1}$},'' \href{http://dx.doi.org/10.1103/PhysRevD.102.065001}{{\em
  Phys. Rev. D} {\bfseries 102} no.~6, (2020) 065001},
  \href{http://arxiv.org/abs/2003.08405}{{\ttfamily arXiv:2003.08405
  [hep-th]}}.

\bibitem{deHaro:2000vlm}
S.~de~Haro, S.~N. Solodukhin, and K.~Skenderis, ``{Holographic reconstruction
  of space-time and renormalization in the AdS / CFT correspondence},''
  \href{http://dx.doi.org/10.1007/s002200100381}{{\em Commun. Math. Phys.}
  {\bfseries 217} (2001) 595--622},
\href{http://arxiv.org/abs/hep-th/0002230}{{\ttfamily arXiv:hep-th/0002230
  [hep-th]}}.
%%CITATION = HEP-TH/0002230;%%.

\bibitem{Bender:1968sa}
C.~M. Bender and T.~T. Wu, ``{Analytic structure of energy levels in a field
  theory model},'' \href{http://dx.doi.org/10.1103/PhysRevLett.21.406}{{\em
  Phys. Rev. Lett.} {\bfseries 21} (1968) 406--409}.

\bibitem{Bender:1969si}
C.~M. Bender and T.~T. Wu, ``{Anharmonic oscillator},''
  \href{http://dx.doi.org/10.1103/PhysRev.184.1231}{{\em Phys. Rev.} {\bfseries
  184} (1969) 1231--1260}.

\bibitem{Klassen:1990dx}
T.~R. Klassen and E.~Melzer, ``{The thermodynamics of purely elastic scattering
  theories and conformal perturbation theory},''
  \href{http://dx.doi.org/10.1016/0550-3213(91)90159-U}{{\em Nucl. Phys. B}
  {\bfseries 350} (1991) 635--689}.

\bibitem{Klassen:1991ze}
T.~R. Klassen and E.~Melzer, ``{Spectral flow between conformal field theories
  in (1+1)-dimensions},''
  \href{http://dx.doi.org/10.1016/0550-3213(92)90422-8}{{\em Nucl. Phys. B}
  {\bfseries 370} (1992) 511--550}.

\bibitem{Pappadopulo:2012jk}
D.~Pappadopulo, S.~Rychkov, J.~Espin, and R.~Rattazzi, ``{OPE Convergence in
  Conformal Field Theory},''
  \href{http://dx.doi.org/10.1103/PhysRevD.86.105043}{{\em Phys. Rev. D}
  {\bfseries 86} (2012) 105043},
  \href{http://arxiv.org/abs/1208.6449}{{\ttfamily arXiv:1208.6449 [hep-th]}}.

\bibitem{Pal:2019yhz}
S.~Pal, ``{Bound on asymptotics of magnitude of three point coefficients in 2D
  CFT},'' \href{http://dx.doi.org/10.1007/JHEP01(2020)023}{{\em JHEP}
  {\bfseries 01} (2020) 023}, \href{http://arxiv.org/abs/1906.11223}{{\ttfamily
  arXiv:1906.11223 [hep-th]}}.

\bibitem{Billo:2016cpy}
M.~Billò, V.~Gonçalves, E.~Lauria, and M.~Meineri, ``{Defects in conformal
  field theory},'' \href{http://dx.doi.org/10.1007/JHEP04(2016)091}{{\em JHEP}
  {\bfseries 04} (2016) 091}, \href{http://arxiv.org/abs/1601.02883}{{\ttfamily
  arXiv:1601.02883 [hep-th]}}.

\bibitem{Giokas:2011ix}
P.~Giokas and G.~Watts, ``{The renormalisation group for the truncated
  conformal space approach on the cylinder},''
  \href{http://arxiv.org/abs/1106.2448}{{\ttfamily arXiv:1106.2448 [hep-th]}}.

\bibitem{Watts:2011cr}
G.~M.~T. Watts, ``{On the renormalisation group for the boundary Truncated
  Conformal Space Approach},''
  \href{http://dx.doi.org/10.1016/j.nuclphysb.2012.01.012}{{\em Nucl. Phys. B}
  {\bfseries 859} (2012) 177--206},
  \href{http://arxiv.org/abs/1104.0225}{{\ttfamily arXiv:1104.0225 [hep-th]}}.

\bibitem{Rychkov:2014eea}
S.~Rychkov and L.~G. Vitale, ``{Hamiltonian truncation study of the $\phi^4$
  theory in two dimensions},''
  \href{http://dx.doi.org/10.1103/PhysRevD.91.085011}{{\em Phys. Rev. D}
  {\bfseries 91} (2015) 085011},
  \href{http://arxiv.org/abs/1412.3460}{{\ttfamily arXiv:1412.3460 [hep-th]}}.

\bibitem{Rychkov:2015vap}
S.~Rychkov and L.~G. Vitale, ``{Hamiltonian truncation study of the $\phi^4$
  theory in two dimensions. II. The $\mathbb Z_2$ -broken phase and the Chang
  duality},'' \href{http://dx.doi.org/10.1103/PhysRevD.93.065014}{{\em Phys.
  Rev. D} {\bfseries 93} no.~6, (2016) 065014},
  \href{http://arxiv.org/abs/1512.00493}{{\ttfamily arXiv:1512.00493
  [hep-th]}}.

\bibitem{Elias-Miro:2015bqk}
J.~Elias-Miro, M.~Montull, and M.~Riembau, ``{The renormalized Hamiltonian
  truncation method in the large $E_T$ expansion},''
  \href{http://dx.doi.org/10.1007/JHEP04(2016)144}{{\em JHEP} {\bfseries 04}
  (2016) 144}, \href{http://arxiv.org/abs/1512.05746}{{\ttfamily
  arXiv:1512.05746 [hep-th]}}.

\bibitem{Elias-Miro:2017tup}
J.~Elias-Miro, S.~Rychkov, and L.~G. Vitale, ``{NLO Renormalization in the
  Hamiltonian Truncation},''
  \href{http://dx.doi.org/10.1103/PhysRevD.96.065024}{{\em Phys. Rev. D}
  {\bfseries 96} no.~6, (2017) 065024},
  \href{http://arxiv.org/abs/1706.09929}{{\ttfamily arXiv:1706.09929
  [hep-th]}}.

\bibitem{Elias-Miro:2017xxf}
J.~Elias-Miro, S.~Rychkov, and L.~G. Vitale, ``{High-Precision Calculations in
  Strongly Coupled Quantum Field Theory with Next-to-Leading-Order Renormalized
  Hamiltonian Truncation},''
  \href{http://dx.doi.org/10.1007/JHEP10(2017)213}{{\em JHEP} {\bfseries 10}
  (2017) 213}, \href{http://arxiv.org/abs/1706.06121}{{\ttfamily
  arXiv:1706.06121 [hep-th]}}.

\bibitem{Cardy:1984bb}
J.~L. Cardy, ``{Conformal Invariance and Surface Critical Behavior},''
  \href{http://dx.doi.org/10.1016/0550-3213(84)90241-4}{{\em Nucl. Phys. B}
  {\bfseries 240} (1984) 514--532}.

\bibitem{Belavin:1984vu}
A.~A. Belavin, A.~M. Polyakov, and A.~B. Zamolodchikov, ``{Infinite Conformal
  Symmetry in Two-Dimensional Quantum Field Theory},''
  \href{http://dx.doi.org/10.1016/0550-3213(84)90052-X}{{\em Nucl. Phys. B}
  {\bfseries 241} (1984) 333--380}.

\bibitem{Cardy:1986gw}
J.~L. Cardy, ``{Effect of Boundary Conditions on the Operator Content of
  Two-Dimensional Conformally Invariant Theories},''
\href{http://dx.doi.org/10.1016/0550-3213(86)90596-1}{{\em Nucl.Phys.}
  {\bfseries B275} (1986) 200--218}.
%%CITATION = NUPHA,B275,200;%%.

\bibitem{Cardy:1989ir}
J.~L. Cardy, ``{Boundary Conditions, Fusion Rules and the Verlinde Formula},''
\href{http://dx.doi.org/10.1016/0550-3213(89)90521-X}{{\em Nucl.Phys.}
  {\bfseries B324} (1989) 581}.
%%CITATION = NUPHA,B324,581;%%.

\bibitem{Cardy:1991tv}
J.~L. Cardy and D.~C. Lewellen, ``{Bulk and boundary operators in conformal
  field theory},''
\href{http://dx.doi.org/10.1016/0370-2693(91)90828-E}{{\em Phys.Lett.}
  {\bfseries B259} (1991) 274--278}.
%%CITATION = PHLTA,B259,274;%%.

\bibitem{Behrend:1999bn}
R.~E. Behrend, P.~A. Pearce, V.~B. Petkova, and J.-B. Zuber, ``{Boundary
  conditions in rational conformal field theories},''
  \href{http://dx.doi.org/10.1016/S0550-3213(99)00592-1}{{\em Nucl. Phys.}
  {\bfseries B570} (2000) 525--589},
  \href{http://arxiv.org/abs/hep-th/9908036}{{\ttfamily arXiv:hep-th/9908036
  [hep-th]}}.
[Nucl. Phys.B579,707(2000)].
%%CITATION = HEP-TH/9908036;%%.

\bibitem{DiFrancesco:1997nk}
P.~Di~Francesco, P.~Mathieu, and D.~Senechal,
  \href{http://dx.doi.org/10.1007/978-1-4612-2256-9}{{\em {Conformal Field
  Theory}}}.
\newblock Graduate Texts in Contemporary Physics. Springer-Verlag, New York,
  1997.

\bibitem{Recknagel:2013uja}
A.~Recknagel and V.~Schomerus,
  \href{http://dx.doi.org/10.1017/CBO9780511806476}{{\em {Boundary Conformal
  Field Theory and the Worldsheet Approach to D-Branes}}}.
\newblock Cambridge Monographs on Mathematical Physics. Cambridge University
  Press, 11, 2013.

\bibitem{Ishibashi:1988kg}
N.~Ishibashi, ``{The Boundary and Crosscap States in Conformal Field
  Theories},'' \href{http://dx.doi.org/10.1142/S0217732389000320}{{\em Mod.
  Phys. Lett. A} {\bfseries 4} (1989) 251}.

\bibitem{Dorey:1999cj}
P.~Dorey, I.~Runkel, R.~Tateo, and G.~Watts, ``{g function flow in perturbed
  boundary conformal field theories},''
  \href{http://dx.doi.org/10.1016/S0550-3213(99)00772-5}{{\em Nucl. Phys. B}
  {\bfseries 578} (2000) 85--122},
  \href{http://arxiv.org/abs/hep-th/9909216}{{\ttfamily arXiv:hep-th/9909216}}.

\bibitem{Cardy:1989fw}
J.~L. Cardy and G.~Mussardo, ``{S Matrix of the Yang-Lee Edge Singularity in
  Two-Dimensions},'' \href{http://dx.doi.org/10.1016/0370-2693(89)90818-6}{{\em
  Phys. Lett. B} {\bfseries 225} (1989) 275--278}.

\bibitem{Zamolodchikov:1995xk}
A.~B. Zamolodchikov, ``{Mass scale in the sine-Gordon model and its
  reductions},'' \href{http://dx.doi.org/10.1142/S0217751X9500053X}{{\em Int.
  J. Mod. Phys. A} {\bfseries 10} (1995) 1125--1150}.

\bibitem{Kramers:1941kn}
H.~A. Kramers and G.~H. Wannier, ``{Statistics of the two-dimensional
  ferromagnet. Part 1.},'' \href{http://dx.doi.org/10.1103/PhysRev.60.252}{{\em
  Phys. Rev.} {\bfseries 60} (1941) 252--262}.

\bibitem{Zamolodchikov:1989fp}
A.~B. Zamolodchikov, ``{Integrals of Motion and S Matrix of the (Scaled)
  $T=T_c$ Ising Model with Magnetic Field},''
  \href{http://dx.doi.org/10.1142/S0217751X8900176X}{{\em Int. J. Mod. Phys. A}
  {\bfseries 4} (1989) 4235}.

\bibitem{Zamolodchikov:1989hfa}
A.~B. Zamolodchikov, ``{Integrable field theory from conformal field theory},''
  {\em Adv. Stud. Pure Math.} {\bfseries 19} (1989) 641--674.

\bibitem{Fonseca:2001dc}
P.~Fonseca and A.~Zamolodchikov, ``{Ising field theory in a magnetic field:
  Analytic properties of the free energy},''
  \href{http://arxiv.org/abs/hep-th/0112167}{{\ttfamily arXiv:hep-th/0112167}}.

\bibitem{Fonseca:2006au}
P.~Fonseca and A.~Zamolodchikov, ``{Ising spectroscopy. I. Mesons at $T <
  T_c$},'' \href{http://arxiv.org/abs/hep-th/0612304}{{\ttfamily
  arXiv:hep-th/0612304}}.

\bibitem{Zamolodchikov:2011wd}
A.~Zamolodchikov and I.~Ziyatdinov, ``{Inelastic scattering and elastic
  amplitude in Ising field theory in a weak magnetic field at $T>T_c$:
  Perturbative analysis},''
  \href{http://dx.doi.org/10.1016/j.nuclphysb.2011.04.005}{{\em Nucl. Phys. B}
  {\bfseries 849} (2011) 654--674},
  \href{http://arxiv.org/abs/1102.0767}{{\ttfamily arXiv:1102.0767 [hep-th]}}.

\bibitem{Zamolodchikov:2013ama}
A.~Zamolodchikov, ``{Ising Spectroscopy II: Particles and poles at $T >
  T_c$},'' \href{http://arxiv.org/abs/1310.4821}{{\ttfamily arXiv:1310.4821
  [hep-th]}}.

\bibitem{Gabai:2019ryw}
B.~Gabai and X.~Yin, ``{On The S-Matrix of Ising Field Theory in Two
  Dimensions},'' \href{http://arxiv.org/abs/1905.00710}{{\ttfamily
  arXiv:1905.00710 [hep-th]}}.

\bibitem{Doyon:2004fv}
B.~Doyon and P.~Fonseca, ``{Ising field theory on a Pseudosphere},''
  \href{http://dx.doi.org/10.1088/1742-5468/2004/07/P07002}{{\em J. Stat.
  Mech.} {\bfseries 0407} (2004) P07002},
\href{http://arxiv.org/abs/hep-th/0404136}{{\ttfamily arXiv:hep-th/0404136
  [hep-th]}}.
%%CITATION = HEP-TH/0404136;%%.

\bibitem{Frohlich:2004ef}
J.~Frohlich, J.~Fuchs, I.~Runkel, and C.~Schweigert, ``{Kramers-Wannier duality
  from conformal defects},''
  \href{http://dx.doi.org/10.1103/PhysRevLett.93.070601}{{\em Phys. Rev. Lett.}
  {\bfseries 93} (2004) 070601},
  \href{http://arxiv.org/abs/cond-mat/0404051}{{\ttfamily
  arXiv:cond-mat/0404051}}.

\bibitem{Bachas:2012bj}
C.~Bachas, I.~Brunner, and D.~Roggenkamp, ``{A worldsheet extension of
  O(d,d:Z)},'' \href{http://dx.doi.org/10.1007/JHEP10(2012)039}{{\em JHEP}
  {\bfseries 10} (2012) 039}, \href{http://arxiv.org/abs/1205.4647}{{\ttfamily
  arXiv:1205.4647 [hep-th]}}.

\bibitem{Kaplan:2009kr}
D.~B. Kaplan, J.-W. Lee, D.~T. Son, and M.~A. Stephanov, ``{Conformality
  Lost},'' \href{http://dx.doi.org/10.1103/PhysRevD.80.125005}{{\em Phys. Rev.
  D} {\bfseries 80} (2009) 125005},
  \href{http://arxiv.org/abs/0905.4752}{{\ttfamily arXiv:0905.4752 [hep-th]}}.

\bibitem{Gorbenko:2018ncu}
V.~Gorbenko, S.~Rychkov, and B.~Zan, ``{Walking, Weak first-order transitions,
  and Complex CFTs},'' \href{http://dx.doi.org/10.1007/JHEP10(2018)108}{{\em
  JHEP} {\bfseries 10} (2018) 108},
  \href{http://arxiv.org/abs/1807.11512}{{\ttfamily arXiv:1807.11512
  [hep-th]}}.

\bibitem{Rutter:2018aog}
D.~Rutter and B.~C. van Rees, ``{Counterterms in Truncated Conformal
  Perturbation Theory},'' \href{http://arxiv.org/abs/1803.05798}{{\ttfamily
  arXiv:1803.05798 [hep-th]}}.

\bibitem{Luscher:1986pf}
M.~Luscher, ``{Volume Dependence of the Energy Spectrum in Massive Quantum
  Field Theories. 2. Scattering States},''
  \href{http://dx.doi.org/10.1007/BF01211097}{{\em Commun. Math. Phys.}
  {\bfseries 105} (1986) 153--188}.

\bibitem{Gary:2009ae}
M.~Gary, S.~B. Giddings, and J.~Penedones, ``{Local bulk S-matrix elements and
  CFT singularities},''
  \href{http://dx.doi.org/10.1103/PhysRevD.80.085005}{{\em Phys. Rev. D}
  {\bfseries 80} (2009) 085005},
  \href{http://arxiv.org/abs/0903.4437}{{\ttfamily arXiv:0903.4437 [hep-th]}}.

\bibitem{Penedones:2010ue}
J.~Penedones, ``{Writing CFT correlation functions as AdS scattering
  amplitudes},'' \href{http://dx.doi.org/10.1007/JHEP03(2011)025}{{\em JHEP}
  {\bfseries 03} (2011) 025}, \href{http://arxiv.org/abs/1011.1485}{{\ttfamily
  arXiv:1011.1485 [hep-th]}}.

\bibitem{Dubovsky:2017cnj}
S.~Dubovsky, V.~Gorbenko, and M.~Mirbabayi, ``{Asymptotic fragility, near
  AdS$_{2}$ holography and $ T\overline{T} $},''
  \href{http://dx.doi.org/10.1007/JHEP09(2017)136}{{\em JHEP} {\bfseries 09}
  (2017) 136}, \href{http://arxiv.org/abs/1706.06604}{{\ttfamily
  arXiv:1706.06604 [hep-th]}}.

\bibitem{Hijano:2019qmi}
E.~Hijano, ``{Flat space physics from AdS/CFT},''
  \href{http://dx.doi.org/10.1007/JHEP07(2019)132}{{\em JHEP} {\bfseries 07}
  (2019) 132}, \href{http://arxiv.org/abs/1905.02729}{{\ttfamily
  arXiv:1905.02729 [hep-th]}}.

\bibitem{Komatsu:2020sag}
S.~Komatsu, M.~F. Paulos, B.~C. Van~Rees, and X.~Zhao, ``{Landau diagrams in
  AdS and S-matrices from conformal correlators},''
  \href{http://dx.doi.org/10.1007/JHEP11(2020)046}{{\em JHEP} {\bfseries 11}
  (2020) 046}, \href{http://arxiv.org/abs/2007.13745}{{\ttfamily
  arXiv:2007.13745 [hep-th]}}.

\bibitem{TK}
T.~H. Koornwinder, ``Dual addition formulas associated with dual product
  formulas,'' \href{http://dx.doi.org/10.1142/9789813228887_0019}{{\em
  Frontiers in Orthogonal Polynomials and q-Series} (2018) 373--392},
  \href{http://arxiv.org/abs/1607.06053}{{\ttfamily arXiv:1607.06053 [math]}}.

\bibitem{Coleman:1977py}
S.~R. Coleman, ``{The Fate of the False Vacuum. 1. Semiclassical Theory},''
  \href{http://dx.doi.org/10.1103/PhysRevD.16.1248}{{\em Phys. Rev. D}
  {\bfseries 15} (1977) 2929--2936}. [Erratum: Phys.Rev.D 16, 1248 (1977)].

\end{thebibliography}\endgroup
 }

\end{document}